\documentclass[11pt]{article}
\usepackage[T1]{fontenc}		
\usepackage[utf8]{inputenc}		
\usepackage{lmodern}			
\usepackage{indentfirst}		
\usepackage{color}			
\usepackage{graphicx}		
\usepackage{float}
\usepackage{tikz}
\usetikzlibrary{positioning}
\usepackage[english]{babel}
\usepackage{amsmath}
\usepackage{amsfonts}
\usepackage{amssymb}
\usepackage{amsthm}
\usepackage{pgfplots} 
\usepackage{physics}
\usepackage{mathtools}
\usepackage{mathrsfs}
\usepackage{dsfont}
\usepackage[top=3cm, left=3cm, right=2cm, bottom=2cm]{geometry}
\usepackage[title, titletoc]{appendix} 
\usepackage{caption}
\usepackage[hyperfootnotes=false]{hyperref}

\newtheorem{definition}{Definition}

\newtheorem{hypothesis}{Hypothesis}
\newtheorem{postulate}{Postulate}

\definecolor{ao}{rgb}{0.0, 0.5, 0.0}
\definecolor{blueink}{RGB}{0,128,128}
\definecolor{darkblueink}{RGB}{0,0,128}
\definecolor{darkredink}{RGB}{128,0,0}
\definecolor{redink}{RGB}{255,42,42}
\definecolor{orangeink}{RGB}{255,162,0}
\definecolor{mathematicablue}{rgb}{0.368, 0.507, 0.710}
\definecolor{mathematicaorange}{rgb}{0.881, 0.505, 0.173}

\usepackage[num, abnt-emphasize=bf, abnt-thesis-year=both, abnt-repeated-author-omit=no, abnt-last-names=abnt, abnt-etal-cite, abnt-etal-list=3, abnt-etal-text=it, abnt-and-type=e, abnt-doi=doi, abnt-url-package=none, abnt-verbatim-entry=no]{abntex2cite} 
\usepackage{cite}

\usepackage{ABNT6023-2018}

\definecolor{blue}{RGB}{41,5,195}

\makeatletter
\hypersetup{
	pdftitle={Light Cone Thermodynamics: Quasi-Local Formalism}, 
	pdfauthor={Matheus G. Barbosa},
	pdfcreator={LaTeX with abnTeX2},
	colorlinks=true,       		
	linkcolor=blue,          	
	citecolor=blue,        		
	filecolor=blue,      		
	urlcolor=blue,
	bookmarksdepth=4,
	pdfstartview={FitH}	
}
\makeatother

\makeindex

\makeatletter
\renewcommand{\thanks}[1]{%
	\footnotetext[0]{#1}%
}
\makeatother

\title{Light Cone Thermodynamics:\\
	 Quasi-Local Formalism\thanks{This work is derived from the author's doctoral thesis in preparation at the S\~{a}o Carlos Institute of Physics, University of S\~{a}o Paulo.}}
\author{Matheus G. Barbosa\\
	\small S\~{a}o Carlos Institute of Physics, University of S\~{a}o Paulo, IFSC -- USP, 13566-590, S\~{a}o Carlos, SP, Brazil. \\
	\small E-mail: matheusgb@ifsc.usp.br \\
	\small ORCID iD: \href{https://orcid.org/0000-0003-0595-9897}{0000-0003-0595-9897}}
\date{}

\begin{document}
	
\maketitle	
	
\begin{abstract}
	Inspired by black hole thermodynamics and other results suggesting that general relativity has some intrinsic connection with thermodynamics, this work aims to establish a quasi-local thermodynamic formalism in general spacetimes. With this purpose, we generalize the Bondi-Sachs formalism to past light cones associated with an arbitrary worldline. From this generalization, we demonstrate the emission of Hawking radiation in a quasi-local context through a few examples, suggesting that it may occur under mild conditions. In addition, we argue that an ``irreversible inequality'' is satisfied in typical gravitational systems, indicating that an analogue of the second law may be established. Together with a definition of internal energy derived from the Landau-Lifshitz pseudotensor, these results are used to construct a postulational framework that generalizes classical thermodynamics in order to include the dynamics of spacetime. As a consequence, it is possible to define the rest-mass density from thermodynamic variables, decoupling it from the dynamics of constituent numbers.
\end{abstract}
	
\pdfbookmark[0]{\contentsname}{toc}
\tableofcontents
\cleardoublepage

\section{Introduction}

Black hole thermodynamics has established a puzzling relation between thermodynamics and gravitation. The analogy between the laws of black hole mechanics and the laws of thermodynamics \cite{Bardeen1973} becomes a physical identity when one considers that the generalized second law is necessary to include black holes in a thermodynamic framework \cite{Bekenstein1972,Bekenstein1973} and that the Hawking temperature has a physical interpretation, which originates from quantum effects \cite{Hawking1974,Hawking1975}. This leads to the idea that general relativity has an intrinsic connection with thermodynamics and quantum theories and that gravitational phenomena could be related to some statistical description of the microscopic degrees of freedom contained in spacetime.

Besides, extensions of some results of black hole thermodynamics to cosmological event horizons \cite{Gibbons1977cosmo}, to the Rindler horizon \cite{Unruh1976}, and even to generalizations of Einstein's theory of gravity \cite{Padmanabhan2010a} suggest that this connection with thermodynamics emerges when one analyses causal horizons. We define a causal horizon as the boundary of the causal past \cite{Hawking1973} of an observer (or family of observers, depending on the context). Since all causal horizons ``hide information from their observers'' and entropy quantifies inaccessible information, we may generalize the reasoning used to attribute an entropy to the event horizon of a black hole \cite{Bekenstein1973} and expect a similar thermodynamic quantity naturally related to any causal horizon \cite{Jacobson1995}. Moreover, the density matrix describing quantum fields in a spacetime with a bifurcate Killing horizon \cite{Wald2001} represents a thermal state when one takes the partial trace over the region hidden by the horizon \cite{Padmanabhan2010a}. Such a partial trace should be performed in the presence of any causal horizon, which would probably produce a mixed state that could be associated with a temperature under conditions compatible with some notion of thermal equilibrium. Therefore, the first hypothesis of this work can be stated as
\begin{hypothesis}\label{hypo1}
	General relativity has intrinsic thermodynamic aspects, whose physical interpretation is associated with causal horizons.
\end{hypothesis}

Following similar ideas, it has been proposed \cite{Jacobson1995,Jacobson2003,Padmanabhan2010a} that the first law of thermodynamics not only is present in local versions of the Rindler horizon but also gives rise to the field equations of the theory. This leads us to an intriguing standpoint, where the seemingly peculiar relation between gravity and thermodynamics is explained by saying that the former is a consequence of the latter. Some questions are left, however, in these proposals to construct a theory of gravitation from thermodynamics. Firstly, the concept of horizons in equilibrium, defined in terms of the shape of the boundary of the system, does not show a clear link with the usual notion of equilibrium in thermodynamics. We also note that the usage of Rindler observers and of the Davies-Unruh temperature seems rather artificial: although there is reason to expect something similar to the Fulling-Davies-Unruh effect in general spacetimes, the relevance of this specific family of observers in a fundamental aspect of the theory is questionable, and the analysis holds when it is taken the limit where the observer's acceleration diverges. Thus, it becomes unclear whether the first law could be verified by a real observer and whether the temperature involved could be measured by them. On account of these remarks, the present work will pursue a reformulation of these ideas.

By taking the phenomenological aspect of classical thermodynamics into consideration, we suppose that all thermodynamic quantities can be observed, through direct measurements or inference, by a real observer. In addition, the thermodynamics of cosmological event horizons and the Fulling-Davies-Unruh effect point to a dependence on the observer of the relevant causal horizon and its temperature. Moreover, in an attempt to base the theory on the way we experience the world, we state that an astronomer on Earth does not have to take into account any hypothetical observer near a star to make observations about it. What really matters are their relative positions and motions. Following this reasoning, we propose that
\begin{hypothesis}\label{hypo2}
	Any observation requires only one observer.
\end{hypothesis}
Thus, we shall abandon the notion of family of observers from the beginning, since it becomes artificial and could obscure the physics. Instead, we take a general observer as the fundamental object of our analysis, represented by a future-directed timelike curve.

We also suppose that a physical law should not depend on the whole history of those who verify it. Then, we ignore the asymptotic future of this observer, avoiding possible teleological issues, as well as their asymptotic past, since some level of independence is expected between what one observes and the path one has taken to the point of observation: it suffices for two observers to observe the same physical parameters that their worldlines coincide in a neighborhood of the points of observation. As a consequence, we shall consider observers following trajectories for a finite interval of proper time, that is, curves contained in a compact subset of spacetime. In the following, we shall describe objects that require a nondegenerate compact to be defined as quasi-local \cite{Szabados2009}.

Then, considering that, in the context of general relativity, there is no minimum time scale to prevent us from shortening the interval of observation indefinitely, we shall assume that
\begin{hypothesis}\label{hypo3}
	Any open interval of proper time during which an observation occurs is sufficient to define the physical parameters of some subset of spacetime.
\end{hypothesis}
Of course, not all definitions may be determined, but we are assuming that there exists some meaningful definition, for each physical parameter, that will satisfy the above hypothesis. We acknowledge the high level of idealization in it, but, in principle, it could be realized in spacetimes that evolve sufficiently slowly, bypassing the limitations of a real measuring device.

From these three hypotheses and considering the limit in which the interval of observation tends to a single point, we conclude that the basic geometric structure in the connection between general relativity and thermodynamics is the past light cone of a point along a worldline. Other structures, like the usual asymptotic horizons, may be thought of as limiting cases of this one, once we take into account possible families of observers (as will be shown in Figures \ref{limit} and \ref{conehole}).

Now, supposing that the observer will be able to observe the variations in the thermodynamic parameters during their lifetime, we are led to consider the same thermodynamic system at two distinct moments in the same spacetime. Therefore, a definition of thermodynamic system should be compatible with some Lie dragging consistent with the time evolution perceived by the observer. If one imagines a thermodynamic system as contained in a three-dimensional slice of time and tries to associate it with a past light cone, then it seems unavoidable to allow this light cone to evolve with the system. For this reason, we shall compare distinct past light cones related to distinct points of the worldline of the observer. Furthermore, we find that the most natural view is to consider these past light cones as slices of time. This formulation of the problem presents a subtle conceptual difference with respect to the results that inspired it. In black hole thermodynamics, for example, one considers that the event horizon is fixed and the variations of the thermodynamic quantities take place in the transition from one spacetime to another \cite{Bardeen1973}. Even though physical interpretation can relate both views, their precise mathematical formulations are not the same. Here, we adopt an explicit time evolution, which in turn requires a comparison of different causal horizons.

Despite our proposal being derived from different reasoning, it is compatible with the idea of ``photographic'' thermodynamics proposed in Reference \citeonline{Dunkel2009}. This could be considered an indication that the supposed thermodynamic aspects of general relativity may merge smoothly with a consistent relativistic thermodynamic theory. It is also worth noting that the ideas and results of the present work are very different from those of the partially homonymous Reference \citeonline{Lorenzo2018}.

In order to analyze the time evolution being proposed, we shall generalize the Bondi-Sachs formalism \cite{Bondi1962} to worldlines and spacetimes limited only by regularity conditions. Part \ref{bondisachs} will be devoted to this generalization, starting with the coordinate system in Section \ref{coorsys}. Section \ref{regcond} analyzes the regularity conditions around the worldline, and Section \ref{EFEs} shows how to construct solutions to the Einstein field equations from a hierarchical integration scheme.

Part \ref{tagr} focuses on the construction of the thermodynamic formalism. Section \ref{qlhre} demonstrates the thermalization of a vacuum state in quasi-local contexts and indicates that this process may be considerably common. In Section \ref{iisec}, we analyze specific models of gravitational collapse, during the process of star formation and prior to supernova explosions, and find indications that an inequality may be satisfied in typical collapses, representing a precursor of a gravitational second law. Section \ref{EC} aims to define the gravitational contribution to the internal energy. First, we derive an analogue of the integral mass formula \cite{Bardeen1973} and argue that it is related to the notion of infinitesimal free energy. The infinitesimal internal energy, entropy, and temperature associated with this concept suggest that a specific normalization of the total energy derived from the Landau-Lifshitz pseudotensor can be used as a definition for the quasi-local internal energy. Based on the previous results, we construct a thermodynamic formalism in Section \ref{lct} that intends to merge classical thermodynamics with the thermodynamic aspects of general relativity into a postulational framework similar to that of Reference \citeonline{Callen1985}. This allows us to give a thermodynamic definition for the rest-mass density. Finally, we relate our results and those of black hole thermodynamics in Section \ref{bbht} and finish with some conclusions.

The software \textit{Wolfram Mathematica} was used to compute many of the results in this work. The suite \textit{xAct} \cite{xAct}, in particular, the packages \textit{xCoba} and \textit{xTensor}, was used to compute curvatures, Christoffel symbols, the equations of Appendix \ref{appfs}, and the Landau-Lifshitz pseudotensor. Curvature calculations were checked with a code kindly provided by D. A. T. Vanzella (personal communication, February 10, 2023). Other calculations done using \textit{Mathematica} were those leading to the expansions \eqref{expTBondi}, the covariant derivative of equation \eqref{tphiQ} and the subsequent solutions to $ \delta $, the analysis of the first example of Subsection \ref{scbhf} and of the models of Section \ref{iisec}, the expansions of Subsection \ref{imf}, the results of Appendix \ref{firstterms}, the equations of Appendix \ref{kinematics} that precede equation \eqref{kappadelrbeta}, and the equations after equation \eqref{Tursol} of Appendix \ref{SSS}.

Figures \ref{z}, \ref{shear}, \ref{analyticity}, \ref{exppeel}, \ref{loop}, \ref{varyingomega}, \ref{limit}, \ref{conehole}, and \ref{branch} were drawn with the software \textit{Inkscape}, Figure \ref{deltheta} with the \textit{LaTeX} package \textit{PGF/TikZ}, and Figures
\ref{muc}, \ref{e2Phi}, \ref{other}, \ref{3d}, \ref{Lie}, \ref{betabound}, and \ref{superiota} with \textit{Mathematica}.

Given the previous motivation, the term \emph{light cone} will refer to past light cones from now on. We use Planck units, $ c = G = \hbar = k_B = 1 $, and the metric signature $(-,+,+,+)$. Points and sets of points in a spacetime $ \mathcal{M} $ are written as capital calligraphic letters. Abstract indices of tensors are represented by the lowercase Latin letters $ \left\{a,\ldots,h\right\} $. Components of tensors are represented by the Greek letters $\lambda, \mu,\nu,\rho,\sigma$. In general, $ \mu,\nu,\ldots\in \left\{0,1,2,3\right\} $, but $ \mu,\nu,\ldots\in \left\{u,r,\theta,\phi\right\} $ in a Bondi-Sachs coordinate system. Non-timelike components of tensors or non-timelike labels of basis vectors are indexed by $ i,j,k \in \left\{1,2,3\right\} $. Capital Latin letters represent angular components, as $A, B, \ldots \in \left\{\theta,\phi\right\}$. The Einstein summation notation is adopted only for pairs of contravariant-covariant indices; otherwise, no summation is implied. $\partial_a$ stands for the partial derivative operator canonically associated with a coordinate system, and $\nabla_a$ for the Levi-Civita connection. The derivative of $ f(x) $ will be denoted by $ \dv*{f}{x} $ or $ D_x f$. The tangent space at $ \mathcal{Q} \in \mathcal{M} $ is written as $ T_{\mathcal{Q}} $. The symbol $ * $ indicates complex conjugation.

\part{A characteristic solution to the Einstein field equations}\label{bondisachs}

\section{The coordinate system}\label{coorsys}

Let $\mathcal{P}\colon \mathbb{R} \supseteq (a,b) \to \mathcal{M}$ be a $C^3$, future-directed timelike curve contained in a convex normal neighborhood $ \mathcal{N} $ \cite{Hawking1973,Chrusciel2011} and parametrized by proper time $ \tau $. That is, $\left. {x^\mu}\right| _{\mathcal{P}}   = {x^\mu}(\tau)$ and the four-velocity $v^\mu \coloneqq \dv*{x^\mu}{\tau}$ satisfies $v^a v_a = -1$. Considering that an observer follows $ \mathcal{P} $, we might have, in addition to a four-acceleration $ a^a \coloneqq v^b\nabla_b v^a $ satisfying $ a^a v_a = 0$, an angular velocity vector $ \Omega^a $ (relative to Fermi-Walker transport \cite[p. 174]{Misner1973}) attributed to this curve, such that $ \Omega^a v_a = 0$. Therefore, given an orthonormal basis $ \left\{ e_{(\bar{\mu})}{}^a \right\} $ at $\mathcal{P}(\tau)$, the most general rule for propagating it along $ \mathcal{P} $ and keep it orthonormal as well as $ e_{(\bar{0})}{}^a = v^a $ corresponds to Fermi-Walker transport plus a rotation \cite{Padmanabhan2010b}:
\begin{equation}\label{tetradtransport}
v^a \nabla_a e_{(\bar{\mu})}{}^b = e_{(\bar{\mu})}{}_a (v^b a^a - v^a a^b +  v_c \Omega_d \epsilon^{cdab}),
\end{equation}
where $ \epsilon_{abcd} $ is the Levi-Civita tensor, with $ \epsilon_{0123} = -\epsilon^{0123} = + 1 $ \cite[p. 87]{Misner1973}.

Now, consider a unitary, spacelike vector $n^a \coloneqq n^{\bar{i}} e_{(\bar{i})}{}^a$ at $\mathcal{P}(\tau)$. We define the functions $ \theta' $ and $ \phi' $ on $ T_{\mathcal{P}(\tau)} $ such that
\begin{equation}\label{n}
n^{\bar{\mu}} \eqqcolon \left(0,\sin(\theta')\cos(\phi'),\sin(\theta')\sin(\phi'),\cos(\theta') \right),
\end{equation}
for all $ n^a $. By keeping the components $ n^{\bar{i}} $ of each $ n^a $ constant along $ \mathcal{P} $, it is possible to extend the definitions of $ \theta' $ and $ \phi' $ to the tangent spaces of the whole curve. For each $ n^a(\theta',\phi') $ with fixed values of $ \theta' $ and $ \phi' $ we construct a past-directed null vector field along $ \mathcal{P} $:
\begin{equation}\label{k}
\left .{k^a(\theta',\phi')}\right|_{\mathcal{P}} \coloneqq \left .{n^a(\theta',\phi')-v^a}\right|_{\mathcal{P}}.
\end{equation}
The null geodesic generated by $ k^a(\theta',\phi') $, satisfying $ k^a \nabla_a k^b =0 $ and starting at $\mathcal{P}(\tau)$, will be labeled by $ \theta', \phi' $ and $ \tau $. Then, each point along this geodesic can be identified by these 3 functions plus the affine parameter $ \lambda' $ for which $ k^\mu = \dv*{x^\mu}{\lambda'} $ and which is set to 0 at $\mathcal{P}(\tau)$.

For each value of $ \tau $ and considering only the sections of positive $ \lambda' $, the set of null geodesics generated by all $ k^a(\theta',\phi') $, i.e., in all directions given by $ \theta' $ and $ \phi' $, constitutes a hypersurface, the past light cone $ \mathcal{C}(\tau) $ of $\mathcal{P}(\tau)$. With that, we can define the function $ u' $ such that $ \left .{u'}\right|_{\mathcal{C}(\tau)} \coloneqq \tau $. 
As a result, we have a coordinate system $ y^{\mu'} = (u', \lambda', \theta', \phi') $ that covers a neighborhood $ \mathcal{N}' \subseteq\mathcal{N} $ of $ \mathcal{P} $.

Denoting the corresponding coordinate basis by $\left\{ \partial_{(\mu')}{}^a \right\} $ and noting that $ \partial_{(\lambda')}{}^a = k^a $, it follows that
\[g_{\lambda' \lambda'} = g_{ab}\partial_{(\lambda')}{}^a\partial_{(\lambda')}{}^b = g_{ab}k^a k^b = 0. \]
Moreover,
\[  k_a k^a = 0 \implies k_a \nabla_b k^a = 0, \]
so
\[ 0 = [\partial_{(A')},\partial_{(\lambda')}]^a = [\partial_{(A')},k]^a, \]
\[ \implies k_a [\partial_{(A')},k]^a = k_a \partial_{(A')}{}^b \nabla_b k^a - k_a k^b \nabla_b \partial_{(A')}{}^a = 0, \]
\[ \implies  k_a k^b \nabla_b \partial_{(A')}{}^a = 0. \]
Given the geodesic equation for $ k^a $, we find that
\begin{equation}\label{ktheta}
k^b \nabla_b (k_a \partial_{(A')}{}^a) = 0.
\end{equation}
Now, consider a coordinate system $ x^{\bar{\mu}} $ in which the coordinate basis coincides with the basis $ \left\{ e_{(\bar{\mu})}{}^a \right\} $ at $\mathcal{P}(\tau)$, where $x^{\bar{\mu}}=0$. Assuming that the coordinates $ x^{\bar{\mu}} $ at $ \mathcal{C}(\tau) $ are real analytic functions of $ \lambda' $, one can expand them as
\[ x^{\bar{\mu}} = \left .{k^{\bar{\mu}}}\right|_{\mathcal{P}(\tau)} \lambda' + \mathcal{O}({\lambda'}^2). \]
Then, equations \eqref{k} and \eqref{n} lead to
\[ \partial_{(A')}{}^{\bar{\mu}} = \partial_{A'} {x^{\bar{\mu}}} = \partial_{A'} \left(  \left .{k^{\bar{\mu}}}\right|_{\mathcal{P}(\tau)} \right) \lambda' + \mathcal{O}({\lambda'}^2) = \partial_{A'} n^{\bar{\mu}} \lambda' + \mathcal{O}({\lambda'}^2), \]
\[ \implies k_{\bar{\mu}} \partial_{(A')}{}^{\bar{\mu}}  = \left[ \left .{k_{\bar{\mu}}}\right|_{\mathcal{P}(\tau)} + \mathcal{O}(\lambda')\right] \partial_{(A')}{}^{\bar{\mu}} =  \mathcal{O}({\lambda'}^2). \]
Since equation \eqref{ktheta} requires $ k_a \partial_{(A')}{}^a $ to be a constant along the null geodesics, the last equation can hold only if
$ k_a \partial_{(A')}{}^a =0 $. As a consequence,
\[ g_{\lambda' A'} = g_{ab}\partial_{(\lambda')}{}^a \partial_{(A')}{}^b = g_{ab}k^a \partial_{(A')}{}^b = 0. \]
We find, then, that $ k^a $ is orthogonal to all vector fields in the tangent bundle of $ \mathcal{C}(\tau) $, including itself, so the past light cone can be called a null hypersurface \cite{Gourgoulhon2006}.

\begin{figure}
	\centering
	\begin{tikzpicture}
	\node[anchor=south west,inner sep=0] (image) at (0,0) {\includegraphics[width=0.6\textwidth]{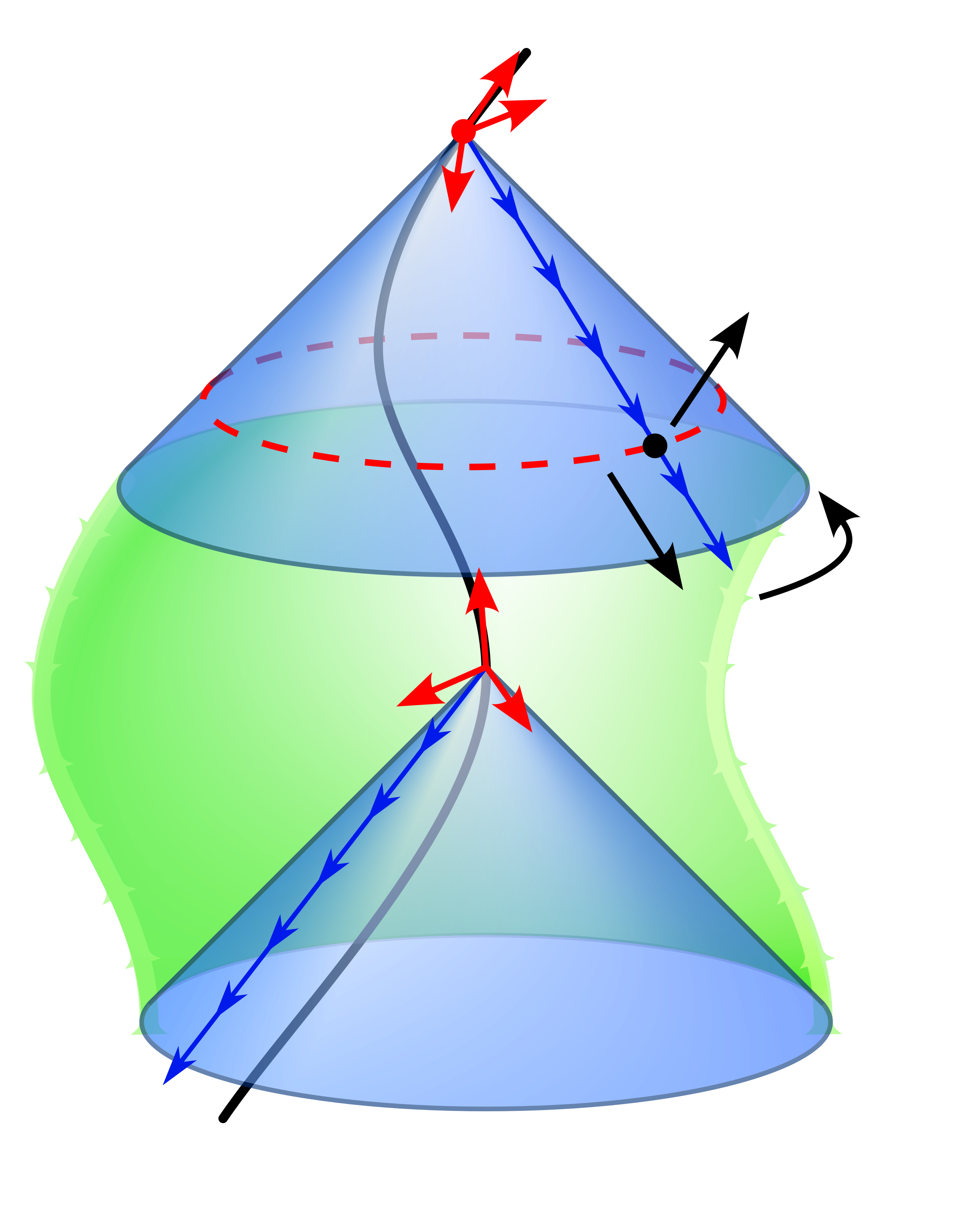}};
	\begin{scope}[x={(image.south east)},y={(image.north west)}]
	\node[anchor=west] at (0.23,0.07) {$\mathcal{P}$};
	\node[anchor=north,text=blue] at (0.13,0.11) {$k^a(\theta,\phi)$};
	\node[anchor=south east,text=red] at (0.46,0.42) {$e_{(\bar{1})}{}^a$};
	\node[anchor=south west,text=red] at (0.53,0.39) {$e_{(\bar{2})}{}^a$};
	\node[anchor=west,text=red] at (0.5,0.5) {$e_{(\bar{0})}{}^a$};
	\node[anchor=west,text=ao] at (0.06,0.425) {$\simeq \mathbb{R} \times S^2$};
	\node[anchor=north west] at (0.68,0.51) {$r$};
	\node[anchor=west] at (0.87,0.55) {$\phi$};
	\node[anchor=south west] at (0.75,0.735) {$u$};
	\node[anchor=north west] at (0.69,0.645) {$\mathcal{Q}$};
	\node[anchor=east,text=red] at (0.2,0.675) {$\mathcal{S}(u,r)$};
	\node[anchor=north,text=red] at (0.47,0.835) {$e_{(\bar{1})}{}^a$};
	\node[anchor=north west,text=red] at (0.52,0.91) {$e_{(\bar{2})}{}^a$};
	\node[anchor=east,text=red] at (0.53,0.96) {$e_{(\bar{0})}{}^a$};
	\node[anchor=east,text=red] at (0.47,0.895) {$\mathcal{P}(\tau)$};
	\node[anchor=east] at (0.32,0.96) {$\boxed{\theta = \text{constant}}$};
	\end{scope}
	\end{tikzpicture}
	\caption[Bondi-Sachs coordinates]{Bondi-Sachs coordinates. The coordinates $ z^\mu $ are constructed around the worldline $ \mathcal{P} $, along which the observer propagates an orthonormal basis $ \left\{ e_{(\bar{\mu})}{}^a \right\} $. Each past light cone is generated by null geodesics, as the ones in blue, which have $k^a(\theta,\phi)$ as the tangent vector field and constant values of $ \theta $ and $ \phi $. At each light cone of constant $ u $, we define a foliation by surfaces $ \mathcal{S}(u,r) \simeq S^2 $ so that their areas equal $ 4\pi r^2 $. Then, a point $ \mathcal{Q} $ at the past light cone of $ \mathcal{P}(\tau) $ will be identified by the coordinates $ z^\mu = (u,r,\theta,\phi) $, with $ u=\tau $. In this picture, $ \theta $ is held constant while the other coordinates increase in the indicated directions. Besides, since our analysis may be limited by the neighborhood $ \mathcal{N} $ to finite values of $ r $, we shall consider only a region of spacetime bounded by some timelike hypersurface $\simeq \mathbb{R} \times S^2$, as the one represented in green}
	\label{z}
\end{figure}

In order to further simplify the metric, we note that each past light cone will be foliated by spacelike surfaces $ \mathcal{S}'(u',\lambda') $, where $ u' $ and $ \lambda' $ are constant, diffeomorphic (at least in $ \mathcal{N} $) to two-dimensional spheres. Therefore, we look for a coordinate system $ z^\mu = (u,r,\theta,\phi) $ in which the induced metric at the surface $ \mathcal{S}(u,r) $ has a determinant with the usual form of the one computed for a sphere embedded in Euclidean space. Then, we define the Bondi-Sachs coordinates $ z^\mu $ (as depicted in Figure \ref{z}), in which $ u \coloneqq u', \theta \coloneqq \theta', \phi \coloneqq \phi' $ and the areal radius $ r $ is such that
\[  \det(g_{AB}) = r^4 \sin[2](\theta).  \]
The metric components transform as
\begin{equation}\label{gtransf}
g_{\mu\nu}(z^\sigma)= \partial_{\mu} y^{\mu'} \partial_{\nu} y^{\nu'} g_{\mu' \nu'} (y^{\rho'}(z^\sigma)),
\end{equation}
so $ g_{\lambda' \lambda'} = g_{\lambda' A'} = 0 $ implies that $ g_{AB}(z^\sigma) = g_{A'B'}(y^{\rho'})  $ and
\begin{equation}\label{r}
r \coloneqq \left(\frac{\det(g_{A'B'}(y^{\rho'}))}{\sin[2](\theta')}\right)^{1/4}
\end{equation}
satisfies the previous condition for $ r $.

Moreover, equation \eqref{gtransf} and $ g_{\lambda' \lambda'} = g_{\lambda' A'} = 0 $ also imply that
\[ g_{rr}=g_{rA}=0. \]
By defining
\begin{equation}\label{h}
h_{AB} \coloneqq r^{-2}g_{AB}, \qq{which has} \det(h_{AB}) = \sin[2](\theta),
\end{equation}
the general form of a Bondi-Sachs metric can be cast as
\[ \dd{s}^2 = g_{uu}\dd{u}^2 + 2g_{ur}\dd{u} \dd{r} + 2g_{uA}\dd{u} \dd{z^A} + r^2 h_{AB}\dd{z^A} \dd{z^B}. \]
For computational convenience, we choose to express this metric (in a way similar to References \citeonline{Maedler2013} and \citeonline{VanderBurg1966}) through the six functions $\Phi,\beta,U,W,\gamma$ and $ \delta $, all of which depend on the coordinates $ u,r,\theta $ and $ \phi $, as
\[ \dd{s}^2 = -e^{2(\Phi + \beta)}\dd{u}^2 + 2e^{2\beta}\dd{u} \dd{r} + r^2 h_{AB}(\dd{z^A} - U^A\dd{u})(\dd{z^B} - U^B\dd{u}), \]
where
\[h_{AB}\dd{z^A} \dd{z^B}= \cosh(2\delta)(e^{2\gamma}\dd{\theta}^2 + e^{-2\gamma}\sin[2](\theta)\dd{\phi}^2) + 2\sin(\theta)\sinh(2\delta)\dd{\theta} \dd{\phi}, \]
\[ U^\theta=U \qq{and} U^\phi=W. \]

Given that $ k^a $ is orthogonal to the light cones of constant $ u $, we must have $ k_a \propto \nabla_a u $, so $ g^{ab}\nabla_a u \nabla_b u = 0 $ and $ g^{uu} = 0 $. In addition, the angular coordinates $ z^A $ are constant along each null generator, meaning that $ k^a \nabla_a z^A = g^{ab}\nabla_a u \nabla_b z^A = 0 $, or $ g^{uA}=0 $. With this, successively using the relation $ g^{\mu\rho}g_{\rho\nu} = \delta^\mu_\nu $ and introducing $h^{AB}$, with $h^{AC}h_{CB}=\delta^A_B$, one can calculate the inverse metric $g^{\mu\nu}$. In summary, one finds that
\begin{gather}\label{g}
\begin{matrix}
g_{\mu\nu} = \mqty(-e^{2(\Phi + \beta)} + r^2 h_{AB}U^A U^B & e^{2\beta} & - r^2 h_{CB}U^C \\ e^{2\beta} & 0 & 0\\ - r^2 h_{CA}U^C & 0 & r^2 h_{AB} ), \\[4ex]
h_{AB} = \mqty(e^{2\gamma}\cosh(2\delta) & \sinh(2\delta)\sin(\theta) \\ \sinh(2\delta)\sin(\theta) & e^{-2\gamma}\cosh(2\delta)\sin[2](\theta)) , \\[4ex]
g^{\mu\nu} = \mqty(0 & e^{-2\beta} & 0 \\ e^{-2\beta} & e^{2(\Phi - \beta)} & e^{-2\beta}U^B\\ 0 & e^{-2\beta}U^A & r^{-2} h^{AB}) , \\[4ex]
h^{AB} = \mqty(e^{-2\gamma}\cosh(2\delta) & -\sinh(2\delta)\csc(\theta) \\ -\sinh(2\delta)\csc(\theta) &  e^{2\gamma}\cosh(2\delta)\csc[2](\theta)) , 
\end{matrix}
\end{gather}
\begin{equation}\label{detg}
\qand* g \coloneqq \det(g_{\mu\nu})=-e^{4\beta}r^4 \sin[2](\theta).
\end{equation}

\section{Regularity conditions}\label{regcond}

\subsection{Central conditions on the metric}

Now, we are going to analyze how the motion of the observer's orthonormal tetrad is expressed as central conditions along $\mathcal{P}$ on the metric. First, we construct what is called the observer's proper reference frame \cite{Misner1973}. As before, the time coordinate is represented by the observer's proper time, $\tau$, and the orthonormal tetrad is transported along $\mathcal{P}$ according to equation \eqref{tetradtransport}. Moreover, for a given point $\mathcal{P}(\tau)$, the spacelike geodesics generated by vectors $n^a$, orthogonal to $e_{(\bar{0})}{}^a=v^a$, will have a definite affine parameter $s$, which equals proper length, once we set $ \left.s\right|_{\mathcal{P}}=0$ and $n^a n_a=1$. As a consequence, in the convex normal neighborhood $ \mathcal{N} $ of $\mathcal{P}$, each event $\mathcal{Q}$ can be completely specified if one knows the value $\tau$ for which a geodesic, starting at $\mathcal{P}(\tau)$ with direction $n^a$, reaches $ \mathcal{Q} $ if extended by a proper length $ s $. For $n^a=n^{\bar{i}} e_{(\bar{i})}{}^a$, the proper coordinates $ x^{\bar{\mu}} $ defined in a neighborhood $ \bar{\mathcal{N}} \subseteq \mathcal{N} $ of $\mathcal{P}$ will be given by
\begin{equation}\label{barcoord}
x^{\bar{\mu}} \coloneqq \left(\tau, s n^{\bar{1}}, s n^{\bar{2}}, s n^{\bar{3}}\right) \text {. }
\end{equation}
Since
$$
\left. \partial_{(\bar{\mu})}{}^a \right|_{\mathcal{P}}=e_{(\bar{\mu})}{}^a
$$
by construction, we have that
$$
\left.g_{\bar{\mu}\bar{\nu}}\right|_{\mathcal{P}}=\eta_{\bar{\mu}\bar{\nu}} .
$$
Furthermore, equation \eqref{tetradtransport} determines some of the Christoffel symbols along $\mathcal{P}$, for
\begin{gather*}
\left.v^{\bar{\mu}} \nabla_{\bar{\mu}} e_{(\bar{\nu})}{}^{\bar{\rho}}\right|_{\mathcal{P}} =\left.\nabla_{\bar{0}} e_{(\bar{\nu})}{}^{\bar{\rho}}\right|_{\mathcal{P}} = \left.\Gamma^{\bar{\rho}}{}_{ \bar{\mu} \bar{0}} e_{(\bar{\nu})}{}^{\bar{\mu}}\right|_{\mathcal{P}} =\left.\Gamma^{\bar{\rho}}{}_{ \bar{\nu}\bar{0}}\right|_{\mathcal{P}},  \\
\therefore \left.\Gamma^{\bar{\rho}}{}_{ \bar{\nu}\bar{0}}\right|_{\mathcal{P}} =\left.e_{(\bar{\nu}) \bar{\mu}}\left(v^{\bar{\rho}} a^{\bar{\mu}}-v^{\bar{\mu}} a^{\bar{\rho}}+v_{\bar{\sigma}} \Omega_{\bar{\lambda}} \epsilon^{\bar{\sigma} \bar{\lambda} \bar{\mu} \bar{\rho}}\right)\right|_{\mathcal{P}}, \\
\text { or }\left.\Gamma^{\bar{\rho}}{}_{ \bar{\nu}\bar{0}}\right|_{\mathcal{P}(\tau)}=\left.\eta_{\bar{\mu} \bar{\nu}}\left(v^{\bar{\rho}} a^{\bar{\mu}}-v^{\bar{\mu}} a^{\bar{\rho}}+v_{\bar{\sigma}} \Omega_{\bar{\lambda}} \epsilon^{\bar{\sigma} \bar{\lambda} \bar{\mu} \bar{\rho}}\right) \right|_{\mathcal{P}(\tau)} \text {. }
\end{gather*}
For $v^{\bar{\mu}}=(1,0,0,0)$ and $a^{\bar{\mu}}=\left(0, a^{\overline{1}}, a^{\overline{2}}, a^{\overline{3}}\right)$, the last equation implies that
$$
\left.\Gamma^{\bar{0}}{}_{ \bar{0}\bar{0}}\right|_{\mathcal{P}(\tau)}=0,\quad \left.\Gamma^{\bar{0}}{}_{ \bar{i}\bar{0}}\right|_{\mathcal{P}(\tau)} =a_{\bar{i}}(\tau), \quad \left.\Gamma^{\bar{i}}{}_{ \bar{0}\bar{0}}\right|_{\mathcal{P}(\tau)}=a^{\bar{i}}(\tau)
$$
$$
\text { and }\quad \left.\Gamma^{\bar{i}}{}_{ \bar{k}\bar{0}}\right|_{\mathcal{P}(\tau)} =\epsilon_{\bar{0} \bar{i} \bar{j} \bar{k}} \Omega^{\bar{j}}(\tau) \text {. }
$$
Each spacelike geodesic used in the construction of the coordinate system can be represented by
$$
x^{\bar{\mu}}=\left(\tau_0, s n^{\overline{1}}, s n^{\overline{2}}, s n^{\overline{3}}\right),
$$
for some $\tau_0$ and $n^{\bar{i}}$ constants. Thus,
$$
\dv{x^{\bar{\mu}}}{s} =n^{\bar{\mu}}, \quad \dv[2]{x^{\bar{\mu}}}{s}=0
$$
and its geodesic equation becomes
$$
\left.\Gamma^{\bar{\mu}}{}_{ \bar{\nu}\bar{\rho}} n^{\bar{\nu}} n^{\bar{\rho}} \right|_{x^{\bar{\sigma}}(s)}=0 \text {. }
$$
So, for $s=0$, this equation holds for any $n^{\bar{i}}$, including $n^{\bar{i}}=\delta^{\bar{i}}{}_{\bar{j}}$, meaning that $\left.\Gamma^{\bar{\mu}}{}_{ \bar{j}\bar{j}}\right|_{\mathcal{P}}=0$. Choosing $n^{\bar{i}}=\delta^{\bar{i}}{}_{\bar{j}}+\delta^{\bar{i}}{}_{\bar{k}}$ then results in
$$
\left.\Gamma^{\bar{\mu}}{}_{ \bar{j}\bar{k}}\right|_{\mathcal{P}}=0,
$$
fixing the remaining Christoffel symbols along $\mathcal{P}$.

Since we are using the Levi-Civita connection,
\[ \nabla_{\bar{\rho}} g_{\bar{\mu} \bar{\nu}} =0, \]
\[ \implies \partial_{\bar{\rho}} g_{\bar{\mu} \bar{\nu}} = g_{\bar{\sigma} \bar{\nu}} \Gamma^{\bar{\sigma}}{}_{ \bar{\mu}\bar{\rho}} +g_{\bar{\mu}\bar{\sigma} } \Gamma^{\bar{\sigma}}{}_{ \bar{\nu}\bar{\rho}} , \]
\[ \therefore \left.  \partial_{\bar{0}} g_{\bar{\mu} \bar{\nu}} \right|_{\mathcal{P}(\tau)} = \left.  \partial_{\bar{k}} g_{\bar{i} \bar{j}} \right|_{\mathcal{P}(\tau)} = 0,\]
\[ \left.\partial_{\bar{i}}g_{\bar{0}\bar{0}}\right|_{\mathcal{P}(\tau)}=-2 a_{\bar{i}}(\tau) \quad\text { and } \quad \left.\partial_{\bar{k}}g_{\bar{0} \bar{i}}\right|_{\mathcal{P}(\tau)}=\epsilon_{\bar{0} \bar{i} \bar{j} \bar{k}} \Omega^{\bar{j}}(\tau)\text {. } \]
With this, we can expand the metric components in proper coordinates around $x^{\bar{\mu}}=(\tau, 0,0,0)$ as
$$
g_{\bar{\mu} \bar{\nu}}\left(x^{\bar{\rho}}\right)=\eta_{\bar{\mu} \bar{\nu}}+\left.\partial_{\bar{\rho}}g_{\bar{\mu} \bar{\nu}}\right|_{\mathcal{P}(\tau)}\left(x^{\bar{\rho}}-\delta^{\bar{\rho}}{}_{\bar{0}} \tau\right)+ \mathcal{O} \left(\left(x^{\bar{\rho}}-\delta^{\bar{\rho}}{}_{\bar{0}} \tau\right)\left(x^{\bar{\sigma}}-\delta^{\bar{\sigma}}{}_{\bar{0}} \tau\right)\right)
$$
and get \cite{Misner1973}
\[
d s^2 = -\left[1+2 a_{\bar{i}}\left(x^{\bar{0}}\right) x^i\right] {d x^{\bar{0}}}^2 +2 \epsilon_{\bar{i} \bar{j} \bar{k}} \Omega^{\bar{j}}\left(x^{\bar{0}}\right) x^{\bar{k}} d x^{\bar{0}} d x^{\bar{i}} +\delta_{\bar{i} \bar{j}} d x^{\bar{i}} d x^{\bar{j}}+ \mathcal{O}\left(x^{\bar{i}} x^{\bar{j}}\right) d x^{\bar{\mu}} d x^{\bar{\nu}},
\]
where $\epsilon_{\bar{i} \bar{j} \bar{k}}$ is the three-dimensional Levi-Civita symbol, with $\epsilon_{\bar{1} \bar{2} \bar{3}}=+1$.

Next, we express each unitary vector $n^a$ using the angles $\theta^{\prime}$ and $\phi^{\prime}$, as in equation \eqref{n}. Moreover, we remind that the vector field $ k^a $ is defined by equation \eqref{k} and by the geodesic equation,
$$
\dv[2]{x^{\bar{\mu}}}{{\lambda'}} +\Gamma^{\bar{\mu}}{}_{ \bar{\nu} \bar{\rho}} \dv{x^{\bar{\nu}}}{{\lambda'}}\dv{x^{\bar{\rho}}}{{\lambda'}}=0 \text {. }
$$
Then, for each geodesic, we have
$$
k^{\bar{\mu}}=\dv{x^{\bar{\mu}}}{\lambda'}
$$
and
$$
\begin{gathered}
\left.\dv[2]{x^{\bar{\mu}}}{{\lambda'}} \right|_{\mathcal{P}(\tau)}=-\left.\Gamma^{\bar{\mu}}{}_{ \bar{\nu} \bar{\rho}} \dv{x^{\bar{\nu}}}{\lambda'}\dv{x^{\bar{\rho}}}{\lambda'}\right|_{\mathcal{P}(\tau)}=-\left.\Gamma^{\bar{\mu}}{}_{ \bar{\nu} \bar{\rho}}\left(n^{\bar{\nu}}-v^{\bar{\nu}}\right)\left(n^{\bar{\rho}}-v^{\bar{\rho}}\right)\right|_{\mathcal{P}(\tau)} , \\
\implies\left.\dv[2]{x^{\bar{0}}}{{\lambda'}} \right|_{\mathcal{P}(\tau)}=2 a_{\bar{k}}(\tau) n^{{\bar{k}}}\left(\theta^{\prime}, \phi^{\prime}\right), \\
\left.\dv[2]{x^{\bar{i}}}{{\lambda'}} \right|_{\mathcal{P}(\tau)}=2 \epsilon_{\bar{i} \bar{j} \bar{k}} \Omega^{\bar{j}}(\tau) n^{{\bar{k}}}\left(\theta^{\prime}, \phi^{\prime}\right)-a^{\bar{i}}(\tau) .
\end{gathered}
$$
Now, suppose that the coordinates $x^{\bar{\mu}}$ are real analytic functions of $\lambda^{\prime}$ in a neighborhood of $\mathcal{P}$. In that case, we may expand them around $\lambda^{\prime}=0$ at $\mathcal{P}(\tau)$ as
\begin{equation}\label{x0}
x^{\bar{0}}\left(u^{\prime}, \lambda^{\prime}, \theta^{\prime}, \phi^{\prime}\right)=u^{\prime}-\lambda^{\prime}+a_{\bar{k}}\left(u^{\prime}\right) n^{\bar{k}}\left(\theta^{\prime}, \phi^{\prime}\right) \lambda^{\prime 2}+\mathcal{O}\left(\lambda^{\prime 3}\right)
\end{equation}
and
\begin{equation}\label{xi}
x^{\bar{i}}\left(u^{\prime}, \lambda^{\prime}, \theta^{\prime}, \phi^{\prime}\right)=n^{\bar{i}}\left(\theta^{\prime}, \phi^{\prime}\right) \lambda^{\prime}+\left[\epsilon_{\bar{i} \bar{j} \bar{k}} \Omega^{\bar{j}}\left(u^{\prime}\right) n^{\bar{k}}\left(\theta^{\prime}, \phi^{\prime}\right)-\frac{a^{\bar{i}}\left(u^{\prime}\right)}{2}\right] \lambda^{\prime 2}+\mathcal{O}\left(\lambda^{\prime 3}\right),
\end{equation}
with the coordinates $y^{\mu^{\prime}}=\left(u^{\prime}, \lambda^{\prime}, \theta^{\prime}, \phi^{\prime}\right)$ defined as before.

From it, we can obtain the partial derivatives with respect to the coordinates $y^{\mu^{\prime}}$ as power series as well:
$$
\begin{aligned}
\partial_{u^{\prime}} x^{\bar{0}} & =1+\dv{a_{\bar{k}}}{u'} n^{\bar{k}} {\lambda'}^2+\mathcal{O}\left({\lambda'}^3\right), \\
\partial_{\lambda^{\prime}} x^{\bar{0}} & =-1+2 a_{\bar{k}} n^{\bar{k}} \lambda'+\mathcal{O}\left({\lambda'}^2\right), \\
\partial_{A^{\prime}} x^{\bar{0}} & =a_{\bar{k}} \partial_{A'} n^{\bar{k}} {\lambda'}^2+\mathcal{O}\left({\lambda'}^3\right), \\
\partial_{u^{\prime}} x^{\bar{i}} & =\left(\epsilon_{\bar{i} \bar{j} \bar{k}} \dv{\Omega^{\bar{j}}}{u'} n^{\bar{k}}-\frac{1}{2} \dv{a^{\bar{i}}}{u'}\right) {\lambda'}^2+\mathcal{O}\left({\lambda'}^3\right), \\
\partial_{\lambda^{\prime}} x^{\bar{i}} & =n^{\bar{i}}+\left(2 \epsilon_{\bar{i} \bar{j} \bar{k}} \Omega^{\bar{j}} n^{\bar{k}}-a^{\bar{i}}\right) \lambda^{\prime}+\mathcal{O}\left({\lambda'}^2\right), \\
\partial_{A^{\prime}} x^{\bar{i}} & =\partial_{A'} n^{\bar{i}} \lambda^{\prime}+\epsilon_{\bar{i} \bar{j} \bar{k}} \Omega^{\bar{j}} \partial_{A'} n^{\bar{k}} {\lambda'}^2+\mathcal{O}\left({\lambda'}^3\right) .
\end{aligned}
$$
Furthermore, the metric components in the coordinates $y^{\mu^{\prime}}$ are given by
$$
g_{\mu^{\prime} \nu^{\prime}}\left(y^{\sigma'} \right)=\partial_{\mu'} x^{\bar{\mu}} \partial_{\nu'} x^{\bar{\nu}} g_{\bar{\mu} \bar{\nu}}\left(x^{\bar{\rho}}\left(y^{\sigma^{\prime}}\right)\right) \text {. }
$$
Since we are looking for general regularity conditions on the metric, from which one may compute the curvature of spacetime after solving the Einstein field equations, it is not reasonable to fix any curvature component from the beginning. Such components appear as second-order corrections to the metric in $x^{\bar{\mu}}$ coordinates \cite{Manasse1963,Ni1978}; hence, we shall omit those terms here. Besides, the angular coordinates do not carry the dimension of length given by the metric tensor. So, each angular index in a metric component increases one order of $\lambda^{\prime}$ in its expansion. As a result, we have
$$
\begin{aligned}
g_{ u^{\prime} u^{\prime}}&=-\left[1+2 a_{\bar{i}} n^{\bar{i}} \lambda^{\prime}\right]+\mathcal{O}	\left({\lambda'}^2\right), \\
g_{u^{\prime} \lambda^{\prime}}&=1+\mathcal{O}\left({\lambda'}^2\right), \\
g_{u^{\prime} A^{\prime}}&=\left(\epsilon_{\bar{i}\bar{j}\bar{k}} \partial_{A'} n^{\bar{i}} \Omega^{ \bar{j}} n^{\bar{k}}-a_{\bar{k}} \partial_{A'} n^{\bar{k}}\right) {\lambda'}^2+\mathcal{O}\left({\lambda'}^3\right) \\
\text {and } \quad  g_{A^{\prime} B^{\prime}}&={\lambda'}^2 \partial_{A'} n^{\bar{i}} \partial_{B'} n_{\bar{i}}+\mathcal{O}\left({\lambda'}^4\right),
\end{aligned}
$$
with $a_{\bar{i}}$ and $\Omega^{\bar{i}}$ evaluated at $\tau=u^{\prime}$, since the supposition that $ \mathcal{P} $ is $ C^3 $ allows us to write
\begin{gather*}
a_{\bar{i}}\left(x^{\bar{0}}\right)=a_{\bar{i}}\left(u^{\prime}\right)+\left. \dv{a_{\bar{i}}}{\tau}\right|_{\tau=u^{\prime}}\left(x^{\bar{0}}-u^{\prime}\right) + \mathcal{O}\left(x^{\bar{0}}-u^{\prime}\right)  =a_{\bar{i}}\left(u^{\prime}\right)+\mathcal{O}\left(\lambda^{\prime}\right) \\
\text {and} \quad  \Omega^{\bar{i}}\left(x^{\bar{0}}\right)=\Omega^{\bar{i}}\left(u^{\prime}\right)+\mathcal{O}\left(\lambda^{\prime}\right) .
\end{gather*}
The quadratic term in $g_{A^{\prime} B^{\prime}}$ equals the usual metric for a sphere of radius $\lambda^{\prime}$, so $\det\left(g_{A^{\prime} B^{\prime}}\right)=\lambda^{\prime 4} \sin^2 (\theta^{\prime})+\mathcal{O}\left({\lambda'}^6\right)$. Given the definition for the areal radius $r$ in equation \eqref{r}, it implies that
$$
\frac{r}{\lambda^{\prime}}=\left[1+\mathcal{O}\left({\lambda'}^2\right)\right]^{1 / 4}=1+\mathcal{O}\left({\lambda'}^2\right) \text {. }
$$
Therefore, the Bondi-Sachs coordinates $z^\mu=(u, r, \theta, \phi)$ will be given by
\begin{equation}\label{ytoz}
u=u^{\prime}, \quad r=\lambda^{\prime}+\mathcal{O}\left({\lambda'}^3\right),\quad \theta=\theta^{\prime}\quad \text {and} \quad \phi=\phi^{\prime} .
\end{equation}
Computing the Jacobian components for this transformation and the metric components as before yields a similar result:
$$
\begin{aligned}
g_{ u u}&=-\left[1+2 a_{\bar{i}} n^{\bar{i}} r\right]+\mathcal{O}	\left(r^2\right), \\
g_{u r}&=1+\mathcal{O}\left(r^2\right), \\
g_{uA}&=\left(\epsilon_{\bar{i}\bar{j}\bar{k}} \partial_{A} n^{\bar{i}} \Omega^{ \bar{j}} n^{\bar{k}}-a_{\bar{k}} \partial_{A} n^{\bar{k}}\right) r^2+\mathcal{O}\left(r^3\right) \\
\text {and} \quad g_{A B}&=r^2 \partial_{A} n^{\bar{i}} \partial_{B} n_{\bar{i}}+\mathcal{O}\left(r^4\right).
\end{aligned}
$$
The last equation determines the matrix $h_{A B}$, defined in equation \eqref{h}:
$$
h_{A B}=\left(\begin{array}{cc}
1 & 0 \\
0 & \sin ^2 (\theta)
\end{array}\right)+\mathcal{O}\left(r^2\right) \text {. }
$$
Now, we are in a position to specify the central conditions on the functions $\Phi, \beta, U, W, \gamma$ and $\delta$. To do so, we assume that they can be
expanded around $r=0$ up to some order $N$ as \footnote{In fact, differentiability up to order $ N $ implies that the remainder is little-o of $ r^N $ \cite[p. 287]{Apostol1967}. Since little-o implies in Big-O, the latter is more usual, and the use of two notations would become cumbersome, we shall assume that functions are differentiable to a sufficient order so that the expansion in question is valid, without explicitly stating it throughout the text.}  
\begin{equation}\label{f}
f\left(z^\mu\right)=\sum_{\xi=0}^N f^{(\xi)}\left(u, z^A\right) \frac{r^{\xi}}{\xi !}+\mathcal{O}\left(r^{N}\right) .
\end{equation}
Then,
$$
\sinh (2 \delta)=\sinh \left(2 \delta^{(0)}\right)+2 \cosh \left(2 \delta^{(0)}\right) \delta^{(1)} r+\mathcal{O}\left(r^2\right),
$$
$$ \cosh (2 \delta)=\cosh \left(2 \delta^{(0)}\right)+2 \sinh \left(2 \delta^{(0)}\right) \delta^{(1)} r+\mathcal{O}\left(r^2\right)
$$
\[\qand* e^{2 \gamma}=e^{2 \gamma^{(0)}}+2 \gamma^{(1)} e^{2 \gamma^{(0)}} r+\mathcal{O}\left(r^2\right) . \]
From $h_{\theta\phi}=\mathcal{O}\left(r^2\right)$, it follows that
$$
\delta^{(0)}=\delta^{(1)}=0,
$$
and $h_{ \theta \theta}=1+\mathcal{O}\left(r^2\right)$ implies that
$$
\gamma^{(0)}=\gamma^{(1)}=0.
$$
Proceeding in a similar way with the components $ g_{uA} $, $ g_{ur} $ and $ g_{uu} $, one obtains
\begin{gather*}
U^{(0)}=a_{\bar{k}} \partial_{\theta} n^{\bar{k}} -\epsilon_{\bar{i}\bar{j}\bar{k}} \partial_{\theta} n^{\bar{i}} \Omega^{ \bar{j}} n^{\bar{k}}, \\
\sin^2 (\theta) W^{(0)}=a_{\bar{k}} \partial_{\phi} n^{\bar{k}} -\epsilon_{\bar{i}\bar{j}\bar{k}} \partial_{\phi} n^{\bar{i}} \Omega^{ \bar{j}} n^{\bar{k}}, \\
\beta^{(0)}=\beta^{(1)}=\Phi^{(0)}=0 \quad\qand\quad \Phi^{(1)}=a_{\bar{i}} n^{\bar{i}} .
\end{gather*}
Writing all sums explicitly, the central conditions on the metric correspond to
\begin{equation}\label{ccg}
\begin{gathered}
\gamma=\mathcal{O}\left(r^2\right),\quad \delta=\mathcal{O}\left(r^2\right), \quad \beta=\mathcal{O}\left(r^2\right), \\
U=\cos( \theta)\left(\cos (\phi) a_{\bar{1}}+\sin (\phi) a_{\bar{2}}\right)-\sin (\theta) a_{\bar{3}}+\sin (\phi) \Omega^{\bar{1}} -\cos(\phi)\Omega^{\bar{2}} + \mathcal{O}(r),\\
W =\csc (\theta)\left(-\sin(\phi)  a_{\bar{1}}+\cos(\phi)a_{\bar{2}}\right)+\cot (\theta)\left(\cos( \phi) \Omega^{\bar{1}}+\sin(\phi)\Omega^{\bar{2}}\right)-\Omega^{\bar{3}} +\mathcal{O}(r)\\
\qand* \Phi=\left[\sin (\theta)\left(\cos (\phi) a_{\bar{1}}+\sin( \phi) a_{\bar{2}}\right)+\cos( \theta) a_{\bar{3}}\right]  r+\mathcal{O}\left(r^2\right) \text {, }
\end{gathered}
\end{equation}
with the terms proportional to the Riemann curvature tensor \cite{Maedler2013} still undetermined, as expected. As we will see, $ U^{(0)} $ and $ W^{(0)} $ are constants of integration. It can also be shown that $ \Phi^{(1)} $ is enforced by the Einstein field equations.

\subsection{Central conditions on the energy-momentum tensor}

The Einstein field equations also require the components of the energy-momentum tensor. So, we consider that these can be written as Taylor polynomials around $x^{\bar{i}}=0$ in proper coordinates:
\begin{equation}\label{expT}
T^{\bar{\mu}\bar{\nu}}(x^{\bar{\rho}}) = T^{\bar{\mu}\bar{\nu}}(x^{\bar{0}},0,0,0) + \mathcal{O}(x^{\bar{i}}).
\end{equation}
To see how this expansion changes with the coordinates, first recall that the functions $\theta'$ and $\phi'$, originally defined on each $ T_{\mathcal{P}(\tau)} $ along $ \mathcal{P} $, were propagated along different geodesics in the coordinate systems $ x^{\bar{\mu}} $ and $ y^{\mu'} $. In order to make this difference clearer, define $\bar{\theta}$ and $\bar{\phi}$ as the extension of the respective functions along the spacelike geodesics defining $ x^{\bar{\mu}} $. Namely, $\bar{\theta}$ and $\bar{\phi}$ are constant along each one of those geodesics. Thus,
\[ \left.\left\{\bar{\theta},\bar{\phi}\right\}\right|_{\mathcal{P}} = \left.\left\{{\theta'},{\phi'}\right\}\right|_{\mathcal{P}} \quad\text{and}\]
\begin{equation}\label{thetaphi}
\left\{{\theta'},{\phi'}\right\}  =  \left.\left\{\bar{\theta},\bar{\phi}\right\}\right|_{\mathcal{P}} + \mathcal{O}(s), 
\end{equation}
with $s$ given by the definition \eqref{barcoord} as
\begin{equation}\label{s}
s=s(x^{\bar{i}})=\frac{x^{\bar{i}}}{n^{\bar{i}}(\bar{\theta},\bar{\phi})}.
\end{equation}
Then, from the last two equations and equation \ref{xi}, we have that 
\begin{equation}\label{lambda}
\lambda'= s + \mathcal{O}(s^2).
\end{equation}
Since $s$, $\bar{\theta}$ and $\bar{\phi}$ do not depend on $x^{\bar{0}}$, the last equations imply that
\begin{equation}\label{d0lambdayA}
\begin{aligned}
\partial_{\bar{0}} \lambda' &=  \mathcal{O}(s^2) = \mathcal{O}({\lambda'}^2) \\
\text{and} \quad \partial_{\bar{0}} y^{A'} &=  \mathcal{O}(s) = \mathcal{O}(\lambda').
\end{aligned}
\end{equation}
Also, from equation \eqref{x0} one finds that
\begin{equation}\label{d0u}
\partial_{\bar{0}} u' = 1 + \mathcal{O}(s^2) = 1 + \mathcal{O}({\lambda'}^2).
\end{equation}

To proceed with the spatial derivatives, we note that $ s $ may be expressed as $ s = \sqrt{\delta_{\bar{i} \bar{j}}x^{\bar{i}}x^{\bar{j}}} $, so its derivatives can be easily calculated:
\begin{equation}\label{dsdxi}
\partial_{\bar{i}} s= \frac{x^{\bar{i}}}{s}.
\end{equation}
Then, it follows from equations \eqref{lambda}, \eqref{s}, and \eqref{thetaphi} that
\begin{equation}\label{dilambda}
\partial_{\bar{i}} \lambda' = n^{\bar{i}}(\theta',\phi') + \mathcal{O}(\lambda'),
\end{equation}
and equation \eqref{x0} yields
\begin{equation}\label{diu}
\partial_{\bar{i}} u' = n^{\bar{i}}(\theta',\phi') + \mathcal{O}(\lambda').
\end{equation}
Now, given their definitions, $\bar{\theta}$ and $\bar{\phi}$ can be written as
\[ \bar{\theta} = \arccos(\frac{x^{\bar{3}}}{s}) \qand \bar{\phi} = \arctan(\frac{x^{\bar{2}}}{x^{\bar{1}}}). \]
Using equations \eqref{thetaphi} and \eqref{dsdxi}, one finds that
\begin{equation}\label{dithetaphi}
\begin{aligned}
\partial_{\bar{i}} \theta' &= \frac{\cot(\theta')n^{\bar{i}}(\theta',\phi') - \csc(\theta')\delta^{\bar{3}}{}_{\bar{i}}}{\lambda'} + \mathcal{O}(1)          \\
\qand* \partial_{\bar{i}} \phi' &= \frac{\csc(\theta')(\cos(\phi')\delta^{\bar{2}}{}_{\bar{i}} - \sin(\phi')\delta^{\bar{1}}{}_{\bar{i}})}{\lambda'} + \mathcal{O}(1),
\end{aligned}
\end{equation}
for $ \bar{\theta} \ne 0,\pi/2 $. Since the transformation between the primed and the Bondi-Sachs coordinates is trivial, it is straightforward to obtain from equations \eqref{d0lambdayA}, \eqref{d0u}, \eqref{dilambda}, \eqref{diu} and \eqref{dithetaphi} the following:
\begin{equation}\label{dbarbondi}
\begin{gathered}
\partial_{\bar{0}} u = 1 + \mathcal{O}(r^2),  \qquad \partial_{\bar{0}} r = \mathcal{O}(r^2), \qquad \partial_{\bar{0}} z^A = \mathcal{O}(r), \\
\partial_{\bar{i}} u = n^{\bar{i}}(\theta,\phi) + \mathcal{O}(r), \qquad \partial_{\bar{i}} r = n^{\bar{i}}(\theta,\phi) + \mathcal{O}(r),\\
\partial_{\bar{1}} \theta = \frac{\cos(\theta)\cos(\phi)}{r} + \mathcal{O}(1), \qquad \partial_{\bar{2}} \theta = \frac{\cos(\theta)\sin(\phi)}{r} + \mathcal{O}(1), \\
\partial_{\bar{3}} \theta = - \frac{\sin(\theta)}{r} + \mathcal{O}(1), \qquad \partial_{\bar{1}} \phi = - \frac{\csc(\theta)\sin(\phi)}{r} + \mathcal{O}(1),  \\
\partial_{\bar{2}} \phi = \frac{\csc(\theta)\cos(\phi)}{r} + \mathcal{O}(1) \quad\qand\quad \partial_{\bar{3}} \phi = \mathcal{O}(1). 
\end{gathered}
\end{equation}

With this, we can calculate the expansion of the contravariant components of the energy-momentum tensor in orders of $ r $ using equation \eqref{expT} and 
\[ T^{\mu\nu}(z^{\sigma})=\partial_{\bar{\mu}} z^{\mu} \partial_{\bar{\nu}} z^{\nu} T^{\bar{\mu}\bar{\nu}}(x^{\bar{\rho}}(z^{\sigma})).\]
Note that each angular index corresponds to the multiplication of the leading terms in equation \eqref{expT} by $\mathcal{O}(1/r)$. Moreover, lowering to an angular index requires components of the metric \eqref{g} which are $\mathcal{O}(r^2)$. The net result is that each covariant angular index raises the order by $\mathcal{O}(r)$, so
\begin{equation}\label{centralT}
T_{uA}{}^{(0)}=T_{rA}{}^{(0)}=T_{AB}{}^{(0)}=T_{AB}{}^{(1)}=0 \text {. }
\end{equation}
For later use, we explicitly give the components
\begin{equation}\label{expTBondi}
\begin{aligned}
T_{ru} = & -{T}^{\bar{0}\bar{0}{(0)}}-\cos (\phi ) \sin (\theta ) {T}^{\bar{0}\bar{1}{(0)}}-\sin (\theta ) \sin (\phi ) {T}^{\bar{0}\bar{2}{(0)}}-\cos (\theta ) {T}^{\bar{0}\bar{3}{(0)}} +\\ 
& \mathcal{O}\left(r\right),  \\
T_{rr} = &\; {T}^{\bar{0}\bar{0}{(0)}}+2 \cos (\phi ) \sin (\theta ) {T}^{\bar{0}\bar{1}{(0)}}+2 \sin (\theta ) \sin (\phi ) {T}^{\bar{0}\bar{2}{(0)}} +\\
&  2 \cos (\theta ) {T}^{\bar{0}\bar{3}{(0)}}+\cos ^2(\phi ) \sin ^2(\theta ) {T}^{\bar{1}\bar{1}{(0)}}+\sin ^2(\theta ) \sin (2 \phi ) {T}^{\bar{1}\bar{2}{(0)}} +\\
&  \cos (\phi ) \sin (2 \theta ) {T}^{\bar{1}\bar{3}{(0)}}+\sin ^2(\theta ) \sin ^2(\phi ) {T}^{\bar{2}\bar{2}{(0)}} +\\
& \sin (2 \theta ) \sin (\phi ) {T}^{\bar{2}\bar{3}{(0)}}+\cos ^2(\theta ) {T}^{\bar{3}\bar{3}{(0)}}+\mathcal{O}\left(r\right),      \\
T_{r\theta} =& \left(\cos (\theta ) \cos (\phi ) {T}^{\bar{0}\bar{1}{(0)}}+\cos (\theta ) \sin (\phi ) {T}^{\bar{0}\bar{2}{(0)}}-\sin (\theta ) {T}^{\bar{0}\bar{3}{(0)}} +\right. \\
& \cos (2 \theta ) \left(\cos (\phi ) {T}^{\bar{1}\bar{3}{(0)}}+\sin (\phi ) {T}^{\bar{2}\bar{3}{(0)}}\right) +\\
& \frac{1}{2} \sin (2 \theta ) \left(\cos ^2(\phi ) {T}^{\bar{1}\bar{1}{(0)}}+\sin (2 \phi ) {T}^{\bar{1}\bar{2}{(0)}}+\sin ^2(\phi ) {T}^{\bar{2}\bar{2}{(0)}} -\right. \\
&\left. \left. {T}^{\bar{3}\bar{3}{(0)}}\right)\right) r + \mathcal{O}\left(r^2\right)  \\
\qand* T_{r\phi} =&\; \sin ^2(\theta ) \left(-\csc (\theta ) \sin (\phi ) {T}^{\bar{0}\bar{1}{(0)}}+\cos (\phi ) \csc (\theta ) {T}^{\bar{0}\bar{2}{(0)}} -\right.\\
&\cos (\phi ) \sin (\phi ) {T}^{\bar{1}\bar{1}{(0)}}+\cos(2\phi ) {T}^{\bar{1}\bar{2}{(0)}}-\cot (\theta ) \sin (\phi ) {T}^{\bar{1}\bar{3}{(0)}} +\\
&\left. \cos (\phi ) \sin (\phi ) {T}^{\bar{2}\bar{2}{(0)}}+\cos (\phi ) \cot (\theta ) {T}^{\bar{2}\bar{3}{(0)}}\right) r+\mathcal{O}\left(r^2\right),
\end{aligned}
\end{equation}
where $T^{\bar{\mu}\bar{\nu}(0)} =   T^{\bar{\mu}\bar{\nu}(0)}(u) \coloneqq T^{\bar{\mu}\bar{\nu}}(x^{\bar{0}},0,0,0) $. As we can see, some of the components of the energy-momentum tensor may not be single-valued functions at the vertex of each light cone (as already mentioned in Reference \citeonline{Penrose1980}). This peculiarity occurs due to the characteristic property of spherical coordinate systems that the direction of maximum increase in $ r $ becomes degenerate at $ r=0 $. Therefore, the angular coordinates must be chosen in order to specify a unique direction onto which a tensor can be projected and have its components fully determined.

\subsection{Axial conditions}

A similar degeneracy of directions occurs along the polar axis, where \(\sin (\theta)=0\). Namely, \(\partial_{(\theta)}{}^a\) is not a single-valued vector when \(\theta\) equals 0 or \(\pi\). Thus, for fixed values of \(u\) and $ r $, \(\partial_{(\theta)}{ }^a(\theta, \phi)\) will keep a dependence on \(\phi\) as $\sin (\theta) \to 0$, and the same applies to $ g_{\theta\theta}$. Moreover, $ g^{\theta\phi} $ and $ g^{\phi\phi} $ diverge, while $ g_{\phi\phi} $ goes to zero as one approaches the axis, meaning that \(\partial_{(\phi)}{}^a \rightarrow 0^a\) and that the coordinate system does not completely describe the geometry of spacetime at points where \(\sin (\theta)=0\). As a consequence, one must impose regularity conditions around the axis in order to keep the geometric invariant quantities well defined.

In particular, one may use Riemann normal coordinates \(w^{\tilde{\mu}}\) \cite{Misner1973,Chrusciel2011} centered at a given point \(\mathcal{Q}_0\) of the axis to calculate the invariant ratio between the circumference of a small circle around the axis and its radius. This circle can be defined in a normal neighborhood \(\mathcal{N}_{\mathcal{Q}_0}\) of \(\mathcal{Q}_0\), in which every point \(\mathcal{Q}\) is connected to \(\mathcal{Q}_0\) through a unique geodesic. Then, the distance $ \Delta s $ between \(\mathcal{Q}_0\) and $ \mathcal{Q} $ is given by integration along this geodesic of
\begin{equation*}
\int_{\mathcal{Q}_0}^{\mathcal{Q}} \sqrt{g_{a b} s^a s^b} \dd{s} \text {.}
\end{equation*}
Given the vector field \(s^a=s^a (\mathcal{Q})\) tangent to the geodesic passing through $ \mathcal{Q}$, affinely parametrized by \(s\) and normalized at \(\mathcal{Q}_0\), we have that
\begin{equation}\label{Deltas}
\Delta s(\mathcal{Q})=\int_{\mathcal{Q}_0}^{\mathcal{Q}}  \dd{s}.
\end{equation}
Furthermore, we shall say that the geodesics passing through \(\mathcal{Q}_1, \mathcal{Q}_2\) and \(\mathcal{Q}_3\) are coplanar with respect to \(\mathcal{Q}_0\) if \(s^a\left(\mathcal{Q}_1\right), s^a\left(\mathcal{Q}_2\right)\) and \(s^a(\mathcal{Q}_3)\) are coplanar at \(\mathcal{Q}_0\). With this, we define the circle \(\mathcal{A}\) of radius $ \alpha $ in the geodesic plane given by \(p_{(1)}{}^a, p_{(2)}{}^a \in T_{\mathcal{Q}_0}\) as
\begin{equation*}
\mathcal{A}\coloneqq \left\{\mathcal{Q} \in \mathcal{N}_{\mathcal{Q}_0}\mid \Delta s(\mathcal{Q})=\alpha, \:\left.s^a(\mathcal{Q})\right|_{\mathcal{Q}_0} \in \operatorname{span}\left(\left\{p_{(1)}{}^a, p_{(2)}{}^a \right\}\right)\right\}.
\end{equation*}

By using orthonormal vectors $ p_{(1)}{}^a $ and $ p_{(2)}{}^a  $ as part of the coordinate basis of \(w^{\tilde{\mu}}\) at $ \mathcal{Q}_0 $, we can write
\begin{equation*}
\left.s^{\tilde{\mu}}\right|_{\mathcal{Q}_0} = \left(0,s^{\tilde{1}}, s^{\tilde{2}}, 0\right)\eqqcolon (0, \cos (\psi), \sin (\psi), 0) ,
\end{equation*}
since \(\left(s^{\tilde{1}}\right)^2+\left(s^{\tilde{2}}\right)^2=1\) at \(\mathcal{Q}_0\). Then, each \(\mathcal{Q} \in \mathcal{A}\) has coordinates of the form
\begin{equation*}
\left.w^{\tilde{\mu}}\right|_{\mathcal{Q}}=(0, \alpha \cos (\psi), \alpha \sin( \psi), 0)
\end{equation*}
and the vector field tangent to \(\mathcal{A}\) can be written as
\begin{equation*}
t_{(\psi)}{}^{\tilde{\mu}}=\dv{w^{\tilde{\mu}}}{\psi}=(0,-\alpha \sin (\psi),\alpha\cos(\psi),0).
\end{equation*}
From that, we define the circumference of \(\mathcal{A}\) as
\begin{equation*}
C\coloneqq \int_0^{2 \pi} \sqrt{g_{a b} t_{(\psi)}{}^a t_{(\psi)}{}^b} \dd{\psi} .
\end{equation*}
For \(g_{ \tilde{\mu} \tilde{\nu}}(w^{\tilde{\rho}}) =\eta_{ \tilde{\mu} \tilde{\nu}}+\mathcal{O}\left(w^{ \tilde{\rho}} w^{\tilde{\sigma}}\right)\), we find that
\begin{equation*}
C=2 \pi \alpha+\mathcal{O}\left(\alpha^3\right)
\end{equation*}
and
\begin{equation}\label{Calpha}
\frac{C}{\alpha}=2 \pi+\mathcal{O}\left(\alpha^2\right) \text {. }
\end{equation}

\begin{figure}
	\centering
	\begin{tikzpicture}
	\draw[blue,line width=0.5mm] (0,3.5) -- (0,-3.5);
	\node[blue,right] at (0,-1.5) {$\sin(\theta)=0$};
	\draw[red,dashed,line width=0.5mm] (0,0) circle (3);
	\node at (0,0) {$\bullet$};
	\node[right] at (0,0) {$r=0$};
	\draw (60:1) arc(60:120:1);
	\node[above] at (0.26,0.97) {$\theta$};
	\node[above] at (-0.26,0.97) {$\theta$};
	\draw[dashed] (0,0) -- (1.5,2.6);
	\draw[dashed] (0,0) -- (-1.5,2.6);
	\node at (1.5,2.6) {$\bullet$};
	\node at (-1.5,2.6) {$\bullet$};
	\draw[-stealth,line width=0.7mm] (1.5,2.6) -- ++(-30:1);
	\draw[-stealth,line width=0.7mm] (-1.5,2.6) -- ++(210:1);
	\node[right] at (1.9,2.6) {$\partial_{(\theta)}{}^a (\phi)$};
	\node[left] at (-1.9,2.6) {$\partial_{(\theta)}{}^a (\phi\pm\pi)$};
	\draw[-stealth,line width=0.7mm] (0,3) -- ++(0:1);
	\draw[-stealth,line width=0.7mm] (0,3) -- ++(180:1);
	\node[blue] at (0,3) {$\bullet$};
	\node[blue,anchor=north west] at (0,3) {$\mathcal{Q}_0$};
	\node[red,right] at (3,0) {$r=$ constant};
	\end{tikzpicture}
	\caption[Differentiability across the polar axis]{Differentiability across the polar axis. The picture represents a coordinate plane, where $ u $ is constant and the azimuthal angle assumes the value $ \phi $ on the right side and $ \phi\pm\pi $ on the left. As one approaches the polar axis (in blue) through curves of constant $ u, r, $ and $ \phi $, the vector field $\partial_{(\theta)}{}^a (\phi)$ must tend to the negative of its opposite counterpart, $\partial_{(\theta)}{}^a (\phi\pm\pi)$, in order to ensure differentiability of the metric with respect to $ \theta $ at points along the axis, like $ \mathcal{Q}_0 $}
	\label{deltheta}
\end{figure}
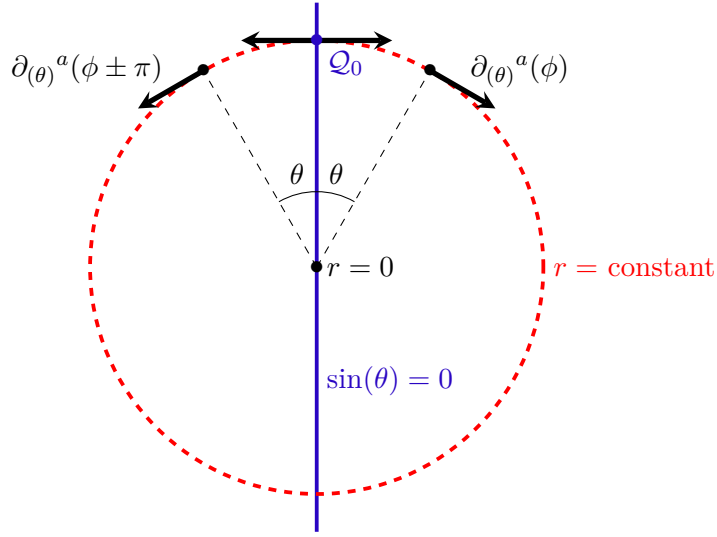

To obtain this result using Bondi-Sachs coordinates, we first suppose that the metric components are free of kinks as the axis is crossed. So we must provide a notion of continuous \(\theta\)-derivative that is well defined in some sense. Recalling the residual dependence on \(\phi\) in \(\partial_{(\theta)}{}^a\) as \(\sin (\theta) \rightarrow 0\), we shall say that \(\partial_\theta\) is continuous along the axis if, for \(u, r\) and \(\phi\) constants,
\begin{equation*}
\lim_{\sin (\theta) \rightarrow 0} \partial_\theta g_{\mu \nu}(\phi)=-\lim_{\sin(\theta) \rightarrow 0} \partial_\theta g_{\mu \nu}(\phi \pm \pi),
\end{equation*}
which implies that
\begin{equation}\label{deltheta02pi}
\left.\partial_{(\theta)}{}^a(\phi)\right|_{\mathcal{Q}_0} \coloneqq \lim_{\sin (\theta) \rightarrow 0} \partial_{(\theta)}{}^a(\phi)=-\lim_{\sin (\theta) \rightarrow 0} \partial_{(\theta)}{}^a(\phi \pm \pi) = - \left.\partial_{(\theta)}{}^a(\phi\pm \pi)\right|_{\mathcal{Q}_0},
\end{equation}
as illustrated in Figure \ref{deltheta}.
Moreover, in case we also require that the surfaces \(\mathcal{S}(u, r)\) (of \(u\) and \(r\) constants) are diffeomorphic to \(S^2\), then \(\left\{\left.\partial_{(\theta)}{}^a\left(\phi_1\right)\right|_{\mathcal{Q}_0}, \left.\partial_{(\theta)}{}^a\left(\phi_2\right)\right|_{\mathcal{Q}_0}\right\} \) will be linearly dependent when \(\sin (\theta) \rightarrow 0\) only if \(\phi_2=\phi_1 \pm \pi\). As a result, the set \(\left\{\partial_{(\theta)}{ }^a(\phi) \in T_{\mathcal{Q}_0} \mid 0 \leq  \phi<\pi\right\}\) spans a single two-dimensional plane, since it must be tangent to \(\mathcal{S}(u, r)\).

Therefore, one may define a circle using
\begin{equation*}
\left.s^a(\phi)\right|_{\mathcal{Q}_0}= \pm\left.\frac{\partial_{(\theta)}{}^a}{\sqrt{g_{\theta \theta}}}(\phi)\right|_{\mathcal{Q}_0},
\end{equation*}
with the sign given by $ \left.\cos(\theta)\right|_{\mathcal{Q}_0} $, and parallelly transporting it by an affine distance \(\alpha\). Because the Christoffel symbols vanish at $ \mathcal{Q}_0$ in the Riemann normal coordinate system, it follows that
\begin{equation*}
\left.s^{\tilde{\mu}} \nabla_{\tilde{\mu}} s^{\tilde{\nu}}\right|_{\mathcal{Q}_0} = \left.s^{\tilde{\mu}} \partial_{\tilde{\mu}} s^{\tilde{\nu}}\right|_{\mathcal{Q}_0}=0,
\end{equation*}
\begin{equation*}
s^{\tilde{\mu}}(\phi, \alpha)= \pm\left.\frac{\partial_{(\theta)}{}^{ \tilde{\mu}}}{\sqrt{g_{\theta \theta}}}(\phi)\right|_{\mathcal{Q}_0}+\mathcal{O}\left(\alpha^2\right)
\end{equation*}
and
\begin{equation*}
\begin{aligned}
s^a \nabla_a \theta & = \pm\left.\frac{\partial_{(\theta)}{}^{ \tilde{\mu}}}{\sqrt{g_{\theta \theta}}}\right|_{\mathcal{Q}_0} \partial_{ \tilde{\mu}} \theta+\mathcal{O}\left(\alpha^2\right) \\
& =\left.\frac{ \pm 1}{\sqrt{g_{ \theta \theta}}}\right|_{\mathcal{Q}_0} \partial_{(\theta)}{ }^{\tilde{\mu}}\partial_{ \tilde{\mu}} \theta+\mathcal{O}(\alpha) .
\end{aligned}
\end{equation*}
As a consequence, along each geodesic generated by a \(s^a(\phi)\),
\begin{equation*}
\partial_s \theta=\left.\frac{ \pm 1}{\sqrt{g_{\theta \theta}}}\right|_{\mathcal{Q}_0}+\mathcal{O}(\alpha)
\end{equation*}
and
\begin{equation*}
\partial_\theta s= \pm \left.\sqrt{g_{ \theta \theta}}\right|_{\mathcal{Q}_0} +\mathcal{O}(\alpha) .
\end{equation*}

It follows from equation \eqref{Deltas} that
\begin{equation}\label{alpha}
\begin{aligned}
\alpha & =\int_{\mathcal{Q}_0}^{\mathcal{Q}} \pm \left[\left.\sqrt{g_{\theta \theta}}\right|_{\mathcal{Q}_0} +\mathcal{O}(\alpha)\right] \dd{\theta} \\
& = \left.r e^{\gamma} \sqrt{\cosh (2 \delta)}\right|_{\mathcal{Q}_0} \Delta \theta+\mathcal{O}\left(\alpha^2\right)\\
&=\left.r e^{\gamma} \sqrt{\cosh (2 \delta)}\right|_{\mathcal{Q}_0} \left.\sin (\theta)\right|_{\mathcal{Q}}+\mathcal{O}\left(\alpha^2\right)\\
&=\left.r e^{\gamma} \sqrt{\cosh (2 \delta)}\sin (\theta)\right|_{\mathcal{Q}}+\mathcal{O}\left(\alpha^2\right),
\end{aligned}
\end{equation}
where \(\Delta \theta=\left| \left.\theta\right|_{\mathcal{Q}} -\left.\theta\right|_{\mathcal{Q}_0} \right| \).

To get an expression for the vector field tangent to the circle, we first note that, even though \(\partial_{(\phi)}{}^a \rightarrow 0\), the limit of \(\partial_{(\phi)}{}^a \dd{\phi}_a\) as \(\sin (\theta) \rightarrow 0\) remains well defined. Thus, we may extend the definition of the \(\phi\)-component of a vector to points along the axis by taking the limit of the contraction with \(\dd{\phi}_a\) as \(\sin (\theta) \rightarrow 0\). Such a process will result in a finite value only for vector fields whose projections onto a unit vector in the direction of \(\partial_{(\phi)}^a\) go to zero as \(\sin( \theta )\rightarrow 0\). This is the case for \(t_{( \psi)}^{ \tilde{\mu}}\), since it vanishes for \(\alpha=0\). It also means that its \(u\)-, \(r\)- and \(\theta\)-components equal zero at the axis, given that \(\partial_{(u)}{}^a, \partial_{(r)}{}^a\) and \(\partial_{(\theta)}{}^a\) differ from \(0^a\) there. So the tangent vector field must be proportional to \(\partial_{(\phi)}{}^\mu\) at \(\mathcal{Q}_0\) and, for convenience, we parametrize it in such a way that
\begin{equation}\label{tphiQO}
\left.t_{(\phi)}{}^\mu\right|_{\mathcal{Q}_0}=\left.\partial_{(\phi)}{}^\mu\right|_{\mathcal{Q}_0}.
\end{equation}

To check the consistency of this limit, we shall expand the components of this vector field as
\begin{equation}\label{tphiexp}
\begin{aligned}
\left.t_{(\phi)}{}^\mu\right|_{\mathcal{Q}}& = \left.\partial_{(\phi)}{}^\mu\right|_{\mathcal{Q}_0} + \mathcal{O}(\alpha)\\
& = \left.\partial_{(\phi)}{}^\mu\right|_{\mathcal{Q}} + \mathcal{O}(\alpha).
\end{aligned}
\end{equation}
Then, for \(\mathcal{Q} \in \mathcal{A}\),
\begin{equation}\label{zphi}
\begin{aligned}
\left.z^\mu(\phi)\right|_{\mathcal{Q}} & = \left.z^\mu(\phi)\right|_{\mathcal{Q}_0}+\left.\dv{z^\mu}{s} \left(\phi\right)\right|_{\mathcal{Q}_0}\alpha + \mathcal{O}\left(\alpha^2\right) \\
& = \left.z^\mu(\phi)\right|_{\mathcal{Q}_0} + \left.s^\mu(\phi)\right|_{\mathcal{Q}_0} \alpha + \mathcal{O}\left(\alpha^2\right) \\
& =\left.z^\mu(\phi)\right|_{\mathcal{Q}_0} \pm \left.\frac{\partial_{(\theta)}{}^\mu}{\sqrt{ g_{\theta\theta} } }(\phi)\right|_{\mathcal{Q}_0} \alpha + \mathcal{O}\left(\alpha^2\right) \\
& =\left.z^\mu(\phi)\right|_{\mathcal{Q}_0} \pm \left.\frac{\partial_{(\theta)}{}^\mu}{\sqrt{ g_{\theta\theta} } }(\phi)\right|_{\mathcal{Q}} \alpha + \mathcal{O}\left(\alpha^2\right)
\end{aligned}
\end{equation}
\begin{equation}\label{deftphi}
\qand* \left.t_{(\phi)}{}^\mu\right|_{\mathcal{Q}} = \left.\dv{\left. z^\mu (\phi)\right|_{\mathcal{A}}}{\phi}\right|_{\mathcal{Q}}.
\end{equation}
Although \(\left.z^\mu(\phi)\right|_{\mathcal{Q}}\) carries a dependence on \(\phi\), in order to ensure continuity of the above expansion, the curve obtained by varying \(\phi\) in this term contains a single point, $ \mathcal{Q}_0 $. It follows that the \(\phi\)-derivative of any single-valued function on $ \mathcal{N}_{\mathcal{Q}_0} $ must vanish along this curve. Equivalently, we can impose that
\begin{equation*}
\dv{ \left.z^\mu(\phi) \right|_{\mathcal{Q}_0}}{\phi}=0^\mu .
\end{equation*}
We distinguish this vector from $ \left.\partial_{(\phi)}{}^\mu\right|_{\mathcal{Q}_0} $ because the order between the operations of taking the \(\phi\)-derivative and the limit as \(\sin (\theta) \rightarrow 0\) is reversed in each case. Consequently, using equations \eqref{deftphi}, \eqref{zphi} and \eqref{tphiexp} yields
\begin{equation}\label{tphiQ}
\begin{aligned}
\left.t_{(\phi)}{}^\mu\right|_{\mathcal{Q}}& = \pm \dv{\phi} \left( \left.\frac{\partial_{(\theta)}{}^\mu}{\sqrt{ g_{\theta\theta} } } \right|_{\mathcal{A}} \right) \alpha + \mathcal{O}(\alpha^2) \\
& = \pm \left[ \left.\partial_{(\phi)}{}^\nu\right|_{\mathcal{Q}} + \mathcal{O}(\alpha) \right] \nabla_{\nu} \left( \left.\frac{\partial_{(\theta)}{}^\mu}{\sqrt{ g_{\theta\theta} } } \right|_{\mathcal{A}} \right) \alpha + \mathcal{O}(\alpha^2) \\
& = \pm \left.\partial_{(\phi)}{}^\nu \nabla_{\nu} \left( \frac{\partial_{(\theta)}{}^\mu}{\sqrt{ g_{\theta\theta} } } \right) \right|_{\mathcal{Q}}  \alpha + \mathcal{O}(\alpha^2).
\end{aligned}
\end{equation}
Next, we evaluate the above covariant derivative, substitute \(\alpha\) by the last line in equation \eqref{alpha}, and take the limit as \(\sin (\theta) \rightarrow 0\), finding that
\begin{equation*}
\left.t_{(\phi)}{}^\mu\right|_{\mathcal{Q}_0} = \left.\left\{\cosh ^2(2 \delta) \mp e^{2 \gamma}\left[\frac{\sinh (4 \delta) \partial_{\phi}\gamma}{2}+\sinh ^2(2 \delta) \partial_{\phi} \delta\right]\right\} \partial_{(\phi)}{}^\mu \right|_{\mathcal{Q}_0} .
\end{equation*}
Therefore, in order to make the limit above equal to the one determined from beginning, in equation \eqref{tphiQO}, we must impose that
\begin{equation}\label{tphiphiQ0}
\left.t_{(\phi)}{}^\phi\right|_{\mathcal{Q}_0} = \left. \left\{\cosh ^2(2 \delta) \mp\left[c_1 \sinh (4 \delta)+c_2 \sinh ^2(2 \delta)\right]\right\} \right|_{\mathcal{Q}_0} = 1,
\end{equation}
\begin{equation*}
\qq*{where} c_1=\left.\frac{e^{2 \gamma} \partial_{\phi} \gamma}{2}\right|_{\mathcal{Q}_0} \quad\text{and}\quad c_2= \left. e^{2 \gamma} \partial_{\phi} \delta \right|_{\mathcal{Q}_0} .
\end{equation*}
For arbitrary \(c_1, c_2 \in \mathbb{R}\), one solution is given by
\begin{equation}\label{sol1}
\left.\delta \right|_{\mathcal{Q}_0}=0,
\end{equation}
while the other is
\begin{equation*}
\left.\delta \right|_{\mathcal{Q}_0}=\frac{1}{4} \ln \left(\frac{\mp 1-2 c_1+c_2}{\mp 1+2 c_1+c_2}\right),
\end{equation*}
provided the argument of the logarithm is positive.

With the aid of condition \eqref{tphiphiQ0}, we can calculate the ratio \(C / \alpha\) using equations \eqref{tphiQ}, \eqref{alpha}, and the metric \eqref{g}:
\begin{equation}\label{CalphaBondi}
\begin{aligned}
\frac{C}{\alpha} & = \frac{1}{\alpha} \int_0^{2 \pi} \left.\sqrt{g_{\mu\nu} t_{(\phi)}{}^\mu t_{(\phi)}{}^\nu}\right|_{\mathcal{A}}  \dd{\phi} \\
& = \int_0^{2 \pi} \frac{1}{\alpha} \sqrt{\left.r^2 e^{-2 \gamma} \sech(2 \delta)\left(t_{(\phi)}{}^{\phi}\right)^2 \right|_{\mathcal{Q}_0}  \sin ^2(\theta) \left[ 1+\mathcal{O}(\sin (\theta))\right]} \dd{\phi} \\
& =\int_0^{2 \pi} \frac{\left.r e^{-\gamma}\sqrt{\sech(2 \delta)} \right|_{\mathcal{Q}_0} \sin (\theta)}{\left.r e^{\gamma}\sqrt{\cosh(2 \delta)} \right|_{\mathcal{Q}_0} \sin (\theta)} \frac{\left[1+\mathcal{O}(\alpha)\right]}{\left[1+\mathcal{O}(\alpha)\right]} \dd{\phi} \\
& = \int_0^{2 \pi} \left.\frac{e^{-2 \gamma}}{\cosh (2 \delta)}\right|_{\mathcal{Q}_0} \dd{\phi} + \mathcal{O}(\alpha).
\end{aligned}
\end{equation}
In case \(\left.\delta \right|_{\mathcal{Q}_0}\) is given by the solution \eqref{sol1} and we assume that \(\left.\gamma \right|_{\mathcal{Q}_0}\) does not depend on \(\phi\), then
\begin{equation*}
\frac{C}{\alpha}=2 \pi \left.e^{-2 \gamma}\right|_{\mathcal{Q}_0} +\mathcal{O}(\alpha)
\end{equation*}
and equation \eqref{Calpha} implies that
\begin{equation*}
\left.\gamma\right|_{\mathcal{Q}_0} = \left.\delta \right|_{\mathcal{Q}_0} =0,
\end{equation*}
what we shall call the \emph{trivial axial condition}. This is compatible with the stronger conditions imposed in References \citeonline{Bondi1962} and \citeonline{Isaacson1983}, where axial symmetry is assumed.

\begin{figure}
	\centering
	\includegraphics[width=0.6\textwidth]{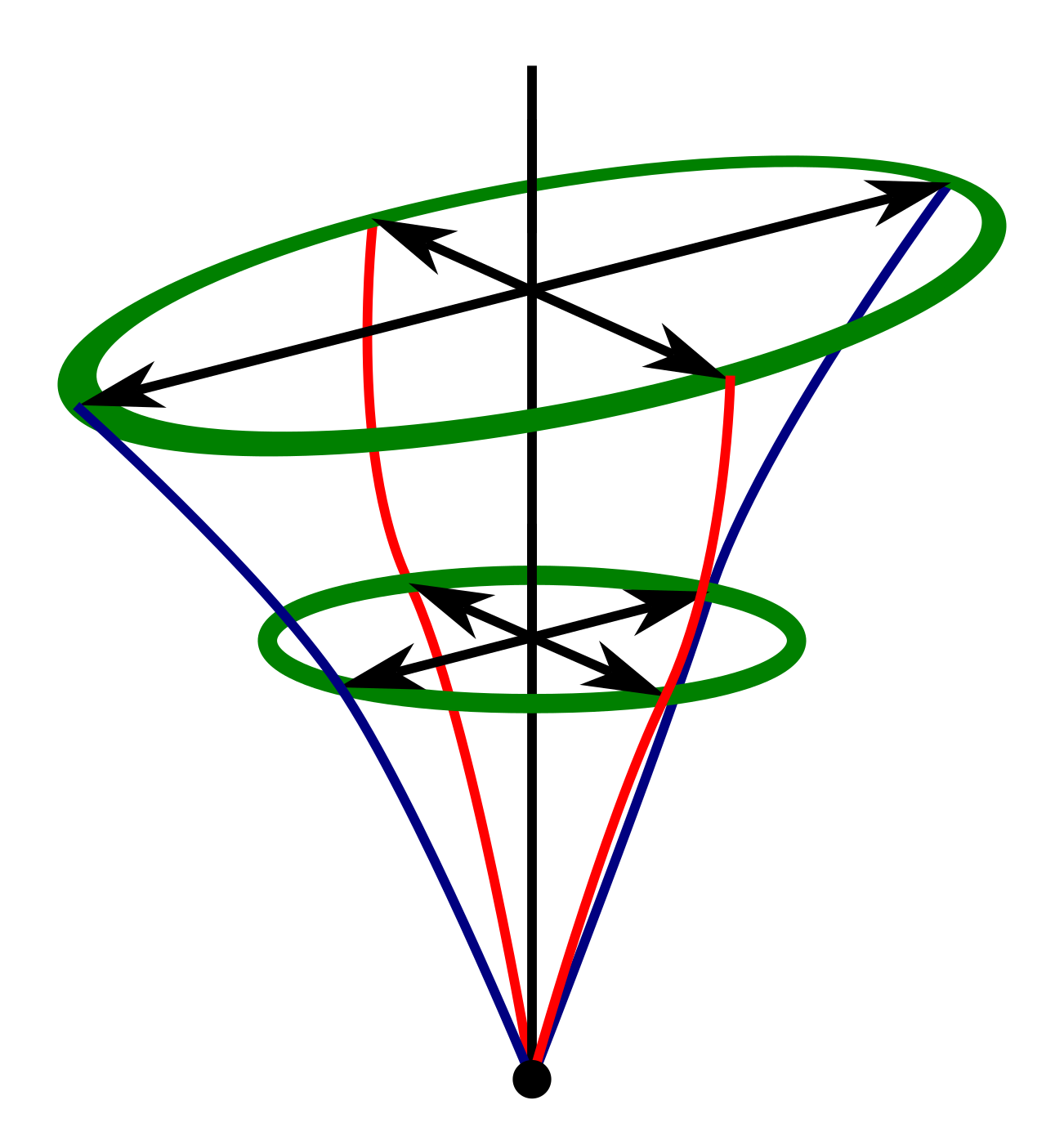}
	\caption[Shear along the polar axis]{Shear along the polar axis. Null geodesics (in red and blue) of constant $ u $ and $ \theta $ emanate from $ r=0 $ (black dot at the bottom) and form an infinitesimal bundle around the axis (vertical black line). Points of constant affine parameter (in green) will constitute a set that approximates a circle (below) near the origin. If there is a non-vanishing shear along the axis, then this circle will be deformed into an ellipse (above). In this case, the vector field \(\partial_{(\theta)}{}^a\) along the axis (black arrows) changes from an initially symmetrical distribution to another one where its ``length'' has a dependence on $ \phi $}
	\label{shear}
\end{figure}

One may consider the possibility to further constrain the functions \(\gamma\) and \(\delta\), so that the above condition could hold in general, for example. However, if we start with the assumption that \(\left.\gamma\right|_{\mathcal{Q}_0}\) is a constant for all values of \(\phi\), then \(c_1=0\) and \(\left.\delta\right|_{\mathcal{Q}_0}=0\), meaning that the trivial axial condition will be satisfied. As a result, \(\partial_r \gamma=\partial_r \delta=0\) along the axis and the double contraction of the covariant shear tensor with its contravariant form necessarily vanishes:
\begin{equation*}
\sigma^{a b} \sigma_{a b}=2\left[\cosh ^2(2 \delta)\left(\partial_r \gamma\right)^2+\left(\partial_r \delta\right)^2\right]=0 .
\end{equation*}
Since this is a coordinate-independent result, it cannot be imposed on every spacetime. Therefore, $ \left.\gamma\right|_{\mathcal{Q}_0} $ will in general fail to be a single-valued function, preserving a dependence on $ \phi $, which emerges from a non-vanishing shear. This can be understood as the effect of shear on the null geodesics integrated from \(r=0\) to \(\left.r \right|_{\mathcal{Q}_0}\). If an infinitesimal bundle of null geodesics around the axis passes through a region of non-vanishing shear, then points of constant affine parameter originally arranged in a circle will be deformed into an ellipse \cite[p. 34]{Poisson2004}. Consequently, the function \(\theta\), defined by these geodesics, will vary at a different rate for different directions pointing outwards from the axis. In other words, \(\partial_{(\theta)}{}^a\) will be stretched or shortened differently for different values of \(\phi\), as represented in Figure \ref{shear}. Thus, $ \left. g_{\theta\theta} \right|_{\mathcal{Q}_0} $ will depend on \(\phi\) and so will $ \left.\gamma\right|_{\mathcal{Q}_0} $ and \(\left.\delta\right|_{\mathcal{Q}_0}\).

Nevertheless, equation \eqref{CalphaBondi} will be enough for our calculations. By comparing it to equation \eqref{Calpha} and expanding the integrand with respect to $ r $, we find that
\[ \int_0^{2 \pi} \left. \left[1-\gamma^{(2)} r^2 + \mathcal{O}(r^3) \right] \right|_{\mathcal{Q}_0} \dd{\phi} = 2\pi. \]
As a consequence, the \emph{general axial condition} must be satisfied:
\begin{equation}\label{gac}
\int_0^{2 \pi} \left. \gamma^{(2)} \right|_{\mathcal{Q}_0} \dd{\phi} = 0.
\end{equation}

\section{Einstein field equations}\label{EFEs}

\subsection{Hierarchical structure of the solution}

The main feature of this coordinate system is the strikingly simple structure presented by the Einstein field equations written in terms of $ z^\mu $, which can be solved by a hierarchical integration scheme. In order to demonstrate this, we first define the tensor
\[ H_{ab} \coloneqq G_{ab} + \Lambda g_{ab} - 8\pi T_{ab}, \]
where $G_{ab}$ is the Einstein tensor, $\Lambda$ the cosmological constant, and $T_{ab}$ the energy-momentum tensor. So, the Einstein field equations for the most general case can be written as
\begin{equation}\label{efe}
H_{ab}=0, 
\end{equation}
which implies that
\begin{equation}\label{he}
H_{r\mu} =0,
\end{equation}
\begin{equation}\label{gee}
H_{AA}- \frac{1}{2}g_{AA}g^{BC}H_{BC}=0,
\end{equation}
\begin{equation}\label{mee}
\qq*{and} \nabla_\nu H^\nu{}_\mu = -8\pi \nabla_\nu T^\nu{}_\mu =0.
\end{equation}

Equations \eqref{he} will be called hypersurface equations (HE$\mu$), since they do not contain any $\partial_u$, equations \eqref{gee} are the gravitational evolution equations (GEE$\theta\theta$ and GEE$\phi\phi$), and equations \eqref{mee} are the matter evolution equations (MEE$\mu$). Introducing the notation $ f(\Psi,\dots) = f(\gamma,\delta,\beta,U,W,\Phi,\dots) $ and keeping in mind that the functions $ f $ below can also depend on first- and second-order partial derivatives of their arguments with respect to non-timelike coordinates, we find that the field equations can be rearranged as
\begin{equation}\label{her}
\qq*{(HE$r$)}  \partial _r \beta = f_\beta(\gamma,\delta,T_{rr}),
\end{equation}
\begin{multline}\label{hetheta}
\qq*{(HE$\theta$)} \partial _r \left(e^{-2 \beta } r^4 \left(e^{2 \gamma } \cosh (2 \delta ) \partial _r U +  \sinh (2 \delta )\sin (\theta )\partial _r W \right)\right) \\
= f_{UW1}(\gamma,\delta,\beta,T_{r\theta}),
\end{multline}
\begin{multline}\label{hephi}
\qq*{(HE$\phi$)} \partial _r \left(e^{-2 \beta } r^4 \left(e^{-2 \gamma }\cosh (2 \delta )  \sin (\theta ) \partial _r W +  \sinh (2 \delta ) \partial _r U \right)\right) \\
= f_{UW2}(\gamma,\delta,\beta,T_{r\phi}),
\end{multline}
\begin{equation}\label{heu}
\qq*{(HE$u$)} \partial _r e^{2 \Phi } + e^{2 \Phi } \frac{ \left(1-4 \pi  r^2 {T}_{{rr}}\right)}{r} = f_\Phi(\gamma,\delta,\beta,U,W,T_{ru},\Lambda),
\end{equation}
\begin{multline}\label{geethetatheta}
\qq*{(GEE$\theta\theta$)} \frac{1}{2} \partial_r \left( r \left(\partial _{u} \gamma \right)\right)  -r \sinh ^2(2 \delta ) \left(\partial _r \gamma \right) \left(\partial _u \gamma \right)  +\\
\tanh (2 \delta ) \left(r \left(\partial _r \delta \right) \left(\partial _u \gamma \right)+\frac{1}{2} \left(\left( 1 +   2 r \left(\partial _r \gamma \right)  \right) \left(\partial _u \delta \right)+r \left(\partial _{{ur}} \delta \right)\right)\right)\\
= f_{\gamma\delta 1}(\Psi,T_{AB}),
\end{multline}
\begin{multline}\label{geephiphi}
\qq*{and (GEE$\phi\phi$)} -\frac{1}{2} \partial_r \left( r \left(\partial _{u} \gamma \right)\right) -r \sinh ^2(2 \delta ) \left(\partial _r \gamma \right) \left(\partial _u \gamma \right)   +\\
\tanh (2 \delta ) \left(- r \left(\partial _r \delta \right) \left(\partial _u \gamma \right)+ \frac{1}{2} \left( \left(1-2 r \left(\partial _r \gamma \right)\right) \left(\partial _u \delta \right)+r \left(\partial _{{ur}} \delta \right)\right) \right)\\
= f_{\gamma\delta 2}(\Psi,T_{AB}).
\end{multline}
Then, \eqref{geethetatheta} $-$ \eqref{geephiphi} and \eqref{geethetatheta} $+$ \eqref{geephiphi} yield, respectively,
\begin{equation}\label{gammaur}
\partial_r \left( r \left(\partial _{{u}} \gamma \right) \right) + 2 \tanh (2 \delta ) \left(\left(\partial _r \delta \right) \left(r \left(\partial _u \gamma \right) \right)+\left(\partial _r \gamma \right) (r \left(\partial _u \delta \right))\right) = f^-(\Psi,T_{\theta\theta},T_{\phi\phi})
\end{equation}
\begin{equation}\label{deltaur}
\qq*{and}   \partial_r \left( r \left(\partial _{{u}} \delta \right) \right) - \sinh (4 \delta ) \left(\partial _r \gamma \right) \left( r\left(\partial _u \gamma \right) \right) = f^+(\Psi,T_{AB}).
\end{equation}

Moreover, from the matter evolution equations one obtains
\begin{equation}\label{meer}
\qq*{(MEE$r$)} \partial _u T_{rr} = f_{T,r}(\Psi,T_{r\mu},T_{AB}),
\end{equation}
\begin{equation}\label{meetheta}
\qq*{(MEE$\theta$)} \partial _u T_{r\theta} = f_{T,\theta}(\Psi,T_{r\mu},T_{AB},T_{u\theta}),
\end{equation}
\begin{equation}\label{meephi}
\qq*{(MEE$\phi$)} \partial _u T_{r\phi} = f_{T,\phi}(\Psi,T_{r\mu},T_{AB},T_{u\phi})
\end{equation}
\begin{equation}\label{meeu}
\qq*{and (MEE$u$)} \partial _u T_{ru} = -\frac{1}{2} {T}_{{rr}} \left(\partial _u e^{2 \Phi }\right) + f_{T,u}(\Psi,T_{\mu\nu},\partial_u \gamma,\partial_u \delta,\partial_u \beta,\partial_u U,\partial_u W).
\end{equation}
In order to solve the last equation, we also need to take the derivative with respect to $u$ of equation \eqref{heu}. Then, substitution of $\partial _u T_{ru}$ by the right-hand side of equation \eqref{meeu} gives
\begin{equation}\label{Phiur}
\partial _r \left(r \left(\partial _u e^{2 \Phi }\right)\right) = f_{\Phi,u}(\Psi,T_{\mu\nu},\Lambda,\partial_u \gamma,\partial_u \delta,\partial_u \beta,\partial_u U,\partial_u W,\partial_u T_{rr}).
\end{equation}
Since the explicit form of most functions $f$ are very long and does not bring much insight into the solution, they were relegated to Appendix \ref{appfs}.

With the above equations, a hierarchical integration scheme can be formulated if $\Lambda$ and the 12 independent (apart from a few constraints) functions of $\left\{ \gamma, \delta, T_{\mu\nu} \right\}$ are given at an initial light cone of $u = u_0$ constant and if the 6 independent functions in $\left\{ T_{AB}, T_{uu}, T_{u\theta}, T_{u\phi} \right\}$ are prescribed for $u > u_0$, through equations of motion or some equation of state, for example. This is accomplished as follows:
\begin{enumerate}
	\item Given the initial data $\left\{ \gamma, \delta, T_{\mu\nu}, \Lambda \right\}$ at $ \mathcal{C}(\tau=u_0) $, one can integrate equation \eqref{her} and then the system of linear equations for $\partial_r U$ and $\partial_r W$, equations \eqref{hetheta} and \eqref{hephi}.
	
	\item The solution to the first-order linear differential equation \eqref{heu} can be written as integrals of its coefficients. At this point, we have found the functions $\beta$, $U$, $W$, and $\Phi$ for $u = u_0$.
	
	\item Solve the system of first-order linear differential equations of \eqref{gammaur} and \eqref{deltaur} for $r (\partial_u  \gamma)$ and $r (\partial_u  \delta)$.
	
	\item Next, equations \eqref{meer}-\eqref{meephi} directly provide $\partial_u T_{rr}$, $\partial_u T_{r\theta}$, and $\partial_u T_{r\phi}$.
	
	\item Take the $u$-derivative of equations \eqref{her}-\eqref{hephi} and integrate in $r$ as before to get
	$\partial_u \beta$, $\partial_u U$, and $\partial_u W$.
	
	\item As mentioned before, take the $u$-derivative of equation \eqref{heu}, substitute $\partial_u T_{ru}$ by the right-hand side of equation \eqref{meeu}, and rearrange the terms to obtain equation \eqref{Phiur}. Then integrate it to find $\partial_u \Phi$.
	
	\item Return to equation \eqref{meeu} and calculate $\partial_u T_{ru}$.
	
	\item From $\partial_u \gamma$, $\partial_u \delta$, $\partial_u T_{r\mu}$ and some suitable prescription for $T_{AB}$, $T_{uu}$, $T_{u\theta}$ and $T_{u\phi}$, one can either numerically calculate the initial data for $u = u_0 + \Delta u$ or keep taking derivatives with respect to $u$ of the previous equations. By repeating the above described order of integration, one may either numerically integrate the spacetime for $u>u_0$ or expand all the functions to some desired order with respect to $u$.
	
	\item Constants of integration that appear in the process are determined by the central conditions \eqref{ccg} along $\mathcal{P}$.
\end{enumerate}
Since the whole scheme depends either on straightforward integration with respect to $r$ or on solutions of systems of first-order linear differential equations, well-posed prescriptions for $T_{AB}$, $T_{uu}$, $T_{u\theta}$ and $T_{u\phi}$ will produce solutions for the other functions that vary continuously with the initial data. In this sense, we can say that this characteristic initial-central value problem is well-posed for real analytic solutions with respect to $ u $.

One can expand all functions with respect to $ r $ and calculate the coefficients using the scheme above. These are shown in Appendix \ref{firstterms} to the first order not determined by the central conditions. It is worth to note that the typical initial-value constraints of the Cauchy formulation are replaced by much simpler ones in the characteristic solution, which derive from the leading-order terms in equations \eqref{gammaur}, \eqref{deltaur}, and MEEs below them. The resulting conditions on coefficients can be put into the form shown in Appendix \ref{firstterms}.

Another interesting property of this solution to the Einstein field equations is that the initial data together with $ \partial_u a^{\bar{i}}, \partial_u \Omega^{\bar{i}} $ and the central conditions determine the first derivative with respect to $u$ of $T_{r\mu}$ and of all functions present in the metric, without making use of any prescription for $T_{AB}$, $T_{uu}$, $T_{u\theta}$ and $T_{u\phi}$ on $u>u_0$. That is to say that, knowing how their four-acceleration and angular velocity are going to change, an observer with access to the initial data at a light cone can calculate to first order in $ \Delta u = u - u_0 $, at a subsequent light cone, the full metric and the four-momentum crossing the hypersurface, regardless of the behavior of matter encoded in the prescription of $T_{AB}$, $T_{uu}$, $T_{u\theta}$ and $T_{u\phi}$ for $u>u_0$. We see, then, that the effects of different interactions of matter enter as second-order corrections in the evolution of the geometry of light cones and of the kinematic parameters of matter. The leading terms generating changes in spacetime and in its four-momentum content may be thought of as given by the determination of the ``present inertial paths'' and by local energy-momentum conservation. Curiously, the hierarchy in the solution brings to light an order of relevance for the dynamics of spacetime, with the generalization of Newton's first law and local energy-momentum conservation above the specification of how the interactions change with configuration of the matter content. This may be seen as an extension of the local universality of the action of gravity for test particles into a quasi-local universality for all kinds of matter to first order in $ \Delta u $.

To make things clearer, $T_{r\mu}$ have been called four-momentum because $T_{r\mu} = g_{r\rho}g_{\mu\nu}T^{\rho\nu} = e^{2\beta}g_{\mu\nu}T^{u\nu} $ fully determines the more usual form for four-momentum, $ T^{u\nu} $. For this reason, the remaining covariant components were attributed to pressures and shear stresses, which are related to momentum exchange between different parts of the matter content. That is, their ``spatial'' derivatives (possibly in null directions) represent forces, or interactions of matter. With this in mind, recall that the equation of geodesic deviation \cite{Misner1973} equates the relative acceleration between neighboring geodesics to a contraction of the Riemann tensor. Thus, the ``gravitational force'' per unit mass is proportional to second derivatives of the metric. Following an analogy with interactions of matter, we could say that terms proportional to $ \partial_i (\partial_\rho g_{\mu\nu}) $ represent the gradient of gravitational stresses, while $ \partial_u (\partial_\rho g_{\mu\nu}) $ characterizes variations in the gravitational momentum. Given the antisymmetry in the indices of the Riemann tensor that originate these second derivatives, the terms involving $ \partial_u (\partial_u g_{\mu\nu}) $ must vanish, and we conclude that the gravitational momentum is proportional to $ \partial_i g_{\mu\nu} $. Since the gravitational degrees of freedom are represented by $ \gamma $ and $ \delta $ in the above formalism, one may expect that the gravitational momentum could be defined in terms of $ \partial_i \gamma $ and $ \partial_i \delta $. In this case, conditions imposed on $ \partial_u (\partial_i \gamma) $ and $ \partial_u (\partial_i \delta) $ could be treated as the gravitational equivalent to local energy-momentum conservation. Therefore, the GEEs, or equations \eqref{gammaur} and \eqref{deltaur}, together with the MEEs \eqref{meer}, \eqref{meetheta}, \eqref{meephi} and \eqref{meeu} may be viewed as the complete law of local energy-momentum conservation. As a result, the previous discussion about the leading terms powering the evolution of the universe can be linked more directly to the grouping we have made with the Einstein field equations. From the initial data, we first determine the ``present inertial paths'' through the HEs; then, it becomes possible to impose the generalized law of local energy-momentum conservation given by the GEEs and the MEEs, thereby obtaining the dominant changes taking place in the quasi-local dynamics.

\subsection{Constants of integration and boundary conditions}\label{coi}

Given the fundamental role that constants of integration play in the original context of the Bondi-Sachs formalism \cite{Bondi1962}, we shall analyze the solutions to the HEs and the GEEs more carefully. First, $ \beta $ is obtained by straightforward integration of HE$r$ \eqref{her} plus a function $C_\beta(u,\theta,\phi)$:
\begin{equation}\label{intbeta}
\beta = \int_{0}^{r} f_{\beta} \dd{r'} + C_{\beta}.
\end{equation}
Then, consistency with central conditions \eqref{ccg} and \eqref{expTBondi} as $ r \to 0 $ implies that $ f_\beta = \mathcal{O}(r) $ and $ C_\beta =0 $. Similarly, HE$ \theta $ \eqref{hetheta} and HE$ \phi $ \eqref{hephi} are integrated as
\begin{equation*}
e^{-2 \beta } r^4 \left(e^{2 \gamma } \cosh (2 \delta ) \partial _r U +  \sinh (2 \delta )\sin (\theta )\partial _r W \right) = \int_{0}^{r} f_{UW1} \dd{r'} + C_{UW1}
\end{equation*}
and
\begin{equation*}
e^{-2 \beta } r^4 \left(e^{-2 \gamma }\cosh (2 \delta )  \sin (\theta ) \partial _r W +  \sinh (2 \delta ) \partial _r U \right) = \int_{0}^{r} f_{UW2} \dd{r'} + C_{UW2}.
\end{equation*}
Once again, we use the central conditions to find that
\begin{equation*}
\mathcal{O}(r^4) \partial _r U +  \mathcal{O}(r^6)\partial _r W  = \int_{0}^{r} \mathcal{O}({r'}^3) \dd{r'} + C_{UW1}
\end{equation*}
and
\begin{equation*}
\mathcal{O}(r^4) \partial _r W +  \mathcal{O}(r^6)\partial _r U  = \int_{0}^{r} \mathcal{O}({r'}^3) \dd{r'} + C_{UW2},
\end{equation*}
meaning that $ C_{UW1} $ and $ C_{UW2} $ also vanish. It follows that
\begin{equation}\label{intU}
U= \int_{0}^{r} \frac{e^{2\beta}}{{r'}^4} \left[ e^{-2\gamma} \cosh(2\delta)\int_{0}^{{r'}} f_{UW1} \dd{r''} - \sinh(2\delta) \int_{0}^{{r'}} f_{UW2} \dd{r''}  \right] \dd{r'} + C_U
\end{equation}
and
\begin{equation}\label{intW}
W= \int_{0}^{r} \frac{e^{2\beta}\csc(\theta)}{{r'}^4} \left[ e^{2\gamma} \cosh(2\delta)\int_{0}^{{r'}} f_{UW2} \dd{r''} - \sinh(2\delta) \int_{0}^{{r'}} f_{UW1} \dd{r''}  \right] \dd{r'} + C_W,
\end{equation}
where the integrands of the outer integrations are $ \mathcal{O} (1)$. Therefore, $ C_U = U^{(0)} $ and $ C_W = W^{(0)} $.

To solve HE$ u $ \eqref{heu}, we need the integrating factor
\begin{equation}\label{FPhi}
F_\Phi (u,r,\theta,\phi) = \exp(\int_{r^*}^{r} \frac{1}{r'} - 4\pi r' T_{rr} \dd{r'} ), 
\end{equation}
where $ r^* > 0 $ is some arbitrary radius in the local coordinate neighborhood. Multiply HE$ u $ by $ F_\Phi $ and integrate to find
\begin{equation}\label{intPhi}
F_\Phi e^{2\Phi} = \int_{0}^{r} F_\Phi f_\Phi \dd{r'} + C_\Phi. 
\end{equation}
Then, by the central conditions, 
\[ F_\Phi =  \exp( \ln(\frac{r}{r^*}) + \mathcal{O}(r^2) + \mathcal{O}(1)) =  \mathcal{O}(r), \]
\[ \mathcal{O}(r)\mathcal{O}(1) = \int_{0}^{r} \mathcal{O}(r') \mathcal{O}\left(\frac{1}{r'}\right) \dd{r'} + C_\Phi =  \mathcal{O}(r) +  C_\Phi\]
and we obtain $ C_\Phi = 0 $.

Similarly, the integrating factor for equation \eqref{gammaur} is
\[ F_\gamma (u,r,\theta,\phi) =  \exp(\int_{0}^{r} 2\tanh(2\delta) \partial_{r'} \delta \dd{r'} ) = \exp(\ln(\cosh(2\delta))) = \cosh(2\delta). \]
Multiply equation \eqref{gammaur} by $ F_\gamma $ and define
\[ G\coloneqq r \cosh(2\delta)\partial_u \gamma \qand D \coloneqq r\partial_u \delta \]
to transform equations \eqref{gammaur} and \eqref{deltaur} into
\begin{equation}\label{GD}
\partial_r G  = -2\sinh(2\delta)(\partial_r \gamma) D + \cosh(2\delta)f^-
\end{equation}
and
\begin{equation}\label{DG}
\partial_r D  = 2\sinh(2\delta)(\partial_r \gamma) G + f^+ .
\end{equation}
The solution to the corresponding homogeneous system, i.e., with $ f^+=f^-=0 $, can be given in terms of
\begin{equation*}
\omega (u,r,\theta,\phi) \coloneqq \int_{0}^{r} 2 \sinh(2\delta) \partial_{r'} \gamma \dd{r'}
\end{equation*}
and
\[
\begin{matrix}
H (u,r,\theta,\phi) \coloneqq \mqty( \cos(\omega) & -\sin(\omega) \\ \sin(\omega) & \cos(\omega) ),
\end{matrix}
 \]
namely
\[
\begin{matrix}
\mqty(G\\D)= H \mqty( C_G \\ C_D ).
\end{matrix}
\]
Therefore, we can write the solution to the inhomogeneous system as
\begin{equation}\label{solGD}
\begin{matrix}\mqty(G\\D) = H \mqty(C_G \\ C_D) \end{matrix} + H \int_{0}^{r} H^{-1} \begin{matrix}\mqty(\cosh(2\delta) f^- \\ f^+)\end{matrix} \dd{r'} ,
\end{equation}
since the central conditions (including \eqref{centralT}) imply that $ f^+ = \mathcal{O}(r) = f^- $ and $ \omega = \mathcal{O}(r^4)$, meaning that $ H(r=0)=\mathds{1}  $ and that the integral is finite and vanishes as $ r\to 0 $. Consequently, $ C_G = G( r=0) = 0 $ and $ C_D = D( r=0)=0 $.

We have found that the only nonzero constants of integration are $ C_U = U^{(0)} (u,\theta,\phi) $ and $ C_W = W^{(0)} (u,\theta,\phi) $, in contrast to $ \partial_u c(u,\theta), M(u,\theta) $ and $ N(u,\theta) $ in the original formulation \cite{Bondi1962}. Apart from the differences related to powers of $ r $ in the terms where they appear, which were expected, given the change from an expansion around infinity to one centered on $ r=0 $, they occur in different functions of the metric and have distinct significance. While $ U^{(0)} $ and $ W^{(0)} $ carry the gauge freedom in choosing the worldline $ \mathcal{P} $ and its orthonormal tetrad from 6 independent functions of $ u $, $ a^{\bar{i}} $ and $ \Omega^{\bar{i}} $, $ M $ and $ N $ are primarily related to the mass and the dipole moment of the isolated system in the interior of spacetime. However, if fiducial values of $ c,\ M, $ and $ N $ are known for a given spacetime, then one may say that $ M $ and $ N $ also carry information about the chosen gauge for the coordinate system, represented by the generalized Lorentz boost $ K(\theta) $ and the supertranslation $ \alpha(\theta) $ between the fiducial gauge and the one being employed. From this remark, we shall compare the symmetry groups of both contexts, in the sense of coordinate transformations that preserve the form of the metric and the suitable limits corresponding to either asymptotic flatness or local flatness.

In the quasi-local version of the formalism, local flatness is guaranteed for any point of any spacetime. As a result, one can freely change the timelike curve $ \mathcal{P} $ and the orthonormal tetrad defined along it in order to reconstruct the coordinate system, with the only restriction given by the necessary differentiability. Since such a curve can be integrated from an arbitrary point, say, $ \mathcal{P}(0) $, if one specifies an initial four-velocity $ v^a(0) $ and the four-acceleration $ a^a $ in an interval of $ \tau $, the complete symmetry group can be represented by possible choices of $ \mathcal{P}(0) $, $ v^a(0) $ and $ a^a(\tau) $ as well as $ \left\{ e_{(\bar{i})}{}^a (0) \right\} $ and $ \Omega^a(\tau) $. Besides, translations of the origin $ \mathcal{P}(0) $ that parallelly transport the orthonormal basis will commute with Lorentz transformations on $ \left\{ e_{(\bar{\mu})}{}^a (0) \right\} $, given that translations act on $ \mathcal{M} $ while these Lorentz transformations act on $ T_{\mathcal{P}(0)} $ in a somewhat independent manner. By noting it, one may naively think that both kinds of transformations constitute a group very similar to the Poincaré group, replacing the semidirect product by a direct product. However, two consecutive translations will lead to an orthonormal basis that, in general, differs from the one obtained by the resulting equivalent translation. That is, the path dependence of parallel transport requires the introduction of a Lorentz transformation in the composition of two translations. Apart from this non-trivial, intertwined construction using translations and Lorentz transformations, we still can choose, in a completely independent way, the 6 functions $ a^{\bar{i}} $ and $ \Omega^{\bar{i}} $. More precisely, if one gives $ a^i(\tau) $ and $ \Omega^i(\tau) $ in an arbitrary coordinate system, then the orthogonality conditions $ a^a v_a = \Omega^a v_a =0 $ eliminate the timelike components of $ a^a $ and $ \Omega^a $ from the differential equations determining the orthonormal tetrad. These functions correspond to differentiable, time-dependent infinitesimal Lorentz transformations acting on the result of parallel transport of the orthonormal basis.

Such time-dependent transformations do not occur in the context of asymptotically flat spacetimes. Hence, we may say that the gauge freedom contained in $ M $ and $ N $ is more closely related to translations of $ \mathcal{P}(0) $ and Lorentz transformations on $ \left\{ e_{(\bar{\mu})}{}^a (0) \right\} $ in the quasi-local context. But, since the metric is expanded around null infinity and not integrated throughout spacetime, $ M $ and $ N $ must also bring information from the inside of spacetime in the form of constants of integration. That is why the gauge in this case must be specified in relation to some fiducial values of $ c,\ M, $ and $ N $. Another difference comes from the fact that null geodesics are tied to infinity, so, for a fixed value of $ u $, these geodesics meet a $ S^2 $ that tends to null infinity. Therefore, the generalized Lorentz transformations \cite{Bondi1962} of this context may be imagined as acting on the $ S^2 $ at $ u=0 $ in the limit of $ r\to \infty $. All the other $ S^2 $ for different $ u $ will be simultaneously defined by imposing geodesic motion on the initial $ S^2 $. Moreover, because each point of this initial $ S^2 $ can be translated in an independent manner, as long as differentiability is preserved, the resulting transformation has an angular dependence and is called a supertranslation. In the limit of $ r\to \infty $, where the elements of the symmetry group become exact isometries of null infinity, translations in the $ \partial_{(r)}{}^a $-direction are suppressed, so we may say that the only relevant translations occur in the $ \partial_{(u)}{}^a $-direction. Given the above dissimilarities, the comparison that can be made between $ \left\{U^{(0)}, W^{(0)} \right\}$ and $ \left\{M,N\right\} $ is that they convey information about the gauge being employed, with the caveat that the meaning of this gauge differs considerably from one context to the other.

\begin{figure}
	\centering
	\begin{tikzpicture}
	\node[anchor=south west,inner sep=0] (image) at (0,0) {\includegraphics[width=0.6\textwidth]{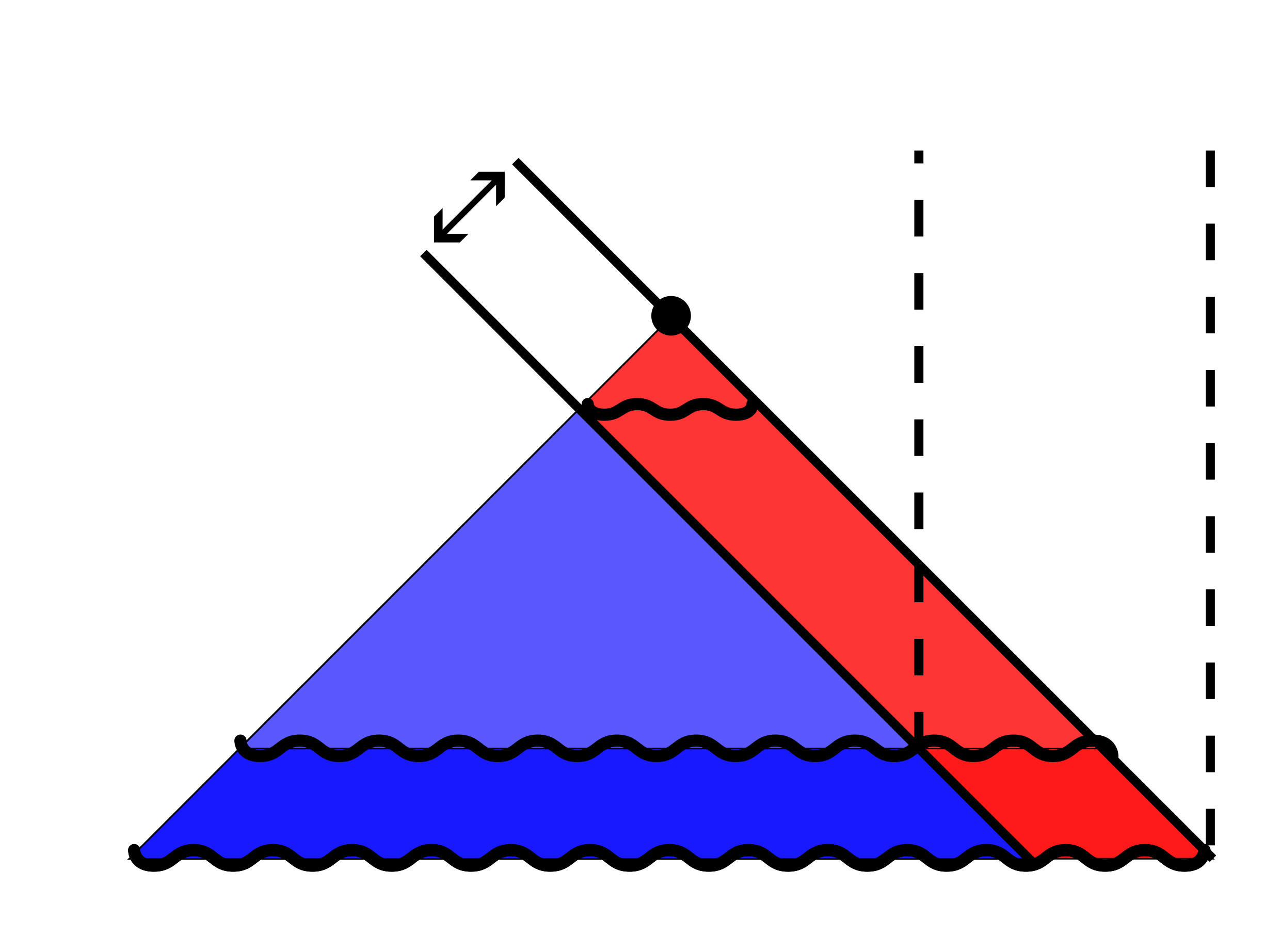}};
	\begin{scope}[x={(image.south east)},y={(image.north west)}]
	\node[anchor=south west] at (.52,.68) {$\mathcal{Q}$};
	\node[anchor=south west] at (.4,.84) {$u$};
	\node[anchor=north east] at (.33,.74) {$u_0$};
	\node[anchor=south east] at (.37,.79) {$\Delta u$};
	\node at (.45,.37) {$\mathcal{K}$};
	\node at (.62,.44) {$\mathcal{U}$};
	\node[below,blue] at (.45,.095) {$\mathcal{H}_K$};
	\node[below,red] at (.87,.095) {$\mathcal{H}_U$};
	\node[left] at (.45,.58) {$\mathcal{H}_0$};
	\node[left] at (.19,.23) {$\mathcal{H}_1$};
	\node[left] at (.1,.1) {$\mathcal{H}_2$};
	\node[above] at (.71,.85) {$r_1$};
	\node[above] at (.95,.85) {$r_2$};
	\end{scope}
	\end{tikzpicture}
	\caption[Characteristic initial value problem]{Characteristic initial value problem. To fully determine the solution for the fields at a point $ \mathcal{Q} $ of the light cone $ u $, one needs initial data placed at some hypersurface bounded by the causal past \cite{Hawking1973} of $ \mathcal{Q} $, $ J^-(\mathcal{Q}) $. Examples of such surfaces are $ \mathcal{H}_0 $, $ \mathcal{H}_1 $, and $ \mathcal{H}_2 $, represented by the wavy lines. Let $ D^+(\mathcal{H}) $ be the future Cauchy development \cite{Hawking1973} of one of these surfaces and $ D^-(\mathcal{C}_0) $ be the past Cauchy development of the light cone $ u_0 $, $ \mathcal{C}_0 $. The blue region $ \mathcal{K} $ corresponds to $ \mathcal{K}\coloneqq D^+(\mathcal{H}) \cap D^-(\mathcal{C}_0) $ and the red one is $ \mathcal{U}\coloneqq D^+(\mathcal{H}) \setminus \mathcal{K} $. Thus, $ \mathcal{H}_K \coloneqq \mathcal{H}\cap \mathcal{K} $ contains the initial data that can always be known from data at $ \mathcal{C}_0 $ and $ \mathcal{H}_U \coloneqq \mathcal{H}\cap \mathcal{U} $ contains the part that may remain unknown. Despite the existence of $ \mathcal{H}_U $, analyticity of the data in $ \mathcal{C}_0 $ can be used to determine the fields at $ \mathcal{H}_0 $ and, consequently, at $ \mathcal{Q} $. That would be possible even for initial data placed at $ \mathcal{C}_0 $ and confined to a finite radius, like $ r_1 $ or $ r_2 $. If analyticity ceases to hold, then the solution at $ \mathcal{Q} $ cannot be determined with such initial data. However, we note that, as the radius increases, it becomes possible to define surfaces $ \mathcal{H} $ such that $ \mathcal{H}_K $ represents a larger part of $ \mathcal{H} $. The same effect occurs as $ \Delta u = u-u_0 $ decreases. Therefore, one can loosely say that the ratio between the volumes of each part of $ \mathcal{H} $, $ V_U/V_K  $, tends to zero as $ \Delta u/ r \to 0 $. When the limit is reached, the only piece of information lacking in the full determination of the fields will be a possible shock wave passing through $ \mathcal{Q} $ tangentially to $ J^-(\mathcal{Q}) $, a piece of information contained in the boundary of $ \mathcal{H} $. This can be interpreted as the expectation that events in the remote past and far away do not have a significant influence on what is observed locally in the short term \cite[Section 13.3]{Ellis1985}. Moreover, it exposes the downside of this characteristic formulation. While the initial data necessary to determine the fields at a compact of spacetime can be placed at a compact spacelike surface, the null case requires a non-compact surface and may still not suffice. We may say that the simplification of the field equations is paid with the demand for an infinite amount of information. But, of course, it could be circumvented by imposing either suitable boundary conditions or a matching scheme \cite{Winicour2012}}
	\label{analyticity}
\end{figure}
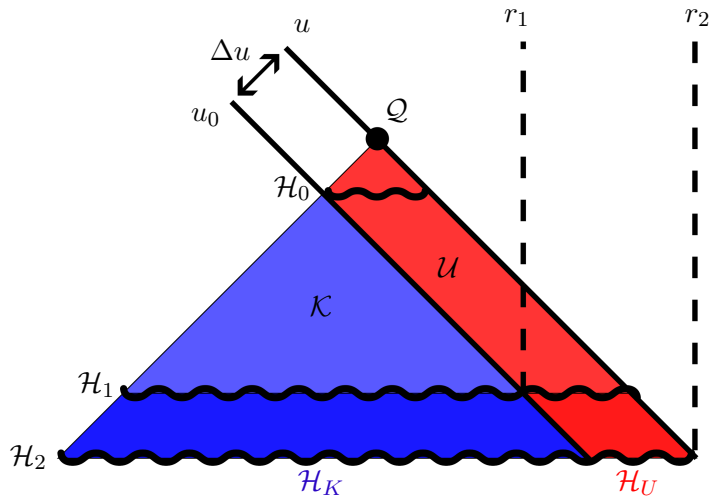

There is still another constant of integration that has no parallel in the quasi-local version of the formalism that we have developed so far, namely the news function $ \partial_u c $. In the original formulation, the Einstein field equations can be integrated if one knows $ \gamma, M $, and $ N $ at the initial hypersurface of constant $ u $, and $ \partial_u c $ is prescribed for all values of $ u $. Thus, the evolution of a neighborhood of null infinity is determined from the initial data apart from the ``news'' brought from the inside of spacetime. That is, $\partial_u c $ dictates the flow of information propagated along null geodesics to infinity about the dynamics taking place in the interior of spacetime. Conversely, in the quasi-local formalism described above, constants of integration follow from central conditions imposed on the metric by assuming that $ g_{\bar{\mu}\bar{\nu}}(x^{\bar{\rho}}) $ are $ C^1 $ functions. Moreover, if one seeks to obtain an exact solution using the hierarchical integration scheme, then it must be iterated ad infinitum, meaning that the solution is assumed to be an analytic function of $ u $. In that case, fields at a point $ \mathcal{Q} $, lying at a light cone $ u $ to the future of $ u_0 $, are determined by information in a Cauchy surface that transcends the causal past of the light cone $ u_0 $, but this information can be determined by the data at $ u_0 $ if $ \Delta u = u-u_0 $ is less than the radius of convergence of the solution (see Figure \ref{analyticity}). That is, the initial data completely determines the solution, without insertion of any data from the outside of the domain of integration in $ r $, as long as the analytic behavior is ensured. We are imposing here what was called the ``no-interference'' condition in Reference \citeonline{Ellis1985}. It basically states that the behavior of a physical system can be extrapolated in time to a sufficient degree, eliminating the possibility of a shock wave suddenly messing up any deterministic prediction. In practice, this is a principle that permeates most of physics and is confirmed by the correctness of its predictions.

Nevertheless, one may still want to model a predictable shock wave, in which case the fields we are interested in should be represented by distributions. When the energy-momentum flux given by such a distribution is transversal to the light cones, the information carried by it should be already contained in the initial data if the equations governing the evolution of the energy-momentum tensor are hyperbolic differential equations with non-spacelike characteristics \cite{Morse1953,Courant1989}. As a consequence, we expect radical changes taking place neither in the hierarchical structure of the solution nor in the constants of integration in this case. It may happen, however, that a Dirac delta function in the components of the energy-momentum tensor propagates tangentially to a light cone $ u $, for example. That would lead to a discontinuity in the transverse extrinsic curvature of the light cone \cite{Barrabes1991}, which may be thought of as the result of an isometric ``gluing'' of two distinct manifolds, $ \mathcal{M}^+ $ and $ \mathcal{M}^- $, at the ``shared'' light cone that bounds each of them. More precisely, one requires that the intrinsic metrics induced on the light cone by the metrics of $ \mathcal{M}^+ $ and $ \mathcal{M}^- $ be equal when written in terms of the same set of intrinsic coordinates. If we include this light cone in a family of hypersurfaces, then we may say that the intrinsic metric induced on them is continuous across the transition from $ \mathcal{M}^- $ to $ \mathcal{M}^+ $. On the other hand, its transversal Lie derivative will not be continuous \cite{Barrabes1991} at all points of the light cone, meaning that $ g_{\bar{\mu}\bar{\nu}} (x^{\bar{\rho}}) $ are no longer $ C^1 $ functions and the central conditions cease to hold. Therefore, we conjecture that the jumps of the first derivatives of the metric may appear in the constants of integration as extra terms not entirely related to $ a^a $ and $ \Omega^a $ but resulting from the $ \delta $-function propagation. Viewed in this way, such constants of integration would play a role similar to the news function.

Another important remark is that proceeding with a pillbox integration in $ u $ (similar to the non-lightlike case \cite[p. 552]{Misner1973}) of equation \eqref{solGD} one finds that a $ \delta $-function in $ T_{AB} $ leads to discontinuities in $ \gamma $ and $ \delta $. Then, the hierarchical scheme may introduce discontinuities in all functions of the metric. Although it may seem strange at first, this result does not conflict with the continuity of the metric mentioned above. The metrics and coordinate systems defined in both regions of spacetime, to the future and to the past of the light cone, are not related a priori. In fact, by imposing isometry at the light cone, one may prevent the soldering of $ \mathcal{M}^+ $ and $ \mathcal{M}^- $ from being ``affinely conciliable'' \cite{Barrabes1991}. That is, affine parameters of null geodesics that are identified in the soldering, calculated from each side of the light cone, cannot be made equal at each point in general, even by a linear transformation. Since the coordinate $ r $ was defined using $ \lambda' $, it may fail to be continuous. Therefore, the functions in the metric with the same name to the future and to the past of the light cone do not need to represent the same mapping from points of spacetime to $ \mathbb{R} $. One could, however, look at the coordinate system and the hierarchical solution without making any connection with the spacetime they represent, regarding the coordinates as well defined. In that case, each function in the metric could be treated as a single entity for all coordinate values, and no condition on its continuity would be imposed. So, we consider the possibility of representing them as distributions, even as $ \delta $-functions, which could be useful in the description of gravitational shock waves. Then, one of the difficulties that should be bypassed would be finding a way to reattach the coordinates to the spacetime manifold. In spite of the interest in this matter, such generalizations will not be pursued in this work. From now on, we shall assume that all functions are continuously differentiable with respect to $ u $.

\subsection[Equivalence between sets of equations]{Equivalence between sets of equations\footnote{This subsection builds upon the proof of Reference \citeonline{Isaacson1983}. The main difference is that here we explicitly use only two GEEs. Thus, it becomes clear that only 10 equations are necessary to ensure that the Einstein field equations are satisfied.}}

So far, we have demonstrated that the Einstein field equations give rise to a set of equations that can be solved in a hierarchical manner, but we still have to show that the converse is also true. Namely, that a solution to this hierarchical scheme necessarily satisfies the Einstein field equations. We start by noting that equation \eqref{he} implies that
\begin{equation}\label{heup}
H^\mu{}_{r} = g^{\mu\nu}H_{\nu r} =0.
\end{equation}
Then, one can express (see details in Appendix \ref{intermediate}) the matter evolution equations as
\begin{equation}\label{div}
\nabla_\nu H^\nu{}_\mu = (-g)^{-1/2} \partial_\nu \left( (-g)^{1/2}  H^\nu{}_\mu \right) + \frac{1}{2} \left( \partial_\mu  g^{\nu \rho} \right) H_{\nu \rho} = 0.
\end{equation}
Take $\mu = r$ and use equation \eqref{heup} to get
\[ \nabla_\nu H^\nu{}_r = \frac{1}{2} \left( \left(  \partial_r g^{u\rho} \right) H_{u\rho} + \left(  \partial_r g^{r\rho} \right) H_{r\rho} + \left(  \partial_r g^{A\rho} \right) H_{A\rho}   \right) = 0.  \]
Since $g^{uu}= g^{uA} = H_{r \rho } = 0$, we find that
\[ \left( \partial_r g^{AB} \right) H_{AB} = 0, \]
\[ \qor*  \left( \partial_r g^{AA} \right) H_{AA} + 2\left( \partial_r g^{\theta\phi} \right) H_{\theta\phi} = 0. \]
Then, equations \eqref{gee} imply
\begin{equation}\label{divHr}
\frac{1}{2}\left( \partial_r g^{AA} \right) g_{AA}g^{BC}H_{BC} + 2\left( \partial_r g^{\theta\phi} \right) H_{\theta\phi} = 0.
\end{equation}

Now, we have that
\[ g^{AB}g_{AB} =  \frac{h^{AB}}{r^2} r^2 h_{AB} = \delta^A{}_A = 2, \]
\begin{equation}\label{gABgAB}
 \implies g^{AA}g_{AA} = 2(1-g^{\theta\phi}g_{\theta\phi} ).
\end{equation}
Moreover,
\[ g^{BC}H_{BC} = g^{AA}H_{AA} + 2 g^{\theta\phi}H_{\theta\phi}, \]
which, by the GEEs, becomes
\[ g^{BC}H_{BC}  = \frac{1}{2}g^{AA}g_{AA}g^{BC}H_{BC} +  2 g^{\theta\phi}H_{\theta\phi}. \]
Using equation \eqref{gABgAB}, we get
\[ g^{BC}H_{BC}  = (1-g^{\theta\phi}g_{\theta\phi} ) g^{BC}H_{BC} +  2 g^{\theta\phi}H_{\theta\phi}, \]
\[ \implies g^{\theta\phi} \left( H_{\theta\phi} - \frac{1}{2} g_{\theta\phi}g^{BC}H_{BC} \right)= 0 . \]
If $ g^{\theta\phi} =0  $, equation \eqref{divHr} can be written as
\begin{equation}\label{divHr2}
\frac{1}{2}\left( \partial_r g^{AB} \right) g_{AB}g^{CD}H_{CD} = 0. 
\end{equation}
Otherwise, 
\[  g^{\theta\phi} \neq 0  \implies   H_{\theta\phi} - \frac{1}{2} g_{\theta\phi}g^{BC}H_{BC} = 0,\]
that is, equation \eqref{divHr} leads to equation \eqref{divHr2} in any case.

From Jacobi's formula, we have that
\[ \left( \partial_r g^{AB} \right) g_{AB} = \frac{\partial_r \det(g^{AB})}{\det(g^{AB})} = \left( \partial_r \frac{1}{r^4 \sin[2](\theta)} \right) r^4 \sin[2](\theta) = -\frac{4}{r} \neq 0. \]
Therefore, equation \eqref{divHr2} implies that
\begin{equation}\label{gCDHCD}
g^{CD}H_{CD} = 0,
\end{equation}
which, by using the GEEs,
\begin{equation}\label{HAA}
\implies H_{AA} =0, 
\end{equation}
\begin{equation}\label{gthetaphiHthetaphi}
\therefore g^{\theta\phi}H_{\theta\phi} = 0.
\end{equation}
If $g^{\theta\phi} \neq 0$, it follows that $ H_{\theta\phi} = 0 $. The case $g^{\theta\phi} = 0$ can happen only if $ \delta =0 $, since $ g^{\theta\phi} = - r^{-2} \csc(\theta) \sinh(2\delta)  $. As seen before, $ \delta $ belongs to the initial data in the hierarchical scheme of integration, and continuous variations in it continuously change the solution of the other functions we have defined. Given that $ H_{\theta\phi} $ depends on these functions, it will also continuously change with $ \delta $. Therefore, as $ H_{\theta\phi} \to 0 $ in the limit of $ \delta \to 0 $, we find that $ H_{\theta\phi} = 0 $ in all cases. So,
\begin{equation}\label{HAB}
H_{AB} =0. 
\end{equation}

Taking into account this last result and the HEs, we have
\[ H^r{}_A = g^{r\nu}H_{\nu A} = g^{r u}H_{u A}, \]
and, for $ g^{uu} = g^{uB} = 0 $,
\[ H^u{}_A = H^B{}_A =0. \]
Using this and the HEs in the $ \mu = A $ component of equation \eqref{div} yields
\[ \left( -g  \right)^{-1/2} \partial_r \left( \left( -g  \right)^{1/2} g^{r u}H_{u A} \right) + \frac{1}{2}\left( \left( \partial_A g^{u\rho} \right) H_{u\rho} + \left( \partial_A g^{B\rho} \right) H_{B\rho} \right) =0, \]
\[ \implies  \partial_r \left( g_{ru} r^2 \sin(\theta) g^{r u}H_{u A} \right) =0,  \]
\[ \implies \partial_r \left( r^2  H_{u A} \right) =0, \]
\[ \therefore H_{uA} = \frac{k}{r^2}, \]
where $ k $ does not depend on $ r $. If we suppose that the domain of $ H_{uA} $ contains the events along $r = 0$, then $ k $ must vanish and
\begin{equation}\label{HuA}
\therefore H_{uA} =0.
\end{equation}
This will happen in the case we are considering, where the covariant components of both the energy-momentum tensor and of the metric are written in terms of positive powers of $ r $, according to the central conditions \eqref{centralT} and \eqref{ccg}, if the metric components are $ C^2 $ functions of $ r $ in a neighborhood of $r = 0$.

Applying the results obtained so far and proceeding in a similar manner as for $ \mu=A $, the component $ \mu=u $ of equation \eqref{div} implies that
\begin{equation}\label{Huu}
H_{uu} =0.
\end{equation}
As a consequence of the HEs and equations \eqref{HAB}, \eqref{HuA} and \eqref{Huu}, we find that
\[ H_{ab}=0. \]
Thus, for a Bondi-Sachs coordinate system in which $ g^{uu} = g^{uB} = 0 $ and $ \det(g_{AB}) =  r^4 \sin[2](\theta)$, the assumption that $ H_{uA} $ and $ H_{uu} $ are defined along $ r=0 $ induces the equivalence
\[ \qq*{HEs}+\qq{GEEs}+\qq{MEEs} \iff \quad H_{ab}=0. \]

\part{Thermodynamic aspects of general relativity}\label{tagr}

\section{Quasi-local Hawking radiation emission}\label{qlhre}

We start analyzing the connection between general relativity and thermodynamics by establishing a notion of temperature and the numerical factor associated with it. To do that, we shall endeavor to demonstrate the emission of Hawking radiation using the formalism so far developed. We first note that, in the original context of black holes, the most important requirement for the emission is an exponential ``peeling'' relation between affine parameters defined on past and future null infinities \cite{Barcelo2011}. Luckily, we already have exactly the same functional form of this relation encoded in equations \eqref{delrlambda}, which can be written as
\begin{equation}\label{peel}
\lambda' = \lambda'_1+ e^{2\beta_1}\int_{r_1}^{r} \exp\left( \int_{r_1}^{r'} \kappa (u_1,r'',\theta_1,\phi_1) \dd{r''}\right) \dd{r'}
\end{equation}
and
\[ r = r_1+ e^{-2\beta_1}\int_{\lambda'_1}^{\lambda'} \exp\left( -\int_{\lambda'_1}^{\lambda''} \kappa (u'_1,\lambda''',\theta'_1,\phi'_1) \dd{\lambda'''}\right) \dd{\lambda''}. \]
Above, we used equation \eqref{kappadelrbeta}, integrated around a point $ (u_1,r_1,\theta_1,\phi_1) $ and employed the notation $ f_1\coloneqq f(u_1,r_1,\theta_1,\phi_1) $, which will be used for any function $ f $ from now on. 

If we use either $ \lambda' $ or $ r $ as the natural coordinate to express the quantized field modes and define a vacuum while the coordinate distance between constant phase surfaces of the modes is kept constant for the other coordinate, then we can proceed as in black hole thermodynamics to demonstrate the emission of Hawking radiation. We shall choose $ \lambda' $ as the natural coordinate, leaving the reasoning that leads to this choice to Subsection \ref{post}.

In order to be able to analyze the evolution of constant phase surfaces, it will be taken into consideration spherically symmetric spacetimes and field modes that propagate transversally to the light cones in the radial direction. In a geometrical optics approximation, the wave vectors will be proportional to the vector field $ {l'}^a $ defined in equation \eqref{lprime}. As noted below this definition, $ {l'}^a $ has the advantage that it maps one light cone into another one, so we can track how the separation between constant phase surfaces evolves between successive light cones.

Pursuing a quasi-local analysis in the spirit of Hypothesis \ref{hypo3} that captures general features of the evolution of the quantized field, we are led to consider how the field evolves between two nearby events, $ \mathcal{Q}_1 $, with coordinates $ (u_1,r_1,\theta_1,\phi_1) $, and $ \mathcal{Q}_2 $, labeled by $ (u_2,r_2,\theta_2,\phi_2) $. If $ r-r_1 $ on one light cone equals $ r-r_2 $ on the successive one, but $ \kappa=0  $ around both events, then equation \eqref{peel} implies a linear relation between $ \lambda'-\lambda'_1 $ and $ \lambda'-\lambda'_2 $, so we lose the exponential ``peeling'' relation. Therefore, the simplest case to analyze happens when $ \kappa\neq 0 $ at one of the events and $ \kappa = 0 $ at the other one, as in the transition from the matter collapsing into a black hole to the exterior region where $ T_{ab}=0 $.

\begin{figure}
	\centering
	\begin{tikzpicture}
	\node[anchor=south west,inner sep=0] (image) at (0,0) {\includegraphics[width=0.5\textwidth]{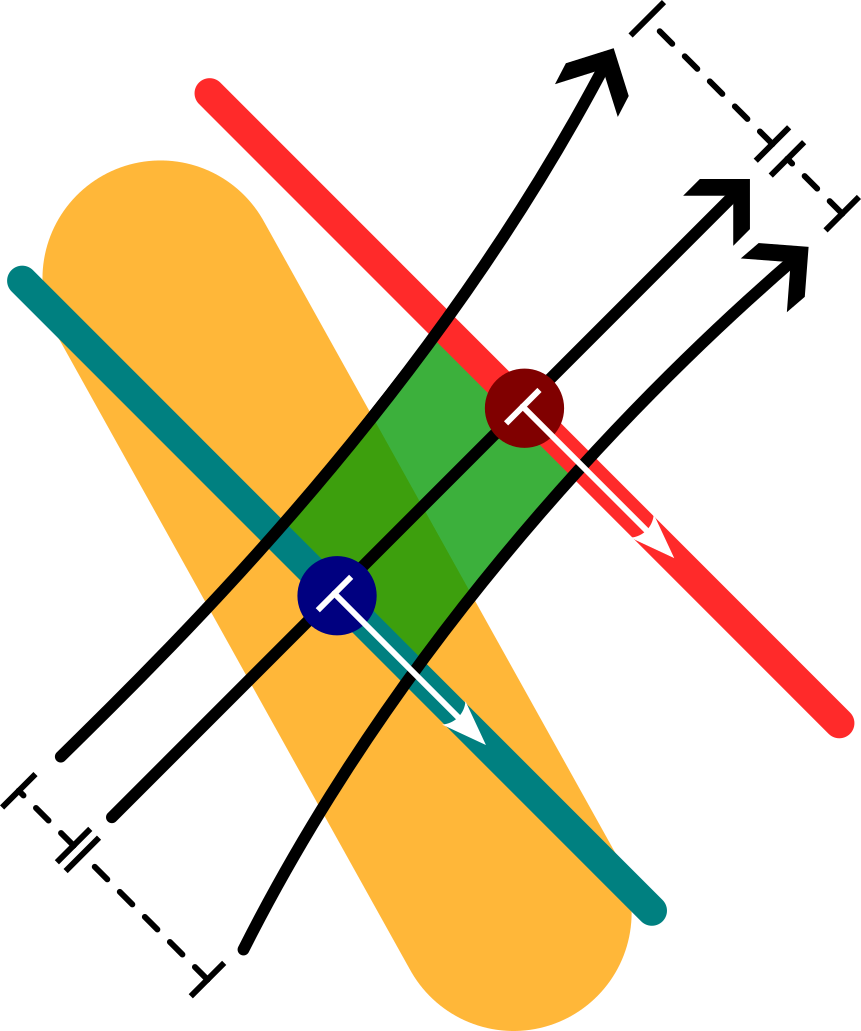}};
	\begin{scope}[x={(image.south east)},y={(image.north west)}]
	\node[anchor=north,text=darkblueink] at (0.39,0.38) {$\mathcal{Q}_1$};
	\node[anchor=south,text=darkredink] at (0.615,0.64) {$\mathcal{Q}_2$};
	\node[anchor=north east,text=orangeink] at (0.6,0) {\textbf{matter}};
	\node[anchor=north west,text=blueink] at (0.77,0.105) {$ u_1 $};
	\node[anchor=north west,text=redink] at (0.99,0.285) {$ u_2 $};
	\node[anchor=north east] at (0.05,0.205) {$ \Delta r_{-} $};
	\node[anchor=north east] at (0.165,0.11) {$ \Delta r_{+} $};
	\node[anchor=south west] at (0.835,0.925) {$ \Delta r_{-} $};
	\node[anchor=south west] at (0.95,0.825) {$ \Delta r_{+} $};
	\node[anchor=north east] at (0.53,0.305) {$ \lambda$};
	\node[anchor=south west] at (0.76,0.48) {$ \tilde{\lambda}$};
	\end{scope}
	\end{tikzpicture}
	\caption[Exponential peeling]{Exponential peeling. A segment of a null generator of the past light cone $ u_1 $ is shown as the blue line. The dark blue dot represents the event $ \mathcal{Q}_1 $. Similarly, the red line and dark red dot are related to the light cone $ u_2 $ and the event $  \mathcal{Q}_2 $, respectively. There is a nonzero flux of matter across $ \mathcal{Q}_1 $, represented by the orange region, while $ T_{ab}=0 $ around $ \mathcal{Q}_2 $. The events $ \mathcal{Q}_1 $ and $ \mathcal{Q}_2 $ are connected by a radial, transversal null geodesic, shown as the central black curve. In case $ {l'}^a $ and $ \partial_{(r)}{}^a $ commute at $ \mathcal{Q}_1 $, the distances measured in $ r $, as $ \Delta r_{+} $ and $ \Delta r_{-} $, between this central curve and its neighboring geodesics will be approximately constant as the geodesics evolve. Then, the $ \lambda' $-distances represented by the white arrows, $ \lambda $ and $ \tilde{\lambda} $ (which are measured from the central geodesic at the cones $ u_1 $ and $ u_2 $, respectively), will be approximately related by an exponential peeling. The green region represents a neighborhood of the central geodesic where the approximations being employed are valid}
	\label{exppeel}
\end{figure}
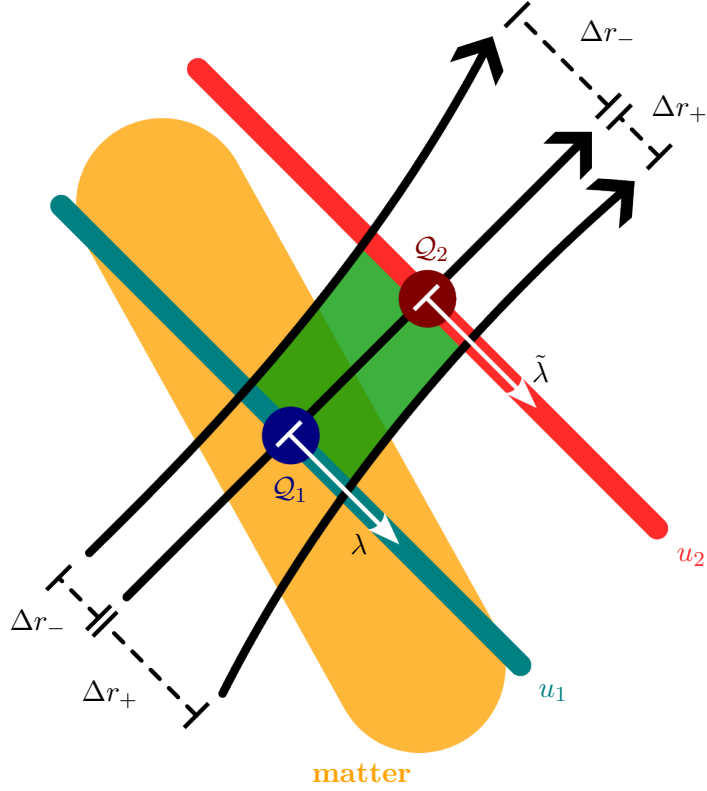

The first cases we shall consider below assume that $ \kappa_1\neq 0 $ and $ \kappa_2=0 $. Besides that, the condition to keep constant the areal distance between constant phase surfaces will be replaced by condition \eqref{commucond}, which implies in the commutation  between $ {l'}^a $ and $ \partial_{(r)}{}^a $. In a sufficiently small neighborhood of an event $ \mathcal{Q}_1 $ where the last condition holds, one can say that the geodesics generated by $ {l'}^a $ preserve their areal separation approximately constant between successive light cones. The general scheme is pictured in Figure \ref{exppeel}.

Then, using the approximation
\begin{equation}\label{kappaapp}
\kappa (u_1,r,\theta_1,\phi_1) \approx \kappa_1
\end{equation}
in equation \eqref{peel}, one gets
\begin{equation}\label{lkr}
\lambda' -\lambda'_1\approx  \frac{e^{2\beta_1}}{\kappa_1}\left[e^{\kappa_1(r-r_1)}-1\right].
\end{equation}
Given that $ \kappa_2=0 $, equation \eqref{peel} calculated at $ u_2 $ yields 
\[ r-r_2= e^{-2\beta_2}(\lambda'-\lambda'_2).\]
Assuming that condition \eqref{commucond} is satisfied, we have that $ r(u_1)-r_1\approx r(u_2)-r_2 $ if $ r(u_1) $ and $ r(u_2) $ are the areal coordinates of the same transversal geodesic at two different times. Defining 
\begin{equation}\label{lltilde}
\lambda\coloneqq\lambda'-\lambda'_1 \qand  \tilde{\lambda}\coloneqq\lambda'-\lambda'_2
\end{equation}
and using the last two equations results in
\begin{equation}\label{lambdas}
\lambda\approx \frac{e^{2\beta_1}}{\kappa_1}\left[e^{\kappa_1 \exp(-2\beta_2)\tilde{\lambda}}-1\right],
\end{equation}
which is the exponential peeling relation between affine distances of two neighboring transversal geodesics at two different light cones.

\subsection{Schwarzschild-de Sitter background}\label{SdS}

Now, we need to find a context where the situation depicted in Figure \ref{exppeel} is realized and the condition \eqref{commucond} is satisfied at $ \mathcal{Q}_1 $. Guided by the coordinate acceleration in equation \eqref{force}, one may expect that condition \eqref{commucond} will be satisfied when the gravitational force changes direction. In a spherically symmetric spacetime, this can occur when the attractive force of a central mass ceases to be dominant and gives place to the repulsive effect of a positive cosmological constant. Therefore, we are led to consider a Schwarzschild-de Sitter background. For the sake of simplicity, we shall perturb this background with a small matter flux around an event $ \mathcal{Q}_1 $ with radius $ r_1 $ given by equation \eqref{r1}, where condition \eqref{commucond} is satisfied. With that, we find a case corresponding to the scheme of Figure \ref{exppeel}.

Then, we shall assume that the characteristic initial-central value problem exposed in Section \ref{EFEs} is well-posed, at least for some classes of initial data. In this case, one may construct a family of spacetimes from a background solution around a light cone with initial data $\left\{ \bar{\gamma}, \bar{\delta}, \bar{T}_{\mu\nu}, \bar{\Lambda} \right\}$ using a parameter $ \tilde{\varepsilon} $ and a perturbation to the initial data proportional to $\left\{ \tilde{\gamma}, \tilde{\delta}, \tilde{T}_{\mu\nu}, \tilde{\Lambda} \right\}$. In other words, one consider as initial data
\[ T_{\mu\nu}= \bar{T}_{\mu\nu}+\tilde{\varepsilon} \tilde{T}_{\mu\nu}, \quad \gamma=\bar{\gamma}+\tilde{\varepsilon}\tilde{\gamma}, \quad \delta=\bar{\delta}+\tilde{\varepsilon}\tilde{\delta} \qand \Lambda=\bar{\Lambda}+\tilde{\varepsilon}\tilde{\Lambda}.\]
In this case, we shall consider that $  \bar{\gamma}= \bar{\delta}=\tilde{\gamma}= \tilde{\delta}=\tilde{\Lambda}=0  $ and that both $ \bar{T}_{\mu\nu}$ and $ \tilde{T}_{\mu\nu}  $ are spherically symmetric. By plugging this initial data in the hierarchical scheme and by expanding the hyperbolic functions with respect to $ \tilde{\varepsilon} $, one finds that all functions of the metric can be written as $ f = \bar{f} + \mathcal{O}(\tilde{\varepsilon}) $, where $ \bar{f} $ is the value of the function in the background solution.

With the spacetime setting defined, we turn our attention to a minimally coupled scalar field of zero mass \cite{Birrell1982}, $ \psi $, which satisfies the wave equation
\[ \Box\psi= \left(-g'\right)^{-1/2} \partial_{\mu'}\left(\sqrt{-g'} g^{\mu' \nu'}\partial_{\nu'}\psi\right)=0. \]
In a general spherically symmetric spacetime, the inverse metric is given by equation \eqref{invgprime}, and the determinant is found above it. Then, it follows that the last equation can be written as
\begin{multline}\label{fieldeq}
\partial_{\lambda'} (r\partial_{u'} \psi) = -\left(\partial_{u'}r+g^{\lambda'\lambda'}\partial_{\lambda'}r  +\frac{r\partial_{\lambda'}g^{\lambda'\lambda'}}{2} 
\right) \partial_{\lambda'}\psi -\frac{r g^{\lambda'\lambda'}}{2}\partial_{\lambda'\lambda'}\psi-\\
\frac{\partial_{\theta'}\left(\sin(\theta')\partial_{\theta'}\psi\right)}{2r\sin(\theta')}-\frac{\partial_{\phi'\phi'}\psi}{2r\sin[2](\theta')}.
\end{multline}

If $ \psi $ is a differentiable function of $ \tilde{\varepsilon} $, then it can be expressed as
$ \psi=\bar{\psi}+\mathcal{O}(\tilde{\varepsilon})  $. Using this expansion and the one for the functions of the metric in the previous equation yields the following condition on the zeroth-order term of $ \psi $:
\begin{multline}
\partial_{\lambda'} (\bar{r}\partial_{u'} \bar{\psi}) = -\left(\partial_{u'}\bar{r}+\bar{g}^{\lambda'\lambda'}\partial_{\lambda'}\bar{r}  +\frac{\bar{r}\partial_{\lambda'}\bar{g}^{\lambda'\lambda'}}{2} 
\right) \partial_{\lambda'}\bar{\psi} -\frac{\bar{r} \bar{g}^{\lambda'\lambda'}}{2}\partial_{\lambda'\lambda'}\bar{\psi}-\\
\frac{\partial_{\theta'}\left(\sin(\theta')\partial_{\theta'}\bar{\psi}\right)}{2\bar{r}\sin(\theta')}-\frac{\partial_{\phi'\phi'}\bar{\psi}}{2\bar{r}\sin[2](\theta')}.
\end{multline}
Since we are considering a Schwarzschild-de Sitter background, the background metric is static and $ \partial_{u'}\bar{r}=0 $. Then, we can use the separation of variables 
\[\bar{\psi}(u',\lambda',\theta',\phi')=\bar{F}(u')\bar{X}(\lambda')\bar{Y}(\theta')\bar{Z}(\phi') \]
to solve the above equation. Introducing the separation constants $ l,\ m $ and $ \omega $, one finds that the angular dependence is given by spherical harmonics,
\[ \bar{Y}(\theta')\bar{Z}(\phi')=Y_{lm}(\theta',\phi'), \]
the time dependent function is
\[ \bar{F}(u')=e^{-i\omega u'}, \]
and $ \bar{X}(\lambda')$ satisfies
\begin{equation}\label{Xbar}
\partial_{\lambda'\lambda'} \bar{X} + \left(\frac{2e^{-2\bar{\beta}_R}}{\bar{r}} +\frac{\partial_{\lambda'}\bar{g}^{\lambda'\lambda'}}{\bar{g}^{\lambda'\lambda'}}-\frac{2i\omega}{\bar{g}^{\lambda'\lambda'}}\right) \partial_{\lambda'} \bar{X} - \left[\frac{2i\omega e^{-2\bar{\beta}_R}}{\bar{r} \bar{g}^{\lambda'\lambda'}}+\frac{l(l+1)}{\bar{r}^2 \bar{g}^{\lambda'\lambda'}}\right]\bar{X}=0,
\end{equation}
where equations \eqref{delrlambda} and \eqref{betaR} were used. It is worth noting that, even though the perturbed energy-momentum tensor is nonzero at $ \mathcal{Q}_1 $, we are analyzing $ \psi $ in the exterior region of the background metric, where $ \bar{T}_{ab}=0 $.

To solve the last equation, we shall perform a geometrical optics approximation around $ \mathcal{Q}_1 $, where the field frequency is high enough so that the coefficients in the equation can be considered as constants as the field oscillates many times around $ \mathcal{Q}_1 $. For this, we introduce another parameter, $ \bar{\varepsilon} $, which modulates the scale of a new coordinate. If we zoom in enough on $ \mathcal{Q}_1 $, then the coefficients of equation \eqref{Xbar} can be treated as slowly varying functions, and the corresponding perturbation method can be applied \cite{Shivamoggi2003}.

Thus, let
\[ \bar{s} \coloneqq \frac{\lambda'-\lambda'_1}{\bar{\varepsilon}} .\]
It follows that, for any function $ f(\lambda'(\bar{s})) $,
\[ \partial_{\bar{s}}f = \bar{\varepsilon}\partial_{\lambda'}f,\]
so we can write
\[ f(\bar{s};\bar{\varepsilon})= f_1+\mathcal{O}(\bar{\varepsilon}) .\]
We expand each coefficient in equation \eqref{Xbar} as above and suppose that $ \bar{X} $ can be written as a formal power series\footnote{Strictly speaking, the concomitant use of Big-O notation and formal power series is an abuse of notation. Here we adopt standard practice and hope that the series will represent a good approximation of some yet unspecified order.} of the form
\[ \bar{X}(\bar{s};\bar{\varepsilon})=\sum_{n=0}^{\infty} \bar{\varepsilon}^n \bar{X}_n(\lambda'(\bar{s})).\]

However, with this procedure, the only term left of the lowest order in $ \bar{\varepsilon} $ is $ \partial_{\bar{s}\bar{s}} \bar{X}_0 $, leading to an unwanted linear dependence on $ \lambda' $. To contour this problem, we shall tie $ \bar{\varepsilon} $ to $ \omega $ (a trick that will be shown to be unnecessary in the next subsection). Let $ K $ be the radius of curvature inside which the coefficients of equation \eqref{Xbar} can be approximated by constants. Then, one expects that this geometric optics approximation will provide reasonable results for modes of the field with a high frequency, such that their wavelengths, $ 2\pi \omega^{-1} $, are much smaller than $ K $. Therefore, defining
\[ \bar{\varepsilon} \coloneqq \frac{1}{\omega K} \]
allows us to substitute $ \omega $ by $ (\bar{\varepsilon} K)^{-1} $ in equation \eqref{Xbar} and obtain
\[ \partial_{\bar{s}\bar{s}} \bar{X}_0 = \frac{2i}{K\bar{g}^{\lambda'\lambda'}_1} \partial_{\bar{s}} \bar{X}_0
 \]
to lowest order in $ \bar{\varepsilon} $, which is solved by
\[ \bar{X}_0 = e^{2i\bar{s}/K\bar{g}^{\lambda'\lambda'}_1} = e^{2i(\lambda'-\lambda'_1)/\bar{\varepsilon}K\bar{g}^{\lambda'\lambda'}_1} = e^{2i\omega(\lambda'-\lambda'_1)/\bar{g}^{\lambda'\lambda'}_1}.\]

As a result, one finds a family of solutions of the form
\[ \psi_{\omega lm}(u',\lambda,\theta',\phi';\bar{\varepsilon},\tilde{\varepsilon})= N_\omega Y_{lm}(\theta',\phi')e^{-i\omega u'} e^{i 2 \omega \lambda/\bar{g}^{\lambda'\lambda'}_1} +\mathcal{O}\left(\bar{\varepsilon}\right) + \mathcal{O}\left(\tilde{\varepsilon}\right) , \]
where $ N_\omega $ is a normalization constant and definition \eqref{lltilde} was used. In this work, we shall not tackle the issue of defining a suitable normalization procedure for the modes $ \psi_{\omega lm} $. Then, we assume that the above family can be used to define the quantum field operator as
\[ \hat{\Psi}\coloneqq\sum_{l,m}\int_{0}^{\infty}\hat{a}_{\omega l m} \psi_{\omega lm} + \hat{a}_{\omega l m}^\dagger \psi_{\omega lm}^* \dd{\omega},\]
where $ \hat{a}_{\omega l m} $ and $ \hat{a}_{\omega l m}^\dagger  $ are annihilation and creation operators satisfying canonical commutation relations. One could do exactly the same procedure around $ \mathcal{Q}_2 $, obtaining another basis for the space of solutions with elements of the form
\[ \tilde{\psi}_{\tilde{\omega }\tilde{l}\tilde{m}}(u',\tilde{\lambda},\theta',\phi';\bar{\varepsilon},\tilde{\varepsilon})= N_{\tilde{\omega }}Y_{\tilde{l}\tilde{m}}(\theta',\phi')e^{-i\tilde{\omega }u'} e^{i 2 \tilde{\omega} \tilde{\lambda} / \bar{g}^{\lambda'\lambda'}_2} +\mathcal{O}\left(\bar{\varepsilon}\right) + \mathcal{O}\left(\tilde{\varepsilon}\right) \]
and a corresponding representation for the field operator,
\[ \hat{\tilde{\Psi}}\coloneqq\sum_{\tilde{l},\tilde{m}}\int_{0}^{\infty}\hat{b}_{\tilde{\omega }\tilde{l}\tilde{m}} \tilde{\psi}_{\tilde{\omega }\tilde{l}\tilde{m}} + \hat{b}_{\tilde{\omega }\tilde{l}\tilde{m}}^\dagger \tilde{\psi}_{\tilde{\omega }\tilde{l}\tilde{m}}^* \dd{\tilde{\omega}}.\]

Following the discussion at the beginning of this section, the affine parameters $ \lambda $ and $\tilde{\lambda}$ on the same constant phase surface will be related by approximation \eqref{lambdas}, since condition \eqref{commucond} is satisfied at $ \mathcal{Q}_1 $. Therefore, the evolution of a mode $ \psi_{\omega lm} $ between the light cones $ u_1 $ and $ u_2 $ will be characterized by
\[ e^{i2\omega\lambda/\bar{g}^{\lambda'\lambda'}_1} \xmapsto{\hspace{1.5em}} e^{i2\omega \exp(2\beta_1)\left[\exp(\kappa_1 \exp(-2\beta_2)\tilde{\lambda})-1\right]/\kappa_1 \bar{g}^{\lambda'\lambda'}_1}. \]
By defining
\[ \chi\coloneqq \frac{2\omega e^{2\beta_1}}{\kappa_1 \bar{g}^{\lambda'\lambda'}_1} \qand \tilde{\kappa}\coloneqq \frac{\kappa_1}{e^{2\beta_2}}, \]
one has that the mode evolves as
\begin{equation}\label{psievol}
\psi_{\omega lm} \xmapsto{\hspace{1.5em}} N_\omega Y_{lm}(\theta',\phi') e^{-i\omega u' }e^{-i\chi }e^{i\chi \exp(\tilde{\kappa}\tilde{\lambda})}+\mathcal{O}\left(\bar{\varepsilon}\right) + \mathcal{O}\left(\tilde{\varepsilon}\right).
\end{equation}

\begin{figure}
	\centering
	\begin{tikzpicture}
	\node[anchor=south west,inner sep=0] (image) at (0,0) {\includegraphics[width=0.5\textwidth]{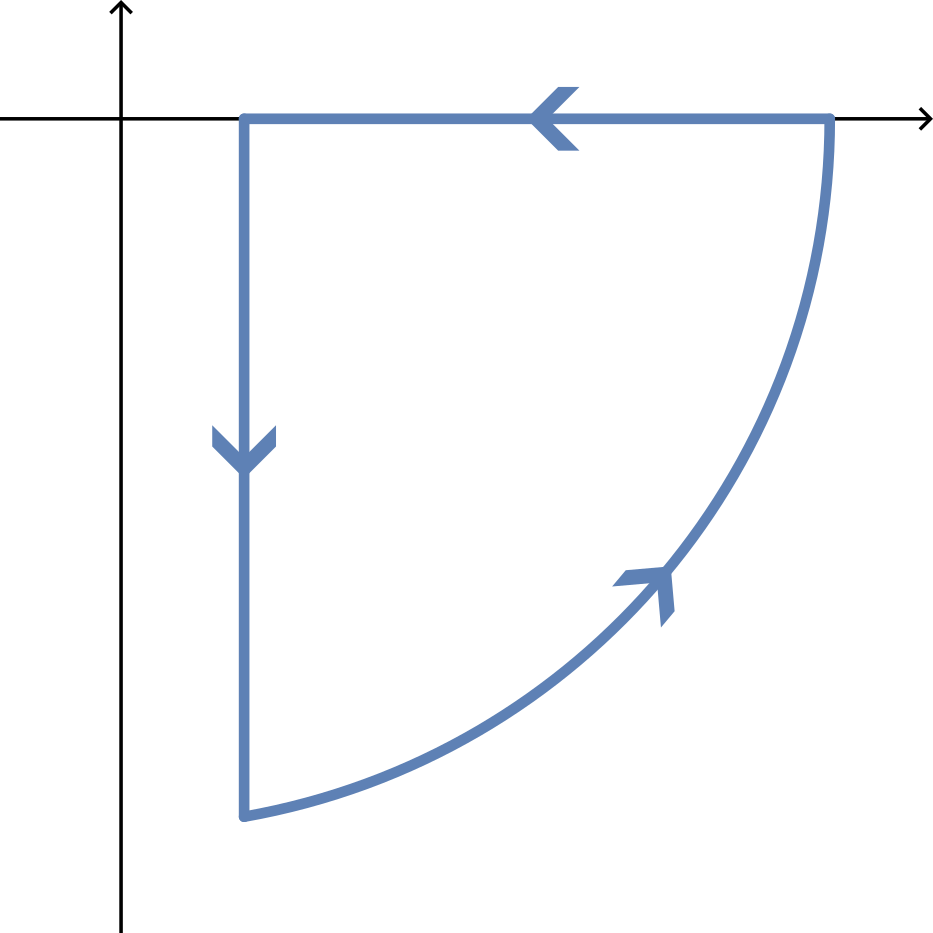}};
	\begin{scope}[x={(image.south east)},y={(image.north west)}]
	\node[anchor=east,text=mathematicablue] at (0.24,0.51) {$P_1$};
	\node[anchor=south,text=mathematicablue] at (0.585,0.9) {$P_3$};
	\node[anchor=north west,text=mathematicablue] at (0.715,0.385) {$P_2$};
	\node[anchor=south] at (0.265,0.875) {$z_{\text{min}}$};			
	\node[anchor=south] at (0.89,0.875) {$z_{\text{max}}$};
	\node[anchor=east] at (0.125,1) {$\operatorname{Im}(z)$};
	\node[anchor=north] at (1,0.875) {$\operatorname{Re}(z)$};
	\end{scope}
	\end{tikzpicture}
	\caption[Closed curve for contour integration]{Closed curve for contour integration. A holomorphic function of $ z $ is integrated along the closed curve $ P=P_1\cup P_2 \cup P_3 $ shown in blue. The line segment $ P_1 $ starts at $z_{\text{min}}\in \mathbb{R} > 0 $ and follows in the direction of $z_{\text{min}} -\infty i $ until $ \abs{z}=z_{\text{max}} $. The circular arc $ P_2 $ keeps $ \abs{z} $ constant, ending at $ z_{\text{max}} $. Lastly, $ P_3 $ goes from $ z_{\text{max}} $ to $ z_{\text{min}} $ along the real line}
	\label{loop}
\end{figure}
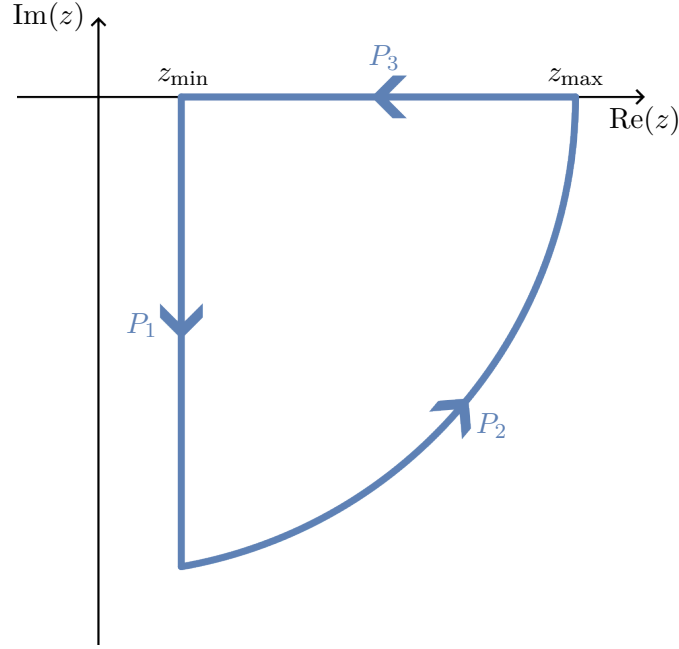

In order to write the mode $ \psi_{\omega lm} $ in terms of the basis $ \left\{\tilde{\psi}_{\tilde{\omega }\tilde{l}\tilde{m}},\tilde{\psi}^*_{\tilde{\omega }\tilde{l}\tilde{m}}\right\} $, we use the Fourier transform and its inverse to express\footnote{The following calculations of this subsection and Figure \ref{loop} were adapted from the unpublished ``Lecture notes'' on quantum field theory in curved spacetimes by D. A. T. Vanzella.}
\[ e^{-i\chi }e^{i\chi \exp(\tilde{\kappa}\tilde{\lambda})}=\frac{1}{2\pi}\int_{-\infty}^{\infty} e^{i\omega'\tilde{\lambda}} \int_{-\infty}^{\infty}e^{-i\omega'z'} e^{-i\chi }e^{i\chi \exp(\tilde{\kappa}z')}\dd{z'}\dd{\omega'}.\]
Then, by defining
\[ z\coloneqq-i\chi e^{\tilde{\kappa}z'},  \]
one obtains
\[ e^{-i\chi }e^{i\chi \exp(\tilde{\kappa}\tilde{\lambda})}=\frac{e^{-i\chi }}{2\pi\tilde{\kappa}}\int_{-\infty}^{\infty} e^{i\omega'\tilde{\lambda}}(-i\chi)^{i\omega'/\tilde{\kappa}} \int_{0}^{-\infty i} z^{-1-i\omega'/\tilde{\kappa}} e^{-z}\dd{z}\dd{\omega'}.\]
The last integrand is holomorphic except along $ \left\{z\in \mathbb{R} \mid z\leq 0 \right\} $, as can be seen by writing the complex exponentiation in terms of the principal branch of the complex logarithm function. Therefore, Cauchy's integral theorem implies that its integral along the loop shown in Figure \ref{loop} equals zero. By taking the limit $ z_{\text{max}}\to \infty $, the integral along the segment $ P_2 $ vanishes (which can be seen using the estimation lemma). Then, one can take the limit $z_{\text{min}}\to 0 $ and find that the inner integral in the above equation equals the gamma function,\footnote{The precise statement is: the gamma function is the analytic continuation of Euler's integral to the punctured imaginary axis $ i\mathbb{R}\setminus \{0\} $.} so that
\[ e^{-i\chi }e^{i\chi \exp(\tilde{\kappa}\tilde{\lambda})}=\frac{e^{-i\chi }}{2\pi\tilde{\kappa}}\int_{-\infty}^{\infty} e^{i\omega'\tilde{\lambda}}e^{\pi\omega'/2\tilde{\kappa}}\chi^{i\omega'/\tilde{\kappa}} \Gamma\left(-\frac{i\omega'}{\tilde{\kappa}}\right)\dd{\omega'}.\]

We split the above integral in positive and negative frequency parts,
\[ e^{-i\chi }e^{i\chi \exp(\tilde{\kappa}\tilde{\lambda})}=\frac{e^{-i\chi }}{2\pi\tilde{\kappa}}\int_{0}^{\infty} e^{i\omega'\tilde{\lambda}}e^{\pi\omega'/2\tilde{\kappa}}\chi^{i\omega'/\tilde{\kappa}} \Gamma\left(-\frac{i\omega'}{\tilde{\kappa}}\right)+e^{-i\omega'\tilde{\lambda}}e^{-\pi\omega'/2\tilde{\kappa}}\chi^{-i\omega'/\tilde{\kappa}} \Gamma\left(\frac{i\omega'}{\tilde{\kappa}}\right)\dd{\omega'},\]
and define a new integration variable, $ \tilde{\omega}\coloneqq \bar{g}^{\lambda'\lambda'}_2 \omega'/2 $. Then, by using the relation
\[ Y^*_{lm}(\theta',\phi') = e^{-2im\phi'}Y_{lm}(\theta',\phi'), \]
which follows from equations (3.54), (3.53) and (3.51) of Reference \citeonline{Jackson1998}, one finds that equation \eqref{psievol} becomes
\begin{multline}
\psi_{\omega lm} \xmapsto{\hspace{1.5em}}\\ \sum_{\tilde{l},\tilde{m}}\delta_{l\tilde{l}}\delta_{m\tilde{m}}\int_{0}^{\infty}\frac{N_\omega}{N_{\tilde{\omega}}} e^{-iu'(\omega -\tilde{\omega}) }\frac{e^{-i\chi }}{\pi\tilde{\kappa} \bar{g}^{\lambda'\lambda'}_2} e^{\pi\tilde{\omega}/\tilde{\kappa}\bar{g}^{\lambda'\lambda'}_2}\chi^{2i\tilde{\omega}/\tilde{\kappa}\bar{g}^{\lambda'\lambda'}_2} \Gamma\left(-\frac{2i\tilde{\omega}}{\tilde{\kappa}\bar{g}^{\lambda'\lambda'}_2}\right)N_{\tilde{\omega }}Y_{\tilde{l}\tilde{m}}e^{-i\tilde{\omega }u'} e^{i 2 \tilde{\omega} \tilde{\lambda} / \bar{g}^{\lambda'\lambda'}_2}+\\
\frac{N_\omega}{N_{\tilde{\omega}}^*}e^{2im\phi'} e^{-iu'(\omega +\tilde{\omega}) }\frac{e^{-i\chi }}{\pi\tilde{\kappa} \bar{g}^{\lambda'\lambda'}_2} e^{-\pi\tilde{\omega}/\tilde{\kappa}\bar{g}^{\lambda'\lambda'}_2}\chi^{-2i\tilde{\omega}/\tilde{\kappa}\bar{g}^{\lambda'\lambda'}_2} \Gamma\left(\frac{2i\tilde{\omega}}{\tilde{\kappa}\bar{g}^{\lambda'\lambda'}_2}\right)N_{\tilde{\omega }}^*Y_{\tilde{l}\tilde{m}}^*e^{i\tilde{\omega }u'} e^{-i 2 \tilde{\omega} \tilde{\lambda} / \bar{g}^{\lambda'\lambda'}_2}\dd{\tilde{\omega}} + \\
\mathcal{O}\left(\bar{\varepsilon}\right) + \mathcal{O}\left(\tilde{\varepsilon}\right),
\end{multline}
where $ \delta_{l\tilde{l}}$ and $\delta_{m\tilde{m}} $ are Kronecker deltas. Next, we treat the terms $ \mathcal{O}\left(\bar{\varepsilon}\right) $ and $ \mathcal{O}\left(\tilde{\varepsilon}\right) $ as constants with respect to $ \tilde{\omega} $, and include them in the integral above as integrands multiplied by a function of $ \tilde{\omega} $ whose integral from $ 0 $ to $ \infty $ equals $ 1 $, like $ e^{-\tilde{\omega}} $. For $ Y_{\tilde{l}\tilde{m}}\neq 0 $, $ \tilde{\psi}_{\tilde{\omega }\tilde{l}\tilde{m}} $ has an inverse,
\[ \tilde{\psi}_{\tilde{\omega }\tilde{l}\tilde{m}}{}^{-1}= \left(N_{\tilde{\omega }}Y_{\tilde{l}\tilde{m}}e^{-i\tilde{\omega }u'} e^{i 2 \tilde{\omega} \tilde{\lambda} / \bar{g}^{\lambda'\lambda'}_2}\right)^{-1} +\mathcal{O}\left(\bar{\varepsilon}\right) + \mathcal{O}\left(\tilde{\varepsilon}\right), \]
and we can multiply the positive frequency part of the integrand above by $ \tilde{\psi}_{\tilde{\omega }\tilde{l}\tilde{m}}{}^{-1}\tilde{\psi}_{\tilde{\omega }\tilde{l}\tilde{m}} $ and the negative one by $ \tilde{\psi}^*_{\tilde{\omega }\tilde{l}\tilde{m}}{}^{-1}\tilde{\psi}^*_{\tilde{\omega }\tilde{l}\tilde{m}} $, obtaining
\begin{multline}
\psi_{\omega lm} \xmapsto{\hspace{1.5em}}\\ \sum_{\tilde{l},\tilde{m}}\int_{0}^{\infty}\left[\frac{\delta_{l\tilde{l}}\delta_{m\tilde{m}}N_\omega e^{\pi\tilde{\omega}/\tilde{\kappa}\bar{g}^{\lambda'\lambda'}_2}}{N_{\tilde{\omega}}\pi\tilde{\kappa} \bar{g}^{\lambda'\lambda'}_2} \Gamma\left(-\frac{2i\tilde{\omega}}{\tilde{\kappa}\bar{g}^{\lambda'\lambda'}_2}\right) e^{-i\left[\chi +u'(\omega -\tilde{\omega})\right] } \chi^{2i\tilde{\omega}/\tilde{\kappa}\bar{g}^{\lambda'\lambda'}_2}  + \mathcal{O}\left(\bar{\varepsilon}\right) + \mathcal{O}\left(\tilde{\varepsilon}\right) \right] \tilde{\psi}_{\tilde{\omega }\tilde{l}\tilde{m}} + \\
\left[\frac{\delta_{l\tilde{l}}\delta_{m\tilde{m}}N_\omega e^{-\pi\tilde{\omega}/\tilde{\kappa}\bar{g}^{\lambda'\lambda'}_2}}{N^*_{\tilde{\omega}}\pi\tilde{\kappa} \bar{g}^{\lambda'\lambda'}_2} \Gamma\left(\frac{2i\tilde{\omega}}{\tilde{\kappa}\bar{g}^{\lambda'\lambda'}_2}\right) e^{-i\left[\chi +u'(\omega +\tilde{\omega})-2m\phi'\right] } \chi^{-2i\tilde{\omega}/\tilde{\kappa}\bar{g}^{\lambda'\lambda'}_2}  + \mathcal{O}\left(\bar{\varepsilon}\right) + \mathcal{O}\left(\tilde{\varepsilon}\right) \right] \tilde{\psi}^*_{\tilde{\omega }\tilde{l}\tilde{m}} 
\dd{\tilde{\omega}}.
\end{multline}

From the last result, one can read the Bogoliubov coefficient relating the modes $ \psi_{\omega lm} $ and $ \tilde{\psi}^*_{\tilde{\omega }\tilde{l}\tilde{m}} $, whose modulus square is
\[ \abs{\beta_{\omega l m\tilde{\omega}\tilde{l}\tilde{m}}}^2= \delta_{l\tilde{l}}\delta_{m\tilde{m}}\abs{\frac{N_\omega}{N_{\tilde{\omega}}^*}}^2\frac{e^{-2\pi\tilde{\omega}/\tilde{\kappa}\bar{g}^{\lambda'\lambda'}_2}}{\left(\pi\tilde{\kappa} \bar{g}^{\lambda'\lambda'}_2\right)^2}  \abs{\Gamma\left(\frac{2i\tilde{\omega}}{\tilde{\kappa}\bar{g}^{\lambda'\lambda'}_2}\right)}^2+ \mathcal{O}\left(\bar{\varepsilon}\right) + \mathcal{O}\left(\tilde{\varepsilon}\right) .\]
Then, by using equation (8.332.1) of Reference \citeonline{Gradshteyn2007}, one finds that
\[ \abs{\beta_{\omega l m\tilde{\omega}\tilde{l}\tilde{m}}}^2= \frac{\delta_{l\tilde{l}}\delta_{m\tilde{m}}\abs{N_\omega}^2}{\pi\tilde{\kappa} \bar{g}^{\lambda'\lambda'}_2 \tilde{\omega}\abs{N_{\tilde{\omega}}^*}^2} \frac{1}{e^{4\pi\tilde{\omega}/\tilde{\kappa}\bar{g}^{\lambda'\lambda'}_2}-1}+ \mathcal{O}\left(\bar{\varepsilon}\right) + \mathcal{O}\left(\tilde{\varepsilon}\right),\]
where one recognizes a Bose-Einstein distribution in the leading-order term with temperature
\begin{equation*}
T=\bar{g}^{\lambda'\lambda'}_2\frac{\tilde{\kappa}}{4\pi}=\frac{\bar{g}^{\lambda'\lambda'}_2}{e^{2\beta_2}} \frac{\kappa_1}{4\pi}. 
\end{equation*}
Therefore, if one has, at $ u_1 $, the vacuum state, $ \ket{0} $, associated with the representation $ \{\psi_{\omega lm},\psi_{\omega lm}^*\} $, such that $ \hat{a}_{\omega l m}\ket{0}=0, \ \forall \ \omega,\ l,\ m, $ then it will evolve to a state with expected number of particles in the mode $ \tilde{\psi}_{\tilde{\omega }\tilde{l}\tilde{m}} $ given by \cite{Hawking1975,Birrell1982}
\begin{equation}\label{expnum}
 \bra{0}\hat{b}_{\tilde{\omega }\tilde{l}\tilde{m}}^\dagger \hat{b}_{\tilde{\omega }\tilde{l}\tilde{m}} \ket{0} = \sum_{l,m}\int_{0}^{\infty}\abs{\beta_{\omega l m\tilde{\omega}\tilde{l}\tilde{m}}}^2 \dd{\omega} = \frac{\int_{0}^{\infty}\abs{N_\omega}^2\dd{\omega}}{4\pi^2 T } \frac{\tilde{\omega}^{-1}\abs{N_{\tilde{\omega}}^*}^{-2}}{e^{\tilde{\omega}/T}-1}+ \mathcal{O}\left(\bar{\varepsilon}\right) + \mathcal{O}\left(\tilde{\varepsilon}\right). 
\end{equation}
In other words, the initial vacuum state becomes, to leading order, a thermal radiation with a well-defined temperature and graybody factor \cite{Parker2009} $ \tilde{\omega}^{-1}\abs{N_{\tilde{\omega}}^*}^{-2} $.

\subsection{Formalizing the method}

The previous derivation was inspired by the original strategy \cite{Hawking1975,Barcelo2011} for demonstrating the emission of Hawking radiation: (i) solve the equation for the scalar field to be quantized in two regions of a background spacetime, (ii) construct a representation for the quantum field operator in each one of them, (iii) make a geometrical optics approximation to relate the bases of each representation using lightlike geodesics, and (iv) find a locus where the coordinate distance between these geodesics in one region is related to the coordinate distance of the same geodesics in the other region by an exponential approximation. We chose to present the former analysis because it is closer to the original formulation, being pedagogical in this sense, and it renders the case analyzed simpler.

However, one can improve the method by noting that the geometrical optics approximation could have been used for any coordinate, not only $ \lambda' $. This would allow us to solve the scalar field equation in any region of a sufficiently smooth spacetime. Moreover, the approximation \eqref{kappaapp} carries a term of $\mathcal{O}(\tilde{\varepsilon})$,
\[ \kappa_1=\tilde{\varepsilon}\left(4\pi r\tilde{T}_{rr}\right), \]
to the temperature $ T $, which can be avoided in a more general geometrical optics approximation. As a result, the steps (i), (iii), and the approximations involved in (iv) can be condensed into a single approximation.

Intuitively, we are assuming that the elements of the relevant basis for the solution space of the wave equation, in any spacetime, look like plane waves locally. Even the low-frequency solutions. They may not entirely fit inside the region where the geometrical optics approximation is valid, but one may consider the ``partial mode'' that fits. Following Hypothesis \ref{hypo3}, the measurements of an ideal detector would be influenced by the variations of the scalar field in the immediate vicinity of the point of observation, regardless of if one consider entire modes or not. But, since we are dealing only with spherically symmetric spacetimes, we shall leave the angular dependence in its exact form, and keep the spherical harmonics in the solutions.

Thus, let
\begin{equation}\label{ts}
t\coloneqq  \frac{u'-u'_1 }{\varepsilon}\qand s\coloneqq \frac{\lambda'-\lambda'_1}{\varepsilon}. 
\end{equation}
Then, for any function $ f(u'(t),\lambda'(s)) $,
\[\partial_t f = \varepsilon\partial_{u'} f  \qand \partial_s f = \varepsilon\partial_{\lambda'} f ,\]
implying that
\[ f(t,s;\varepsilon)=f_1+\mathcal{O}\left(\varepsilon\right).\]
Now, return to equation \eqref{fieldeq}, write each coefficient as above, and consider its solutions as formal power series of the form 
\begin{equation}\label{psiformal}
\psi(t,s,\theta',\phi';\varepsilon)=Y(\theta')Z(\phi')\sum_{n=0}^{\infty} \varepsilon^n \psi_n(u'(t),\lambda'(s)).	
\end{equation}
After multiplying this equation by $ \varepsilon^2(r_1YZ)^{-1} $, one finds that the resulting zeroth-order terms are
\[ \partial_{st}\psi_0+\frac{g^{\lambda'\lambda'}_1}{2}\partial_{ss}\psi_0=0. \]

We separate variables once again and assume that
\[ \psi_0(t,s) = F(t)X(s),  \]
obtaining
\[ \frac{\partial_tF}{F}=-\frac{g^{\lambda'\lambda'}_1}{2}\frac{\partial_{ss}X}{\partial_{s}X}\eqqcolon-i\varepsilon\omega. \]
As a result, one has solutions of the form
\[ F= e^{-i\omega u'} \qand X= e^{i2\omega(\lambda'-\lambda'_1)/g^{\lambda'\lambda'}_1}. \]
Applying separation of variables to equation \eqref{fieldeq} without expanding any of its terms with respect to $ \varepsilon $ leads to spherical harmonics:
\[ Y(\theta')Z(\phi')=Y_{lm}(\theta',\phi'). \]
Consequently, one finds a family of solutions of the form
\[ \psi_{\omega lm}(u',\lambda,\theta',\phi';\varepsilon)= N_\omega Y_{lm}(\theta',\phi')e^{-i\omega u'} e^{i 2 \omega\lambda/g^{\lambda'\lambda'}_1}+\mathcal{O}\left(\varepsilon\right). \]

From the last result, one can define the wave vector
\[ k_{\omega lm}^{\mu'}\coloneqq -ig^{\mu'\nu'}\nabla_{\nu'} \ln \psi_{\omega lm} .\]
By factoring the zeroth-order term of $ \psi_{\omega lm} $, expanding the logarithm, and using equation \eqref{lprimeprime}, one finds that
\[ k_{\omega lm}^{{\mu'}} = \frac{2\omega}{g^{{\lambda'\lambda'}}}{l'}^{{\mu'}}-i\left(0,0,g^{{\theta'\theta'}}\partial_{{\theta'}} \ln Y_{lm},g^{{\phi'\phi'}}\partial_{{\phi'}} \ln Y_{lm}\right)+ \mathcal{O}\left(\varepsilon\right).\]
Therefore, the only radial transverse modes left in the limit $ \varepsilon\to 0 $ will be the ones with $ l=m=0 $, so that the derivatives of the spherical harmonics vanish. It becomes clear then that our analysis is restricted to a subspace of the space of solutions to the wave equation, which is spanned by the modes $ \psi_{\omega}\coloneqq \psi_{\omega 00} $ with wave vector
\[ k_{\omega}^{{\mu'}} \coloneqq k_{\omega 00}^{{\mu'}} = \frac{2\omega}{g^{\lambda'\lambda'}}{l'}^{\mu'}+ \mathcal{O}\left(\varepsilon\right).\]
This restriction is analogous to the division in the original calculation \cite{Hawking1975} between modes that enter the black hole, modes that are scattered by the collapsing matter, and modes that travel through the collapsing matter and follow to the future null infinity.

The last equation implies that all modes $\psi_{\omega}$ follow lightlike geodesics to zeroth order. As a consequence, one can exchange the terms $ \mathcal{O}\left(\bar{\varepsilon}\right) $ and $ \mathcal{O}\left(\tilde{\varepsilon}\right) $ by $ \mathcal{O}\left(\varepsilon\right) $ and the background metric by the spacetime metric in the calculations of the previous subsection to obtain similar results for these modes. By assuming that $ \beta_{\omega l m\tilde{\omega}\tilde{0}\tilde{0}} $ is proportional to $ \delta_{l\tilde{0}}\delta_{m\tilde{0}} $, equation \eqref{expnum} becomes valid for $ \tilde{l}=\tilde{m}=0 $, and we recover the same expression for the zeroth-order term with a temperature
\begin{equation}\label{T}
T=\frac{g^{\lambda'\lambda'}_2}{e^{2\beta_2}} \frac{\kappa_1}{4\pi}. 
\end{equation}

We conclude that the result of the previous subsection could have been demonstrated in a fully dynamical spacetime, with no reference to a background or to modes of high frequency. The only difference would be a small change in the position of the event $ \mathcal{Q}_1 $ as an effect of the perturbation $ \tilde{\varepsilon} \tilde{T}_{\mu\nu} $.

The geometrical optics approximation allows us to solve the wave equation, which enables the construction of a quantum field operator. Together with condition \eqref{commucond}, it also implies in an exponential relation between the variables that define each one of the families of solutions in consideration. As a result, one finds that the subspace of radial transverse modes thermalizes an initial vacuum state. We also note that the infinite redshift caused by the event horizon, which justifies the geometrical optics approximation in the original derivation, becomes unnecessary in the present context, when one considers the limit where the events $ \mathcal{Q}_1 $ and $ \mathcal{Q}_2 $ get arbitrarily close.

\subsection{Spherical collapse and black hole formation}\label{scbhf}

From the last subsection, one sees that the requirement for Hawking radiation emission in a spherically symmetric spacetime is the existence of a point where condition \eqref{commucond} is satisfied and the component $ T_{rr} $ differs from zero, but quickly vanishes along future-directed radial transverse lightlike geodesics. As a concrete example where we shall compute the metric explicitly, we take the spherical collapse of matter that precedes the formation of a black hole, bridging the gap between the present formalism and the results that inspired it.

For simplicity, we consider a spacetime filled with dust up to some radius $ R $, from which one has $ T_{\mu\nu} =0$. Besides, we assume that $ \Lambda=0 $ and that the observer stays at the center of symmetry. As shown in the second subsection of Appendix \ref{SSS}, this spacetime is characterized by the initial values of the proper energy density, $ \mu(u,r) $, and of the observed redshift, $ z(u,r) $. We assume that the initial profile of $ \mu(u,r) $ on $ u=u_1 $ is given by
\[ \mu(u_1,r)= 
\begin{cases}
\mu_0 & \qif* r< 1,       \\
100 \mu_0(1.1-r)^2 & \qif* 1 \leq r < 1.1,   \\
0 & \text{otherwise},
\end{cases} \]
where $ \mu_0 $ is a positive constant. We also assume that $ z(u_1,r)=z_0 $ is initially constant. Even though this is not in conformity with the initial-value constraint \eqref{ivcmuz}, one may consider that $ \mu(u_1,r) $ is an approximation of a distribution that is constant outside a small neighborhood of $ r=0 $ and decreases very fast towards it, vanishing there. Therefore, we are assuming that the behavior of $ \mu(u_1,r) $ near $ r=0 $ will not have a strong effect on the fields near the surface of the matter distribution, which is the region we are interested in.

The values of $ \mu_0 $ and $ z_0 $ shall be determined so that the fluid be in a collapse process and its outermost layers be about to cross the event horizon. As a sign of the formation and of the approximate localization of the latter, we shall take the emergence of a canonical dynamical horizon \cite{Ashtekar2003}. Such a horizon is foliated by closed two-dimensional surfaces, which, by symmetry, correspond to spheres of constant radius $ r $. Each one of these leaves is orthogonal to a pair (aside from possible rescalings) of future-directed lightlike vector fields, such that the expansion scalar of one of them is null, while the other one is strictly negative. For any sphere of constant radius $ r $ in this spacetime, these normal fields are given by $ - \partial_{(r)}{}^a $ and $ {l'}^a $. By construction, $ - \partial_{(r)}{}^a $ has a negative expansion scalar in the coordinate neighborhood under consideration. For this reason, one must search for events where $ {l'}^{a} $ points to regions that keep $ r $ constant, that is, where $ {l'}^{a} \propto \partial_{(u)}{}^a$ and
\begin{equation}\label{condi}
e^{2 \Phi }=0. 
\end{equation} 
This same condition can be obtained by calculating the expansion scalar of $ {l'}^{a} $. From equations \eqref{x}, \eqref{xidef}, and \eqref{Theta}, it follows that $ \Xi_{(l)}=e^{2(\Phi-\beta)}/r $. Then, definition \eqref{lprime} and the rescaling rules in Table 1 of Reference \citeonline{Gourgoulhon2006} (where $ \theta' $ is the function of interest) imply that
\[ \Xi_{(l')}=e^{2\beta}\Xi_{(l)}=\frac{e^{2\Phi}}{r} . \]
One notes that $ \Phi $ will not be defined in the regions where $ e^{2 \Phi }\leq 0  $. However, the function $ e^{2 \Phi } $ still can be obtained through the hierarchical solution, being, in this sense, more fundamental than $ \Phi $.

We want to model the case in which the horizon forms close to $ r=1 $, with the quadratic decay of $ \mu(u_1,r) $ representing the last layers necessary for the formation of the horizon and contributing with a negligible portion of the mass of the black hole. Therefore, the value of $ \mu_0 $ must be slightly smaller than the critical density $ \mu_c $ for which $ e^{2 \Phi(u_1,1) }=0  $. Thus, we use $ \mu(u_1,r)=\mu_0  $ in equation \eqref{phisol} to calculate $ e^{2 \Phi } $ for an arbitrary radius $ r $, and, by introducing the variables
\[ x\coloneqq 2\sqrt{\pi \mu_0}(1+z_0)r \qand \zeta\coloneqq 1 + 4 z_0 + 2 z_0^2 ,\]
condition \eqref{condi} becomes
\begin{equation}\label{condix}
\frac{x\sqrt{1-x^2}+\zeta\arcsin(x)}{2x(1+z_0)^2}=0. 
\end{equation}

In principle, one considers that $ 0\leq x$, since $ \mu_0\geq0 $ and inequality \eqref{vu} holds. In addition, $ x\leq 1  $, since equation \eqref{betasol} implies that $ e^{-2 \beta(u_1,x) }=\sqrt{1-x^2}  $ for $ \mu(u_1,r)=\mu_0  $ and since $ \partial_{\lambda'}r = e^{-2\beta}  $, indicating that the coordinate system $ z^\mu $ is not well defined for $ x>1 $. In fact, the Raychaudhuri equation \cite[eq. 6.9]{Gourgoulhon2006} implies that $ \partial_{\lambda'}\Theta'\leq 0 $ if the fluid satisfies the null energy condition \cite{Poisson2004}, where $ \Theta' $ is the expansion scalar associated with $ \partial_{(\lambda')}{}^a $.
Given that $ \partial_{(\lambda')}{}^a=e^{-2\beta}\partial_{(r)}{}^a $, one can see from the rescaling rules of Reference \citeonline{Gourgoulhon2006} that $ \Theta' $ vanishes in the limit $ x\to 1 $. Therefore, its subsequent values will be non-positive, and the area of the spheres that foliate the light cone cannot increase anymore. In other words, the coordinate $ r $ ceases to be strictly increasing in the direction of $ \partial_{(\lambda')}{}^a $.

However, the above condition is not satisfied in the limit $ x\to 0 $, which can be evaluated using l'Hôpital's rule. The case $ x=1 $ is satisfied only for $ \zeta=0 $, leading to $ {l'}^{a} =\partial_{(u)}{}^a$. But, for $ \zeta=0 $, one can use equations \eqref{gsphe}, \eqref{betasol}, and \eqref{phisol} with $ \mu(u_1,r)=\mu_0  $ to find that $ g_{uu}=-0.5 (1 + z_0)^{-2} $. As a consequence, $ {l'}^{a} $ ceases to be lightlike and to define a dynamical horizon in the pathological limit $ x\to 1 $. This leads to the restriction $ 0<x<1 $ for the location of this horizon.

\begin{figure}
	\centering
	\begin{tikzpicture}
	\node[anchor=south west,inner sep=0] (image) at (0,0) {\includegraphics[width=0.6\textwidth]{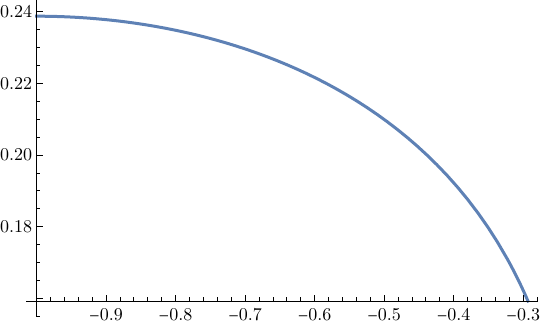}};
	\begin{scope}[x={(image.south east)},y={(image.north west)}]
	\node[anchor=east] at (0,0.53) {$ \mu_c $};
	\node[anchor=north] at (0.53,0) {$ z_0 $};
	\end{scope}
	\end{tikzpicture}
	\caption[$ \mu_c $ as a function of $ z_0 $]{$ \mu_c $ as a function of $ z_0 $}
	\label{muc}
\end{figure}

Defining the new variable
\[ y\coloneqq 2 \arcsin(x),  \]
one concludes that condition \eqref{condi} is equivalent to Kepler's equation:
\[  y+\zeta^{-1} \sin (y)=0. \]
In the interval $ 0<y<\pi $, this equation will be satisfied only if $ \zeta<0 $. Consequently, $ -1<z_0<-1+0.5\sqrt{2} $. Besides, $ \zeta=2(1+z_0)^2-1>-1 $ coincides with the lower bound imposed by the last equation. Solving equation \eqref{condix} using the function \textit{Solve} of the software \textit{Wolfram Mathematica}, one obtains $ \mu_c $ as a function of $ z_0 $ by fixing $ r=1 $, which is shown in Figure \ref{muc}. In the allowed interval of $ z_0 $, the range of $ \mu_c $ is
\[ 0.159\approx\mu_{\text{min}}<\mu_c<\mu_{\text{max}}\approx0.239 .\]

To completely fix the spacetime, we chose the values $ z_0=-0.7 $ and $ \mu_0 = 0.2044555 < \mu_c(z_0=-0.7)\approx 0.230	$. The value of $ \mu_0 $ was chosen after successive trials and determined by the interval $ \Delta u $ for which $ T_{rr}(u_1,r_1)+\partial_u T_{rr}(u_1,r_1) \Delta u =0$, where $ r_1 $ is defined by the equation $ \partial_r e^{2 \Phi(u_1,r_1) }=0 $. For this value of $ \Delta u $, one can say that, to first order in $ u $, all the fluid has already crossed the surface of radius $ r_1 $. In case the canonical dynamical horizon is found in a sphere of radius $ r\geq r_1 $ at the light cone $ u_2\coloneqq u_1+\Delta u $, there will no longer be a matter flux through the horizon, and it will stop expanding, reaching a stable maximum radius \cite{Ashtekar2003}. By choosing the specific value of $ \mu_0 $ given above, the horizon stabilizes, with a sufficiently high degree of precision, at the radius $ r_1 $. This degree of precision is given by the radial variation $ \Delta r $, also calculated to first order in $ u $, of a lightlike geodesic tangent to $ {l'}^a $ at $ \mathcal{Q}_1$ during the interval $ \Delta u $. For $ {l'}^\mu = \dv*{z^\mu}{w} $, one has that $ \Delta z^\mu \approx {l'}^\mu \Delta w $, and equation \eqref{lprime} implies that $ \Delta u\approx\Delta w $ and $ \Delta r \approx e^{2\Phi_1}\Delta u/2$.

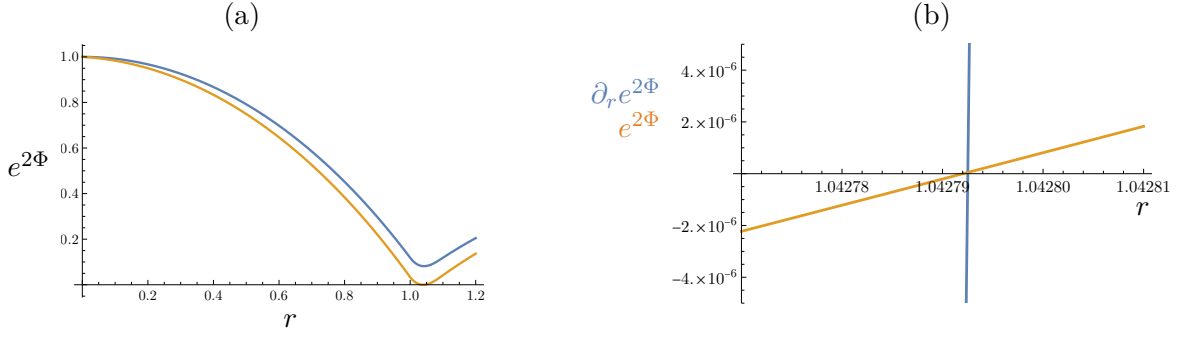
\begin{figure}
	\centering
	\begin{minipage}[t]{.45\textwidth}
		\vspace{0pt} 
		\centering
		\begin{tikzpicture}
		\node at (0.5,1) {(a)};
		\end{tikzpicture}
	\end{minipage}%
	\hfill
	\begin{minipage}[t]{.45\textwidth}
		\vspace{0pt} 
		\centering
		\begin{tikzpicture}
		\node at (0.5,1) {(b)};
		\end{tikzpicture}
	\end{minipage}%

	\begin{minipage}[t]{.45\textwidth}
		\vspace{0pt} 
		\centering
	\begin{tikzpicture}
	\node[anchor=south west,inner sep=0] (image) at (0,0) {\includegraphics[height=0.15\textheight]{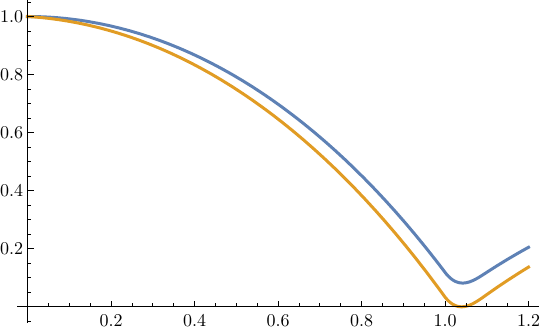}};
	\begin{scope}[x={(image.south east)},y={(image.north west)}]
	\node[anchor=east] at (0,0.54) {$ e^{2 \Phi } $};
	\node[anchor=north] at (0.54,0) {$ r $};
	\end{scope}
	\end{tikzpicture}
	\end{minipage}%
\hfill
	\begin{minipage}[t]{.45\textwidth}
		\vspace{0pt}
		\centering
		\begin{tikzpicture}
		\node[anchor=south west,inner sep=0] (image) at (0,0) {\includegraphics[height=0.15\textheight]{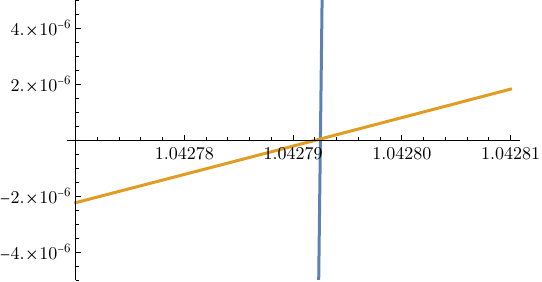}};
		\begin{scope}[x={(image.south east)},y={(image.north west)}]
		\node[anchor=east,text=mathematicablue,overlay] at (0,0.81) {$ \partial_r e^{2 \Phi } $};
		\node[anchor=east,text=mathematicaorange,overlay] at (0,0.69) {$ e^{2 \Phi } $};
		\node[anchor=north] at (0.94,0.43) {$ r $};
		\end{scope}
		\end{tikzpicture}
	\end{minipage}
		\hspace{0pt}
	\caption[Evolution of $ e^{2 \Phi } $]{Evolution of $ e^{2 \Phi } $. \textbf{(a)} $ e^{2 \Phi(u_1,r) } $ is represented in blue. In orange, one has the first order approximation: $ e^{2 \Phi(u_2,r) }\approx e^{2 \Phi(u_1,r) }+\partial_u e^{2 \Phi(u_1,r)} \Delta u $. For computational convenience, $ \partial_u e^{2 \Phi(u_1,r)}  $ was approximated by $ A r $, for $ r\leq 1 $, and by $ -A r + 2A $, for $ r>1 $, with $ A=\partial_u e^{2 \Phi(u_1,1) }\approx- 13.5057$. The maximum relative error in a total of 12 equally spaced sampling points of $ \partial_u e^{2 \Phi(u_1,r)}  $ in the interval of interest was approximately $ 0.04 $. \textbf{(b)} In blue, the function $ \partial_r e^{2 \Phi(u_1,r) } $ identifies the minimum of $ e^{2 \Phi(u_1,r) } $ at $ r_1\approx1.042792 $. In orange, the outermost zero of the approximation for $ e^{2 \Phi(u_2,r) } $ localizes the canonical dynamical horizon near to $ r_1 $}
	\label{e2Phi}
\end{figure}

From the values given above and the results shown in the second subsection of Appendix \ref{SSS}, it was calculated $ r_1\approx1.042792 $, $ \Delta u \approx 0.006 $ and $ \Delta r \approx 0.0003 $. The evolution of the function $ e^{2 \Phi } $ is shown in Figure \ref{e2Phi} (a), where it is possible to note the global minimum at $ r_1 $ and the existence of zeros at $ u_2 $, which indicates the presence of a canonical dynamical horizon. In Figure \ref{e2Phi} (b), a vicinity of $ r_1 $ is magnified, showing its value with greater precision at the point where condition \eqref{commucond} is satisfied. One can also verify that the canonical dynamical horizon, where condition \eqref{condi} is satisfied, is found very close to the surface of radius $ r_1 $ at the cone $ u_2 $. It is possible to see that the lightlike geodesic tangent to  $ {l'}^a $ at $ \mathcal{Q}_1 $ covers a distance $ \Delta r $ much larger than the separation between the horizon and $ r_1 $. We conclude that, for small errors in this first order approximation, this geodesic will stay outside the black hole and follow to future null infinity, reaching far away observers.

\begin{figure}
	\centering
	\hspace{-5pt}
	\begin{minipage}[t]{.45\textwidth}
		\vspace{0pt} 
		\centering
		\begin{tikzpicture}
		\node at (0.5,1) {(a)};
		\end{tikzpicture}
	\end{minipage}%
	\hfill
	\begin{minipage}[t]{.45\textwidth}
		\vspace{0pt} 
		\centering
		\begin{tikzpicture}
		\node at (0.5,1) {(b)};
		\end{tikzpicture}
	\end{minipage}%

	\begin{minipage}[t]{.45\textwidth}
		\vspace{0pt} 
		\centering
		\begin{tikzpicture}
		\node[anchor=south west,inner sep=0] (image) at (0,0) {\includegraphics[height=0.15\textheight]{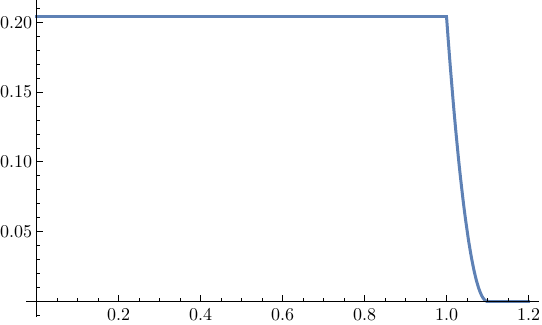}};
		\begin{scope}[x={(image.south east)},y={(image.north west)}]
		\node[anchor=east] at (0,0.52) {$ \mu $};
		\node[anchor=north] at (0.52,0) {$ r $};
		\end{scope}
		\end{tikzpicture}
	\end{minipage}%
\hfill
	\begin{minipage}[t]{.45\textwidth}
		\vspace{0pt}
		\centering
		\begin{tikzpicture}
		\node[anchor=south west,inner sep=0] (image) at (0,0) {\includegraphics[height=0.15\textheight]{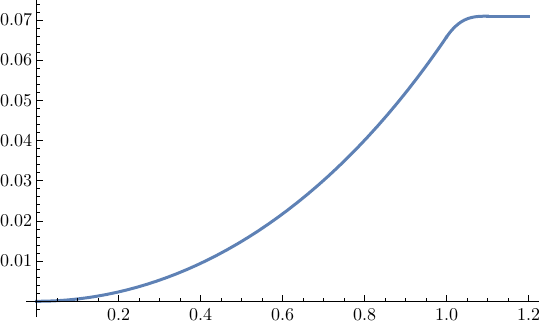}};
		\begin{scope}[x={(image.south east)},y={(image.north west)}]
		\node[anchor=east] at (0,0.52) {$ \beta $};
		\node[anchor=north] at (0.52,0) {$ r $};
		\end{scope}
		\end{tikzpicture}
	\end{minipage}

\vspace{10pt}

	\begin{minipage}[t]{.45\textwidth}
	\vspace{0pt} 
	\centering
	\begin{tikzpicture}
	\node at (0.5,1) {(c)};
	\end{tikzpicture}
\end{minipage}%
\hfill
\begin{minipage}[t]{.45\textwidth}
	\vspace{0pt} 
	\centering
	\begin{tikzpicture}
	\node at (0.5,1) {(d)};
	\end{tikzpicture}
\end{minipage}%

	\begin{minipage}[t]{.45\textwidth}
	\vspace{0pt} 
	\centering
	\begin{tikzpicture}
	\node[anchor=south west,inner sep=0] (image) at (0,0) {\includegraphics[height=0.15\textheight]{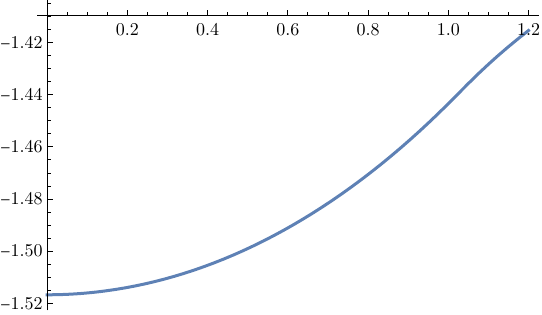}};
	\begin{scope}[x={(image.south east)},y={(image.north west)}]
	\node[anchor=east] at (0,0.46) {$ v^r $};
	\node[anchor=north] at (0.53,.88) {$ r $};
	\end{scope}
	\end{tikzpicture}
\end{minipage}%
\hfill
\begin{minipage}[t]{.45\textwidth}
	\vspace{0pt}
	\centering
	\begin{tikzpicture}
	\node[anchor=south west,inner sep=0] (image) at (0,0) {\includegraphics[height=0.15\textheight]{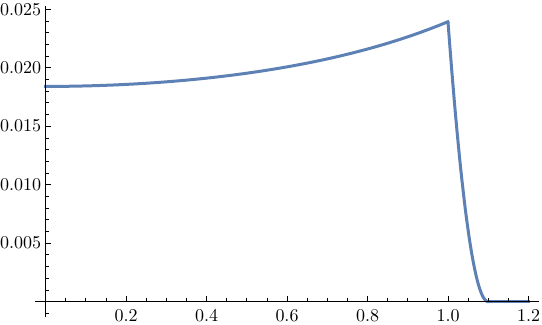}};
	\begin{scope}[x={(image.south east)},y={(image.north west)}]
	\node[anchor=east] at (0,0.53) {$ T_{rr}$};
	\node[anchor=north] at (0.53,0) {$ r $};
	\end{scope}
	\end{tikzpicture}
\end{minipage}

\vspace{10pt}

	\begin{minipage}[t]{.45\textwidth}
	\vspace{0pt} 
	\centering
	\begin{tikzpicture}
	\node at (0.5,1) {(e)};
	\end{tikzpicture}
\end{minipage}%
\hfill
\begin{minipage}[t]{.45\textwidth}
	\vspace{0pt} 
	\centering
	\begin{tikzpicture}
	\node at (0.5,1) {(f)};
	\end{tikzpicture}
\end{minipage}%

	\begin{minipage}[t]{.45\textwidth}
	\vspace{0pt} 
	\centering
	\begin{tikzpicture}
	\node[anchor=south west,inner sep=0] (image) at (0,0) {\includegraphics[height=0.15\textheight]{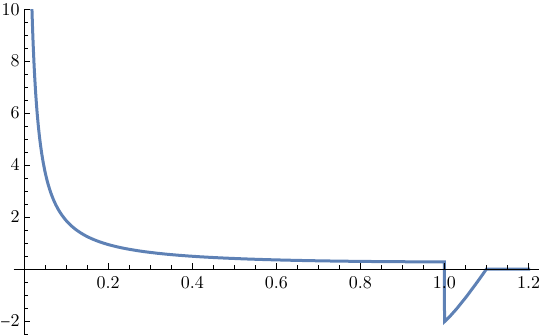}};
	\begin{scope}[x={(image.south east)},y={(image.north west)}]
	\node[anchor=east] at (0,0.57) {$ \partial_u T_{rr}$};
	\node[anchor=north] at (0.51,0.14) {$ r $};
	\end{scope}
	\end{tikzpicture}
\end{minipage}%
\hfill
\begin{minipage}[t]{.45\textwidth}
	\vspace{0pt}
	\centering
	\begin{tikzpicture}
	\node[anchor=south west,inner sep=0] (image) at (0,0) {\includegraphics[height=0.15\textheight]{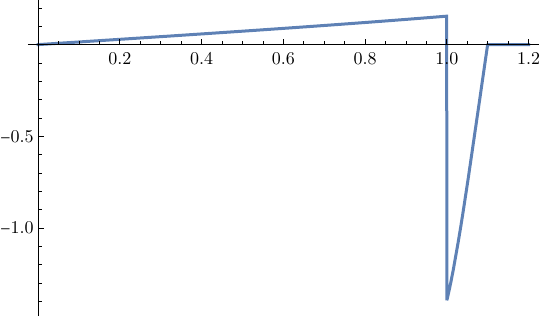}};
	\begin{scope}[x={(image.south east)},y={(image.north west)}]
	\node[anchor=east] at (0,0.42) {$ \partial_u \mu $};
	\node[anchor=north] at (0.52,0.79) {$ r $};
	\end{scope}
	\end{tikzpicture}
\end{minipage}
\hfill

\vspace{10pt}

	\begin{minipage}[t]{.45\textwidth}
	\vspace{0pt} 
	\centering
	\begin{tikzpicture}
	\node at (0.5,1) {(g)};
	\end{tikzpicture}
\end{minipage}

	\begin{minipage}[t]{.45\textwidth}
	\vspace{0pt} 
	\centering
	\begin{tikzpicture}
	\node[anchor=south west,inner sep=0] (image) at (0,0) {\includegraphics[height=0.15\textheight]{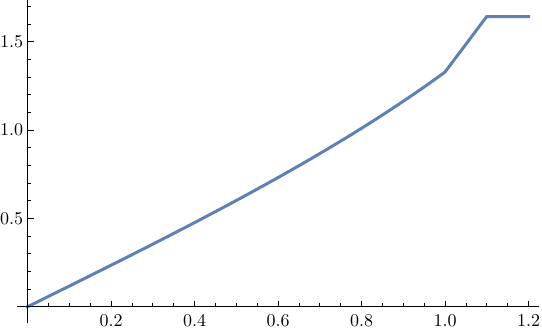}};
	\begin{scope}[x={(image.south east)},y={(image.north west)}]
	\node[anchor=east] at (0,0.54) {$ \partial_u \beta $};
	\node[anchor=north] at (0.52,0) {$ r $};
	\end{scope}
	\end{tikzpicture}
\end{minipage}
	\caption[Functions describing the collapse]{Functions describing the collapse. It is shown the radial dependence of other relevant functions computed for $ z_0=-0.7 $ and $ \mu_0=0.2044555 $ at $ u=u_1 $. In (e), one can see a divergence as $ r\to 0 $, which is due to the violation of the initial-value constraint \eqref{ivcmuz} and the terms proportional to $ r^{-1} $ in equation \eqref{deluTrrsphe}}
	\label{other}
\end{figure}
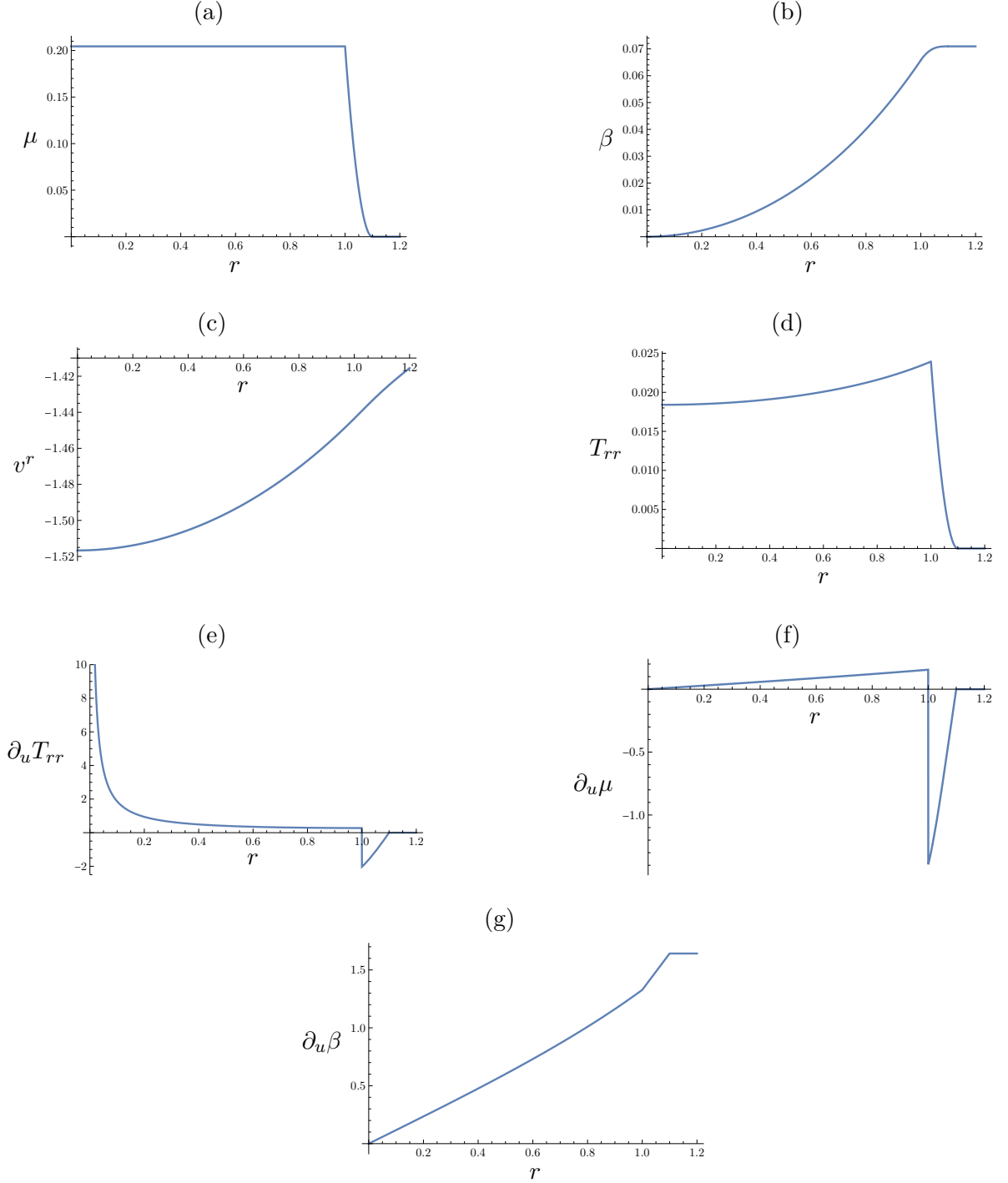

For the sake of transparency, other relevant functions calculated for this spacetime are shown in Figure \ref{other}. Especially, one can see from Figure \ref{other} (c) that $ v^r $ is negative, meaning that this model indeed describes a collapse process. Moreover, Figures \ref{other} (e) and (f) indicate that all matter will cross the horizon shortly after $ u_1 $.

In sum, we have found an event, $ \mathcal{Q}_1 $, where condition \eqref{commucond} is satisfied and $ T_{rr}\neq 0 $. Moreover, $ T_{rr} = 0 $ at a point $ \mathcal{Q}_2 $ on $ u_2 $ with radius $ r_2 \approx r_1+\Delta r $, which can be reached from $ \mathcal{Q}_1 $ along a radial transverse lightlike geodesic. As a consequence, the results from the last subsection apply to this example, and we can say that if one has a vacuum state initially, then the radial transverse modes of the quantum field will evolve to a thermal state with a temperature given by equation \eqref{T}. Assuming that the emitted radiation does not change its thermal character while propagating through regions devoid of matter, it will reach future null infinity with a non-vanishing temperature. Therefore, one recovers the original Hawking radiation from a quasi-local analysis, at least in a qualitative manner.

From equations \eqref{kappadelrbeta}, \eqref{betasol}, \eqref{Tursol}, and \eqref{M} one perceives that the relation between $ \kappa_1 $ and the mass of the black hole is likely very complicated. Even for a constant proper energy density, the dependence of the Schwarzschild radius on $ \mu_0 $ and $ z_0 $ is given implicitly by the transcendental equation \eqref{condix}. For this reason, we shall resort to an extremely simplified model of collapse in order to compare the temperature observed at future null infinity with a more quantitative result.

Let $ z(u_1,r)=z_0 $ and the proper energy density profile be described by a bump function with an almost constant plateau and a very steep, but sufficiently smooth, lateral decay. In practice, we shall consider an approximation where $ \mu(u_1,r)=\mu_0 $ for $ r_{\text{in}}\leq r \leq r_{\text{out}} $, vanishing outside this interval. We shall demand that the integrand of equation \eqref{phisol} be less than 1 for some radius in the interval above so that $ e^{2\Phi(u_1,r)}<1 $. First we note that, in the coordinate neighborhood, one must have $ 1-4\pi\mu_0(1+z_0)^2(r^2-r_{\text{in}}^2)>0 $ in order for equations \eqref{betasol} and \eqref{phisol} to be well-defined. In case $ 1-4\pi\mu_0 r^2>0 $, the above condition on the integrand implies that
\[ r<(8\pi\mu_0)^{-1/2}\sqrt{2-(1+z_0)^2 + \sqrt{[2-(1+z_0)^2]^2+16\pi\mu_0(1+z_0)^2r_{\text{in}}^2}}. \]
Then, the requirement that $ r_{\text{in}}<r $ leads to $ r_{\text{in}}<(2\pi\mu_0)^{-1/2} $, which is automatically satisfied in this case. Given that the case $ 1-4\pi\mu_0 r^2\leq 0 $ already implies that the integrand in consideration is less than 1, there will exist some subinterval of $ (r_{\text{in}},r_{\text{out}}) $ for which $ e^{2\Phi(u_1,r)}<1 $.

Since $ e^{2\Phi(u_1,r)}=1 $ for $ r\leq r_{\text{in}} $, $\partial_r e^{2\Phi(u_1,r)} $ must be negative at some point inside the matter distribution. In addition, the $ r $-derivative of equation \eqref{e2phi} is strictly positive for $ r\geq r_{\text{out}} $. Therefore, if $ e^{2\Phi(u_1,r)} $ is a $ C^1 $ function of $ r $, then condition \eqref{commucond} will be satisfied at some point, $ \mathcal{Q}_1 $, inside the fluid. For a sufficiently small $ d\coloneqq r_{\text{out}}-r_{\text{in}} $, the radial transverse lightlike geodesic starting at this point will leave the fluid sufficiently fast, and the previous analysis will hold for this example too.

We suppose that $ d/r_{\text{out}}\ll 1 $, so that the integral in equation \eqref{M} can be approximated by its integrand evaluated at $ r_{\text{out}} $ times $ d $. For a shell that has just crossed the canonical dynamical horizon, condition \eqref{condi} will be satisfied at $ r_{\text{out}} $, and equation \eqref{e2phi} leads to the usual relation for the Schwarzschild radius, $ r_{\text{out}} =2M $. Using this relation, equation \eqref{Tur}, and condition \eqref{condi} in the approximation for $ M $ yields
\[ \mu_0\approx\frac{1}{8\pi M d}. \]
From equations \eqref{Trr}, \eqref{beta}, \eqref{kappadelrbeta}, and \eqref{T}, it follows that
\begin{equation}\label{TH}
T\approx (1+z_0)^2 g^{\lambda'\lambda'}_2 e^{2(2\beta_1-\beta_2)} \frac{r_1}{d}\frac{1}{8\pi M},
\end{equation}
which is proportional to the well-known result $ T_{\mathscr{I}^+}=(8\pi M )^{-1}$ \cite{Hawking1975}.

Assuming that the approximation for $\mu_0 $ holds if one shifts the matter distribution a little to the outside of the horizon so that the radial transverse lightlike geodesic starting at $ \mathcal{Q}_1 $ can reach future null infinity, one obtains a result that is quite similar to the original Hawking radiation. We interpret the proportionality coefficient as a consequence of this specific choice of model and of different redshift effects related to different observers. It is not clear whether an equality between both results could be achieved.

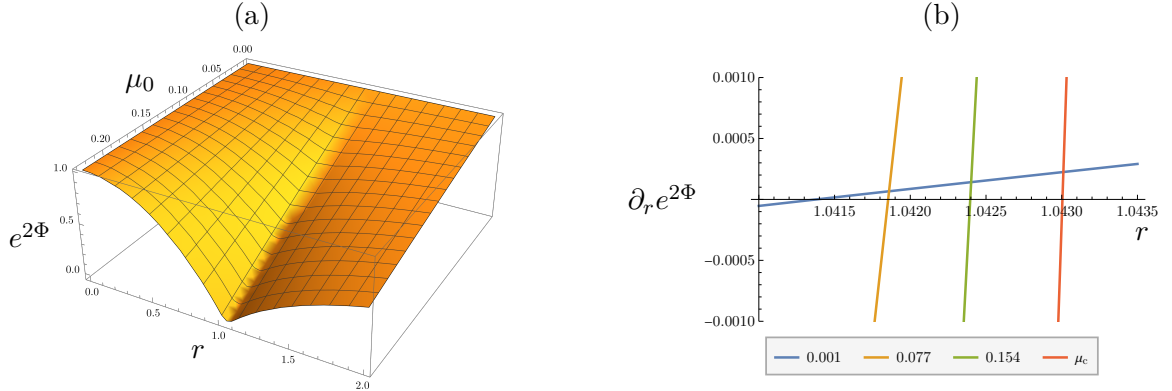
\begin{figure}
	\centering
	\begin{minipage}[t]{.45\textwidth}
		\vspace{0pt} 
		\centering
		\begin{tikzpicture}
		\node at (0.5,1) {(a)};
		\end{tikzpicture}
	\end{minipage}%
	\hfill
	\begin{minipage}[t]{.45\textwidth}
		\vspace{0pt} 
		\centering
		\begin{tikzpicture}
		\node at (0.5,1) {(b)};
		\end{tikzpicture}
	\end{minipage}%

	\begin{minipage}[t]{.45\textwidth}
		\vspace{0pt} 
		\centering
		\begin{tikzpicture}
		\node[anchor=south west,inner sep=0] (image) at (0,0) {\includegraphics[height=0.2\textheight]{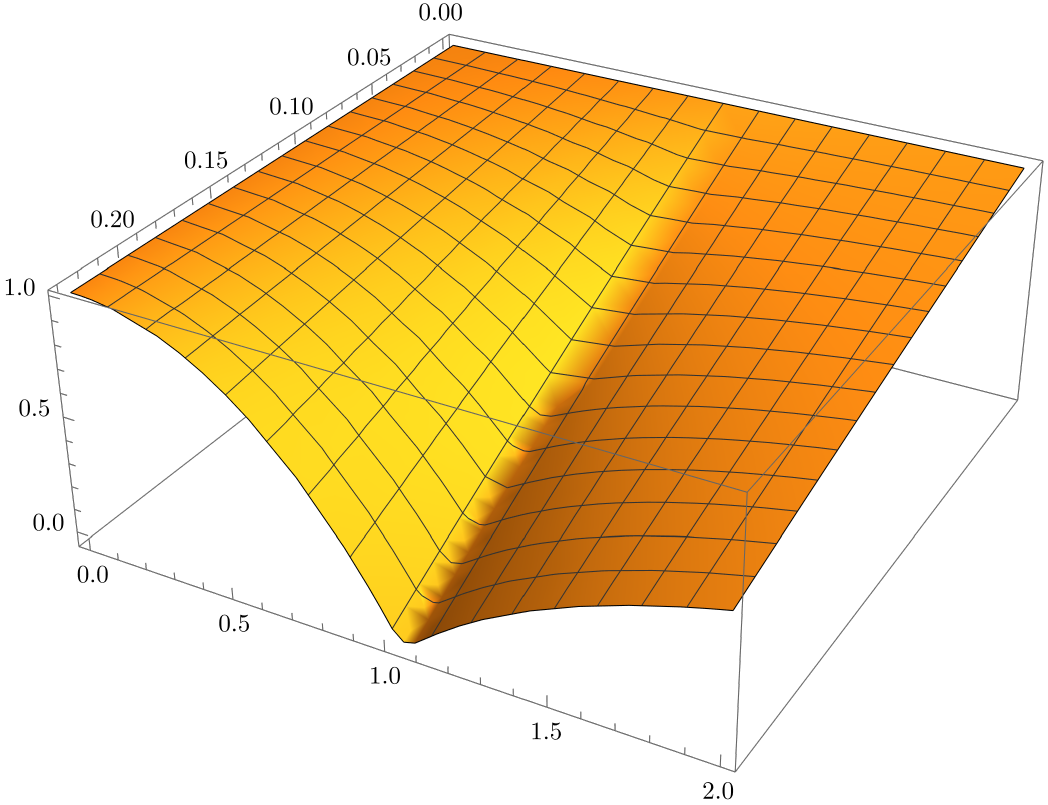}};
		\begin{scope}[x={(image.south east)},y={(image.north west)}]
		\node[anchor=east] at (0.02,0.45) {$ e^{2 \Phi } $};
		\node[anchor=north east] at (0.36,0.16) {$ r $};
		\node[anchor=south east] at (0.25,0.83) {$ \mu_0 $};
		\end{scope}
		\end{tikzpicture}
	\end{minipage}%
	\hfill
	\begin{minipage}[t]{.45\textwidth}
		\vspace{9pt}
		\centering
		\begin{tikzpicture}
		\node[anchor=south west,inner sep=0] (image) at (0,0) {\includegraphics[height=0.18\textheight]{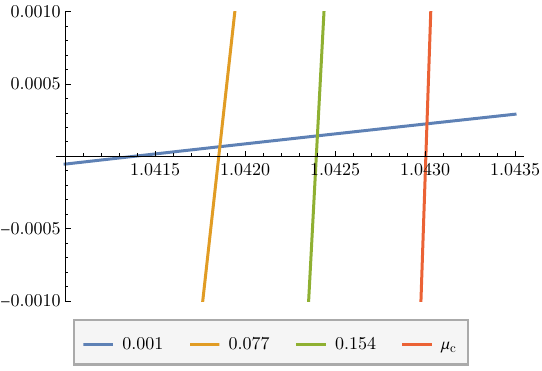}};
		\begin{scope}[x={(image.south east)},y={(image.north west)}]
		\node[anchor=east,overlay] at (0.01,0.58) {$ \partial_r e^{2 \Phi } $};
		\node[anchor=north] at (0.96,0.52) {$r$};
		\end{scope}
		\end{tikzpicture}
	\end{minipage}
	\hspace{0pt}
	\caption[Dependence of $ e^{2 \Phi } $ on $\mu_0$]{Dependence of $ e^{2 \Phi } $ on $\mu_0$. \textbf{(a)} $ e^{2 \Phi } $ as a function of $\mu_0$ and $ r $ for $ z_0=-0.7 $ at $ u=u_1 $. The range of $\mu_0$ is $ [0,\mu_c] $, with $ \mu_c(z_0=-0.7)\approx 0.230$. \textbf{(b)} The $ r $-derivative of the previous function is shown for a few values of $ \mu_0 $, which are indicated by color}
	\label{3d}
\end{figure}

So far, we have discussed only collapse examples where the formation of a black hole is imminent. But this is not a requirement for the emission of Hawking radiation, as already noted in the literature \cite{Barcelo2011}. In fact, the only reasons for the strong restrictions on the values of $\mu_0 $ and $ z_0 $ in the first example were to ensure the formation of a horizon and the escape of radiation to future null infinity. As can be seen in Figure \ref{3d} (a), condition \eqref{commucond} will be satisfied at some radius for all values of $ \mu_0 $ below $ \mu_c $ by keeping $ z_0 $ constant. Therefore, as long as $ v^r $ is negative, the same analysis will hold for all these cases. Furthermore, Figure \ref{3d} (b) shows that the radius where condition \eqref{commucond} is satisfied is practically indifferent to the value of $\mu_0$, and determined mainly by the profile of $ \mu $.

If $ \mu\neq 0 $ at $ r=0 $, then one can generalize the above conclusions for other profiles of $ \mu $. By expanding $ \mu $ and $ z $ with respect to $ r $ (as in expansion \eqref{f}) in equation \eqref{phisol}, one finds that
\[\partial_r e^{2\Phi} = - \left[2-\left(1+z^{(0)}\right)^2\right]\frac{4\pi\mu^{(0)}}{3}r + \mathcal{O}\left(r^2\right). \]
Consequently, if $ 2-\left(1+z^{(0)}\right)^2>0 $, then $ \partial_r e^{2\Phi}<0 $ for sufficiently small $ r $, and the continuity of this derivative implies that it must vanish inside the fluid, given that it is positive outside the matter distribution. Therefore, it is possible to repeat the previous analysis for matter distributions that are collapsing sufficiently fast. In case condition \eqref{commucond} is satisfied deep inside the fluid or the collapse is not quick enough, the radial transverse lightlike geodesic leaving $ \mathcal{Q}_1 $ will stay longer inside the matter, and
$ \kappa_2 $ will be different from zero. Thus, approximation \eqref{lambdas} may be adapted, which will be analyzed in the next subsection.

\subsection{Generalizing $ \kappa_1 $ and $ \kappa_2 $}\label{k1k2}

We first consider the case where $ \kappa_1 $ and $ \kappa_2 $ are different from zero. Then, approximation \eqref{lkr} is valid around both $ \mathcal{Q}_1 $ and $ \mathcal{Q}_2 $, leading to
\begin{equation}\label{k1k2neq0}
\lambda\approx \frac{e^{2\beta_1}}{\kappa_1}\left[e^{\kappa_1 \ln(1+\kappa_2\exp(-2\beta_2)\tilde{\lambda})/\kappa_2}-1\right].
\end{equation}
Consequently, for sufficiently small $ \tilde{\lambda} $, one recovers approximation \eqref{lambdas} and the thermalization process.

There remains the case with $ \kappa_1=0 $ and $ \kappa_2\neq 0 $. By exchanging the labels 1 and 2 in the calculation that led to approximation \eqref{lambdas}, one finds that
\[ \tilde{\lambda}\approx \frac{e^{2\beta_2}}{\kappa_2}\left[e^{\kappa_2 \exp(-2\beta_1)\lambda}-1\right]. \]
From this, one could evolve the quantum field backwards in time with condition \eqref{commucond} holding at $ \mathcal{Q}_2 $. By doing that, one finds that a vacuum at $ \mathcal{Q}_2 $ would have been a thermal state with respect to the radial transverse modes at $ \mathcal{Q}_1 $. Therefore, one has the contrapositive statement that a vacuum state at $ \mathcal{Q}_1 $ cannot be a vacuum state at $ \mathcal{Q}_2 $.

The above results show that one can relax the condition of rapid decrease of $\kappa$ between $ \mathcal{Q}_1 $ and $ \mathcal{Q}_2 $. As long as condition \eqref{commucond} is satisfied and approximation \eqref{lambdas} is valid, an emission process from a vacuum state will occur whenever there is a gradient in the values of $ \kappa $ along radial transverse lightlike geodesics.

Therefore, if one considers a matter distribution with a very low density, so that spacetime can be approximated by Minkowski space, and the quantum field is found in its vacuum state, then any point where condition \eqref{commucond} is satisfied may give rise to particle production as matter collapses and increases the value of $\kappa$. As a consequence, the assumption of an initial vacuum state inside the collapsing matter about to form a black hole may be erroneous, even if the ingoing modes coming from past null infinity are in a vacuum state. We shall come back to this discussion in Section \ref{bbht}.

\subsection{The vertex and an overlooked condition}\label{infgen}

A natural candidate for a point that allows a generalization of the above analysis is the vertex of a light cone. Indeed, the central conditions \eqref{ccg} imply that condition \eqref{commucond} is satisfied there in case $ a^a=0 $. However, an issue that is more evident in this case is the fact that, by using equations \eqref{delrlambda} and \eqref{lprime} as well as the central conditions \eqref{ccg},
\begin{equation}\label{commullambda}
\left[l',\partial_{(\lambda')}\right]^\mu= \left[l',e^{-2\beta}\partial_{(r)}\right]^\mu=\left(  \partial_u e^{-2\beta} + \frac{e^{2\Phi}}{2}\partial_r e^{-2\beta} -\frac{\partial_r e^{2\Phi}}{2} \right) \partial_{(r)}{}^\mu
\end{equation}
also equals zero at $ r=0 $. Being so, $ \lambda\approx\tilde{\lambda} $ in a first approximation, which is incompatible with approximation \eqref{lambdas}. Therefore, we have found another condition for the emission of Hawking radiation that has been overlooked so far. Namely, one has to guarantee that $ r(u_1)-r_1\approx r(u_2)-r_2 $ while $ \lambda $ differs sufficiently from $ \tilde{\lambda} $ so that approximation \eqref{lambdas} makes sense.

One could check this by analyzing higher-order terms. The relevance of condition \eqref{commucond} can be seen by the contraction
\[ \left[l',\partial_{(r)}\right]^c\nabla_c r =- \partial_{(r)}{}^b \nabla_b \left(  {l'}^c\nabla_c r \right),  \] 
which vanishes where this condition is satisfied. As a result, the variations in $ r $ along $ {l'}^a $ are, to first order, constant in the $ \partial_{(r)}{}^a $-direction, justifying the approximation $ r(u_1)-r_1\approx r(u_2)-r_2 $. In case $ \lambda=\tilde{\lambda} $ to first order, then one may consider the corrections given by 
\begin{equation}\label{delrr}
\partial_{(r)}{}^a \nabla_a\left(  \partial_{(r)}{}^b \nabla_b \left(  {l'}^c\nabla_c r \right)\right) \qand \partial_{(r)}{}^a \nabla_a\left(  \partial_{(r)}{}^b \nabla_b \left(  {l'}^c\nabla_c \lambda' \right)\right).
\end{equation}

From equations \eqref{lprime}, \eqref{delrlambda}, \eqref{Phi2}, and \eqref{beta2}, the restrictions on the initial data shown at the beginning of Appendix \ref{SSS}, and the central conditions \eqref{ccg} one finds that 
\[ \eval{\partial_{(r)}{}^a \nabla_a\left(  \partial_{(r)}{}^b \nabla_b \left(  {l'}^c\nabla_c r \right)\right)}_{r=0}=\Phi^{(2)}= 2 \pi T_{rr}{}^{(0)} + \frac{8\pi}{3}T_{ur}{}^{(0)} - \frac{\Lambda}{3} \quad \text{and}  \]
\[ \eval{\partial_{(r)}{}^a \nabla_a\left(  \partial_{(r)}{}^b \nabla_b \left(  {l'}^c\nabla_c \lambda' \right)\right)}_{r=0}=\Phi^{(2)} +\beta^{(2)}=4\pi T_{rr}{}^{(0)}+\frac{8\pi}{3}T_{ur}{}^{(0)}-\frac{\Lambda}{3} . \]
Thus, the latter is indeed greater than the first, but not considerably.

The comparison can be made more precise in the case of a dusty spacetime with $ T_{rr}{}^{(0)}\neq 0 $ and a negligible cosmological constant, where one can use the initial-value constraint \eqref{ivcTrrTur} to show that
\[ \eval{\partial_{(r)}{}^a \nabla_a\left(  \partial_{(r)}{}^b \nabla_b \left(  {l'}^c\nabla_c r \right)\right)}_{r=0} \approx  -\frac{2\pi}{3}T_{rr}{}^{(0)} \quad \text{and}\]
\[ \eval{\partial_{(r)}{}^a \nabla_a\left(  \partial_{(r)}{}^b \nabla_b \left(  {l'}^c\nabla_c \lambda' \right)\right)}_{r=0}\approx \frac{4\pi}{3}T_{rr}{}^{(0)} . \]
Since the modulus of the corrections differ only by a factor of 2, the question of whether an exponential approximation like \eqref{lambdas} is valid near the vertex remains open. A possible way to further investigate it would be to analyze the behavior of lightlike geodesics near the vertex more in depth, which will be left to future work.

Although the above analysis has been inconclusive, it offers a basis to check the validity of approximation \eqref{lambdas} in the previous examples. In the case of subsection \ref{SdS}, one can use the background metric of the first subsection of Appendix \ref{SSS} to show that equation \eqref{commullambda}  also vanishes to zeroth order in $ \tilde{\varepsilon} $ at $ r_1 $. Then, the leading-order terms to be considered are the ones in \eqref{delrr}, which become
\[ \eval{\partial_{(r)}{}^a \nabla_a\left(  \partial_{(r)}{}^b \nabla_b \left(  {l'}^c\nabla_c \lambda' \right)\right)}_{r=r_1}=-e^{4\beta_R}\Lambda = e^{2\beta_R}\eval{\partial_{(r)}{}^a \nabla_a\left(  \partial_{(r)}{}^b \nabla_b \left(  {l'}^c\nabla_c r \right)\right)}_{r=r_1} . \]
Therefore, if $ e^{2\beta_R}\gg 1 $, then approximation \eqref{lambdas} will be valid in a sufficiently small neighborhood of $ \mathcal{Q}_1 $. From definition \eqref{betaR}, one can expect this to occur when the total mass of the central body is large enough. In addition, the computation of the full metric without splitting it in a background component and a perturbation might render the Lie derivative of $ \partial_{(\lambda')}{}^a $ with respect to $ {l'}^a $ non-vanishing, but we shall not carry out such an analysis in this work.

\begin{figure}
	\centering
	\begin{tikzpicture}
	\node[anchor=south west,inner sep=0] (image) at (0,0) {\includegraphics[width=0.6\textwidth]{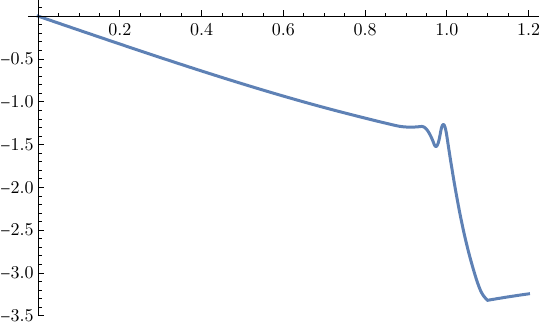}};
	\begin{scope}[x={(image.south east)},y={(image.north west)}]
	\node[anchor=east] at (0,0.49) {$ \left[l',\partial_{(\lambda')}\right]^r $};
	\node[anchor=north] at (0.525,.88) {$ r $};
	\end{scope}
	\end{tikzpicture}
	\caption[{$ \left[l',\partial_{(\lambda')}\right]^r $ in black hole formation}]{$ \left[l',\partial_{(\lambda')}\right]^r $ in black hole formation. From the results of the first example of subsection \ref{scbhf}, it was computed $ \left[l',\partial_{(\lambda')}\right]^r $ as a function of $ r $. As before, $ \Lambda=0 $, $ z_0= - 0.7 $, $ \mu_0=0.2044555 $, and $ u=u_1 $}
	\label{Lie}
\end{figure}

Fortunately, in the first example of subsection \ref{scbhf}, one can compute the right-hand side of equation \eqref{commullambda} and find that its $ r $-component is approximately $ -2.5 $ at $ r_1 $, as shown in Figure \ref{Lie}, validating approximation \eqref{lambdas} and making any further analysis unnecessary.

\section{Irreversible inequality}\label{iisec}

Another result that will guide the definitions of thermodynamic variables, especially their signs, is given below. Pursuing an inequality that may lead to an analogue of the second law of thermodynamics, we consider a purely gravitational irreversible process, namely, the spherical collapse of dust. Then, we take the $ u $-derivative of equation \eqref{betasphe}, use equation \eqref{deluTrrsphe}, integrate by parts the terms containing $ r' $-derivatives, use equation \eqref{delrre2Phi} and the $ r $-derivative of equation \eqref{betasphe}, and, in the boundary terms, employ equations \eqref{Trr} and \eqref{vr} to obtain
\begin{equation}\label{delubeta}
\partial_u \beta = -\frac{2\pi rT_{rr}}{1+z}v^r+\pi\int_{0}^{r}e^{2\beta}\left(  \mu-T_{rr} + \Lambda T_{rr}{r'}^2  \right) \dd{r'},
\end{equation}
where all functions are evaluated at $ u $, the ones outside the integral symbol are evaluated at $ r $, and the ones inside at $ r' $.

From equations \eqref{Trr} and \eqref{betasol}, one can see that the term
\begin{equation}\label{muTrr}
\mu-T_{rr}= \mu\left[1-e^{4\beta}(1+z)^2\right]= \mu\frac{1-\int_{0}^{r'} 8\pi r'' \mu (1+z)^2 \dd{r''}-(1+z)^2}{1-\int_{0}^{r'} 8\pi r'' \mu (1+z)^2 \dd{r''}}
\end{equation}
will be positive as long as $ z<0 $ and the integrands above are sufficiently small. In this case, the assumptions that $ v^r<0 $ and $ \Lambda\geq 0 $ imply that $ \partial_u \beta >0 $ for $ r\neq 0 $, since the null energy condition \cite{Poisson2004},
\begin{equation}\label{nec}
T_{rr} = T_{ab}\partial_{(r)}{}^a \partial_{(r)}{}^b \geq 0,
\end{equation}
and inequality \eqref{vu} hold.

The mentioned smallness of the integrands leads us to consider a weak-field regime and impose some condition on the active gravitational mass. By adding and subtracting $ e^{2\Phi}e^{2\beta}(1+z)^2 $, one can use equation \eqref{vr} to show that
\begin{equation}\label{1e4b}
1-e^{4\beta}(1+z)^2= -2e^{2\beta}(1+z)v^r+ e^{2\beta}(1+z)^2\left(e^{2\Phi}-e^{2\beta}\right).
\end{equation} 
Equation \eqref{e2P} implies that equation \eqref{e2phi} is valid inside the fluid if one replaces $ R $ and $ r $ by $ r' $. From this and equation \eqref{Trr}, one can evaluate equations \eqref{1e4b}, \eqref{muTrr}, and \eqref{delubeta} to find that
\begin{equation}\label{delubeta2}
\partial_u \beta = -\frac{2\pi rT_{rr}}{1+z}v^r+2\pi\int_{0}^{r}T_{rr}e^{2\beta}\left\{ \frac{\Lambda}{3}{r'}^2 - \frac{e^{-2\beta}v^r}{(1+z)}\left[1+\frac{(1+z)M(u,r')}{e^{-2\beta}v^r r'}\right] \right\} \dd{r'}.
\end{equation}
Therefore, the conditions
\[ \abs{\frac{(1+z)M(u,r')}{e^{-2\beta}v^r r'}}\leq 1, \quad v^r<0, \qand \Lambda\geq 0 \implies \partial_u \beta >0 \qif r\neq 0, \]
since the presence of a fluid implies that $ T_{rr}\neq 0 $, as it can be seen in equation \eqref{Trr}, and inequality \eqref{Mpositive} holds, i.e., the dominant energy condition is implicit.

The condition $ v^r<0 $ conveys the information that the fluid is collapsing, as desired. The other relevant condition states that the field strength, $ M(u,r')/r' $, cannot be greater than the ``kinetic parameter'' $ {e^{-2\beta}v^r}/{(1+z)} $. That the field strength is small in many systems is an easily acceptable, representative supposition. But that the kinetic parameter is not even smaller is an assumption that requires more care. First, we note that the collapse condition applied to equation \eqref{vr} and multiplied by $ 2e^{2\Phi}(1+z) $ yields
\[ e^{4\Phi}(1+z)^2<e^{2\Phi}e^{-2\beta}, \]
where we are assuming a collapse prior to the formation of a dynamical horizon so that condition \eqref{condi} is not satisfied anywhere and $ e^{2\Phi}>0 $. Then, by taking the square of equation \eqref{vr} and using the above inequality, one finds that 
\begin{equation}\label{ineq1}
2 \abs{v^r}^2<\abs{\frac{e^{-2\beta}v^r}{1+z}}. 
\end{equation}
As a result, one can consider the stronger condition
\[ \abs{\frac{(1+z)M(u,r')}{e^{-2\beta}v^r r'}}<\frac{M(u,r')}{2 \abs{v^r}^2r'}\leq 1. \]

It is convenient to introduce the \emph{gravitational irreversibility},
\begin{equation}\label{gi}
\iota(u,r)\coloneqq -2v^r(u,r)\sqrt{\frac{r}{r_S(u,r)}},
\end{equation}
where $ r_S(u,r)=2M(u,r) $ is the Schwarzschild radius, since the collapse condition and the last inequality are equivalent to a single condition on $ \iota $:
\begin{equation}\label{gi1}
\iota\geq 1.
\end{equation}
With definition \eqref{gi}, one can use equations \eqref{delubeta2} and \eqref{Trr}, the conditions $ v^r<0 $ and $ \Lambda\geq 0 $, and inequality \eqref{ineq1} to show that 
\[ \partial_u \beta>2\pi\int_{0}^{r}\mu e^{4\beta}(1+z)\abs{v^r}\left(1-\frac{1}{\iota^2}\right) \dd{r'}. \]
Consequently, we can estimate $ \partial_u \beta $ in a Newtonian system, where the metric is nearly flat and the velocities are low, by approximating $ \beta\approx z \approx 0 $ in the above integral, leading to
\begin{equation}\label{delubetaineq}
\partial_u \beta>2\pi\int_{0}^{r}\mu \abs{v^r}\left(1-\frac{1}{\iota^2}\right) \dd{r'}.
\end{equation}

As an example, we consider the process of star formation
inside molecular cloud cores. Specifically, we shall analyze the ``inside-out collapse'' model \cite{Shu1977}, since it is compatible with the above model (aside from the presence of pressure) and is widely employed for describing the formation of low-mass stars \cite{Gao2010}. Taking into account that this category represents the vast majority of the stellar population \cite{Kroupa2001,Bastian2010}, such an analysis offers a reliable benchmark for the behavior of $ \beta $.

Then, by identifying $ v^r $ and $ M $ with the Newtonian fluid velocity and total mass, respectively, restoring dimensional constants, and using definitions (8) of Reference \citeonline{Shu1977} and \eqref{gi}, one obtains
\begin{equation}\label{iotax}
\iota(x) = \abs{v(x)}\sqrt{\frac{2x}{m(x)}}.
\end{equation}
Here, $ x=r/at $ is the similarity variable, $ r $ and $ t $ are the radial and time Newtonian coordinates, respectively, and $ a $ is the isothermal speed of sound of the cloud core.

\begin{figure}
	\centering
	\begin{minipage}[t]{.45\textwidth}
		\vspace{0pt} 
		\centering
		\begin{tikzpicture}
		\node at (0.5,1) {(a)};
		\end{tikzpicture}
	\end{minipage}%
	\hfill
	\begin{minipage}[t]{.45\textwidth}
		\vspace{0pt} 
		\centering
		\begin{tikzpicture}
		\node at (0.5,1) {(b)};
		\end{tikzpicture}
	\end{minipage}%

	\begin{minipage}[t]{.45\textwidth}
		\vspace{0pt} 
		\centering
		\begin{tikzpicture}
		\node[anchor=south west,inner sep=0] (image) at (0,0) {\includegraphics[height=0.15\textheight]{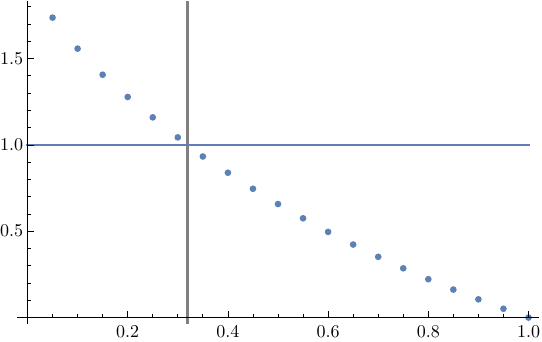}};
		\begin{scope}[x={(image.south east)},y={(image.north west)}]
		\node[anchor=east] at (0,0.54) {$ \iota $};
		\node[anchor=north] at (0.52,0) {$ x $};
		\node[anchor=north] at (0.35,0) {$ x_c $};
		\end{scope}
		\end{tikzpicture}
	\end{minipage}%
	\hfill
	\begin{minipage}[t]{.45\textwidth}
		\vspace{0pt}
		\centering
		\begin{tikzpicture}
		\node[anchor=south west,inner sep=0] (image) at (0,0) {\includegraphics[height=0.15\textheight]{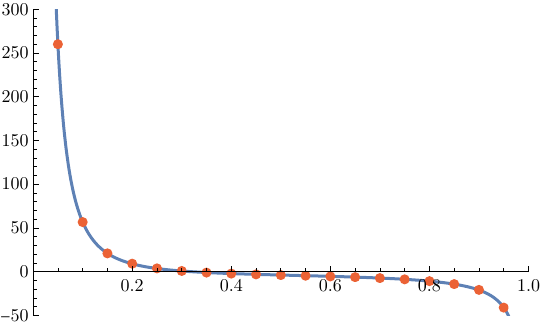}};
		\begin{scope}[x={(image.south east)},y={(image.north west)}]
		\node[anchor=east] at (0,0.54) {$ f $};
		\node[anchor=north] at (0.52,0.1) {$ x $};
		\end{scope}
		\end{tikzpicture}
	\end{minipage}
	\hspace{0pt}
	\caption[$ \iota(x) $ and $ f(x) $ in star formation]{$ \iota(x) $ and $ f(x) $ in star formation. \textbf{(a)} The dots represent the values of $ \iota(x) $ calculated from Table 2 of Reference \citeonline{Shu1977}. One can see that $ \iota(x)\leq 1 $ for $ x\geq x_c \approx 0.32  $. \textbf{(b)} The red dots represent the integrand of inequality \eqref{beta01}, also calculated from Table 2 of Reference \citeonline{Shu1977}. We used the function \textit{NonlinearModelFit} of the software \textit{Wolfram Mathematica} to fit a function of $ x $ to these values. The functional form assumed was $ f(x)=(c_0 + c_1 x + c_2 x^2) x^{\epsilon_0} (1 - x)^{\epsilon_1} $, together with the initial guesses $ c_0=1, \ c_1=-2, \ c_2=1, \ \epsilon_0=-2, \ \text{and} \ \epsilon_1=-1 $, obtained after some trials and an analysis of the behavior of $ v(x) $ as $ x\to 1 $. The best-fit parameters were $ c_0\approx 0.70, \ c_1\approx-1.89, \ c_2\approx-0.96, \ \epsilon_0\approx-2.01, \ \text{and} \ \epsilon_1\approx-0.98 $, which were used to evaluate the function $ f(x) $ represented by the blue line. The maximum absolute error obtained was $ \approx 0.18 $, the maximum relative error was $ \approx 0.07 $, and the root mean square error was $ \approx 0.08 $}
	\label{betabound}
\end{figure}

Firstly, we focus on the ``expansion-wave collapse solution'' of Reference \citeonline{Shu1977}, which represents the evolution of an initially hydrostatic cloud core. Using equation \eqref{iotax} and the values of Table 2 of Reference \citeonline{Shu1977}, which describe the infall region of the cloud core, one can plot the function $ \iota(x) $ and find that $ \iota(x)\leq 1 $ for $ x\geq x_c \approx 0.32  $, as shown in Figure \ref{betabound} (a).
Therefore, we must resort to inequality \eqref{delubetaineq} in order to demonstrate the positivity of $ \partial_u \beta $.

To compute it, we restore dimensional constants, use definitions (8) of Reference \citeonline{Shu1977} to evaluate $ \mu=c^2\rho $ and $ v^r $, where $ \rho $ is the Newtonian mass density, change the integration variable to $ x $, and integrate up to the infall radius, $ x=1 $, since the integral decreases from $ x_c $, obtaining
\begin{equation}\label{beta01}
\partial_u \beta>\frac{a^2}{2c^2t}\int_{0}^{1}\alpha(x) \abs{v(x)}\left(1-\frac{1}{\iota^2(x)}\right) \dd{x}.
\end{equation}
Again, we use Table 2 of Reference \citeonline{Shu1977} to compute the integrand, which is plotted in Figure \ref{betabound} (b). To these results, we fit a function $ f(x) $ that diverges, approximately, as $ x^{-2} $ as $ x\to 0 $ and as $ -(1-x)^{-1} $ as $ x\to 1 $ (details are explained in the caption of Figure \ref{betabound}). Thus, one can say that the resulting integral diverges positively.

A more satisfying bound for $\partial_u \beta$ can be given when one considers the fact that the model in question is singular. What is called the ``core'', at $ r=0 $, in Reference \citeonline{Shu1977} can be interpreted as the protostar, which has a finite size in reality. Therefore, we can introduce a lower cutoff in the integral of inequality \eqref{beta01} to accommodate this physical requirement, as well as an upper cutoff for computational convenience. Setting the last one to $ 0.999 $, one can vary the lower cutoff and find that the maximum value of it for which the integral remains positive is $ x_+\approx 0.038 $. Then, we can compare this value with observed ones.

A good candidate to be used as an example is the protostar inside the Bok globule B335. The infalling matter around it can extend down to radii of the order of $ 10\operatorname{au} $; its inferred age is $ 5.0 \times 10^4 \operatorname{years} $, and the effective speed of sound of the envelope is $ a_{\text{eff}}=0.233\operatorname{km} \operatorname{s}^{-1} $ \cite{Evans2015}, implying a lower cutoff of $ x_{\text{B335}}\approx 0.004 \approx 0.1 x_+$. Therefore, we conclude that $ \partial_u \beta>0 $ along at least $ 99.9\% $ of the infall radius of B335.

We can generalize the above conclusion by analyzing the behavior of the ``collapse solutions without critical points'' near the origin. From equations (17) of Reference \citeonline{Shu1977} and \eqref{iotax}, one finds that $ \iota(x)\to 2 $ as $ x\to 0 $ and condition \eqref{gi1} is satisfied. Consequently, $ \partial_u \beta>0 $ near the origin in all solutions describing star formation from an isothermal sphere that was initially in a state of nearly hydrostatic equilibrium. Given the supersonic flow near the origin, the infalling matter approaches free fall, the fluid pressure becomes negligible, and the model becomes more and more similar to a description of dust. From this reasoning, we infer that if all the other forces were suddenly ``turned off'', then gravity would compel typical systems to proceed an evolution in which $ \partial_u \beta>0 $ most of the time.

One can interpret the above conclusion as follows. The initial data of a dusty spacetime could in principle be arranged in such a way that the fluid is either expanding or collapsing with $ \partial_u \beta<0 $. However, in the case of star formation, the early marginal hydrostatic equilibrium of the isothermal cloud core is perturbed by the gravitational collapse of the central regions, leading to an initial density profile similar to that of the ``singular isothermal sphere'' \cite{Shu1977}. In general terms, the thermodynamic evolution prior to the collapse phase favors classes of initial data that yield an evolution dominated by the condition $ \partial_u \beta>0 $ during the collapse. As a consequence, if we associate $ \partial_u \beta $ with the variation of some form of gravitational entropy, then we may say that collapse becomes the natural, entropy-increasing continuation of the previous thermodynamic evolution once it reaches a certain threshold, similarly to what happens in a phase transition.

Based on this dominance of the condition $ \partial_u \beta>0 $ and on equation \eqref{delrlambda}, we shall assume that
\begin{equation}\label{ii}
\partial_u \lambda'=2\int_{0}^{r}e^{2\beta}\partial_u \beta\dd{r'}> 0
\end{equation}
characterizes a gravitational irreversible inequality. That is, in (i) typical irreversible processes that are (ii) dominated by gravity, there will be a (iii) tendency for the inequality above to be satisfied, and, once it has been established, it will be valid until the end of the process. This might seem a little vague at first, but our aim here is to account for (i) the statistical, empirical nature of the second law of thermodynamics, (ii) other forces that may be balancing the gravitational irreversibility, and (iii) possible transient effects. To verify such a hypothesis with a high level of confidence would probably require exhaustive numerical simulations based on accurate astronomical data, something that is far beyond the scope of this work. Nonetheless, a few rough estimates can still be given to support the idea.

Another typical process where collapse plays a major role is a supernova event. We searched the literature and found two references in which the parameters necessary to calculate the gravitational irreversibility were available for the same simulation. In both, the values represent the spherically symmetric collapse of the iron core inside massive stars that precedes type II supernovae. Reference \citeonline{Siebel2002} uses the Bondi-Sachs formalism and future light cones with vertices at the center of the core. Since the resulting spacetime does not deviate much from Minkowski spacetime, with a redshift factor between the proper times at the origin and at future null infinity of $ \approx 1.12 $ \cite{Siebel2002}, the values for the radial velocity and the Bondi mass will not differ much from what would be calculated using past light cones. In this simulation, the core is modeled by a hybrid polytropic equation of state.

\begin{figure}
	\centering
	\begin{minipage}[t]{.45\textwidth}
		\vspace{0pt} 
		\centering
		\begin{tikzpicture}
		\node at (0.5,1) {(a)};
		\end{tikzpicture}
	\end{minipage}%
	\hfill
	\begin{minipage}[t]{.45\textwidth}
		\vspace{0pt} 
		\centering
		\begin{tikzpicture}
		\node at (0.5,1) {(b)};
		\end{tikzpicture}
	\end{minipage}%

	\begin{minipage}[t]{.45\textwidth}
		\vspace{0pt} 
		\centering
		\begin{tikzpicture}
		\node[anchor=south west,inner sep=0] (image) at (0,0) {\includegraphics[height=0.15\textheight]{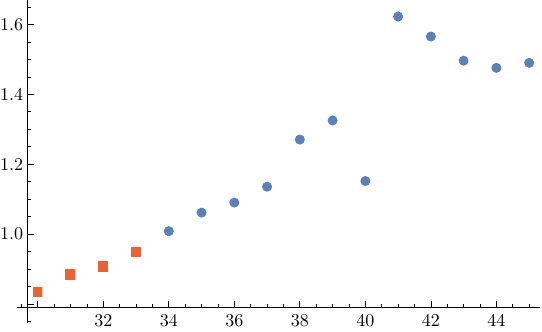}};
		\begin{scope}[x={(image.south east)},y={(image.north west)}]
		\node[anchor=east] at (0,0.54) {$ \iota $};
		\node[anchor=north] at (0.52,0) {$ u_B (\operatorname{ms}) $};
		\end{scope}
		\end{tikzpicture}
	\end{minipage}%
	\hfill
	\begin{minipage}[t]{.45\textwidth}
		\vspace{0pt}
		\centering
		\begin{tikzpicture}
		\node[anchor=south west,inner sep=0] (image) at (0,0) {\includegraphics[height=0.15\textheight]{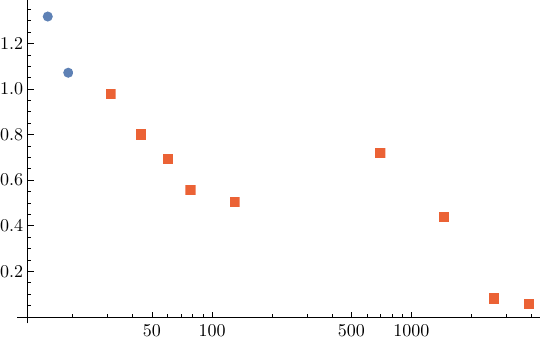}};
		\begin{scope}[x={(image.south east)},y={(image.north west)}]
		\node[anchor=east] at (0,0.54) {$ \iota $};
		\node[anchor=north] at (0.52,0) {$ r (\operatorname{km}) $};
		\end{scope}
		\end{tikzpicture}
	\end{minipage}
	\hspace{0pt}
	\caption[$ \iota $ in core-collapse supernovae]{$ \iota $ in core-collapse supernovae. In both plots, the blue dots satisfy condition \eqref{gi1}, while the red squares do not. \textbf{(a)} $ \iota $ is shown as a function of Bondi time, $ u_B $, calculated at the points of maximum infall velocity of each snapshot reported in the simulation of Reference \citeonline{Siebel2002}. The shock wave forms between $ 40 \operatorname{ms} $ and $ 41 \operatorname{ms} $. \textbf{(b)} $ \iota $ as a function of $ r $ at the moment of bounce of the collapsing core. A few convenient points of the infall region outside the inner core were chosen to represent the behavior of $ \iota $ during the gravitational collapse that precedes a supernova explosion. The data were taken from the ``model SH'' of Reference \citeonline{Sumiyoshi2005}}
	\label{superiota}
\end{figure}

We shall compute $ \iota $ using Figures 7.3 and 7.2 of Reference \citeonline{Siebel2002}, where one finds snapshots of radial velocity profiles and the worldtubes of mass shells, respectively. The plots were imported to the software \textit{Inkscape}, where grid lines and lines parallel to the light cones of Figure 7.2 were drawn in order to read the values more cleanly. At each snapshot, we read the values for the points where the infall velocity attains a maximum, since they lie above the homologous inner core and are the closest representatives to a free-fall state at any given moment. There was a large uncertainty in this process, especially in reading the enclosed mass, which varies at large steps of $ 0.2M_\odot $. We chose the value for the enclosed mass based on the visual proximity between the point of interest and its neighboring shells as well as their midpoint. The readings are given in Table \ref{table} of Appendix \ref{super}, and the calculate $ \iota $, using equation \eqref{gi}, is plotted in Figure \ref{superiota} (a).

A similar analysis was done with data from Reference \citeonline{Sumiyoshi2005}, which models the collapse of a $ 15 M_\odot $ star using a general relativistic Cauchy evolution based on the Misner-Sharp formalism \cite{Misner1964} and neutrino radiation hydrodynamics. The velocity profile is shown exclusively at bounce, and the relation between the baryon mass coordinate and radius is restricted to the ``model SH''. We use the values of the baryon mass coordinate and of the reported infall velocity as an estimate for the Bondi mass and the velocity measured in the Bondi-Sachs coordinate system of the present work, respectively. From Figures 6 and 1 of Reference \citeonline{Sumiyoshi2005}, we read the values of $ M $, $ v^r $, and $ r $ at bounce for a set of different radii, shown in Table \ref{table2}. As before, there is a large uncertainty in the relation between $ M $ and $ r $. Even with small steps between the mass shells, the plot becomes overcrowded at bounce, and makes the readings difficult, especially around $ r=100\operatorname{km} $. From these data, we calculated $ \iota $ as a function of $ r $, as shown in Figure \ref{superiota} (b).

Despite the uncertainties in the calculation of $ \iota $, Figures \ref{superiota} (a) and (b) provide complementary information about the qualitative behavior of $ \iota $ during the collapse phase prior to a supernova explosion. The first plot shows the tendency of the collapse process to satisfy condition \eqref{gi1} and yield a positive contribution to $ \partial_u \beta $ as the collapsing matter approaches a free-fall state and loses its ``memory'' of the previous influence of pressure. In other words, as gravity becomes the dominant force, the gain in infall velocity overcomes the loss given by the compactification of matter, increasing $ \iota $ and favoring the irreversible inequality \eqref{ii}.

By comparing Figure \ref{superiota} (b) with Figure \ref{betabound} (a), one can expect that the high concentration of matter in the innermost components of the infalling matter may give enough weight to the region satisfying condition \eqref{gi1} so that the integrand of inequality \eqref{delubetaineq} implies a positive contribution to $ \partial_u \beta $ in a large portion of the infall region. Unfortunately, a more precise statement would require a dedicated simulation to calculate the relevant parameters.

Taking the above estimates together with the positive $ \partial_u \beta $ shown in Figure \ref{other} (g) (which results from a simulation that is far from being realistic, but serves as an example), one can say that there is evidence to support the idea that the irreversible inequality \eqref{ii} will be favored during the collapse processes in the life of a star, from birth to death. It may not be attained completely, but one could in principle relate this fact to other irreversible processes, such as heating, nuclear fusion, and supernova explosions. Therefore, one can assume that the irreversible inequality \eqref{ii} typically holds when gravity is the main force.

\section{Quasi-local energy}\label{EC}

\subsection{The analogue of the integral mass formula}\label{imf}

In order to establish a thermodynamic formalism, it is necessary to introduce the concept of internal energy. Our strategy will be to start analyzing a definition similar to the one used in black hole thermodynamics and then proceed to a quasi-local definition. Firstly, we define a mass density as given by the source of the gravitational field. In the Newtonian limit, one finds that \cite[p. 415]{Misner1973}
\[ g_{uu} \simeq  -1 - 2\Phi_N, \]
where $ \Phi_N $ is the gravitational potential. We shall generalize this notion and define
\[ \Phi_N \coloneqq \frac{-1-g_{uu}}{2}. \]
Now, we expand the metric with respect to $ r $ and find that 
\[g_{uu} = -1-2  \Phi ^{{(1)}} r + \left(( U^{{(0)}} )^2+\sin ^2(\theta ) ( W^{{(0)}} )^2 -2 ( \Phi ^{{(1)}} )^2-\beta ^{{(2)}} -\Phi ^{{(2)}}\right) r^2 +\mathcal{O}\left(r^3\right). \]
Therefore,
\[ \Phi_N = \Phi ^{{(1)}} r + \frac{1}{2} \left( \Phi ^{(2)} + 2 ( \Phi ^{(1)} )^2+\beta ^{(2)} - (U^{{(0)}} )^2-\sin ^2(\theta ) ( W^{{(0)}} )^2\right) r^2 +\mathcal{O}\left(r^3\right). \]
The first term on the right side represents inertial forces and is present also in the Newtonian theory \cite[p. 295]{Misner1973}. In turn, the terms $ \mathcal{O} (r^3) $ may be viewed as corrections similar to the more usual multipole moments. Thus, they are more closely related to how the total mass is distributed than to its total value. For example, consider a mass distribution in the Newtonian context given by
\[ \rho = \rho_0 + \mathcal{O}(r). \]
In a sufficiently small neighborhood of the origin, we may approximate Poisson's equation by its spherically symmetric part \cite[p. 28]{PoissonWill2014}:
\[ \frac{1}{r^2} \partial_r \left( r^2 \partial_r \Phi_N \right) = 4\pi \rho_0  + \mathcal{O}(r). \]
Then, we find that
\[ \Phi_N = \frac{2\pi}{3} \rho_0 r^2 + \mathcal{O}(r^3), \]
showing that the terms $ \mathcal{O}(r^3) $ are related to deviations in $ \rho $ from its central value.
Because we are interested in the value of $ \rho  $ at $ r=0 $, we shall define it as
\[ \rho \coloneqq  \frac{3\Phi_N^{(2)}}{2\pi}. \]

By expanding the functions of the metric with respect to $ r $ and solving the hierarchical scheme for the first orders after the central conditions (given in Appendix \ref{firstterms}), we find that
\[ 
\begin{aligned}
	\rho = & \; 3 T_{{rr}}{}^{{(0)}}+2 T_{{ur}}{}^{{(0)}} -\frac{1}{4} \left(\csc ^2(\theta ) \partial _{\phi \phi } T_{{rr}}{}^{{(0)}}+\cot (\theta ) \partial _{\theta } T_{{rr}}{}^{{(0)}}+\partial _{\theta \theta } T_{{rr}}{}^{{(0)}}\right) + \\
	& \; \frac{3}{16 \pi } \left(\csc ^2(\theta ) \partial _{\phi \phi } \gamma ^{{(2)}}  -3 \cot (\theta ) \partial _{\theta } \gamma ^{{(2)}} -2 \csc (\theta ) \partial _{\theta \phi } \delta ^{{(2)}} -\partial _{\theta \theta } \gamma ^{{(2)}}- \right .\\
	& \; \left. 2 \cot (\theta ) \csc (\theta ) \partial _{\phi } \delta ^{{(2)}} +2 \gamma ^{{(2)}}\right) +\frac{5}{2} \left(T_{{r\theta }}{}^{{(1)}} \cot (\theta )+\partial _{\theta } T_{{r\theta }}{}^{{(1)}}+\csc ^2(\theta ) \partial _{\phi } T_{{r\phi }}{}^{{(1)}}\right) -  \\
	&  \; \frac{\Lambda }{4 \pi } -\frac{3}{4 \pi } \left(( U^{{(0)}})^2+\sin^2(\theta)( W^{{(0)}} )^2\right).
\end{aligned}
 \]
Imposing the central conditions \eqref{expTBondi} on the components of the energy-momentum tensor yields
\begin{equation}\label{rhobar}
\begin{aligned}
\rho = & \; {T}^{\bar{0}\bar{0}{(0)}}+\frac{1}{4}  \left(5+3 \cos (2 \theta )-6 \sin ^2(\theta ) \cos (2 \phi )\right){T}^{\bar{1}\bar{1}{(0)}} +\\
&\; \frac{1}{4}  \left(5+3 \cos (2 \theta ) + 6 \sin ^2(\theta ) \cos (2 \phi )\right){T}^{\bar{2}\bar{2}{(0)}}+\frac{1}{2} (1-3 \cos (2 \theta )) {T}^{\bar{3}\bar{3}{(0)}} 	-\\
&\; 3 \left(\sin ^2(\theta ) \sin (2 \phi ){T}^{\bar{1}\bar{2}{(0)}} +\sin (2 \theta ) \left( \cos (\phi ) {T}^{\bar{1}\bar{3}{(0)}} + \sin (\phi ){T}^{\bar{2}\bar{3}{(0)}}\right)\right) +\\
&\; \frac{3}{16 \pi } \left(2 \gamma ^{{(2)}} + \csc ^2(\theta ) \partial _{\phi \phi } \gamma ^{{(2)}} -3 \cot (\theta ) \partial _{\theta } \gamma ^{{(2)}} -\partial _{\theta \theta } \gamma ^{{(2)}}-2 \csc (\theta ) \partial _{\theta \phi } \delta ^{{(2)}} - \right .\\
&\; \left .2 \cot (\theta ) \csc (\theta ) \partial _{\phi } \delta ^{{(2)}} \right)-\frac{\Lambda }{4 \pi }  -\frac{3}{4 \pi } \left(( U^{{(0)}} )^2+\sin ^2(\theta ) ( W^{{(0)}} )^2\right) .
\end{aligned}
\end{equation}

In order to eliminate the angular dependence and obtain the monopole term in the source of the gravitational potential, we shall define a mean value for the expressions above. Although a simple integration over the angular coordinates could provide such a mean (as it was done in the original formulation \cite{Bondi1962}), we consider that a three-dimensional mean is more suitable to the notion of a thermodynamic system contained in a light cone. However, there is no canonical scalar surface element for a null hypersurface. The vector surface element compatible with Stokes' theorem for a general hypersurface is given by \cite{Poisson2004,Wald1984}
\[\dd{\Sigma_\mu} = \epsilon_{\mu\alpha\beta\gamma}\partial_{(1)}{}^{\alpha}\partial_{(2)}{}^{\beta}\partial_{(3)}{}^{\gamma} \dd{x^1} \dd{x^2} \dd{x^3}. \]
Therefore, for a non-null hypersurface, one can split the above as
\[  \dd{\Sigma_\mu} = \varepsilon n_{\mu} \dd{\Sigma}, \]
where $ n^{\mu} $ is the vector field orthonormal to the surface and $ \varepsilon = n^a n_a $. Then, the scalar surface element can be calculated as 
\begin{equation}\label{surfelem}
\dd{\Sigma} = n^{\mu} \dd{\Sigma_\mu}.
\end{equation}
In the case of a light cone covered by Bondi-Sachs coordinates, we have that
\[ \dd{\Sigma_\mu} =  \partial_{(r)}{}_{\mu} r^2 \sin(\theta) \dd{r} \dd{\theta} \dd{\phi},\]
but there is no unique transversal vector field to provide a canonical scalar surface element. As a consequence, we must choose one to play the role of $ n^a $ in equation \eqref{surfelem}. One of the natural choices is $ \partial_{(u)}{}^a $, which leads to
\begin{equation}\label{dSigma}
\dd{\Sigma} \coloneqq  \partial_{(u)}{}^{\mu} \dd{\Sigma_\mu} = e^{2\beta}r^2 \sin(\theta) \dd{r} \dd{\theta} \dd{\phi}.
\end{equation}
Using this surface element, one finds that the four-dimensional volume element can be written as
\[ \dd{v} = \epsilon_{\mu\alpha\beta\gamma}\partial_{(u)}{}^{\mu}\partial_{(r)}{}^{\alpha}\partial_{(\theta)}{}^{\beta}\partial_{(\phi)}{}^{\gamma}\dd{u}\dd{r} \dd{\theta} \dd{\phi} = \dd{u} \dd{\Sigma}, \]
so we may attribute to $ u $ an integration measure that is equal to $ 1 $. Such a split in the integration measure mimics the invariant behavior of time in Newtonian mechanics. Moreover, the affine parameter is related to the areal radius by equation \eqref{delrlambda}, allowing us to write
\begin{equation}\label{newmeasure}
\dd{\Sigma} = \left( \sqrt{\det(g_{A'B'})}\dd{\theta'} \dd{\phi'}  \right) \dd{\lambda'}.
\end{equation}
Given that the vector field associated with $ \lambda' $, $ k^a $, is ``calibrated by the observer'' along $ \mathcal{P} $ and parallelly transported, we find that $ \lambda' $ is one of the most natural representations of a distance along the null generators of a light cone. Thus, we see once again a special split in which we may attribute a unit integration measure to $ \lambda' $, similarly to integration measures of linear coordinates in the Newtonian theory. In this sense, we may say that definition \eqref{dSigma} is the ``most Newtonian surface element'' for a light cone, in which the differences between the Newtonian theory and general relativity are concentrated in $ \det(g_{A'B'}) $. Since we are trying to translate concepts from Newtonian gravity to general relativity, definition \eqref{dSigma} will be our choice for integrating scalar functions over light cones.

Now, we define the mean mass density inside a ball $ \mathcal{B}_R $ of radius $ r=R $ and volume $ V_R $ as
\[ \bar{\rho} (u,R) \coloneqq \frac{1}{V_R} \int_{\mathcal{B}_R} \rho \dd{\Sigma}. \]
Then, the \emph{local mean density} will be defined as
\[\bar{\rho}^{(0)} \coloneqq  \bar{\rho} (u,0) = \lim_{R\to 0} \frac{3}{4\pi R^3} \int_{\mathcal{B}_R} \rho r^2 \sin(\theta) \dd{r} \dd{\theta} \dd{\phi}  + \mathcal{O}(R^2) = \frac{1}{4\pi} \int_{S^2} \rho \sin(\theta)  \dd{\theta} \dd{\phi}. \]
Using expression \eqref{rhobar} for $ \rho $, one finds that
\begin{equation}\label{rhotraceT}
\bar{\rho}^{(0)} =  {T}^{\bar{0}\bar{0}{(0)}} + {T}^{\bar{1}\bar{1}{(0)}}+ {T}^{\bar{2}\bar{2}{(0)}} + {T}^{\bar{3}\bar{3}{(0)}} + \dots,
\end{equation}
where the omitted terms are not related to the energy-momentum tensor. One sees that this expression is compatible with the source of the analogue of the Newtonian potential in the interior Schwarzschild solution for perfect fluids \cite[p. 127]{Wald1984}, indicating that our definition is consistent with well established results, at least in some cases.

To make the analogy between the local mean density and the integral mass formula of black hole thermodynamics apparent, we first define some tensors that are useful in the kinematics \cite{Gourgoulhon2006} of light cones. There is a preferred transversal null vector field,
\begin{equation}\label{ldef}
l^a, \qq{such that} l_al^a=0, \quad l_a \partial_{(r)}{}^a=1 \qand l_a \partial_{(A)}{}^a=0,
\end{equation}
with which we can define the orthogonal projector onto the two-surfaces $ \mathcal{S}(u,r) $, 
\[ q^{a}{}_{b} \coloneqq \delta^{a}{}_{b} - \partial_{(r)}{}^a l_b- l^a \partial_{(r)}{}_b. \]
Next, the normal curvature $ \Theta_{ab} $, the expansion scalar $ \Theta $, the shear tensor $ \sigma_{ab} $, the Hájíček one-form $ \Omega^H{}_a $, the transversal deformation rate $ \Xi_{ab} $, the transversal expansion scalar $ \Xi $, and the inaffinity are defined as
\[ \Theta_{ab} \coloneqq  q^{c}{}_{a}q^{d}{}_{b} \nabla_c \partial_{(r)}{}_d, \quad \Theta \coloneqq g^{ab} \Theta_{ab} = \frac{2}{r}, \quad \sigma_{ab} \coloneqq \Theta_{ab} - \frac{\Theta}{2} q_{ab}, \]
\[ \Omega^H{}_a \coloneqq  q^{c}{}_{a}l^{d} \nabla_c \partial_{(r)}{}_d, \quad \Xi_{ab} \coloneqq q^{c}{}_{a}q^{d}{}_{b} \nabla_c l_d , \quad \Xi \coloneqq g^{ab} \Xi_{ab},\]
\begin{equation}\label{kappa}
\qand*\kappa \coloneqq \partial_{(r)}{}^a l^b \nabla_a \partial_{(r)}{}_b .
\end{equation}
All the previous definitions are well known in the geometry of null hypersurfaces \cite{Gourgoulhon2006}. In addition, we introduce the \emph{reduced transversal expansion scalar},
\begin{equation}\label{xidef}
\xi \coloneqq 2\Xi - \Theta,
\end{equation}
in order to eliminate the trivial divergent behavior of $ \Xi $ present in all cases.

Then, one can write $T_{ur}{}^{(0)}, T_{rr}{}^{(0)}, T_{r\theta}{}^{(1)}, \partial_{\theta} T_{r\theta}{}^{(1)}$ and $\partial_{\phi} T_{r\phi}{}^{(1)} $ in terms of the above definitions (see Appendix \ref{kinematics}, especially the comments below equation \eqref{kT0}) and find that
\begin{equation}\label{rhokine}
\begin{aligned}
\frac{\rho}{3} = &\; \frac{1}{8 \pi }\left((\kappa \Theta) ^{{(0)}}+\xi ^{(1)}\right) +\frac{1 }{2 \pi }\left(\cot (\theta ) \Omega ^H{}_{\theta }{}^{{(2)}} +\partial _{\theta } \Omega ^H{}_{\theta }{}^{{(2)}}+\csc ^2(\theta ) \partial _{\phi } \Omega ^H{}_{\phi }{}^{{(2)}}\right) -\\
&\; \frac{1}{2} \left(\csc ^2(\theta ) \partial _{\phi \phi } T_{{rr}}{}^{{(0)}}+\cot (\theta ) \partial _{\theta } T_{{rr}}{}^{{(0)}}  +\partial _{\theta \theta } T_{{rr}}{}^{{(0)}}\right) -\frac{1}{4 \pi }\left(( U^{{(0)}} )^2+\sin ^2(\theta ) ( W^{{(0)}} )^2\right). 
\end{aligned}
\end{equation}
Integrating the above equation to obtain the local mean density using the central conditions \eqref{ccg} for $ U^{{(0)}} $ and $ W^{{(0)}} $ yields a term that depends on $ a^2 = a_{\bar{i}} a^{\bar{i}}$ and $ \Omega^2 = \Omega_{\bar{i}} \Omega^{\bar{i}}$. Besides, by imposing conditions \eqref{expTBondi}, the integral of the terms involving the energy-momentum tensor components vanishes. Moreover, the terms containing $ \Omega ^H{}_{\theta }{}^{{(2)}} $ and its derivative can be integrated in $ \theta $, as $ \partial_\phi \Omega ^H{}_{\phi }{}^{{(2)}} $ can be integrated in $ \phi $, yielding
\begin{equation}\label{lmd}
\begin{aligned}
\frac{4\pi}{3}\bar{\rho}^{(0)} = &\; \int_{S^2} \frac{(\kappa \Theta) ^{{(0)}}+\xi ^{(1)}}{8 \pi } \sin(\theta) \dd{\theta} \dd{\phi} -\frac{4\pi}{3}\frac{a^2 +\Omega^2}{2\pi}+\\
&\; \int_{0}^{2\pi} \left(\left. \frac{\sin(\theta)\Omega ^H{}_{\theta }{}^{{(2)}}}{2\pi}\right|_{\theta=0}^{\theta=\pi}\right) \dd{\phi} + \int_{0}^{\pi} \left(\left. \frac{\Omega ^H{}_{\phi }{}^{{(2)}}}{2\pi\sin(\theta)}\right|_{\phi=0}^{\phi=2\pi}\right) \dd{\theta}.
\end{aligned}
\end{equation}
The integrand of the last integral vanishes identically. The other integral related to the Hájíček one-form equals zero if we impose the central conditions \eqref{expTBondi} on equation \eqref{Omegatheta}, suppose that $ \partial_\theta \gamma^{(2)} $ is finite along the axis, and impose the general axial condition \eqref{gac} on the terms that do not vanish after integration. Consequently,\footnote{This is another way of writing equation (8) of Box 17.2 of Reference \citeonline{Misner1973}. It is also an analogue of equation (37) of Reference \citeonline{Chrusciel2014}.}
\begin{equation}
\frac{4\pi}{3}\bar{\rho}^{(0)} = \int_{S^2} \frac{(\kappa \Theta) ^{{(0)}}+\xi ^{(1)}}{8 \pi } \sin(\theta) \dd{\theta} \dd{\phi} -\frac{4\pi}{3}\frac{a^2 +\Omega^2}{2\pi}.
\end{equation}

Since the expansion scalar quantifies the rate of change in area of the surfaces $ \mathcal{S}(u,r) $ along the null generators, we can relate the above expression to the integral mass formula for a non-rotating black hole \cite{Bardeen1973} by identifying the first term at the right side with the area term for a black hole. Then, the integral of the reduced transversal expansion can be linked to the integral related to the matter outside the event horizon, in which the integrand is projected onto the Killing vector field. For a non-rotating black hole, this projection becomes transversal to the event horizon at the intersection between the hypersurface of integration and the horizon. Apart from these common terms, the proper acceleration and the proper angular frequency appear explicitly in the previous equation in what we shall call the \emph{fictitious mass density},
\[ \rho_{fic} \coloneqq - \frac{a^2 +\Omega^2}{2\pi}. \]

To clarify the relation between the last equation and the integral mass formula, we define the following variable:
\[ Q \coloneqq \frac{3}{8\pi} \int_{\mathcal{B}_R} \kappa \Theta \dd{\Sigma} .\]
By using the definition \eqref{dSigma} and equations \eqref{kappadelrbeta} and \eqref{Theta} from Appendix \ref{kinematics}, one gets
\[ Q = \frac{3}{2\pi} \int_{S^2}\int_{0}^{R} (\partial_r \beta) e^{2\beta} r \sin(\theta) \dd{r} \dd{\theta} \dd{\phi}. \]
Expanding the integrand with respect to $ r $ and using the central conditions \eqref{ccg}, it follows that
\[ \begin{aligned}
Q & = \frac{3}{2\pi} \int_{S^2}\int_{0}^{R} \left[\beta^{(2)}r^2+ \mathcal{O}(r^3)\right]\sin(\theta) \dd{r} \dd{\theta} \dd{\phi}\\
& = \frac{4\pi R^2}{4\pi} \frac{1}{4\pi} \int_{S^2} 2\beta^{(2)}R\sin(\theta) \dd{\theta} \dd{\phi}  + \mathcal{O}(R^4). \end{aligned}\]
Then, by expanding equation \eqref{kappadelrbeta} as
\[\kappa= 2\beta^{(2)}r +\mathcal{O}(r^2)  \]
and using the average per solid angle
\[ \hat{\kappa} (u,r) \coloneqq \frac{1}{4\pi} \int_{S^2} \kappa(u,r,\theta,\phi) \sin(\theta) \dd{\theta} \dd{\phi}  , \]
one can show that
\[ Q = \frac{\hat{\kappa}(u,R)A_R}{4\pi}  + \mathcal{O}(R^4),  \]
where $ A_R $ is the area of $ \mathcal{S}(u,R) \simeq S^2 $. As a result, the associated density
\[ \bar{\rho}_\parallel \coloneqq \frac{Q}{V_R} \]
leads to
\begin{equation}\label{rhoparallel}
\frac{4\pi}{3}\bar{\rho}_\parallel{}^{(0)} = \int_{S^2} \frac{(\kappa \Theta) ^{{(0)}} }{8 \pi } \sin(\theta)\dd{\theta} \dd{\phi}  = \lim_{R\to 0} \frac{\hat{\kappa}(u,R)A_R}{4\pi}   \frac{1}{R^3}.
\end{equation}

For notational convenience, let
\[ \frac{4\pi}{3}\bar{\rho}_\perp{}^{(0)}\coloneqq  \int_{S^2} \frac{\xi ^{(1)}}{8 \pi } \sin(\theta) \dd{\theta} \dd{\phi}. \]
Thus,
\begin{equation}\label{rhosplit}
\frac{4\pi}{3}\bar{\rho}^{(0)} \dd{r}^3 = \frac{4\pi}{3} \left( \bar{\rho}_\parallel{}^{(0)}  + \bar{\rho}_\perp{}^{(0)} +  \rho_{fic} \right) \dd{r}^3
\end{equation}
may be seen as a representation for the total mass inside an infinitesimal ball around the observer. In this case, it is convenient to express the above limit using the expansion
\begin{equation}\label{lim}
\frac{\hat{\kappa}(u,r)A_r}{4\pi}   \frac{1}{r^3} = \lim_{R\to 0} \frac{\hat{\kappa}(u,R)A_R}{4\pi}   \frac{1}{R^3}     +\mathcal{O}(r)
\end{equation}
so that
\[ \frac{4\pi}{3}\bar{\rho}^{(0)} \dd{r}^3 = \frac{\hat{\kappa}(u,\dd{r})A_{\dd{r}}}{4\pi}  +\frac{4\pi}{3} \left( \bar{\rho}_\perp{}^{(0)} +  \rho_{fic} \right) \dd{r}^3 + \mathcal{O}(\dd{r}^4). \]
Accordingly, we shall call the lowest-order approximation in the above equation the \emph{infinitesimal gravitational mass}:
\begin{equation}\label{igm}
M_{\dd{r}}(u) \coloneqq \frac{\hat{\kappa}(u,\dd{r})A_{\dd{r}}}{4\pi}  +\frac{4\pi}{3} \left( \bar{\rho}_\perp{}^{(0)} +  \rho_{fic} \right) \dd{r}^3.
\end{equation}
This is the analogue of the integral mass formula for a non-rotating black hole. By definition, both are related to the coefficient of the leading-order term in a multipole expansion of the equivalent of the Newtonian gravitational potential constructed from the metric. Consequently, both can be interpreted as the gravitational mass of the system, i.e., they are the source of the gravitational field. Besides, they are split into a term arising from a vector field that is tangent to the causal horizon and another term related to structures that are transversal to it. What is truly remarkable in this analogy is the appearance of the ``tangent term'', or area term, since it has exactly the same form in both cases, including the same numerical factor. Given that this term is the one that connects black hole mechanics to thermodynamics, we shall pursue a thermodynamic analogy through an interpretation of the last equation that is adapted to our model of system and observer and by providing a link between the area term and the temperature and entropy of the system.

\subsection{Infinitesimal free energy}

To start exploring the possible thermodynamic meaning of equation \eqref{igm}, we first analyze the physical interpretation of the ``transversal term''. Using equation \eqref{xi1} of Appendix \ref{kinematics}, one finds that the terms containing components of the Hájíček one-form can be expressed as $ -1/6 $ times the last two integrals of equation \eqref{lmd}, so they vanish by the same arguments. Therefore,
\[ 
\begin{aligned}
\frac{4\pi}{3}\bar{\rho}_\perp{}^{(0)} = &\;
\frac{2}{3}\int_{S^2} T_{ur}{}^{(0)} \sin(\theta) \dd{\theta}\dd{\phi} -\frac{\Lambda}{12\pi} \int_{S^2} \sin(\theta) \dd{\theta}\dd{\phi} +\\
&\; \frac{1}{24\pi}\int_{S^2} \partial_\phi \left( 2\cot(\theta)\delta^{(2)} - \csc(\theta)\partial_\phi \gamma^{(2)} + 2 \partial_\theta \delta^{(2)} \right) \dd{\theta}\dd{\phi} +\\
&\; \frac{1}{24\pi}\int_{S^2} \partial_\theta \left( 2\cos(\theta)\gamma^{(2)} +  \sin(\theta)\partial_\theta \gamma^{(2)} \right) \dd{\theta}\dd{\phi}.
\end{aligned}
 \]
The third integral vanishes identically, and the last one equals zero if we consider that $ \partial_\theta \gamma^{(2)} $ is finite along the axis and use the general axial condition \eqref{gac}. Integration of the other two integrals employing the central conditions \eqref{expTBondi} yields
\[ \frac{4\pi}{3}\bar{\rho}_\perp{}^{(0)} = -\frac{8\pi}{3}\left( {T}^{\bar{0}\bar{0}{(0)}} + \frac{\Lambda}{8\pi}\right). \]

Before we proceed, it is worth checking once again the consistency of the definitions we have made so far. By calculating the ``tangent term'' \eqref{rhoparallel} using equation \eqref{kT0} and the central conditions \eqref{expTBondi}, one gets
\begin{equation}\label{parallel}
\frac{4\pi}{3}\bar{\rho}_\parallel{}^{(0)} = \frac{4\pi}{3} \left( 3{T}^{\bar{0}\bar{0}{(0)}} + {T}^{\bar{1}\bar{1}{(0)}} + {T}^{\bar{2}\bar{2}{(0)}} + {T}^{\bar{3}\bar{3}{(0)}}\right).
\end{equation}
Thus, we can write equation \eqref{rhosplit} as
\[ \frac{4\pi}{3}\bar{\rho}^{(0)} \dd{r}^3 = \frac{4\pi}{3} \left[ {T}^{\bar{0}\bar{0}{(0)}} + \frac{\Lambda}{8\pi} - \frac{a^2}{2\pi} + \sum_{\bar{i}} {T}^{\bar{i}\bar{i}{(0)}} - \frac{\Lambda}{8\pi} - \frac{\left(\Omega^{\bar{i}}\right)^2}{2\pi}  \right] \dd{r}^3.\]
With this result, we see that almost all the information contained in $ \bar{\rho}^{(0)} $ was already shown in equation \eqref{rhotraceT}. What one could do to complete that equation would be to include the contributions of $ \Lambda $, $ a^2 $, and $ \Omega^2 $ in the energy-momentum tensor, as if they were treated as matter fields with the corresponding energy density and pressures present in the last equation. However, since only the cosmological constant can be modeled in this way in the whole theory (not only in this equation) and any energy associated with $ \rho_{fic} $ ceases to exist if $ a^2 $ and $ \Omega^2 $ become zero, we shall treat on the same footing $ \pm \Lambda/8\pi $ and $ {T}^{\bar{\mu}\bar{\mu}{(0)}} $, but not $ \rho_{fic} $.

With this in mind and returning to equation \eqref{igm}, we consider that the ``real'' gravitational mass, capable of transferring energy to other systems, is given by
\[ M_{\dd{r}}(u) - \frac{4\pi}{3}\rho_{fic} \dd{r}^3= \frac{\hat{\kappa}(u,\dd{r})A_{\dd{r}}}{4\pi}  +\frac{4\pi}{3} \bar{\rho}_\perp{}^{(0)}\dd{r}^3.\]
Then, denoting by
\[ m_{fic}(u,\dd{r}) \coloneqq \frac{4\pi}{3}\rho_{fic}(u) \dd{r}^3 \]
the \emph{fictitious mass} and by
\[ E(u,\dd{r})
\coloneqq \frac{4\pi}{3}\left( {T}^{\bar{0}\bar{0}{(0)}} (u) + \frac{\Lambda}{8\pi}\right)\dd{r}^3 \]
the \emph{infinitesimal energy} inside the infinitesimal ball as measured by the observer's orthonormal basis $ \left\{ e_{(\bar{\mu})}{}^a \right\} $, we have that
\[ M_{\dd{r}}(u) - m_{fic}(u,\dd{r})= \frac{\hat{\kappa}(u,\dd{r})A_{\dd{r}}}{4\pi}  - 2 E(u,\dd{r}).\]

Following the results of black hole thermodynamics and based on the analogy with the integral mass formula, we shall define a temperature proportional to $ \hat{\kappa} $, which is proportional to an integral of $ {T}_{{rr}} $, and an entropy proportional to $ A_{\dd{r}} $. In case we neglect $ \sigma^{a b} \sigma_{a b} $ and consider a thermodynamic process in which this temperature is kept constant, then there should be a change in some ``thermodynamic energy'' while the angular average of $ {T}_{{rr}} $ is kept constant, which seems to be too restrictive for general physical processes. Moreover, the variations of physical quantities that are naturally related (in this model) to what the observer could measure are those given by $ u $-derivatives. Therefore, we are led to treat the temperature as the thermodynamic variable, instead of entropy, suggesting that the variations of the above equation are related to some form of free energy.

However, taking the $ u $-derivative of the area term does not change its numerical factor as the variation considered for the integral mass formula does, resulting in the differential mass formula of Reference \citeonline{Bardeen1973}. Thus, the product of temperature and entropy differs from that of black hole thermodynamics by a factor of two. In addition, the term $  - 2 E $ also contains what seems to be an extra factor of two, and its sign complicates the interpretation of the last equation as a thermodynamic potential that measures variations in the system's energy. To address these issues, we shall define the \emph{infinitesimal free energy} as
\begin{equation}\label{ife}
F_{\dd{r}}(u)\coloneqq E(u,\dd{r})+\frac{M_{\dd{r}}(u) - m_{fic}(u,\dd{r})}{2}= \frac{\hat{\kappa}(u,\dd{r})A_{\dd{r}}}{8\pi}  .
\end{equation}

To understand the meaning of this definition, we first note that the energy transfer that can be primarily associated with an active gravitational mass is the work done by the gravitational force it generates. In order to have this work transferred to another system, some mass $ m $ must be ``lowered'' in the direction of the active gravitational mass $ M_{\dd{r}} $. But it may happen that one lowers it so much that $ m $ passes through an event horizon, in which case the energy contained in $ m $ will be lost, and no more work will be available to be transferred to regions of spacetime other than the one inside the event horizon. Consequently, this work can be transferred only in a region where the areal radius $ r $ is greater than or equal to the Schwarzschild radius
\[ r_S= 2(M_{\dd{r}}+m). \]
First, consider that $ r_S>\dd{r} $. If we regard $ m $ as a point mass and use Newtonian gravity to compute the gravitational work, $ W_g $, done from infinity to $ r_S $, we find that
\[ W_g = \frac{M_{\dd{r}}m}{r_S} = \frac{M_{\dd{r}}}{r_S}\left( \frac{r_S}{2}-M_{\dd{r}}\right) =\frac{M_{\dd{r}}}{2} +\mathcal{O}(\dd{r}^5). \]
In case $ r_S\leq\dd{r} $, one may lower $ m $ down to $ \dd{r} $. But, past this point, we shall consider $ m $ as a part of the system. Then, a transfer of work may be thought of as a transfer of matter fields, related to a change in $ E $. For this reason, the maximum work that is not computed by $ E $ is
\[ W_g = \frac{M_{\dd{r}}m}{\dd{r}} = \frac{M_{\dd{r}}}{\dd{r}}\left( \frac{r_S}{2}-M_{\dd{r}}\right) \leq \frac{M_{\dd{r}}}{\dd{r}}\left( \frac{\dd{r}}{2}-M_{\dd{r}}\right) = \frac{M_{\dd{r}}}{2} +\mathcal{O}(\dd{r}^5). \]
Therefore, we can interpret the infinitesimal free energy as the sum of the total energy that can be exchanged by fluxes of matter fields (including a field that represents the cosmological constant) plus the lowest-order term of the maximum ``real'' gravitational work that can be extracted from the system without matter fluxes. All the other forces will be mediated by fields whose energies are already computed by $ E $, while the gravitational work is not. In this sense, $ F_{\dd{r}} $ may be viewed as the maximum amount of energy that can be transferred to other systems, which is suitable to a notion of thermodynamic potential.

\subsection{Inner and outer systems}

In order to generalize the thermodynamic potentials to systems of finite radius and to show that they can be understood as conserved quantities, we introduce a splitting of the light cone into inner and outer parts. Let
\[ \mathcal{S}_{\text{i}}(\tau, L)\coloneqq \left\{  \mathcal{Q}\in \mathcal{C}(\tau)\cap\mathcal{N}'\mid \eval{\lambda'}_{\mathcal{Q}}\leq L \right\} \qand \mathcal{S}_{\text{o}}(\tau, L)\coloneqq \mathcal{C}(\tau)\setminus \mathcal{S}_{\text{i}}(\tau, L) \]
be called the \emph{inner} and \emph{outer systems}, respectively (remind that $ \mathcal{N}' $ is the neighborhood covered by the coordinate system $ y^{\mu'} $).

Then, based on the sign and the numerical factor associated with the temperature \eqref{T} and on the indication of a possible infinitesimal generalization shown in Subsection \ref{infgen}, we are led to define the \emph{temperature} and the \emph{entropy} of $ \mathcal{S}_{\text{i}}(u, \dd{L}(\dd{r})) $ as
\begin{equation}\label{TS}
T(u,\dd{r})\coloneqq \frac{\hat{\kappa}(u,\dd{r})}{4\pi} \qand S(\dd{r})\coloneqq -\frac{A_{\dd{r}}}{2}=-2\pi \dd{r}^2 ,
\end{equation}
respectively. From these definitions, it follows that the variation in the infinitesimal free energy of $ \mathcal{S}_{\text{i}}(u, \dd{L}(\dd{r})) $ is given by
\begin{equation}\label{dF}
\dd{F}(u,\dd{r}) \coloneqq \partial _u F_{\dd{r}}\dd{u} = -S(\dd{r}) \partial _u T(u,\dd{r}) \dd{u}.
\end{equation}
The strange minus sign in the definition of $ S $ will be discussed in Section \ref{lct}.

\subsection{Free energy conservation}

To support the idea that the above defined free energy is associated with a conservation law, we shall analyze an example using the Landau-Lifshitz pseudotensor, which provides conserved charges by definition. Surprisingly, the total energy derived from it assumes a very simple form, shown in equation \eqref{Pu}. A first interesting remark that we can make is that it remains constant if the boundary of the system follows the $ \partial_{(u)}^a $ vector field. Therefore, we shall analyze how this energy changes with respect to $ u' $.

We consider a spherically symmetric spacetime and the inner system $ \mathcal{S}_{\text{i}}(u_1, L) $. Following the results of Appendix \ref{SSS} and equation \eqref{kappadelrbeta}, neither $ \beta $ nor $ \kappa $ will depend on angular coordinates, meaning that $ \hat{\kappa}=\kappa $. Then, we imagine that the initial matter distribution inside $ \mathcal{S}_{\text{i}} $ is held fixed while a fluid enters from $ \mathcal{S}_{\text{o}}(u_1, L) $, superimposing its contribution to the energy-momentum tensor on the initial component. This process starts at $ u=u_1 $, and the external fluid gradually occupies regions described by $ r \geq \tilde{R}(u')  $, with $ \tilde{R}(u')< R_1 $ for $ u'>u'_1 $, $ R_1\coloneqq r(u'_1,L) $, and $ \tilde{R}(u'_1)= R_1 $. We assume a constant contribution of the external fluid for the inaffinity, $ \Delta\kappa $, which vanishes for $ r < \tilde{R}(u')  $. As a result, one has, for $ u>u_1 $, $ \kappa(u,r) =\kappa(u_1,r) $ for $ r < \tilde{R}(u')  $ and $ \kappa(u,r) =\kappa(u_1,r)+\Delta\kappa $ for $ r \geq \tilde{R}(u')  $.

From equations \eqref{delrlambda} and \eqref{kappadelrbeta}, one finds that
\[ R(u')\coloneqq r(u',L)= \int_{0}^{L}\exp\left( -\int_{0}^{r(u',\lambda'')}\kappa(u,r') \dd{r'} \right)\dd{\lambda''},\]
where $ u=u' $. It follows that
\begin{multline*}
R(u') = \int_{0}^{\lambda'(u,\tilde{R})}\exp\left( -\int_{0}^{r(u',\lambda'')}\kappa(u_1,r') \dd{r'} \right)\dd{\lambda''} + \\ \int_{\lambda'(u,\tilde{R})}^{L}\exp\left( -\int_{0}^{\tilde{R}(u')}\kappa(u_1,r') \dd{r'} -\int_{\tilde{R}(u')}^{r(u',\lambda'')}\kappa(u_1,r')+\Delta\kappa \dd{r'}\right)\dd{\lambda''},
\end{multline*}
and, since $ r(u',\lambda'')=r(u_1',\lambda'') $ for $ \lambda''\leq \lambda'(u,\tilde{R}) $,
\begin{multline*}
R(u') = \int_{0}^{\lambda'(u,\tilde{R})}\exp\left( -\int_{0}^{r(u_1',\lambda'')}\kappa(u_1,r') \dd{r'} \right)\dd{\lambda''}+ \\ \int_{\lambda'(u,\tilde{R})}^{L}\exp\left( -\int_{0}^{r(u_1',\lambda'')}\kappa(u_1,r') \dd{r'} +\int_{r(u',\lambda'')}^{r(u_1',\lambda'')}\kappa(u_1,r')\dd{r'}\right)e^{-\Delta\kappa\left (r(u',\lambda'')-\tilde{R}(u')\right )}\dd{\lambda''}.
\end{multline*}

Then, we define
\[ l(u',\lambda'') \coloneqq \lambda'' - \lambda'(u(u'),\tilde{R}(u')), \]
and expand $ r(u_1',\lambda'') $ and $ r(u',\lambda'') $ around $ \lambda'(u(u'),\tilde{R}(u'))=\lambda'(u_1,\tilde{R}(u')) $ as
\[ r(u_1',\lambda'')=\tilde{R}(u')+e^{-2\beta(u_1',\lambda'(u_1,\tilde{R}))}l(u',\lambda'')+\mathcal{O}\left (l^2(u',\lambda'')\right )\]
\[  \qand* r(u',\lambda'')=\tilde{R}(u')+e^{-2\beta(u_1',\lambda'(u_1,\tilde{R}))}l(u',\lambda'')+\mathcal{O}\left (l^2(u',\lambda'')\right ), \]
given that $\beta(u',\lambda'(u_1,\tilde{R})) = \beta(u_1',\lambda'(u_1,\tilde{R})) $. Thus, $ r(u_1',\lambda'')-r(u',\lambda'')= \mathcal{O}\left (l^2(u',\lambda'')\right )$, and, by expanding $ \kappa(u_1,r') $ around $ r(u',\lambda'') $, it follows that
\[ \int_{r(u',\lambda'')}^{r(u_1',\lambda'')}\kappa(u_1,r')\dd{r'}= \mathcal{O}\left(r(u_1',\lambda'')-r(u',\lambda'')\right)=\mathcal{O}\left (l^2(u',\lambda'')\right ).\]

Using these results and expanding the exponentials related to them in the last equation for $ R $, we obtain
\[ R(u') = R_1-  \int_{\lambda'(u,\tilde{R})}^{L}e^{-2\beta(u_1,r(u_1',\lambda''))}\left(\Delta\kappa e^{-2\beta(u_1,\tilde{R})}l(u',\lambda'') +\mathcal{O}\left (l^2(u',\lambda'')\right )\right)\dd{\lambda''}. \]
We expand the remaining exponential around $ \lambda'(u(u'),\tilde{R}(u'))$ and change the integration variable to $ l(u',\lambda'') $ to get
\[ R(u') = R_1 - \Delta\kappa e^{-4\beta(u_1,\tilde{R})}\frac{l^2(u',L)}{2}+\mathcal{O}\left (l^3(u',L)\right ). \]
By defining $ v_{\tilde{R}}(u')\coloneqq \partial_{u'}l(u',L) $ and expanding $ e^{-4\beta(u_1,\tilde{R}(u'))} $ around $ u'_1 $, we have that
\[  l(u',L)=v_{\tilde{R}}(u'_1)(u'-u'_1) +\mathcal{O}\left((u'-u'_1)^2\right) \]
\[\qand* R(u') = R_1 - v_{\tilde{R}}^2(u'_1)\Delta\kappa e^{-4\beta(u'_1,L)}\frac{(u'-u'_1)^2}{2}+\mathcal{O}\left ((u'-u'_1)^3\right ). \]
From the last result, one can read the first right derivatives
\[  \eval{\partial_{u'}R(u')}_{u'_1}=0 \qand \eval{\partial_{u'u'}R(u')}_{u'_1} = - v_{\tilde{R}}^2(u'_1) \Delta\kappa  e^{-4\beta(u'_1,L)}.\]

It follows from equation \eqref{Pu} and the above derivatives that
\begin{equation}\label{PuuL}
P^u(u')= P^u(R(u'))=P^u(u'_1)+\frac{3\pi}{8}v_{\tilde{R}}^2(u'_1) R_1^2 \Delta\kappa  e^{-4\beta(u'_1,L)}(u'-u'_1)^2+\mathcal{O}\left ((u'-u'_1)^3\right ).
\end{equation}
Then, we define
\begin{equation}\label{DF}
\Delta F(R_1) \coloneqq \int_{u_0}^{u_2} \dd{F}(u,R_1)=\frac{R_1^2}{2}\Delta\kappa
\end{equation}
for $ u_0<u_1<u_2 $, where the $ u $-derivative in definition \eqref{dF} is to be understood as a distributional derivative in this case. Therefore,
\[ P^u(u')= P^u(u'_1)+\frac{3\pi}{4}v_{\tilde{R}}^2(u'_1) \Delta F(R_1)  e^{-4\beta(u'_1,L)}(u'-u'_1)^2+\mathcal{O}\left ((u'-u'_1)^3\right ).\]
If the total energy of the composite system $\mathcal{C}(u_1)= \mathcal{S}_{\text{i}}(u_1, L)\cup \mathcal{S}_{\text{o}}(u_1, L) $ is conserved, then its $ u' $-derivatives will be equal to zero to all orders, and the coefficients of the above expression will represent the contribution of the inner system to the variation of conserved quantities. Given that $ v_{\tilde{R}}^2(u'_1) e^{-4\beta(u'_1,L)} $ is a factor common to both inner and outer systems, we conclude that $ F $ is conserved for the composite system and $ \Delta F(R_1)  $ is the contribution of the inner system to the variation of $ F $.

Moreover, by expanding the exponential and using the central conditions \eqref{ccg} in the previous equation, one finds that
\begin{equation}\label{PuR}
P^u(u')-P^u(u'_1) = \frac{3\pi}{4}v_{\tilde{R}}^2(u'_1) \Delta F(R_1)  (u'-u'_1)^2+\mathcal{O}\left ((u'-u'_1)^3\right )+\mathcal{O}\left (R_1^4(u'-u'_1)^2\right ).
\end{equation}
Consequently, one can say that, for $ R_1\ll 1 $, the leading-order term of the variation in total energy is given by a kinematic parameter of the model, $ v_{\tilde{R}}^2(u'_1) $, and a dynamical factor, $ \Delta F(R_1) $.

\subsection{Internal energy and generating functions}

Now that we have a suitable definition for the free energy of the inner system, it becomes desirable to define its internal energy through the usual Legendre transformation. However, it is necessary that this definition be associated with some conservation law in order to keep it in accordance with classical thermodynamics. Therefore, we shall adapt the previous analysis of the free energy to a case in which the temperature \eqref{TS} is kept constant. Since the variations in internal energy will be given by variations in entropy, and, therefore, in the areal radius, the problem can be viewed either as a virtual process where the boundary of the outer system is moved on a fixed light cone $ u $ or a real process in a static spacetime. For simplicity, we adopt the first view and suppress the $ u $-dependence of all functions, which is tantamount to considering a static spherically symmetric spacetime.

Let $ \kappa $ be a function of $ r $ for $ r<\bar{R} $ and a constant, $ \bar{\kappa} $, for $ r\geq \bar{R} $, where $ \bar{R}\coloneqq r(\bar{L}) $ and $ \bar{L} $ is some value for $ \lambda' $. The boundary of the inner system is at $ L(R) $, where $ R $ is a function of $ \lambda' $ to be expanded around $ \bar{L} $. From equations \eqref{delrlambda} and \eqref{kappadelrbeta}, it follows that
\begin{equation}\label{RlL}
R(\lambda';\bar{L})= \bar{R}+e^{-2\beta(\bar{L})}(\lambda'-\bar{L})-e^{-4\beta(\bar{L})}\bar{\kappa}\frac{(\lambda'-\bar{L})^2}{2}+\mathcal{O}\left((\lambda'-\bar{L})^3\right). 
\end{equation}

Then, we use definition \eqref{TS} to write equation \eqref{Pu} as
\[ P^u=-\frac{[-S(R)]^{3/2}}{8\sqrt{2\pi}}, \]
take its $ S $-derivative, express it in terms of $ R $, expand the exponentials, and use the central conditions \eqref{ccg} and equations \eqref{ytoz} to find that
\begin{equation}\label{DSPu}
D_S P^u =\frac{3}{16}R,
\end{equation}
or
\begin{multline}\label{dPdS}
D_S P^u(\lambda',\bar{L}) =  \frac{3}{16}\lambda'-\frac{3\pi}{4} \frac{\bar{\kappa}}{4\pi}\frac{(\lambda'-\bar{L})^2}{2}+\\ \mathcal{O}\left(\bar{L}^3\right)+\mathcal{O}\left(\bar{L}^2(\lambda'-\bar{L})\right)+\mathcal{O}\left((\lambda'-\bar{L})^3\right)+\mathcal{O}\left(\bar{L}^2(\lambda'-\bar{L})^2\right).
\end{multline}
If the total energy of the composite inner-plus-outer system is conserved, the negative of this expression will correspond to the derivative with respect to $ S $ of the total energy of the outer system, and each coefficient can be viewed as the variation of some conserved quantity. By identifying the temperature as the first non-trivial, information carrying coefficient, it becomes natural to define this coefficient as the variation of internal energy with respect to entropy.

Before we do that, it is desirable to give some thermodynamic meaning to this expression. Since this $ S $-derivative carries information about temperature, we shall seek to define some generalized form of internal energy. Therefore, we shall normalize the total energy in such a way that the first term in the right-hand side of the above equation is eliminated, as well as the factor $ -3\pi/4 $ that multiplies the temperature in the total energy variation of the inner system. From these requirements, we define the \emph{generating internal energy} as
\begin{equation}\label{GIE}
U_G(R)\coloneqq -\frac{4}{3\pi}P^u(R)+\int_{0}^{S(R)}\frac{\lambda'(S')}{4\pi}\dd{S'}=\frac{1}{\pi^2}\int_{0}^{2\pi}\int_{0}^{\pi}\int_{0}^{R}(r-\lambda'(r))r\sin[2](\theta) \dd{r}\dd{\theta}\dd{\phi}.
\end{equation}

The conservation of total energy implies that $ U_G $ is a conserved charge with respect to variations in $ S $, since these variations only change how one splits the above integral into inner and outer parts when $ U_G $ is calculated for the composite system. Moreover, the above definition and equation \eqref{DSPu} imply that
\begin{equation}\label{dUdS}
D_{S(R)} U_G(R)= \frac{\lambda'(R)-R}{4\pi}.
\end{equation}
Now, from equations \eqref{delrlambda}, \eqref{beta} and \eqref{shear2}, we have that
\begin{equation}\label{lambdaint}
\lambda'(R)=\int_{0}^{R}\exp\left(\int_{0}^{r}r'\left(\frac{\sigma^2}{2}+4\pi T_{rr}\right)\dd{r'}\right) \dd{r}.
\end{equation}
Since $ \sigma^2 \geq 0$ by definition and the null energy condition \eqref{nec} is assumed, it follows that $ \lambda'(R)\geq R $,
\begin{equation}\label{TUineq}
D_{S(R)} U_G(R)\geq 0, \qand U_G(R)\leq 0. 
\end{equation}
In particular, equalities will hold in Minkowski, de Sitter, and anti-de Sitter spacetimes. Consequently, if one takes $ U_G $ as a generalization of internal energy with $ S $ as a thermodynamic variable, the interpretation of $ S $ as entropy is kept and one gets a positive \emph{generating temperature},
\begin{equation}\label{GT}
T_G(R)\coloneqq \frac{\lambda'(R)-R}{4\pi}.
\end{equation}

In addition, one can use equations \eqref{GIE} and \eqref{dPdS} to show that
\[ D_{S} U_G(\lambda',\bar{L}) =  \frac{\bar{\kappa}}{4\pi}\frac{(\lambda'-\bar{L})^2}{2}+\\ \mathcal{O}\left(\bar{L}^3\right)+\mathcal{O}\left(\bar{L}^2(\lambda'-\bar{L})\right)+\mathcal{O}\left((\lambda'-\bar{L})^3\right)+\mathcal{O}\left(\bar{L}^2(\lambda'-\bar{L})^2\right). \]
If the above function is analytic around the origin, then
\[ \eval{\partial_{\lambda'\lambda'}D_{S} U_G(\lambda',\bar{L})}_{(0,0)} =\frac{\bar{\kappa}}{4\pi}  \]
and we are led to introduce the variations in the \emph{infinitesimal internal energy} of $ \mathcal{S}_{\text{i}}(u, \dd{L}(\dd{r})) $ as
\begin{equation}\label{dU}
\dd{U_{\dd{r}}}(u)\coloneqq T(u,\dd{r})\dd{S(\dd{r})}.
\end{equation}
In the example above, $ \dd{r}=R $ and $ T(u,\dd{r}) = \bar{\kappa}/4\pi $, implying that
\begin{equation}\label{delSU}
\eval{\partial_{\lambda'\lambda'}D_{S(\lambda',\bar{L})} U_G(\lambda',\bar{L})}_{(0,0)} =D_{S(\dd{r})} U_{\dd{r}}(u).
\end{equation}
One can generalize the above procedure for non-analytic functions at the expense of some abuse of notation, namely, by defining the action of derivatives on generating functions evaluated at some point as selecting their coefficients as they would do in the analytic case.

The construction of $ U_G $ and $ T_G $ also indicates that a similar quantity may be related to the free energy. Thus, we define the \emph{generating free energy} via the usual Legendre transform \cite{Callen1985}:
\begin{equation}\label{GFE}
F_G(R)\coloneqq U_G(R)-T_G(R)S(R).
\end{equation}
From definitions \eqref{GIE}, \eqref{GT}, and \eqref{TS}, one can change the integration variable to $ r $, integrate by parts, and use equations \eqref{delrlambda} and \eqref{Pu} to obtain
\begin{equation}\label{FG}
F_G(R)=\frac{2}{3\pi}P^u(R)+\frac{1}{2}\int_{0}^{R}e^{2\beta}r^2\dd{r} = \frac{1}{2\pi^2}\int_{0}^{2\pi}\int_{0}^{\pi}\int_{0}^{R}(e^{2\beta}-1)r^2\sin[2](\theta) \dd{r}\dd{\theta}\dd{\phi}.
\end{equation}
Equations \eqref{beta}, \eqref{ccg}, and \eqref{nec} imply that
\[ \beta\geq 0 \qand F_G(R)\geq 0, \]
and, again, equalities are attained in Minkowski, de Sitter, and anti-de Sitter spacetimes.

One could also have changed the integration variable in the previous equation to $ \lambda' $, leading to
\[ F_G(u',L)=\frac{2}{3\pi}P^u(u',L)+\frac{1}{2}\int_{0}^{L}r^2(\lambda')\dd{\lambda'},  \]
where $ L $ is the affine boundary used in the previous subsection. Then, it follows from equations \eqref{PuuL}, \eqref{ytoz}, and \eqref{ccg} that
\begin{multline*}
F_G(u',L)=\frac{2}{3\pi}P^u(u'_1,L)+\frac{L^3}{6}+\frac{v_{\tilde{R}}^2(u'_1) \Delta \kappa}{4} L^2 (u'-u'_1)^2+\\
\mathcal{O}\left ((u'-u'_1)^3\right )+\mathcal{O}\left (L^5\right ) +\mathcal{O}\left (L^4(u'-u'_1)^2\right ),
\end{multline*}
and equation \eqref{DF} can be used to show that
\begin{equation}\label{delFG}
\eval{\partial_{LLu'u'}F_G(u',L)}_{(u'_1,0)}\dd{r}^2 =2v_{\tilde{R}}^2(u'_1) \Delta F(\dd{r}).
\end{equation}

Similarly, equations \eqref{GT}, \eqref{RlL}, \eqref{ccg}, and \eqref{ytoz} can be used to expand $ T_G $ with respect to $ \lambda' $ and $ \bar{L} $:
\begin{equation}\label{TGlL}
T_G(\lambda',\bar{L}) = \frac{\bar{\kappa}}{4\pi}\frac{(\lambda'-\bar{L})^2}{2}+ \mathcal{O}\left(\bar{L}^3\right)+\mathcal{O}\left(\bar{L}^2(\lambda'-\bar{L})\right)+\mathcal{O}\left((\lambda'-\bar{L})^3\right)+\mathcal{O}\left(\bar{L}^2(\lambda'-\bar{L})^2\right).
\end{equation}
Therefore,
\[ \eval{\partial_{\lambda'\lambda'}T_G(\lambda',\bar{L})}_{(0,0)} =\frac{\bar{\kappa}}{4\pi} =  T(u,\dd{r}), \]
and we justify the adjective ``generating'' in the name of the functions we have defined.

\subsection{Euler relations}

Other interesting results can be derived from the previous definitions as follows. Using definitions \eqref{GIE}, \eqref{GT}, and \eqref{TS}, one finds that
\[ U_G(R)= -\int_{0}^{R} T_G(r)4\pi r\dd{r}= \int_{0}^{S(R)} T_G(S')\dd{S'} .\]
Then, by ignoring the fact that equations \eqref{ytoz} imply that $ T_G(0)=0 $, one can consider $ T_G $ as a positive constant, or a virtual independence between $ T_G $ and $ S $, and recover the Euler relation \cite{Callen1985}
\[ U_G(R)= T_G S(R). \]

In addition, definition \eqref{GT} and equation \eqref{delrlambda} yield
\[ D_r T_G(r)= \frac{e^{2\beta}-1}{4\pi}.\]
If there exists an interval $ (0,r) $ in which $ \beta(r)\neq 0 $, then $ T_G(r) $ is invertible and one can use equation \eqref{FG} and definition \eqref{TS} to show that
\[ F_G(R)=- \int_{0}^{R} S(r) D_r T_G(r) \dd{r}=- \int_{0}^{T_G(R)} S(T'_G)\dd{T'_G}.\]
Again, to consider a virtual independence between $ T_G $ and $ S $ leads to another Euler relation:
\[ F_G(R)= - S T_G(R) . \]

Since this virtual independence will not actually hold, it seems prudent to distinguish the usual Euler relations from the \emph{generating Euler relations}:
\[  U_G(S)= \int_{0}^{S} T_G(S')\dd{S'} \qand F_G(T_G) = - \int_{0}^{T_G} S(T'_G)\dd{T'_G}. \]

\section{Light cone thermodynamics}\label{lct}

\subsection{Prelude}

In order to establish a clear connection with classical thermodynamics, we first remind that the only dynamical equations considered so far are the Einstein field equations. Consequently, even if the previous definitions for the free and internal energies represent the ``total energy'' in some sense, they do not convey all the information about the evolution of the system. Then, it becomes reasonable to assume that the present thermodynamic description contains information only about gravitational phenomena, which do not relate directly to the usual thermodynamic properties of matter, like the thermodynamic pressure.

In addition, the analysis of Section \ref{iisec} indicates that the entropy \eqref{TS} cannot account for the second law alone, since an expanding fluid together with a negligible cosmological constant cannot satisfy the irreversible inequality \eqref{ii}. As a result, one has to consider ordinary entropy too, and the most obvious candidate to represent the total entropy is the sum of $ S $ and ordinary entropy. In turn, this suggests that the total internal energy is given by the sum of $ U_G $ and ordinary internal energy.

If that is the case, then both the total entropy and the ``overall temperature'' will depend on the ordinary thermodynamic variables, which are clearly not included in $ T_G $. It follows that this ``overall temperature'' cannot be identically equal to $ T_G $, and we are led to consider equilibrium states with two distinct concepts of temperature and of entropy. In other words, the variation of the total internal energy will contain two $ T\dd{S} $ terms that cannot be unified.

To interpret this situation, we first note that the results of Section \ref{qlhre} suggest that Hawking radiation emission can occur under mild conditions. Moreover, since the temperature associated with it is determined by the inaffinity, a similar physical interpretation could be attributed to this scalar field wherever it is defined. The indication of a possible infinitesimal generalization presented in Subsection \ref{infgen} reinforces this view. Then, we conjecture that the temperature \eqref{TS} is associated with a Bose-Einstein distribution that contributes to the spectrum of the vacuum excited by the matter fields on top of it and that the generating temperature \eqref{GT} may contain additional information about this spectrum. Following this line of reasoning, we shall attribute the meaning of temperature to $ T $ and $ T_G $ and of entropy to $ S $ when they are viewed as properties of the underlying vacuum.

If one does not know that $ T_G $ and $ S $ have interpretations of temperature and entropy, respectively, then one could treat a variation in $ U_G $ as a work term related to the matter fields. One can argue that this duality of interpretations is inherited from the Einstein field equations, since all the quantities in them can be understood as properties of matter and of spacetime concomitantly. Then, one could say that the work done by matter in changing the area of constant $ \lambda' $ surfaces, or in generating the gravitational field, or on spacetime itself, can be interpreted, at the same time, as heat transferred to the vacuum.\footnote{If one regards changes in the area of constant $\lambda'$ surfaces as a sort of ``metrical elasticity'' associated with the vacuum energy, then one can recognize interesting echoes of Sakharov's ideas \cite{Sakharov1991}.} When all the terms in the total internal energy are associated with matter, the ``overall temperature'' and the total entropy become naturally associated with the matter fields.

It appears that matter and vacuum can be regarded as two distinct thermodynamic systems separated by an adiabatic wall, but coexisting in equilibrium at the same location. From that point of view, the usual zeroth law of thermodynamics seems to be inappropriate to tackle these two unusual temperatures. To circumvent this problem, we shall adopt a postulational approach similar to that of Reference \citeonline{Callen1985}, in which equilibrium is defined by the state of the system itself.

With this purpose, it is interesting to first understand which new piece of information $ U $ and $ S $ could provide in a more general thermodynamic theory, and this can be more easily seen in spherically symmetric dusty spacetimes. In this case, equations \eqref{TS}, \eqref{kappadelrbeta}, \eqref{betasphe}, \eqref{Trr}, \eqref{ccg}, and \eqref{ivcmuz}, together with the assumption that $ \mu^{(0)}\neq 0 $, imply that the leading-order coefficient of the temperature is $ T^{(1)} = T_{rr}{}^{(0)}=\mu^{(0)}$. Thus, the infinitesimal internal energy determines $ \mu^{(0)} $ through equation \eqref{dU} in the same way that ordinary internal energy determines ordinary intensive parameters.

In the general case, one can use equations \eqref{rhoparallel}, \eqref{lim}, and \eqref{parallel} to show that
\[ T^{(1)} = {T}^{\bar{0}\bar{0}{(0)}} + \frac{{T}^{\bar{1}\bar{1}{(0)}} + {T}^{\bar{2}\bar{2}{(0)}} + {T}^{\bar{3}\bar{3}{(0)}}}{3}. \]
Therefore, if the mean pressure,
\[ \bar{p} \coloneqq \frac{{T}^{\bar{1}\bar{1}{(0)}} + {T}^{\bar{2}\bar{2}{(0)}} + {T}^{\bar{3}\bar{3}{(0)}}}{3},\]
is determined by ordinary thermodynamics, then we still can say that $ U $ and $ S $ determine the local value of the energy density.

{\emergencystretch=1em In a remarkable congruence of physical meanings, all the eigenvalues of the energy-momentum tensor of a perfect fluid become thermodynamic quantities derived from intensive parameters. Furthermore, since the proper energy density of a perfect fluid is the sum of rest-mass energy density and ordinary, or \emph{classical internal energy density}, $ u_{\text{c}} $, the above conclusion suggests that we can define the rest-mass density from thermodynamic parameters:
\begin{equation}\label{rest}
\rho_{\text{r}}\coloneqq T^{(1)}-\bar{p}-u_{\text{c}}. 
\end{equation}
Then, we can generalize this thermodynamic interpretation of rest mass by assuming that, with knowledge of the four-velocity and of the ordinary thermodynamics associated with the fluid, one could determine the rest-mass density of any matter field through a thermodynamic analysis. \par}

We conclude that the new information brought by this extension of classical thermodynamics concerns the dynamical aspects of rest mass. It ceases to be fixed by constituent numbers (and ordinary internal energy) and acquires an active status with a new pair of thermodynamic conjugated variables associated with it. In other words, rest-mass density becomes dissociated, at least in principle, from conservation laws of constituent numbers, like baryon and lepton numbers; it becomes a ``gravitational potential'' (in the same sense as that of the chemical potential, at least in dusty spacetimes) that is determined by the work it does on spacetime.

Despite these similarities between the proposed thermodynamic extension and ordinary thermodynamics, a structural difference appears when one notes that area is not proportional to either volume or ordinary constituent number. Therefore, one cannot construct a density from $ S $ and use it in some local form of thermodynamics as it is usually done in relativistic contexts. This can be seen as a natural consequence of the fact that gravity does not exist locally and justifies the somewhat cumbersome use of infinitesimals in $ U_{\dd{r}} $, $ T $, and $ S $. If $ S $ is supposed to represent the entropy associated with gravitational collapse and this process loses its meaning locally, then $ S $ should do the same.

Then, one is led to consider a quasi-local formulation for this thermodynamic extension. Here, another issue appears, namely, the fact that gravity turns homogeneity into an unnatural property of physical systems. Thus, we shall not impose such a restriction, in contrast to ordinary thermodynamics. As a consequence, we have to content ourselves with a thermodynamic description that yields less information about the system. Specifically, we shall use quasi-local definitions to determine local quantities in the limit $ r\to 0 $.

\subsection{Postulates}\label{post}

In order to propose a theory that is as general as possible, we first generalize the definition \eqref{GIE} to dynamical, asymmetrical spacetimes by including the dependence of $\lambda'$ on $ u, \ \theta, \ \text{and} \ \phi $:
\begin{equation}\label{gGIE}
U_G(u,R)\coloneqq\int_{0}^{R} \left\{ \frac{1}{\pi^2}\int_{0}^{2\pi}\int_{0}^{\pi}[r-\lambda'(u,r,\theta,\phi)]\sin[2](\theta) \dd{\theta}\dd{\phi}\right\} r\dd{r}.
\end{equation}
It is important to note that only inner systems of the form $ \mathcal{S}_{\text{i}}(u, \lambda'(u,R,\theta,\phi)) $ will be considered, where $ R $ is a fixed radius. This determines how one varies the boundaries of successive inner systems in the limit $ R\to0 $ and makes $ S $-derivatives well-defined operators. Additionally, we suppose that, for each $ u $, there exists a value $ R_{\text{th}} $, called the \emph{thermodynamic radius}, inside which all the concepts to be constructed are well defined. Hence, all inner systems to be considered from now on are those lying inside a thermodynamic radius, and we shall say that an inner system of the form above is \emph{centered} at $ (u,0,0,0) $.

Using definition \eqref{TS}, one can write the above one as
\begin{equation}\label{UG}
U_G(u,S)=\int_{0}^{S} \left\{ \frac{1}{\pi^2}\int_{0}^{2\pi}\int_{0}^{\pi}\frac{\lambda'(u,S',\theta,\phi)-r(S')}{4\pi}\sin[2](\theta) \dd{\theta}\dd{\phi}\right\} \dd{S'}.
\end{equation}
It follows from the arguments above inequalities \eqref{TUineq} that
\begin{equation}\label{TG}
\partial_S U_G = \frac{1}{\pi^2}\int_{0}^{2\pi}\int_{0}^{\pi}\frac{\lambda'(u,S,\theta,\phi)-r(S)}{4\pi}\sin[2](\theta) \dd{\theta}\dd{\phi} \eqqcolon T_G(u,S) \geq 0 \qand  U_G\leq0. 
\end{equation}

From the above definition for $ T_G $, one can express $ \lambda' $ as the integral of equation \eqref{delrlambda}, expand $ \beta $ with respect to $ r $, use the central conditions \eqref{ccg}, equation \eqref{beta2}, and the central conditions \eqref{expTBondi}, divide the resulting equation by $ r^3(S) $, invert definition \eqref{TS} for $ r $, and take the limit $ S\to0 $ to find that
\[ \lim_{S\to 0} 6(2\pi)^{3/2} \frac{T_G(u,S)}{(-S)^{3/2}}= {T}^{\bar{0}\bar{0}{(0)}}(u) + \frac{3{T}^{\bar{1}\bar{1}{(0)}}(u) + 3{T}^{\bar{2}\bar{2}{(0)}}(u) + 2{T}^{\bar{3}\bar{3}{(0)}}(u)}{8}. \]
In the case of a perfect fluid at rest in the proper reference frame of the observer with classical internal energy density $ u_{\text{c}} $ and isotropic pressure $ p $, the last equality can be used to calculate the rest-mass density in a way alternative to equation \eqref{rest}:
\begin{equation}\label{restlim}
\rho^{(0)}_{\text{r}}(u)=\lim_{S\to 0} 6(2\pi)^{3/2} \frac{T_G(u,S)}{(-S)^{3/2}}-p^{(0)}(u)-u^{(0)}_{\text{c}}(u).
\end{equation}
As before, if $ p^{(0)} $ and $ u^{(0)}_{\text{c}} $ are determined by ordinary thermodynamics, then the proposed extension completes the description of the system based on its gravitational effects.

But to restore ordinary thermodynamics and make it compatible with the above ideas, we have to prescribe how to attribute a volume to the inner systems under consideration. Conveniently, this was already done in defining the scalar surface element \eqref{dSigma}, and the volume of $ \mathcal{S}_{\text{i}}(u, \lambda'(u,R,\theta,\phi)) $ is given by
\[ V(u,R)=\int_{\mathcal{S}_{\text{i}}}\dd{\Sigma}=\int_{0}^{2\pi}\int_{0}^{\pi}\int_{0}^{R} e^{2\beta(u,r,\theta,\phi)}r^2 \sin(\theta) \dd{r} \dd{\theta} \dd{\phi}=\frac{4\pi}{3}R^3+\mathcal{O}\left(R^5\right). \]
From the last equality, one sees that $ V $ plays a role similar to that of the previous generating functions, containing the ``ordinary volume'' $ 4\pi R^3/3 $ in its leading-order term.

It becomes clear that, in the limit $ R\to0 $, the higher-order terms, which contain the degrees of freedom associated with $ \kappa $, and, thus, with $ T_{rr} $, are suppressed faster than the ``ordinary volume''. Therefore, we can say that, locally, $ S $ and $ V $ lose their mutual independence and the dependence on $ S $ is effectively suppressed from the manifold of equilibrium states, while $ V $ preserves a local meaning in the densities of extensive variables. We also note that, close to the vertex of the light cone, the properties of isotropy and homogeneity may be recovered to a sufficient approximation. As a result, it becomes reasonable to expect that ordinary thermodynamics can provide the usual information, like isotropic pressure\footnote{Possibly through expressions like $ p=\lim_{R\to0}Nk_B T/V $.}, in the limit $ R\to0 $, with the additional information about the rest-mass density being retained by the proposed extension, as shown in equation \eqref{restlim}.

To formalize these ideas, we first translate the concept of ``macroscopic properties of matter determined by classical thermodynamics'' to the present context. Since they are represented by intrinsic, nonkinematic parameters measured in the rest frame of the system, we can say that these properties are characterized by the scalar fields associated with the local matter content.

Next, we establish a few concepts, namely, the \emph{classical internal energy},
\[ U_C(u,R) \coloneqq \int_{\mathcal{S}_{\text{i}}} u_{\text{c}} \dd{\Sigma}, \]
the \emph{total internal energy},
\[ U_T(u,R)\coloneqq U_C(u,R)+U_G(u,R),\]
and
\begin{definition}
Independent thermodynamic variables: the set of functions of inner systems composed of $ U_T $, $ S $, $ V $, and other possible extensive parameters obtained by integration of their densities with respect to the surface element \eqref{dSigma} over the inner system; these functions are evaluated by an observer locally at rest with respect to all the matter fields.
\end{definition}

From that, we specify the notion of equilibrium as
\begin{definition}
	Local equilibrium: the attribute given to a subset of spacetime in which the matter fields are completely characterized by a four-velocity and a set of scalar fields whose values are determined by functions of the independent thermodynamic variables in the limit $ S\to 0 $.
\end{definition}
Hence, if one accepts the idea that the results of classical thermodynamics can be recovered in the limit proposed and recognizes the existence of systems that are described by these results under certain circumstances, then it becomes evident that
\begin{postulate}
	There exist subsets of spacetimes in local equilibrium.
\end{postulate}

As suggested before, we consider that the full theory will be dictated by the concept of \emph{total entropy}, $ S_T $, which we conceive as the sum of ordinary, or \emph{classical entropy}, $ S_C $, and $ S $. In addition, the interpretation of $ S $ as an entropy seems to be better suited for the view of $ S $ as a property of the vacuum. Thus, we define the \emph{vacuum entropy},
\[ S_V(S)\coloneqq S, \]
to separate the roles of independent thermodynamic variable and of entropy attributed to this parameter.

To make clear that $ S_V $ does satisfy an inequality analogue to the second law in, at least, some typical gravitational collapses, we use the irreversible inequality \eqref{ii}, definition \eqref{TG}, equation \eqref{UG}, and the sign of $ S $ to show that
\[ \partial_u T_G (u,S) > 0 \qand \partial_u U_G (u,S) < 0, \quad \forall S\neq 0.\]
Besides, it follows from the first inequality in \eqref{TG} that $ U_G $ increases monotonically with $ S $. Therefore, the requirement that $ U_G $ keeps its value constant implies that $ S $ has to increase. At this point, we justify the sign of definition \eqref{TS} with the standard inequality of the second law, rejecting the existence of two distinct arrows of time\cite{Bonnor1987} and keeping the proposed temperatures positive.

Then, one can generalize the results of Section \ref{iisec} and the above conclusion and suppose that, for any sequence of events in local equilibrium along a fixed worldline tangent to the unique four-velocity of the matter fields, the surfaces of constant independent thermodynamic variable $ U_G $ centered at these events will have values of $ S_V $ whose variation in the future direction obey the inequality
\begin{equation}\label{2law}
\Delta S_V\geq 0,
\end{equation}
whenever gravitational collapse is the main process inside these surfaces. Extending the generalization even further, one can consider that this supposition merges smoothly with the usual principle of entropy maximization, turning the following into a reasonable proposition:
\begin{postulate}
	There exists a function, which is to be called the total entropy, $ S_T $, of the independent thermodynamic variables of any inner system centered at an event in local equilibrium that determines the values of these variables by a maximization procedure. Namely, for any pair of inner systems with the same value of $ U_T $ that are centered at two different events in local equilibrium connected by an integral curve of the unique four-velocity of the matter fields, the values of the remaining independent thermodynamic variables of the inner system with the greater value of $ u $ will be those that maximize the total entropy subjected to a set of constraints compatible with the independent thermodynamic variables of the inner system of lower $ u $.
\end{postulate}

Perhaps the most convincing argument in favor of the above proposition is the fact that it can be viewed as a generalization of the recently published ``maximum entropy conjecture for black hole mergers'' \cite{Ramirez2026}. The analogue Kerr black hole considered in this conjecture is described by the first multipole moments of a negative-power expansion far away from the system, i.e., its mass and angular momentum. In a purely gravitational process, the relevant independent thermodynamic variable is $ U_G $, which, in the limit $ S\to 0 $, is effectively replaced by the first nonvanishing multipole moment, $ U_{\dd{r}} $, of a positive-power expansion associated with an observer immersed in the system. If one maps one of these notions of first multipole moments into the other one and replaces the constraints mentioned in the postulate by the balance laws of Reference \citeonline{Ramirez2026}, then the analogy between both proposals becomes apparent as well as the generalization to processes that include non-gravitational effects.

In classical thermodynamics, one has, at least in principle, complete control of the constraints imposed on the system, and this is what makes the parameters in question easily measurable and the theory fundamentally phenomenological. However, the usual, simple constraints found in ordinary thermodynamics look much more artificial in the present context, where one expects that the evolution of the system will be governed solely by the laws of physics, including the Einstein field equations. In fact, this is exactly what motivated the irreversible inequality \eqref{ii} and the gravitational second law \eqref{2law}: typical gravitational processes constrained by the Einstein field equations leading to a maximum permissible increase in $ S_V $. From this point of view, the above principle of entropy maximization seems to merge in some aspects with the principle of least action, aligning with the idea that entropy, free energy and action are closely related concepts in the context of general relativity \cite{Gibbons1977action,Hawking1996,Padmanabhan2010a,Chakraborty2020}. An interesting question that follows from it is whether a similar relation between the action and the entropy of classical matter fields could be found. By replacing the usual constraints of ordinary thermodynamics with the laws of physics, one could conceive of a ``truly dynamic'' thermodynamics, possibly incorporating aspects of gravity, relativistic hydrodynamics, and thermal physics into a unified variational principle.\footnote{Another interesting parallel appears if one replaces the word ``thermostat'' in de Broglie's work on the equivalence between the principles of entropy maximization and of least action by ``vacuum''. In this case, the real trajectory of a relativistic particle minimizes its action while maximizing the entropy of the vacuum \cite{Pires2023}.}

This perspective on the previous postulate helps us to identify which notion of additivity the total entropy should encompass. First, we note that disjoint sets of spacetime evolve independently. If two sets are connected by a boundary, then one can match all the fields on the boundary and treat the boundary conditions as constraints for each set independently. What really matters are the fields on the boundary; the bulk of one system does not interfere directly on the evolution of the other. Furthermore, the thermodynamic analysis being proposed will be completed only in the limit $ R\to0 $, meaning that, ultimately, we are effectively disconnecting every inner system. Therefore, additivity with respect to spatially separated subsystems becomes completely irrelevant, and we shall consider only subsystems inside the same inner system.

Besides, by adding the assumption that $ T^a{}_b\partial_{(r)}{}^b = T^{ab}\nabla_b u $ is timelike at $ r=0 $ to the ones that led to the first inequality in \eqref{TG}, one finds that
\[ \partial_S U_G = T_G(u,S) > 0, \quad \forall S\neq0. \]
If we write $ S_T(U_T(U_C,U_G),\dots) $ as $ S_T(U_C,U_G,\dots) $, then it follows from the last inequality and the usual relation between $ S_C $ and $ U_C $ that $ \partial_{U_C} S_T =\partial_{U_C} S_C  $ and $ \partial_{U_G} S_T =\partial_{U_G} S_V$ will be strictly positive. Once again, we generalize our results and suppose that the \emph{thermodynamic temperature}
\[ T_{\text{th}}\coloneqq  \partial_{S_T} U_T>0, \quad \forall S\neq0.\]
Taking the above arguments into account and extending the results of Section \ref{qlhre} to the vacuum states of all conceivable matter fields, it becomes reasonable to state that
\begin{postulate}
	The total entropy is additive over all possible matter fields, including those in vacuum states, inside a given inner system. This function is differentiable for all $ S \neq 0 $ and increases monotonically with the total internal energy.
\end{postulate}

The next postulate follows from the previous discussion about the independence between $ V $ and $ S $. As $ S_T $ loses its explicit dependence on $ S $ in the limit $ R\to 0 $, the dynamics associated with the rest-mass density either ceases to exist or is incorporated into the dynamics of constituent numbers. If one fixes these numbers, then $ U_G $ will not vary independently anymore, which, in turn, implies that $\partial_{U_T}S_T \to \partial_{U_C}S_T = \partial_{U_C}S_C$. In other words, the thermodynamic temperature tends to the ordinary temperature of classical thermodynamics. Therefore, as both temperatures go to zero close to the vertex, $ S_C \to 0$ by the extended version of the third law of thermodynamics \cite{Callen1985}. Since $ S_T $ is supposed to be the sum of $ S_C $ and $ S_V $, $ S_T/V $ is not well defined at the vertex, and we restrict ourselves to stating that
\begin{postulate}
	If $ T_{\text{th}}\to0 $ in the limit $ S\to 0 $, then $ S_C/V\to0 $ as $ S\to 0 $.
\end{postulate}

With this postulate, we complete the analogy between some of the aspects of general relativity and classical thermodynamics. Beyond that, our point of view is that it represents a natural extension of classical thermodynamics, providing the information about rest-mass density in an independent, dynamical way. This compels us to take the propositions above seriously and to believe that they could be related to measurements, at least in some idealized sense. In particular, the temperature \eqref{T} should be measurable. Therefore, this extension of thermodynamics requires the additional prescription for which basis for the space of solutions to the wave equation of the quantum field should be considered. In other words, we have to define the vacuum.

\begin{figure}
	\centering
	\begin{tikzpicture}
	\node[anchor=south west,inner sep=0] (image) at (0,0) {\includegraphics[width=0.6\textwidth]{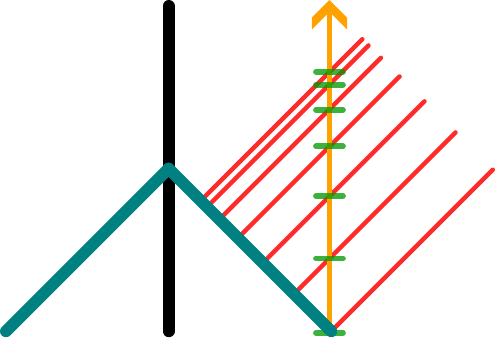}};
	\begin{scope}[x={(image.south east)},y={(image.north west)}]
	\node[anchor=south] at (.67,1) {$ \tau_d $};
	\end{scope}
	\end{tikzpicture}
	\caption[Inaffinity and varying frequencies]{Inaffinity and varying frequencies. For a sufficiently small interval of proper time, the worldline of the observer (black) looks like a straight line. A detector follows a trajectory (orange) with zero coordinate velocity. The constant phase surfaces (red) associated with plane waves with a lightlike coordinate that is also the parameter of an accelerating vector field start at a light cone (blue) and reach the detector. The resulting signal, as a function of the proper time of the detector, $ \tau_d $, will display a varying frequency, as suggested by the intersections in green}
	\label{varyingomega}
\end{figure}

For this, we suppose that some notion of almost instantaneous measurement could be constructed, in alignment with Hypothesis \ref{hypo3}. Additionally, we shall assume that detectors at rest with respect to some coordinate system associated with the observer and close to they should agree on the vacuum, at least for a very brief amount of time. As shown in Section \ref{qlhre}, the nonangular part of the basis modes will look like plane waves close to the initial position of a detector, and the surfaces of constant phase will follow lightlike geodesics in a first approximation. Then, we can consider that the lightlike coordinate of these plane waves is associated with a vector field with a nonvanishing inaffinity. As a result, the intersection between the constant phase surfaces with the worldline of a nearby detector will correspond to a signal with a varying frequency in terms of the proper time of the detector, $ \tau_d $, as pictured in Figure \ref{varyingomega}.

The Fourier transform of a signal with a varying frequency will, in general, contain tails, which stretch out into positive and negative frequencies, the so-called ``concomitant frequencies'' \cite{Ahmadzadegan2018}. That is, the ``acceleration of the constant phase surfaces'' in the immediate neighborhood of the worldline of the detector may force the detected signal to deviate from a plane wave, mixing positive and negative frequency modes. Thus, it seems that the safest way for the observer to state that their immediate neighborhood is in an ``instantaneous vacuum state'' is to use modes related to an affine parameter. We could, in principle, extend this argument from detector to detector all the way to the past on the light cone. Consequently, it appears that the coordinate system $ y^{\mu'} $ will have some preference in any attempt to construct a localized definition of vacuum. Together with the method developed in Section \ref{qlhre}, this idea leads us to propose a new, complementary, and final postulate:
\begin{postulate}
	The invariantly affine vacuum state, whose basis for the solution space of the field equations under consideration is a set of formal power series that are functions of a scale parameter with scale-independent terms composed of eigenfunctions of the operator $ i\partial_{u'} $, belongs to the class of states that are compatible with the physical vacuum.
\end{postulate}

\subsection{Overview}

Once all concepts are established for the proposed extension for thermodynamics, we can proceed to the question of its possible origins. Unlike black hole thermodynamics, it is not yet clear that we are tracing out any degrees of freedom. In addition, the boundaries of the inner systems are somewhat arbitrary, and one considers the limit in which they collapse at the vertex. Thus, the original hypothesis that the information hidden by the horizon leads to a thermodynamic description seems to disappear from the theory.

To amend this apparent loss of a reasonable origin, we advance a perspective on the roots of thermodynamics that can explain its connections with general relativity. Namely, we take \emph{predictability} as the cornerstone of thermodynamics. Heuristically, predictability implies that things cannot be created out of nowhere, but they can be destroyed in a predictable way. In other words, nothing can happen without a preexisting cause, but some things can cease to be the cause of everything that follows. If one associates some form of energy with everything that exists, then one can impose conservation of energy by storing things that vanish in a black box called entropy. Since things can only be destroyed, entropy can only increase. Temperature appears as a quantifier of the rate of energy being lost forever.

In an attempt to clarify the perspective a little more, we could say that as a thermodynamic system reaches equilibrium, most of the information about its past is lost \cite{Callen1985}. Therefore, one ceases to know how to use a part of the energy of the system in a way that its interactions with other systems could attest that this energy still exists. Even though one is unable to confirm that this energy is still there, one refuses to deny the existence of its carrier and is forced to devise the concepts of heat, temperature, and entropy.

In its purest form, predictability requires the system to be isolated. One could improve the theory at some point and include boundary conditions for open systems, but such conditions would have to be justified by analyzing some other isolated system. Thus, we can say that the main concern of thermodynamics is to predict the behavior of a system from its initial data. This means that, at a fundamental level, thermodynamics requires \emph{trivial boundary conditions}, which are compatible with an isolated system.

On top of predictability, we place the choice of the minimum set of variables required to describe a system. Together with the trivial boundary conditions, this naturally leads us to characteristic formulations. To see why, we first note that hyperbolic equations are defined by the presence of distinct families of real characteristic curves \cite{Morse1953}. These curves split the number of degrees of freedom of the initial data in equal parts, one for each family. Therefore, a characteristic evolution that is not based in a combination of families of characteristics, like the one in this work, requires boundary conditions a priori, since the characteristic surface does not contain all the degrees of freedom of the theory. But the supposition that these boundary conditions are trivial turns their explicit usage unnecessary. That is to say that we can work with less degrees of freedom. This is what we have been doing in supposing an analytic characteristic evolution without boundary conditions. Trivial boundary conditions are another way to refer to the ``no-interference'' condition of Subsection \ref{coi}.

While the Cauchy problem of general relativity also does not require boundary conditions, it requires the double of initial data, which is then constrained by elliptic equations. Thus, it complicates the thermodynamic description. A characteristic evolution provides a clear division between the true, freely specifiable initial data and the boundary conditions.

Despite constrained initial data being particular to general relativity, one can extend the above argument to more general hyperbolic equations. Take the classical one-dimensional wave equation as an example,
\[ \partial_{tt}  \phi_w = \partial_{xx}\phi_w.\]
The initial data on a Cauchy surface is composed of $ \phi_w(0,x) $ and $ \partial_{t}\phi_w(0,x) $, while two arbitrary functions can be specified along two arbitrary characteristic curves, one of each family, $ x-t =\text{constant}$ and $ x+t =\text{constant}$. In the Cauchy problem, one cannot dismiss one of the functions as a trivial boundary condition. Meanwhile, in the characteristic formulation, it is possible to take the values along one characteristic curve as initial data and suppose that the remaining function is equal to zero. By doing this, we completely determine the evolution of the wave from a characteristic surface and the assumption of trivial boundary conditions, reducing the number of degrees of freedom to its minimum. The initial surface determines the transversal family of characteristics, just like the initial values of a past like cone determine the future behavior of the radial transverse modes.

In addition, one could say that the difference between thermodynamics and dynamics is, to a great extent, the presence of $ \partial_t  \phi_w $ in the initial data. This idea fits nicely with the notion that thermodynamics is the analysis of systems that have lost their memories \cite{Callen1985}, especially of their initial velocity profiles. Equilibrium thermodynamics does not accept time derivatives as variables. It integrates out kinematic degrees of freedom.

We could summarize the above discussion by noting that nothing prevents a lightlike spherical shell from collapsing at the position of some observer and forming a black hole at any given moment, without any presage of its coming. But nature is kind to us and never turned anyone into a black hole ``out of the blue'', at least on Earth. This is the reason why we have faith in predictability. Even though the theory allows unpredictable results based solely on initial values, the experimental fact that the universe is indeed predictable to some extent makes us comfortable in ignoring any possible unpredictability. Thus, we assume that a true prediction can be made, reject the necessity of the concept of a domain of prediction \cite{Geroch1977}, and accept the fact that every statement that pretends to be a prediction contains a little of faith.\footnote{This prompts the following question: ``Could we relate this unavoidable amount of faith to the intrinsic probabilistic nature of the world?'' In the future, ahead of our present (past light cone), there lies a sea of uncertainties. Do any aspects of quantum theories arise from it?} In sum, ``nothing is created'' cannot account for lightlike shock waves, and this leads to analytic\footnote{Presumably, the word ``analytic'' automatically implies that the theory is ``hole-free'' \cite{Geroch1977}.} causal horizons, since these are the physically relevant characteristic surfaces. ``Nothing is lost'' requires the existence of entropy and the associated concept of temperature.

As an addendum to this discussion, we note that the lack of the notion of a domain of dependence in Newtonian mechanics \cite{Geroch1977} precludes us from making a clear division between initial values and boundary conditions like the one mentioned above. Therefore, to properly include long-range interactions in a thermodynamic framework, it seems mandatory to consider their relativistic, causal generalizations. One could extrapolate this reasoning and state that the minimal requirement for the existence of a thermodynamic theory is a set of hyperbolic differential equations in a causal spacetime. From this, we conjecture that an extension for thermodynamics similar to the one proposed could be constructed for any long-range interaction described by hyperbolic differential equations, with electromagnetism being the first obvious candidate to be analyzed.

Now, we turn our attention to the thermodynamic variables. Following the discussion of Reference \citeonline{Callen1985}, these variables can be traced back to symmetry principles that are the source of conservation laws and Goldstone excitations. If one thinks of ``Goldstone modes of infinite wavelength'' as the properties that are homogeneous throughout the system, the excitations that do not vary from point to point, then one could relate them to the monopoles of each point of the system. In the case of a homogeneous system, these terms contain, at each point, the information that is shared by the whole system.

Similarly, by associating symmetries with the things that do not vary and represent the whole system, one is also led to first moments of the system. If we restrict our attention to first moments, then we are effectively mimicking the conditions of homogeneity and isotropy found in classical thermodynamics. In a sense, we restore the spatial symmetries. Interestingly, the energy does not seem to be related to first moments in the same way. It does have a deep connection in asymptotic expansions in negative powers, but, for positive-power expansions, we cannot find the same explicit relation. In our proposal, energy conservation is more of a definition than anything else, even though we had defined $ U_G $ from the total energy \eqref{Pu}. The surfaces of constant energy $ U_T $ act as intangible boundaries that define an isolated system. Nonetheless, it is forced to be represented by the leading-order term $ U_{\dd{r}} $ only for the sake of consistency with the rest of the theory, which is valid in the limit $ R\to 0 $.

The entropy $ S $ appears as a completely different quantity. It does not represent any leading-order term; it represents the ``spatial'' coordinate itself. One could, in principle, imagine that it has something to do with the approximate local symmetries of spacetime. However, since we already have related the principle of entropy maximization to the principle of least action, we see this qualitative difference between $ S $ and the other variables as a consequence of the distinctive feature of general relativity that its action requires boundary terms. Hence, $ S $ could represent the information about the boundary, which is necessary to normalize the action \cite{Hawking1996}.

Except for the exceptional character of $ S $, the present perspective can be summarized by saying that thermodynamics is, at its core, the theory of leading-order dynamics with trivial boundary conditions. Since, typically, we do physics order by order, we could say that, in a sense, everything starts with thermodynamics. This reinforces our view that the principle of entropy maximization can be seen as the most primitive form of the principle of least action. The transition occurs by replacing the simple constraints on the first moments and the trivial  boundary conditions by complex boundary conditions and transport equations encoded in Lagrangians.

Here, we finally find out what we have been ``integrating out'' all along: higher-order terms and all possible complex boundary conditions (including those that are disguised as initial values in Cauchy surfaces). In a completely unexpected manner, regularity conditions emerge to play the central role in the connection between general relativity and thermodynamics, from the expansion that gives the leading-order terms to the definition of analytic horizons.

Lastly, if one adopts the point of view we have been discussing, remembers that stationary black holes are described only by the first multipole moments \cite{Wald1984}, and recognizes that the main geometric object of a spacetime containing a black hole is an analytic horizon, then the fact that the thermodynamic aspects of general relativity were discovered in what would be called black hole thermodynamics ceases to be a surprise.

\section{Back to black hole thermodynamics}\label{bbht}

A possible way of comparing our results with the ones of black hole thermodynamics is to imagine that we have turned an asymptotically flat spacetime containing a black hole inside out. The relevant boundary in the context of black holes is at the center of spacetime, the event horizon. In the context of light cones, it becomes an outer boundary that defines the inner system under consideration. The family of observers at the asymptotically flat region becomes one single observer, and the regularity conditions around the vertex follow from the local flatness of spacetime. More importantly, all the expansions in negative powers of the radius become expansions in positive powers, and the identification of the leading-order terms in both cases can be used to establish an analogy between their respective thermodynamic variables.

The temperatures obtained in both contexts show the same dependence on the mass of a black hole, as can be seen in equation \eqref{TH}, at least in one model of collapse. By assuming that external observers in the far future will measure a temperature that is compatible with the one measured by the observer at the center of the collapse, we can say that the present derivation shows how Hawking radiation is produced in loco, as matter falls into the dynamical horizon.

With respect to the integral mass formula \cite{Bardeen1973}, we believe that, in the negative power expansion, the active gravitational mass could represent internal energy, since the black hole looks like a point particle from far away, and the famous mass-energy equivalence could be invoked to connect both concepts. But in a positive power expansion, the observer immersed in the system has access to the internal degrees of freedom of the system, like pressure, and it seems necessary to make a distinction between the active gravitational mass and the thermodynamic potentials. This is the reason we have argued in favor of the infinitesimal free energy \eqref{ife}. This access to internal degrees of freedom also explains why the distinction between the thermodynamic temperature and $ T_G $ appears more explicitly in the present context. By immersing the observer in the system, we have a more complete picture of the relation between thermodynamics and general relativity.

Now, the second law \eqref{2law} seems to have a somewhat distinct character of that of the generalized second law of thermodynamics \cite{Bekenstein1973}, which is more closely related to entropy flux, even though there is an excess of entropy being produced by this flux. In the present context, we are explicitly associating the entropy increase with gravitational collapse at fixed internal energy; thus, we have suppressed the entropy flow to a greater degree (but it is still possible that some entropy flux may occur).

An important distinction appears in the derivation of the area theorem \cite{Hawking1971} from a local energy condition. Since spatial cross sections of the event horizon represent spatial one-way membranes by definition, the local positivity of energy can be used to derive a second law that relates to matter fluxes. This notion does not generalize to typical systems, i.e., there are no spatial one-way membranes like these near general spacetime events. Therefore, it seems that the only gravitational second law that can be generalized is the one related to gravitational collapse. This process concerns the evolution of the matter distribution, making it fundamentally nonlocal. Hence, local energy conditions cannot constrain the matter distribution and its velocity profile to the point that a second law is enforced. For this reason, the argument of typicality seems to be unavoidable in this case.

\begin{figure}
	\centering
	\begin{minipage}[t]{.45\textwidth}
		\vspace{0pt} 
		\centering
		\begin{tikzpicture}
		\node at (0.5,1) {(a)};
		\end{tikzpicture}
	\end{minipage}%
	\hfill
	\begin{minipage}[t]{.45\textwidth}
		\vspace{0pt} 
		\centering
		\begin{tikzpicture}
		\node at (0.5,1) {(b)};
		\end{tikzpicture}
	\end{minipage}%

	\begin{minipage}[t]{.45\textwidth}
		\vspace{0pt} 
		\centering
		\begin{tikzpicture}
		\node[anchor=south west,inner sep=0] (image) at (0,0) {\includegraphics[height=0.15\textheight]{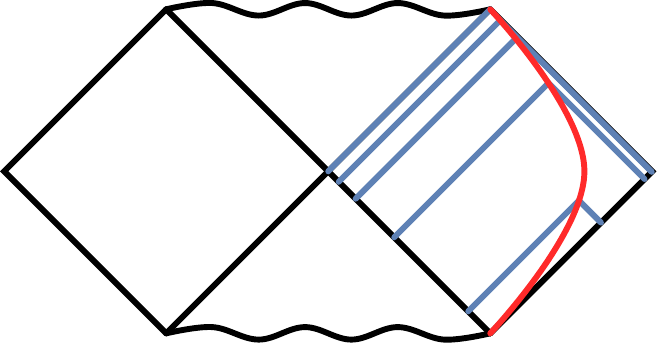}};
		\begin{scope}[x={(image.south east)},y={(image.north west)}]
		\end{scope}
		\end{tikzpicture}
	\end{minipage}%
	\hfill
	\begin{minipage}[t]{.45\textwidth}
		\vspace{0pt}
		\centering
		\begin{tikzpicture}
		\node[anchor=south west,inner sep=0] (image) at (0,0) {\includegraphics[height=0.15\textheight]{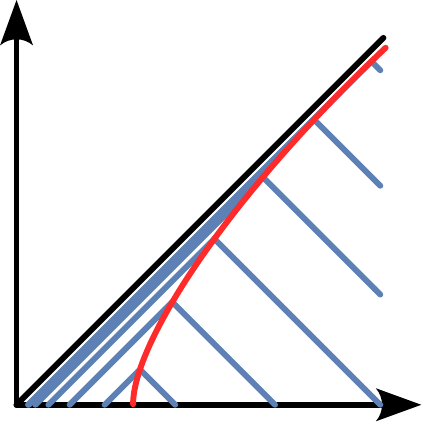}};
		\begin{scope}[x={(image.south east)},y={(image.north west)}]
		\node[anchor=east] at (0,0.5) {$ t $};
		\node[anchor=north] at (0.5,0) {$ x $};
		\end{scope}
		\end{tikzpicture}
	\end{minipage}
	\hspace{0pt}
	\caption[Light cones in the asymptotic future]{Light cones in the asymptotic future. \textbf{(a)} The Penrose diagram of the maximally extended Schwarzschild spacetime \cite{Hawking1973}. The red curve represents a worldline that follows to future timelike infinity. Past light cones associated with this worldline are represented in blue. In the asymptotic future, at least one generator of the light cone will coincide with a generator of the event horizon. \textbf{(b)} The sector with positive time and space Cartesian coordinates, $ t $ and $ x $, respectively, of the two-dimensional Minkowski spacetime is shown, where the black diagonal line represents the Rindler horizon. A Rindler observer is represented in red and its light cones in blue, which approach the Rindler horizon in the asymptotic future}
	\label{limit}
\end{figure}

\begin{figure}
	\centering
	\begin{minipage}[t]{.45\textwidth}
		\vspace{0pt} 
		\centering
		\begin{tikzpicture}
		\node at (0.5,1) {(a)};
		\end{tikzpicture}
	\end{minipage}%
	\hfill
	\begin{minipage}[t]{.45\textwidth}
		\vspace{0pt} 
		\centering
		\begin{tikzpicture}
		\node at (0.5,1) {(b)};
		\end{tikzpicture}
	\end{minipage}%

	\begin{minipage}[t]{.45\textwidth}
		\vspace{0pt} 
		\centering
		\includegraphics[width=\textwidth]{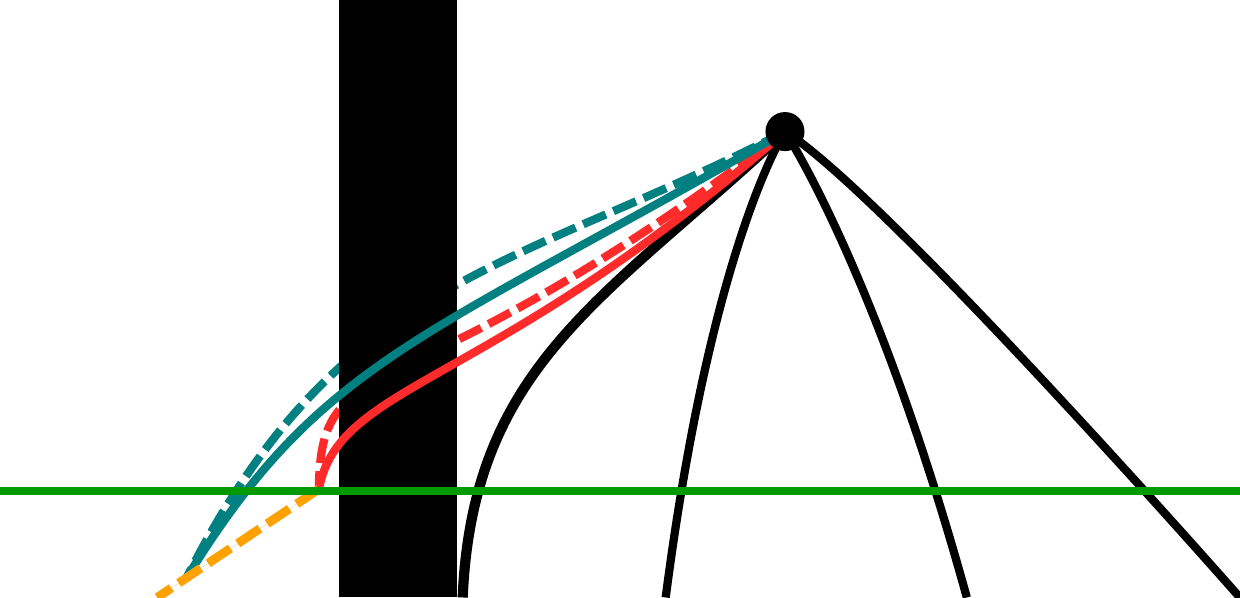}
	\end{minipage}%
	\hfill
	\begin{minipage}[t]{.45\textwidth}
		\vspace{0pt}
		\centering
		\includegraphics[width=\textwidth]{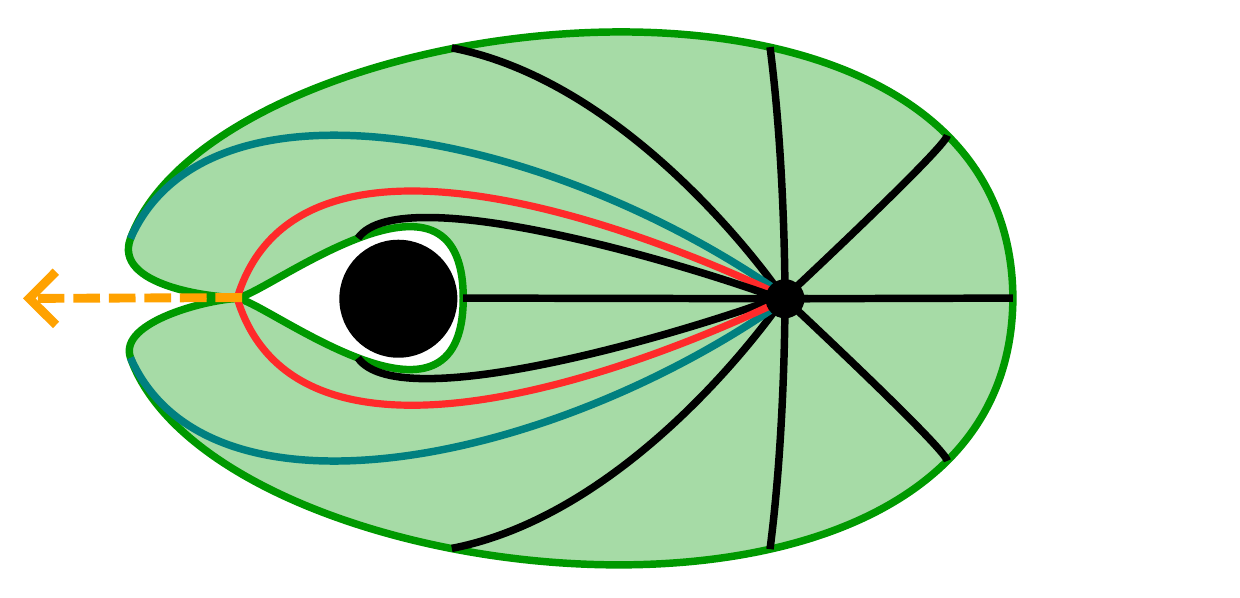}
	\end{minipage}

	\vspace{10pt}

	\begin{minipage}[t]{.45\textwidth}
		\vspace{0pt} 
		\centering
		\begin{tikzpicture}
		\node at (0.5,1) {(c)};
		\end{tikzpicture}
	\end{minipage}%
	\hfill
	\begin{minipage}[t]{.45\textwidth}
		\vspace{0pt} 
		\centering
		\begin{tikzpicture}
		\node at (0.5,1) {(d)};
		\end{tikzpicture}
	\end{minipage}%
	
	\begin{minipage}[t]{.45\textwidth}
		\vspace{0pt} 
		\centering
		\includegraphics[width=\textwidth]{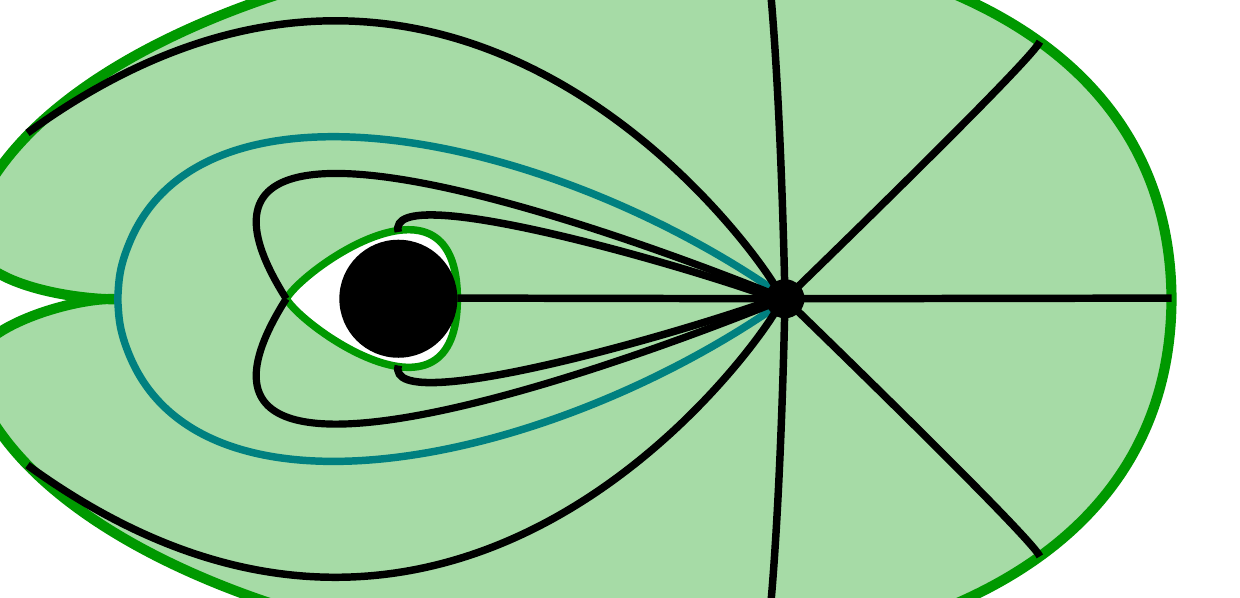}
	\end{minipage}%
	\hfill
	\begin{minipage}[t]{.45\textwidth}
		\vspace{0pt}
		\centering
		\includegraphics[width=\textwidth]{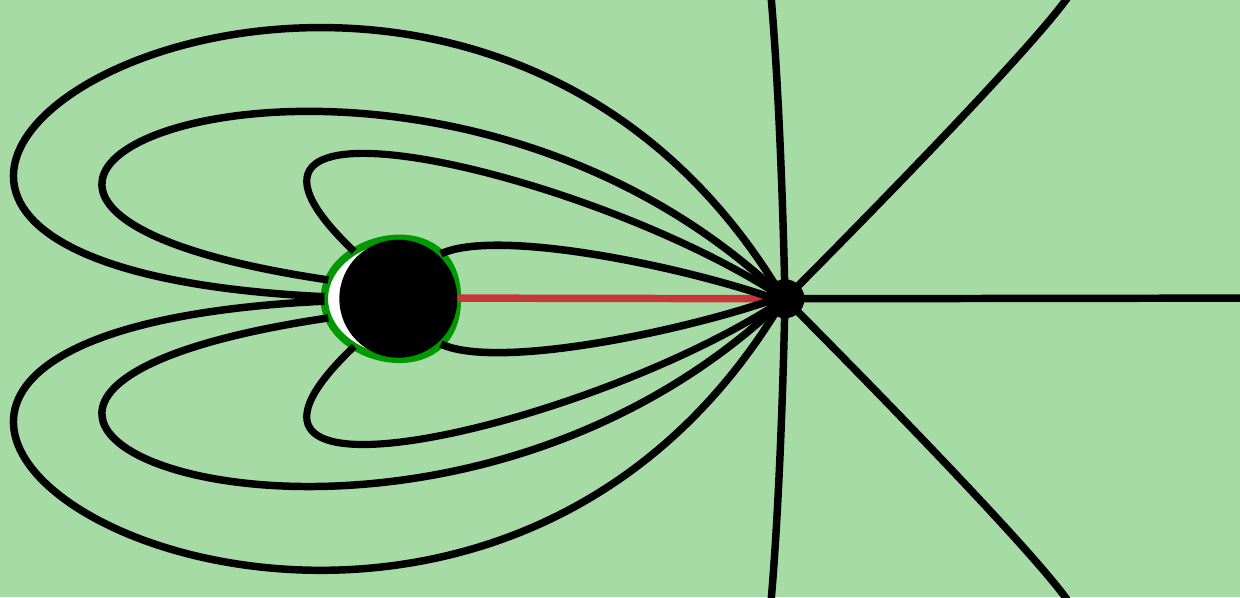}
	\end{minipage}
	\caption[Light cone near a black hole]{Light cone near a black hole. \textbf{(a)} A Schwarzschild black hole is represented by the black region. The black point belongs to the worldline of an observer. The black lines are generators of a light cone and follow into the past. The red and blue lines are generators that are curved around the black hole and meet their symmetric counterpart, which pass behind the black hole and are represented by dashed lines, at a caustic along the orange dashed line. The green line represents a spacelike hypersurface. \textbf{(b)} The intersection between this hypersurface and the light cone is represented by the green contour. The light green region is the interior of this contour inside the hypersurface. The intersection between the black hole and the hypersurface is represented by the black disk. The projection of the black point and the generators in (a) is represented in the light green region. As the hypersurface moves into the past (or the black point moves into the future), caustics are ``sewn'' in the direction of the orange arrow. \textbf{(c)} The hypersurface is moved into the past and does not capture the red generators anymore. \textbf{(d)} In the limit in which the hypersurface goes to past timelike infinity, the terracotta generator adheres to the event horizon. It is not clear whether other generators do the same}
	\label{conehole}
\end{figure}

Nonetheless, we could relate the light cone second law to the generalized second law and explain the differences in the sign of the entropy by regarding the event horizon as the limit of a sequence of past light cones associated with a static observer as their proper time goes to infinity, as represented in Figure \ref{limit} (a). If we restrict our attention to a single observer, some of the past-directed lightlike geodesics that pass nearby the black hole will bend towards it in the past direction, getting closer and closer to the horizon as the affine parameter increases until a caustic is formed. One can picture (and perhaps define) the black hole as a hole in the past light cone of an external observer, as shown in Figure \ref{conehole}.

\begin{figure}
	\centering
	\begin{tikzpicture}
	\node[anchor=south west,inner sep=0] (image) at (0,0) {\includegraphics[height=0.15\textheight]{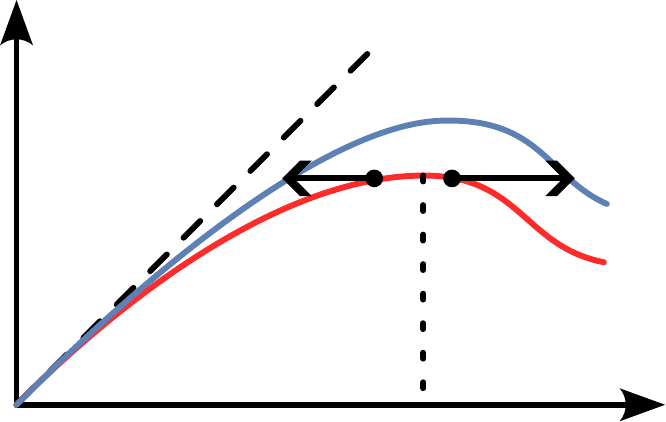}};
	\begin{scope}[x={(image.south east)},y={(image.north west)}]
	\node[anchor=east] at (0,0.5) {$ r $};
	\node[anchor=north] at (0.5,0) {$ \lambda' $};
	\node[anchor=north] at (0.65,0) {$ \lambda_b $};
	\end{scope}
	\end{tikzpicture}
	\caption[Branching of the areal radius]{Branching of the areal radius. Areal radius as a function of affine parameter along a generator that gets close to an event horizon as $ \lambda'\to\infty $. At $ u_1 $, the behavior of $ r $ changes at $ \lambda_b $, as represented by the red curve. If matter falls into the black hole between $ u_1 $ and $ u_2 $, $ r(\lambda') $ approaches the identity function (diagonal traced line) and can be represented by the blue curve at $ u_2 $. The arrows connecting the red and blue curves represent $ \partial_u\lambda' $ in each branch}
	\label{branch}
\end{figure}

The important feature of this light cone is that the expansion of its generators will become negative near the black hole. As they start pointing at the black hole and converging towards it in the past direction, the area of the bundles of generators will start decreasing. Thus, the areal radius will certainly cease to be a monotonic function of the affine parameter if the generators are sufficiently extended into the past. The metric will no longer be well defined in the Bondi-Sachs coordinates, but one still can consider these coordinates as functions that are well defined before caustics form. One just has to keep in mind that the areal radius will be split into two branches: one that increases with $ \lambda' $ and one that decreases. The branching value will be denoted by $ \lambda_b $, so that $ r $ is monotonically increasing in the inner branch, $ r\leq r(\lambda_b) $, and monotonically decreasing in the outer branch, $ r> r(\lambda_b) $. Since $ \lambda'\geq r $, with equality being satisfied in Minkowski spacetime, one sees that, by decreasing $ T_{\lambda'\lambda'} $, $ r(\lambda') $ approaches the identity function. This could represent the process of matter accretion by the black hole between the light cones $ u_1 $ and $ u_2 $, as shown in Figure \ref{branch}.

Since the coordinates are well defined, we could assume that the integrand in definition \eqref{GIE} remains valid in the outer branch. But, to keep $ U_G $ decreasing in the past direction, we should replace $ \dd{r} $ by $ \dd{\lambda'} $ times a positive function. By doing this, $ U_G $ ceases to be a monotonic function of $ S $. In other words, something has to change: either $ U_G $ changes its behavior as one selects inner systems in the outer (past) direction, or we change the sign of $ S $, or deal with a negative temperature. By changing the sign of $ S $, we recover the second law \eqref{2law}.

If $ T_{\lambda'\lambda'} $ decreases in the future direction, then $\partial_u \lambda'>0$ in the outer branch, as represented in Figure \ref{branch}. Assuming that the value of $ U_G $ is more strongly influenced by the matter fluxes in the outer branch, near the black hole, it follows that if we fix the radius $ R_1 $ of the inner system under consideration, then the redefined $ U_G $, which decreases in the past direction of a fixed light cone, will decrease between $ u_1 $ and $ u_2 $. In order to keep its value fixed, we have to shrink the domain of integration, which implies a new radius $ R_2>R_1 $. If the constant $ U_G $ surfaces can get arbitrarily close to the event horizon in the past direction, then we can say that the area theorem is recovered in the limit $ \lambda'\to \infty $. Redefining the sign of the entropy in the outer branch ensures that the second law will be satisfied and brings our definition closer to black hole entropy. The factor of 2 between them remains unexplained, but it could be amended by explicitly calculating the temperature associated with Hawking radiation produced at the vertex (as proposed in Subsection \ref{infgen}). The formalization of the above arguments will be left to future investigations.

As a final remark, we return to the discussion at the end of Subsection \ref{k1k2}. The results of Section \ref{qlhre} indicate that the phenomenon of particle production out of a vacuum state is much more general than the one found in the collapse process forming a black hole. Even if the exponential peeling is not exactly satisfied, the main mechanism seems to be the variation in the way that bundles of lightlike geodesics are distorted by successive layers of matter. In general, this radiation will not exhibit a perfectly thermal spectrum, but it is reasonable to expect that there will be some emission of particles. Therefore, the assumption that ``[...] massless fields are completely determined by their data on $ \mathscr{I}^- $ [...]'' \cite[p. 206]{Hawking1975} may not be valid in general, since these fields may be interacting with timelike matter in the whole process of collapse. In other words, there may be relevant initial data at $ i^- $.

Following the reasoning of Reference \citeonline{Ahmadzadegan2018}, which shows how one can reconstruct an arbitrary worldline from the complete spectrum of the radiation that results from an inertial vacuum, we conjecture that the complete spectrum of Hawking radiation carries a great deal of information regarding the collapsing matter about to form a black hole. Specifically, if particles are being produced in very general matter distributions as we have been suggesting, it is possible that the complete spectrum can determine the class of matter distributions that would result in the same spacetime curvature produced by the collapsing matter.

\section{Conclusions}

We derived a set of results that support the thermodynamic formalism constructed, which represents an extension of classical thermodynamics that includes the dynamics of spacetime. It remains to explore the consequences of the proposed ideas and define their domain of validity.

Probably, the most interesting step in this direction would be the analysis of the degrees of freedom that were ignored. The somewhat unexpected conclusion that they are composed of higher-order terms and complex boundary conditions becomes appealing in the context of black holes when one considers an event horizon as a mere feature of spacetime, which is described by the metric tensor field. In this case, all the relevant degrees of freedom must be contained in this field, and one possible way to treat them is to sort them into multipole moments and degrees of freedom associated with boundary conditions. It is possible that the correct identification and quantification of the elements in the latter group leads to a statistical description of black hole entropy, given its relation with the boundary terms of the gravitational action.

In the more general context of light cones, this conclusion seems to lead to far-reaching consequences. Even though we still can consider all higher-order terms of the quantities defined on a light cone, the future boundary conditions are never known in practice. A first approach to this problem is to hope that the solution to all fields will be analytic and the non-analytic behavior contained in the boundary conditions can be neglected. However, statistical mechanics suggests that a better method may be followed.

The microscopic degrees of freedom of an ordinary thermodynamic system could in principle be treated in a completely deterministic way, but the impossibility of doing that in reality makes the introduction of probabilities a necessity. In fact, it proves to be a much more natural manner to deal with the system, since the statistical predictions are more closely related to how one measures the thermodynamic parameters. Then, we are led to question whether the causal structure of spacetime is inviting us to describe the universe statistically. If so, the resulting theory would have a much more fundamental nature than that of statistical mechanics, affecting every physical system. The first guess is that it could represent a quantum theory.

Together with the fact that gravity cannot be shielded, these ideas could be interpreted as an indication of the existence of an unavoidable noisy background. Such a concept has already been used to derive quantum mechanics \cite{Nelson1966} and even to interpret this theory as a consequence of the structure of spacetime \cite{Lindgren2019}. The relation between diffusion coefficient and mass in these proposals could serve as a further indication that the required stochastic process originates from general relativity. Finally, we may have found the reason to introduce some noise.

\section*{Acknowledgments}

The author would like to thank Professor Daniel A. T. Vanzella for presenting him with a fascinating problem, providing valuable feedback, and being supportive during this project. He also thanks Professor Carlos R. Ordó\~{n}ez and Professor Lucas C. Céleri for helpful references and comments. This study was financed in part by the Coordenação de Aperfeiçoamento de Pessoal de Nível Superior - Brasil (CAPES) - Finance Code 001.

\section*{Data Availability}

All data generated or analyzed during this study are included in this article.

\section*{Conflict of interest}

The author declares that there is no conflict of interest.

\appendix

\begin{appendices}

\section{Complete field equations}\label{appfs}

\begin{equation}\label{beta}
\partial _r \beta =\frac{r}{2}  \left(\cosh ^2(2 \delta ) \left(\partial _r \gamma \right){}^2+\left(\partial _r \delta \right){}^2+4 \pi  {T}_{{rr}}\right),
\end{equation}

\begin{equation}\label{UW1}
\begin{aligned}
&\partial _r \left(e^{-2 \beta } r^4 \left(e^{2 \gamma } \cosh (2 \delta ) \left(\partial _r U\right)+  \sinh (2 \delta )\sin (\theta )\left(\partial _r W\right)\right)\right)\\
&\; = 4 r \partial _{\theta } \beta + r^2 \left(\left(\partial _r \gamma \right) \left(-2 e^{2 \gamma } \csc (\theta ) \left(\sinh (4 \delta ) \left(\partial _{\phi } \gamma \right)+2 \cosh (4 \delta ) \left(\partial _{\phi } \delta \right)\right)+\right .\right .\\
&\; \phantom{=} \; \left .4 \cosh ^2(2 \delta ) \left(\cot (\theta )-\partial _{\theta } \gamma \right)+4 \sinh (4 \delta ) \left(\partial _{\theta } \delta \right)\right)-4 \left(\partial _{\theta } \delta \right) \left(\partial _r \delta \right)+\\
&\; \phantom{=} \; \; e^{2 \gamma } \csc (\theta ) \left(4 \left(\partial _{\phi } \gamma \right) \left(\partial _r \delta \right)-\sinh (4 \delta ) \left(\partial _{{r\phi }} \gamma \right)+2 \left(\partial _{{r\phi }} \delta \right)\right)-2 \left(\partial _{{r\theta }} \beta \right)+\\
&\; \phantom{=} \; \left .2 \cosh ^2(2 \delta ) \left(\partial _{{r\theta }} \gamma \right)-16 \pi  {T}_{{r\theta }}\right),
\end{aligned}
\end{equation}

\begin{equation}\label{UW2}
\begin{aligned}
& \partial _r \left(e^{-2 \beta } r^4 \left(e^{-2 \gamma }\cosh (2 \delta )  \sin (\theta ) \left(\partial _r W\right)+  \sinh (2 \delta ) \left(\partial _r U\right)\right)\right)\\
& \; = e^{-2 \gamma } r^2 \left(4 \left(\partial _r \delta \right) \left(\cot (\theta )-\partial _{\theta } \gamma \right)+2 \left(\partial _r \gamma \right) \left(\sinh (4 \delta ) \left(\cot (\theta )-\partial _{\theta } \gamma \right)+2 \cosh (4 \delta ) \left(\partial _{\theta } \delta \right)\right)+\right .\\
&\; \phantom{=}\;\left . \sinh (4 \delta ) \left(\partial _{{r\theta }} \gamma \right)+2 \left(\partial _{{r\theta }} \delta \right)\right)-2 r \csc (\theta ) \left(r \left(2 \left(\partial _{\phi } \delta \right) \left(\sinh (4 \delta ) \left(\partial _r \gamma \right)+\partial _r \delta \right)+\right .\right .\\
&\; \phantom{=} \left .\left .2 \cosh ^2(2 \delta ) \left(\partial _{\phi } \gamma \right) \left(\partial _r \gamma \right)+\partial _{{r\phi }} \beta +\cosh ^2(2 \delta ) \left(\partial _{{r\phi }} \gamma \right)+8 \pi  {T}_{{r\phi }}\right)-2 \left(\partial _{\phi } \beta \right)\right),
\end{aligned}
\end{equation}

\begin{equation}\label{Phi}
\begin{aligned}
& \partial _r e^{2 \Phi } + e^{2 \Phi } \frac{\left(1-4 \pi  r^2 {T}_{{rr}}\right)}{r} \\
&= r \left(U \left(\partial _{{r\theta }} \gamma +2 \left(\partial _r \gamma \right) \left(\cot (\theta )-\partial _{\theta } \gamma \right)\right)+W \left(-\left(\partial _{{r\phi }} \gamma \right)-2 \left(\partial _r \gamma \right) \left(\partial _{\phi } \gamma \right)\right)\right) \cosh ^2(2 \delta )+\\
&\phantom{=} \frac{1}{4r} e^{-2 (\beta +\gamma )} \cosh (2 \delta ) \left(\left(\sin ^2(\theta ) \left(2 W \left(2 \partial _r W \left(r \left(\partial _r \beta +\partial _r \gamma \right)-2\right)-r \partial _{{rr}} W\right)-r \left(\partial _r W\right){}^2\right) -\right .\right .\\
&\phantom{=}  4 e^{2 \gamma } r \left(W \left(\partial _r U\right)+U \left(\partial _r W\right)\right) \left(\partial _r \delta \right) \sin (\theta )+\\
&\phantom{=} \left . e^{4 \gamma } \left(2 U \left(2 \left(\partial _r U\right) \left(r \left(\partial _r \beta -\partial _r \gamma \right)-2\right)-r \left(\partial _{{rr}} U\right)\right)-r \left(\partial _r U\right){}^2\right)\right) r^3+\\
&\phantom{=} 4 e^{4 \beta } \left(-e^{4 \gamma } \left(\left(\partial _{\phi } \beta \right){}^2+2 \left(\partial _{\phi } \gamma \right) \left(\partial _{\phi } \beta \right)+2 \left(\partial _{\phi } \gamma \right){}^2+2 \left(\partial _{\phi } \delta \right){}^2+\partial _{\phi \phi } \beta +\partial _{\phi \phi } \gamma \right) \csc ^2(\theta )+\right .\\
&\phantom{=} 2 e^{2 \gamma } \left(\left(\partial _{\phi } \beta +\partial _{\phi } \gamma \right) \left(\partial _{\theta } \delta \right)+\partial _{\theta \phi } \delta +\left(\cot (\theta )+\partial _{\theta } \beta -\partial _{\theta } \gamma \right) \left(\partial _{\phi } \delta \right)\right) \csc (\theta )-\left(\partial _{\theta } \beta \right){}^2-\\
&\phantom{=} 2 \left(\partial _{\theta } \gamma \right){}^2-2 \left(\partial _{\theta } \delta \right){}^2-\cot (\theta ) \left(\partial _{\theta } \beta \right)-\partial _{\theta \theta } \beta +3 \cot (\theta ) \left(\partial _{\theta } \gamma \right)+2 \left(\partial _{\theta } \beta \right) \left(\partial _{\theta } \gamma \right)+\partial _{\theta \theta } \gamma +\\
&\phantom{=} \left .\left .1\right)\right)-e^{2 \gamma }\cosh (2 \delta ) r U \csc (\theta ) \left(\partial _{{r\phi }} \gamma \right) \sinh (2 \delta ) -2 \left(\partial _{\theta } U\right)-\\
&\phantom{=} \frac{1}{2} r \left(2 e^{2 \beta } \Lambda +\cot (\theta ) \left(\partial _r U\right)+\partial _{{r\theta }} U+\partial _{{r\phi }} W\right)-2 \left(\partial _{\phi } W\right)+2 W \left(\partial _{\phi } \beta \right)+U \left(-2 \cot (\theta )+\right .\\
&\phantom{=} \left .2 \left(\partial _{\theta } \beta \right)+r \left(-\left(\partial _{{r\theta }} \beta \right)+e^{2 \gamma } \csc (\theta ) \left(2 \left(\partial _{\phi } \gamma \right) \left(\partial _r \delta \right)+\partial _{{r\phi }} \delta \right)-2 \left(\partial _r \delta \right) \left(\partial _{\theta } \delta \right)\right)\right)-\\
&\phantom{=} 2 e^{-2 \gamma } r \cosh (4 \delta ) \left(\partial _r \gamma \right) \left(e^{4 \gamma } U \csc ^2(\theta ) \left(\partial _{\phi } \delta \right)-W \left(\partial _{\theta } \delta \right)\right) \sin (\theta )+\\
&\phantom{=} e^{-2 \gamma } r W \left(-e^{2 \gamma } \left(\partial _{{r\phi }} \beta \right)+\left(\partial _{{r\theta }} \delta \right) \sin (\theta )+2 \left(\partial _r \delta \right) \left(\cos (\theta )-e^{2 \gamma } \left(\partial _{\phi } \delta \right)-\left(\partial _{\theta } \gamma \right) \sin (\theta )\right)\right)+\\
&\phantom{=} \frac{1}{2} e^{-2 (\beta +\gamma )} \frac{1}{r} \left(-2 W \left(\partial _r W\right) \left(\partial _r \delta \right) \sin ^2(\theta ) r^4-\right .\\
&\phantom{=} 2 e^{4 \beta } \left(\left(3 \cot (\theta )+2 \left(\partial _{\theta } \beta \right)-4 \left(\partial _{\theta } \gamma \right)\right) \left(\partial _{\theta } \delta \right)+\partial _{\theta \theta } \delta \right)+\\
&\phantom{=} 2 e^{4 \gamma } \left(-U \left(\partial _r U\right) \left(\partial _r \delta \right) r^4-e^{4 \beta } \csc ^2(\theta ) \left(2 \left(\partial _{\phi } \beta +2 \left(\partial _{\phi } \gamma \right)\right) \left(\partial _{\phi } \delta \right)+\partial _{\phi \phi } \delta \right)\right)+\\
&\phantom{=} e^{2 \gamma } \left(\left(-r U \partial _{{rr}} W+\partial _r W\left(2 U \left(r \partial _r \beta-2\right)-r \partial _r U\right)+W \left(2 \partial _r U \left(r \partial _r \beta -2\right)-r \partial _{{rr}} U\right)\right) r^3+\right .\\
&\phantom{=}\left .\left . 4 e^{4 \beta } \csc ^2(\theta ) \left(\partial _{\theta \phi } \beta +\left(\partial _{\theta } \beta \right) \left(\partial _{\phi } \beta \right)+2 \left(\partial _{\theta } \delta \right) \left(\partial _{\phi } \delta \right)\right)\right) \sin (\theta )\right) \sinh (2 \delta )+\\
&\phantom{=} \frac{1}{2} r \left(e^{-2 \gamma } W \left(\left(\partial _{{r\theta }} \gamma \right) \sin (\theta )+2 \left(\partial _r \gamma \right) \left(\cos (\theta )-2 e^{2 \gamma } \left(\partial _{\phi } \delta \right)-\left(\partial _{\theta } \gamma \right) \sin (\theta )\right)\right)-\right .\\
&\phantom{=}\left . 2 U \left(\partial _r \gamma \right) \left(e^{2 \gamma } \csc (\theta ) \left(\partial _{\phi } \gamma \right)-2 \left(\partial _{\theta } \delta \right)\right)\right) \sinh (4 \delta )+8 \pi  r {T}_{{ur}},
\end{aligned}
\end{equation}

\[ 
\begin{aligned}
& \partial_r \left( r \left(\partial _{{u}} \gamma \right) \right) + 2 \tanh (2 \delta ) \left(\left(\partial _r \delta \right) \left(r \left(\partial _u \gamma \right) \right)+\left(\partial _r \gamma \right) (r \left(\partial _u \delta \right))\right)\\
&= \frac{1}{8} \frac{1}{r} e^{-2 (\beta +\gamma )} {\sech}(2 \delta ) \left(-8 e^{4 \beta +2 \gamma } \csc (\theta ) \left(\left(\partial _{\theta } \beta \right) \left(\partial _{\phi } \delta \right)-\left(\partial _{\phi } \beta \right) \left(\partial _{\theta } \delta \right)\right)+\right .\\
&\phantom{=} e^{4 \gamma } \left(4 e^{4 \beta } \csc ^2(\theta ) \left(\left(\partial _{\phi } \beta \right){}^2+\partial _{\phi \phi } \beta +4 \pi  {T}_{\phi \phi }\right)-r^4 \left(\partial _r U\right){}^2\right)+r^4 \sin ^2(\theta ) \left(\partial _r W\right){}^2-\\
&\phantom{=} \left .4 e^{4 \beta } \left(\left(\partial _{\theta } \beta \right){}^2-\cot (\theta ) \left(\partial _{\theta } \beta \right)+\partial _{\theta \theta } \beta +4 \pi  {T}_{\theta \theta }\right)\right)+\\
&\phantom{=} \tanh (2 \delta ) \left(\frac{1}{4} e^{-2 \gamma } \left(-8 r e^{2 (\gamma +\Phi )} \left(\partial _r \gamma \right) \left(\partial _r \delta \right)-\sin (\theta ) \left(r \left(\partial _{{r\theta }} W\right)+2 \left(\partial _{\theta } W\right)\right)+\right .\right .\\
&\phantom{=} \left . e^{4 \gamma } \csc (\theta ) \left(r \left(\partial _{{r\phi }} U\right)+2 \left(\partial _{\phi } U\right)\right)\right)+\\
&\phantom{=} r \left(\left(\partial _r \delta \right) \left(U \left(\cot (\theta )-2 \left(\partial _{\theta } \gamma \right)\right)-\left(\partial _{\theta } U\right)-2 W \left(\partial _{\phi } \gamma \right)+\partial _{\phi } W\right)-\right .\\
&\phantom{=} \left .\left .2 \left(\partial _r \gamma \right) \left(U \left(\partial _{\theta } \delta \right)+W \left(\partial _{\phi } \delta \right)\right)\right)\right)+\\
&\phantom{=} \frac{1}{4} \left(-4 e^{2 \Phi } \left(\partial _r \gamma \right)+2 U \left(-2 \left(\partial _{\theta } \gamma \right)+\cot (\theta ) \left(1-r \left(\partial _r \gamma \right)\right)-2 r \left(\partial _{{r\theta }} \gamma \right)\right)+\right .\\
&\phantom{=} r \left(-2 e^{2 \Phi } \left(2 \left(\partial _r \gamma \right) \left(\partial _r \Phi \right)+\partial _{{rr}} \gamma \right)-2 \left(\partial _r \gamma \right) \left(\partial _{\theta } U\right)+\right .\\
&\phantom{=} \left .4 e^{-2 \gamma } \sin (\theta ) \left(\partial _r \delta \right) \left(e^{4 \gamma } \csc ^2(\theta ) \left(\partial _{\phi } U\right)-\partial _{\theta } W\right)-\left(\partial _{{r\theta }} U\right)-4 W \left(\partial _{{r\phi }} \gamma \right)+\partial _{{r\phi }} W\right)+\\
&\phantom{=} \left .r \partial _r U \left(\cot (\theta )-2 \partial _{\theta } \gamma \right)+\left(2-2 r \partial _r \gamma \right) \left(\partial _{\phi } W\right)-2 \partial _{\phi } \gamma \left(r \partial _r W+2 W\right)-2 \partial _{\theta } U\right),
\end{aligned}
 \]

\[ 
\begin{aligned}
& \partial_r \left( r \left(\partial _{{u}} \delta \right) \right) - \sinh (4 \delta ) \left(\partial _r \gamma \right) \left( r\left(\partial _u \gamma \right) \right) \\
&\; = \frac{1}{8} \frac{1}{r} e^{-2 (\beta +\gamma )} \sinh (2 \delta ) \left(4 e^{4 \beta } \left(\left(\partial _{\theta } \beta \right){}^2-\cot (\theta ) \left(\partial _{\theta } \beta \right)+\partial _{\theta \theta } \beta +\right .\right .\\
&\; \phantom{=}\left .\left . e^{4 \gamma } \csc ^2(\theta ) \left(\left(\partial _{\phi } \beta \right){}^2+\partial _{\phi \phi } \beta +4 \pi  {T}_{\phi \phi }\right)+4 \pi  {T}_{\theta \theta }\right)-r^4 \left(e^{4 \gamma } \left(\partial _r U\right){}^2+\sin ^2(\theta ) \left(\partial _r W\right){}^2\right)\right)+\\
&\; \phantom{=} \frac{1}{4} e^{-2 \beta } \frac{1}{r} \cosh (2 \delta ) \left(r^4 \sin (\theta ) \left(-\left(\partial _r U\right)\right) \left(\partial _r W\right)-\right .\\
&\; \phantom{=}\left . 4 e^{4 \beta } \csc (\theta ) \left(-\left(\partial _{\theta } \beta \right) \left(\partial _{\phi } \gamma \right)+\left(\partial _{\phi } \beta \right) \left(\partial _{\theta } \beta +\partial _{\theta } \gamma -\cot (\theta )\right)+\partial _{\theta \phi } \beta +4 \pi  {T}_{\theta \phi }\right)\right)+\\
&\; \phantom{=} \frac{1}{4} e^{-2 \gamma } \left(-2 e^{2 \gamma } \left(r \left(\left(\partial _r \delta \right) \left(\partial _{\theta } U+U \cot (\theta )+\partial _{\phi } W\right)+\left(\partial _{\phi } \delta \right) \left(\partial _r W\right)+2 U \left(\partial _{{r\theta }} \delta \right)\right)+\right .\right .\\
&\; \phantom{=} \left .2 W \left(\partial _{\phi } \delta +r \left(\partial _{{r\phi }} \delta \right)\right)+\left(\partial _{\theta } \delta \right) \left(r \left(\partial _r U\right)+2 U\right)\right)+\sin (\theta ) \left(-r \left(\partial _{{r\theta }} W\right)-2 \left(\partial _{\theta } W\right)\right)-\\
&\; \phantom{=} \left .2 e^{2 (\gamma +\Phi )} \left(2 \left(\partial _r \delta \right) \left(r \left(\partial _r \Phi \right)+1\right)+r \left(\partial _{{rr}} \delta \right)\right)-e^{4 \gamma } \csc (\theta ) \left(r \left(\partial _{{r\phi }} U\right)+2 \left(\partial _{\phi } U\right)\right)\right)+\\
&\; \phantom{=} \frac{1}{2} r \sinh (4 \delta ) \left(\partial _r \gamma \right) \left(e^{2 \Phi } \left(\partial _r \gamma \right)+U \left(2 \left(\partial _{\theta } \gamma \right)-\cot (\theta )\right)+\partial _{\theta } U+2 W \left(\partial _{\phi } \gamma \right)-\partial _{\phi } W\right)-\\
&\; \phantom{=} e^{-2 \gamma } r \cosh ^2(2 \delta ) \sin (\theta ) \left(\partial _r \gamma \right) \left(e^{4 \gamma } \csc ^2(\theta ) \left(\partial _{\phi } U\right)-\partial _{\theta } W\right),
\end{aligned}
 \]

\begin{equation}\label{deluTrr}
\begin{aligned}
\partial _u {T}_{{rr}}= &\; e^{2 \beta } \frac{1}{r^3} \sinh (2 \delta ) \left(r \left(-e^{-2 \gamma } \left({T}_{\theta \theta } \left(\partial _r \delta \right)+2 {T}_{{r\theta }} \left(\partial _{\theta } \delta \right)\right)-\right .\right .\\
&\; e^{2 \gamma } \csc ^2(\theta ) \left({T}_{\phi \phi } \left(\partial _r \delta \right)+2 {T}_{{r\phi }} \left(\partial _{\phi } \delta \right)\right)+\\
&\; \left .\left .\csc (\theta ) \left(2 {T}_{{r\theta }} \left(\partial _{\phi } \beta \right)+\partial _{\phi } {T}_{{r\theta }}+2 {T}_{{r\phi }} \left(\partial _{\theta } \beta \right)+\partial _{\theta } {T}_{{r\phi }}\right)\right)-2 \csc (\theta ) {T}_{\theta \phi }\right)+\\
&\; \frac{1}{r^3} e^{2 (\beta -\gamma )} \cosh (2 \delta ) \left(r {T}_{{r\theta }} \left(-2 \left(\partial _{\theta } \beta \right)+2 e^{2 \gamma } \csc (\theta ) \left(\partial _{\phi } \delta \right)+2 \left(\partial _{\theta } \gamma \right)-\cot (\theta )\right)-\right .\\
&\; r \left(\partial _{\theta } {T}_{{r\theta }}\right)-e^{4 \gamma } \csc ^2(\theta ) \left(r \left(2 {T}_{{r\phi }} \left(\partial _{\phi } \beta +\partial _{\phi } \gamma \right)+\partial _{\phi } {T}_{{r\phi }}\right)+{T}_{\phi \phi } \left(r \left(\partial _r \gamma \right)-1\right)\right)+\\
&\; \left .2 e^{2 \gamma } r \csc (\theta ) \left({T}_{\theta \phi } \left(\partial _r \delta \right)+{T}_{{r\phi }} \left(\partial _{\theta } \delta \right)\right)+{T}_{\theta \theta } \left(r \left(\partial _r \gamma \right)+1\right)\right)+\\
&\; \frac{1}{r} \left(2 r \left(\partial _r \beta \right) \left(U {T}_{{r\theta }}+W {T}_{{r\phi }}+{T}_{{ur}}\right)-2 {T}_{{r\theta }} \left(r \left(\partial _r U\right)+U\right)-\right .\\
&\; r \left(U \left(\partial _r {T}_{{r\theta }}+\partial _{\theta } {T}_{{rr}}\right)+e^{2 \Phi } \left(\partial _r {T}_{{rr}}\right)+W \left(\partial _r {T}_{{r\phi }}\right)+2 {T}_{{r\phi }} \left(\partial _r W\right)+\partial _r {T}_{{ur}}\right)+\\
&\; W \left(-r \left(\partial _{\phi } {T}_{{rr}}\right)-2 {T}_{{r\phi }}\right)+\\
&\; \left . {T}_{{rr}} \left(e^{2 \Phi } \left(r \left(\partial _r \beta -3 \left(\partial _r \Phi \right)\right)-2\right)-r \left(\partial _{\theta } U+U \cot (\theta )+\partial _{\phi } W\right)\right)-2 {T}_{{ur}}\right),
\end{aligned}
\end{equation}

\[ 
\begin{aligned}
\partial _u {T}_{{r\theta }}= &\;  \frac{e^{2 \beta }}{r^2} \sinh (2 \delta ) \left(-\csc (\theta ) \left({T}_{\theta \phi } \left(\cot (\theta )-2 \left(\partial _{\theta } \beta \right)\right)-2 {T}_{\theta \theta } \left(\partial _{\phi } \beta \right)-\partial _{\theta } {T}_{\theta \phi }-\left(\partial _{\phi } {T}_{\theta \theta }\right)\right)-\right .\\
&\; \left .3 e^{-2 \gamma } {T}_{\theta \theta } \left(\partial _{\theta } \delta \right)-e^{2 \gamma } \csc ^2(\theta ) \left({T}_{\phi \phi } \left(\partial _{\theta } \delta \right)+2 {T}_{\theta \phi } \left(\partial _{\phi } \delta \right)\right)\right)+\\
&\; \frac{1}{r^2} e^{2 (\beta -\gamma )} \cosh (2 \delta ) \left({T}_{\theta \theta } \left(-2 \left(\partial _{\theta } \beta \right)+3 \left(\partial _{\theta } \gamma \right)-\cot (\theta )\right)+\right .\\
&\; e^{4 \gamma } \csc ^2(\theta ) \left(-2 {T}_{\theta \phi } \left(\partial _{\phi } \beta +\partial _{\phi } \gamma \right)+{T}_{\phi \phi } \left(\cot (\theta )-\partial _{\theta } \gamma \right)-\left(\partial _{\phi } {T}_{\theta \phi }\right)\right)+\\
&\; \left . 2 e^{2 \gamma } \csc (\theta ) \left(2 {T}_{\theta \phi } \left(\partial _{\theta } \delta \right)+{T}_{\theta \theta } \left(\partial _{\phi } \delta \right)\right)-\partial _{\theta } {T}_{\theta \theta }\right)+\\
&\; \frac{1}{r} \left({T}_{{r\theta }} \left(-r \left(U \left(\cot (\theta )-2 \left(\partial _{\theta } \beta \right)\right)+2 \left(\partial _{\theta } U\right)+\partial _{\phi } W\right)-2 e^{2 \Phi } \left(r \left(\partial _r \Phi \right)+1\right)\right)+\right .\\
&\;  r \left(W \left(-\partial _r {T}_{\theta \phi }-\left(\partial _{\phi } {T}_{{r\theta }}\right)+2 {T}_{{r\phi }} \left(\partial _{\theta } \beta \right)\right)+U \left(-\partial _r {T}_{\theta \theta }-\left(\partial _{\theta } {T}_{{r\theta }}\right)\right)+\right .\\
&\; \left .e^{2 \Phi } \left({T}_{{rr}} \left(\partial _{\theta } \beta -\partial _{\theta } \Phi \right)-\partial _r {T}_{{r\theta }}\right)-\partial _r {T}_{{u\theta }}-{T}_{\theta \phi } \partial _r W-{T}_{{r\phi }} \partial _{\theta } W+2 {T}_{{ur}} \partial _{\theta } \beta \right)+\\
&\; \left .{T}_{\theta \theta } \left(-r \left(\partial _r U\right)-2 U\right)-2 \left({T}_{{u\theta }}+W {T}_{\theta \phi }\right)\right),
\end{aligned}
 \]

\[ 
\begin{aligned}
\partial _u {T}_{{r\phi }}= &\; e^{2 \beta } \frac{1}{r^2} \sinh (2 \delta ) \left(2 {T}_{\theta \phi } \left(\csc (\theta ) \left(\partial _{\phi } \beta \right)-e^{-2 \gamma } \left(\partial _{\theta } \delta \right)\right)-\right .\\
&\; \csc (\theta ) \left({T}_{\phi \phi } \left(3 e^{2 \gamma } \csc (\theta ) \left(\partial _{\phi } \delta \right)-2 \left(\partial _{\theta } \beta \right)\right)-\partial _{\theta } {T}_{\phi \phi }\right)-e^{-2 \gamma } {T}_{\theta \theta } \left(\partial _{\phi } \delta \right)+\\
&\; \left .\csc (\theta ) \left(\partial _{\phi } {T}_{\theta \phi }\right)\right)+\frac{1}{r^2} e^{2 (\beta -\gamma )} \cosh (2 \delta ) \left({T}_{\theta \phi } \left(-2 \left(\partial _{\theta } \beta \right)+2 \left(\partial _{\theta } \gamma \right)-\cot (\theta )\right)-\right .\\
&\; e^{4 \gamma } \csc ^2(\theta ) \left({T}_{\phi \phi } \left(2 \partial _{\phi } \beta +3 \partial _{\phi } \gamma \right)+\partial _{\phi } {T}_{\phi \phi }\right)+2 e^{2 \gamma } \csc (\theta ) \left({T}_{\phi \phi } \partial _{\theta } \delta +2 {T}_{\theta \phi } \partial _{\phi } \delta \right)+\\
&\; \left . {T}_{\theta \theta } \left(\partial _{\phi } \gamma \right)-\partial _{\theta } {T}_{\theta \phi }\right)+\frac{1}{r} \left(r \left(e^{2 \Phi } \left({T}_{{rr}} \left(\partial _{\phi } \beta -\partial _{\phi } \Phi \right)-\partial _r {T}_{{r\phi }}\right)+\right .\right .\\
&\; U \left(-\partial _r {T}_{\theta \phi }-\left(\partial _{\theta } {T}_{{r\phi }}\right)\right)+W \left(-\partial _r {T}_{\phi \phi }-\left(\partial _{\phi } {T}_{{r\phi }}\right)\right)-\left(\partial _r {T}_{{u\phi }}\right)-{T}_{\phi \phi } \left(\partial _r W\right)+\\
&\; \left .{T}_{{r\theta }} \left(2 U \left(\partial _{\phi } \beta \right)-\partial _{\phi } U\right)+2 {T}_{{ur}} \left(\partial _{\phi } \beta \right)\right)+\\
&\; {T}_{{r\phi }} \left(-r \left(U \cot (\theta )+\partial _{\theta } U-2 W \left(\partial _{\phi } \beta \right)+2 \left(\partial _{\phi } W\right)\right)-2 e^{2 \Phi } \left(r \left(\partial _r \Phi \right)+1\right)\right)+\\
&\; \left . {T}_{\theta \phi } \left(-r \left(\partial _r U\right)-2 U\right)-2 \left({T}_{{u\phi }}+W {T}_{\phi \phi }\right)\right),
\end{aligned}
 \]

\begin{equation}\label{deluTur}
\begin{aligned}
\partial _u {T}_{{ur}}=&\; \frac{e^{2 (\beta -\gamma )} }{r^2} \cosh (2 \delta ) \left({T}_{\theta \theta } \partial _u \gamma -e^{4 \gamma } \csc ^2(\theta ) \left({T}_{\phi \phi } \partial _u \gamma +2 {T}_{{u\phi }} \left(\partial _{\phi } \beta +\partial _{\phi } \gamma \right)+\partial _{\phi } {T}_{{u\phi }}\right)+\right .\\
&\; 2 e^{2 \gamma } \csc (\theta ) \left({T}_{\theta \phi } \left(\partial _u \delta \right)+{T}_{{u\phi }} \left(\partial _{\theta } \delta \right)\right)+\\
&\;\left . {T}_{{u\theta }} \left(-2 \left(\partial _{\theta } \beta \right)+2 e^{2 \gamma } \csc (\theta ) \left(\partial _{\phi } \delta \right)+2 \left(\partial _{\theta } \gamma \right)-\cot (\theta )\right)- \partial _{\theta } {T}_{{u\theta }}\right)+\\
&\; e^{2 \beta } \frac{1}{r^2} \sinh (2 \delta ) \left(-e^{-2 \gamma } \left({T}_{\theta \theta } \left(\partial _u \delta \right)+2 {T}_{{u\theta }} \left(\partial _{\theta } \delta \right)\right)-\right .\\
&\; e^{2 \gamma } \csc ^2(\theta ) \left({T}_{\phi \phi } \left(\partial _u \delta \right)+2 {T}_{{u\phi }} \left(\partial _{\phi } \delta \right)\right)+\\
&\;\left . \csc (\theta ) \left(2 {T}_{{u\theta }} \left(\partial _{\phi } \beta \right)+\partial _{\phi } {T}_{{u\theta }}+2 {T}_{{u\phi }} \left(\partial _{\theta } \beta \right)+\partial _{\theta } {T}_{{u\phi }}\right)\right)+\\
&\; \frac{1}{r} \left(r \left(U \left(-\partial _r {T}_{{u\theta }}+2 {T}_{{r\theta }} \left(\partial _u \beta \right)-\left(\partial _{\theta } {T}_{{ur}}\right)\right)-e^{2 \Phi } \left(\partial _r {T}_{{ur}}-{T}_{{rr}} \left(\partial _u \beta \right)\right)+\right .\right .\\
&\; W \left(-\partial _r {T}_{{u\phi }}+2 {T}_{{r\phi }} \left(\partial _u \beta \right)-\left(\partial _{\phi } {T}_{{ur}}\right)\right)-\left(\partial _r {T}_{{uu}}\right)-{T}_{{u\phi }} \left(\partial _r W\right)-{T}_{{r\theta }} \left(\partial _u U\right)-\\
&\;\left . {T}_{{r\phi }} \partial _u W\right)+{T}_{{ur}} \left(-r \left(-2 \partial _u \beta +\partial _{\theta } U+U \cot (\theta )+\partial _{\phi } W\right)-2 e^{2 \Phi } \left(r \partial _r \Phi +1\right)\right)+\\
&\; \left . {T}_{{u\theta }} \left(-r \left(\partial _r U\right)-2 U\right)-2 \left({T}_{{uu}}+W {T}_{{u\phi }}\right)\right)-\frac{1}{2} {T}_{{rr}} \left(\partial _u e^{2 \Phi }\right).
\end{aligned}
\end{equation}

\section{First terms of the solution}\label{firstterms}

\begin{equation}\label{beta2}
\beta^{(2)}=2 \pi  T_{{rr}}{}^{{(0)}},
\end{equation}

\begin{equation}\label{U1}
U^{(1)}=-4 \pi  T_{{r\theta }}{}^{{(1)}}+\frac{1}{2} \left(\partial _{\theta } \gamma ^{{(2)}}+\csc (\theta ) \left(\partial _{\phi } \delta ^{{(2)}}\right)\right)+\gamma ^{{(2)}} \cot (\theta ),
\end{equation}

\begin{equation}\label{W1}
W^{(1)}=\csc ^2(\theta ) \left(-4 \pi  T_{{r\phi }}{}^{{(1)}}+\frac{1}{2} \left(\sin (\theta ) \left(\partial _{\theta } \delta ^{{(2)}}\right)-\partial _{\phi } \gamma ^{{(2)}}\right)+\delta ^{{(2)}} \cos (\theta )\right),
\end{equation}

\begin{equation}\label{Phi2}
\begin{aligned}
\Phi^{(2)}& =\frac{1}{3} \left(\pi  \left(\cot (\theta ) \left(10 T_{{r\theta }}{}^{{(1)}}-\partial _{\theta } T_{{rr}}{}^{{(0)}}\right)+\csc ^2(\theta ) \left(10\partial _{\phi } T_{{r\phi }}{}^{{(1)}}-\partial _{\phi \phi } T_{{rr}}{}^{{(0)}}\right)- \right .\right .\\
&\phantom{=}\; \left .\left . \partial _{\theta \theta } T_{{rr}}{}^{{(0)}}+6 T_{{rr}}{}^{{(0)}}+8 T_{{ur}}{}^{{(0)}}+10 \left(\partial _{\theta } T_{{r\theta }}{}^{{(1)}}\right)\right)-\Lambda \right)+\\
&\phantom{=}\; \frac{1}{2} \left(\gamma ^{{(2)}}+\csc (\theta ) \left(-\cot (\theta ) \left(\partial _{\phi } \delta ^{{(2)}}\right)-\left(\partial _{\theta \phi } \delta ^{{(2)}}\right)\right)\right)+\\
&\phantom{=}\; \frac{1}{4} \left(\csc ^2(\theta ) \left(\partial _{\phi \phi } \gamma ^{{(2)}}\right)-3 \cot (\theta ) \left(\partial _{\theta } \gamma ^{{(2)}}\right)-\partial _{\theta \theta } \gamma ^{{(2)}}\right)-2 ( \Phi ^{{(1)}} )^2,
\end{aligned}
\end{equation}

\begin{equation}\label{gamma2u}
\begin{aligned}
\partial_{u}\gamma^{(2)}&=\frac{1}{45} \left(-15 \gamma ^{{(2)}} \left(\partial _{\theta } U^{{(0)}}+U^{{(0)}} \cot (\theta )+\partial _{\phi } W^{{(0)}}+8 \Phi ^{{(1)}}\right)-45 U^{{(0)}} \left(\partial _{\theta } \gamma ^{{(2)}}\right)+\right .\\
&\quad 45 \delta ^{{(2)}} \csc (\theta ) \left(\partial _{\phi } U^{{(0)}}\right)-45 W^{{(0)}} \left(\partial _{\phi } \gamma ^{{(2)}}\right)-45 \delta ^{{(2)}} \sin (\theta ) \left(\partial _{\theta } W^{{(0)}}\right)+\\
&\quad 6 \pi  \csc ^2(\theta ) \left(\partial _{\phi \phi } T_{{rr}}{}^{{(1)}}\right)+6 \pi  \cot (\theta ) \left(\partial _{\theta } T_{{rr}}{}^{{(1)}}\right)-6 \pi  \left(\partial _{\theta \theta } T_{{rr}}{}^{{(1)}}\right)-24 \pi  T_{{r\theta }}{}^{{(2)}} \cot (\theta )+\\
&\quad 24 \pi  \left(\partial _{\theta } T_{{r\theta }}{}^{{(2)}}\right)-24 \pi  \csc ^2(\theta ) \left(\partial _{\phi } T_{{r\phi }}{}^{{(2)}}\right)-3 \csc ^2(\theta ) \left(\partial _{\phi \phi } \gamma ^{{(3)}}\right)+\\
&\quad 12 \gamma ^{{(3)}} \left(\csc ^2(\theta )-3\right)-3 \cot (\theta ) \left(\partial _{\theta } \gamma ^{{(3)}}\right)-3 \left(\partial _{\theta \theta } \gamma ^{{(3)}}\right)+12 \cot (\theta ) \csc (\theta ) \left(\partial _{\phi } \delta ^{{(3)}}\right)-\\
&\quad \left . 10 \pi  T_{\theta \theta }{}^{{(3)}}+10 \pi  T_{\phi \phi }{}^{{(3)}} \csc ^2(\theta )\right),
\end{aligned}
\end{equation}

\begin{equation}\label{delta2u}
\begin{aligned}
\partial_{u}\delta^{(2)}=\;& 
\frac{1}{45} \left(-15 \delta ^{{(2)}} \left(\partial _{\theta } U^{{(0)}}+U^{{(0)}} \cot (\theta )+\partial _{\phi } W^{{(0)}}+8 \Phi ^{{(1)}}\right)-45 \gamma ^{{(2)}} \csc (\theta ) \left(\partial _{\phi } U^{{(0)}}\right)-\right .\\
& 45 U^{{(0)}} \left(\partial _{\theta } \delta ^{{(2)}}\right)+45 \gamma ^{{(2)}} \sin (\theta ) \left(\partial _{\theta } W^{{(0)}}\right)-45 W^{{(0)}} \left(\partial _{\phi } \delta ^{{(2)}}\right)-\\
& 12 \pi  \csc (\theta ) \left(\partial _{\theta \phi } T_{{rr}}{}^{{(1)}}\right)+12 \pi  \cot (\theta ) \csc (\theta ) \left(\partial _{\phi } T_{{rr}}{}^{{(1)}}\right)+24 \pi  \csc (\theta ) \left(\partial _{\phi } T_{{r\theta }}{}^{{(2)}}\right)-\\
& 48 \pi  T_{{r\phi }}{}^{{(2)}} \cot (\theta ) \csc (\theta )+24 \pi  \csc (\theta ) \left(\partial _{\theta } T_{{r\phi }}{}^{{(2)}}\right)-12 \cot (\theta ) \csc (\theta ) \left(\partial _{\phi } \gamma ^{{(3)}}\right)-\\
& 3 \csc ^2(\theta ) \left(\partial _{\phi \phi } \delta ^{{(3)}}\right)+12 \delta ^{{(3)}} \left(\csc ^2(\theta )-3\right)-3 \cot (\theta ) \left(\partial _{\theta } \delta ^{{(3)}}\right)-3 \left(\partial _{\theta \theta } \delta ^{{(3)}}\right)-\\
&\left . 20 \pi  T_{\theta \phi }{}^{{(3)}} \csc (\theta )\right),
\end{aligned}
\end{equation}

\begin{equation}\label{Trru}
\begin{aligned}
\partial_{u}T_{rr}{}^{(0)}= \;&
-3 U^{{(0)}} T_{{r\theta }}{}^{{(1)}}-W^{{(0)}} \left(\partial _{\phi } T_{{rr}}{}^{{(0)}}+3 T_{{r\phi }}{}^{{(1)}}\right)-\\
& \left(T_{{rr}}{}^{{(0)}} \left(\partial _{\theta } U^{{(0)}}+U^{{(0)}} \cot (\theta )+\partial _{\phi } W^{{(0)}}+7 \Phi ^{{(1)}}\right)\right)-U^{{(0)}} \left(\partial _{\theta } T_{{rr}}{}^{{(0)}}\right)-\\
& 3 T_{{rr}}{}^{{(1)}}-3 T_{{ur}}{}^{{(1)}}+\frac{1}{2} \left(T_{{r\theta }}{}^{{(2)}} (-\cot (\theta ))-\partial _{\theta } T_{{r\theta }}{}^{{(2)}}-\csc ^2(\theta ) \left(\partial _{\phi } T_{{r\phi }}{}^{{(2)}}\right)\right)+\\
& \frac{1}{6} \left(T_{\theta \theta }{}^{{(3)}}+T_{\phi \phi }{}^{{(3)}} \csc ^2(\theta )\right),
\end{aligned}
\end{equation}

\begin{equation}\label{Trthetau}
\begin{aligned}
\partial_{u}T_{r\theta}{}^{(1)}=\;& 
-2 T_{{r\theta }}{}^{{(1)}} \left(\partial _{\theta } U^{{(0)}}+4 \Phi ^{{(1)}}\right)-6 U^{{(0)}} \left(\partial _{\theta } T_{{r\theta }}{}^{{(1)}}\right)-4 W^{{(0)}} \left(\partial _{\phi } T_{{r\theta }}{}^{{(1)}}\right)-\\
& T_{{r\theta }}{}^{{(1)}} \left(\partial _{\phi } W^{{(0)}}\right)+T_{{rr}}{}^{{(0)}} \left(-\left(\partial _{\theta } \Phi ^{{(1)}}\right)\right)+U^{{(0)}} \csc ^2(\theta ) \left(\partial _{\phi } T_{{r\phi }}{}^{{(1)}}\right)-\\
& 3 W^{{(0)}} \left(\partial _{\theta } T_{{r\phi }}{}^{{(1)}}\right)+T_{{r\phi }}{}^{{(1)}} \left(6 W^{{(0)}} \cot (\theta )-\partial _{\theta } W^{{(0)}}\right)+\\
& \frac{3}{8 \pi } \left(U^{{(0)}} \left(\csc ^2(\theta ) \partial _{\phi \phi } \gamma ^{{(2)}}+\gamma ^{{(2)}} \left(-\cot ^2(\theta )-3 \csc ^2(\theta )+5\right)+\cot (\theta ) \partial _{\theta } \gamma ^{{(2)}}+\right .\right .\\
&\left . \partial _{\theta \theta } \gamma ^{{(2)}}-4 \cot (\theta ) \csc (\theta ) \partial _{\phi } \delta ^{{(2)}}\right)+W^{{(0)}} \left(4 \cot (\theta ) \partial _{\phi } \gamma ^{{(2)}}+\sin (\theta ) \partial _{\theta \theta } \delta ^{{(2)}}+\right .\\
&\left .\left . \csc (\theta ) \left(\partial _{\phi \phi } \delta ^{{(2)}}\right)-\left(\delta ^{{(2)}} (3 \cos (2 \theta )+1) \csc (\theta )\right)+\cos (\theta ) \left(\partial _{\theta } \delta ^{{(2)}}\right)\right)\right)-\\
& 4 U^{{(0)}} \left(T_{{rr}}{}^{{(0)}}+T_{{ur}}{}^{{(0)}}\right)-\frac{1}{2} U^{{(0)}} \csc ^2(\theta ) \left(\partial _{\phi \phi } T_{{rr}}{}^{{(0)}}\right)-\frac{1}{2} U^{{(0)}} \cot (\theta ) \partial _{\theta } T_{{rr}}{}^{{(0)}}+\\
& \frac{1}{2} U^{{(0)}} \partial _{\theta \theta } T_{{rr}}{}^{{(0)}}-W^{{(0)}} \cot (\theta ) \partial _{\phi } T_{{rr}}{}^{{(0)}}+W^{{(0)}} \partial _{\theta \phi } T_{{rr}}{}^{{(0)}}-2 T_{{r\theta }}{}^{{(2)}}-2 T_{{u\theta }}{}^{{(2)}}+\\
& \frac{1}{6} \left(T_{\theta \theta }{}^{{(3)}} (-\cot (\theta ))-\partial _{\theta } T_{\theta \theta }{}^{{(3)}}-\csc ^2(\theta ) \left(\partial _{\phi } T_{\theta \phi }{}^{{(3)}}\right)+T_{\phi \phi }{}^{{(3)}} \cot (\theta ) \csc ^2(\theta )\right),
\end{aligned}
\end{equation}

\begin{equation}\label{Trphiu}
\begin{aligned}
\partial_{u}T_{r\phi}{}^{(1)}=\;& 
T_{{r\theta }}{}^{{(1)}} \left(-\partial _{\phi } U^{{(0)}}-\frac{5}{2} W^{{(0)}} \sin (2 \theta )\right)-3 U^{{(0)}} \left(\partial _{\phi } T_{{r\theta }}{}^{{(1)}}\right)+\\
& W^{{(0)}} \sin ^2(\theta ) \left(\partial _{\theta } T_{{r\theta }}{}^{{(1)}}\right)+T_{{rr}}{}^{{(0)}} \left(-\left(\partial _{\phi } \Phi ^{{(1)}}\right)\right)-4 U^{{(0)}} \left(\partial _{\theta } T_{{r\phi }}{}^{{(1)}}\right)+\\
& T_{{r\phi }}{}^{{(1)}} \left(-\partial _{\theta } U^{{(0)}}+5 U^{{(0)}} \cot (\theta )-2 \left(\partial _{\phi } W^{{(0)}}\right)-8 \Phi ^{{(1)}}\right)-6 W^{{(0)}} \left(\partial _{\phi } T_{{r\phi }}{}^{{(1)}}\right)+\\
& \frac{1}{16 \pi }\left(6 U^{{(0)}} \left(4 \cot (\theta ) \left(\partial _{\phi } \gamma ^{{(2)}}\right)+\sin (\theta ) \left(\partial _{\theta \theta } \delta ^{{(2)}}\right)+\csc (\theta ) \left(\partial _{\phi \phi } \delta ^{{(2)}}\right)-\right .\right .\\
& \left .\left(\delta ^{{(2)}} (3 \cos (2 \theta )+1) \csc (\theta )\right)+\cos (\theta ) \left(\partial _{\theta } \delta ^{{(2)}}\right)\right)+3 W^{{(0)}} \left(-2 \sin ^2(\theta ) \left(\partial _{\theta \theta } \gamma ^{{(2)}}\right)+\right .\\
& \left .\left .\gamma ^{{(2)}} (6 \cos (2 \theta )+2)-\sin (2 \theta ) \left(\partial _{\theta } \gamma ^{{(2)}}\right)-2 \left(\partial _{\phi \phi } \gamma ^{{(2)}}\right)+8 \cos (\theta ) \left(\partial _{\phi } \delta ^{{(2)}}\right)\right)\right)-\\
& 4 W^{{(0)}} \sin ^2(\theta ) \left(T_{{rr}}{}^{{(0)}}+T_{{ur}}{}^{{(0)}}\right)-U^{{(0)}} \cot (\theta ) \left(\partial _{\phi } T_{{rr}}{}^{{(0)}}\right)+U^{{(0)}} \left(\partial _{\theta \phi } T_{{rr}}{}^{{(0)}}\right)-\\
& \frac{1}{2} W^{{(0)}} \sin ^2(\theta ) \left(\partial _{\theta \theta } T_{{rr}}{}^{{(0)}}\right)+\frac{1}{4} W^{{(0)}} \sin (2 \theta ) \left(\partial _{\theta } T_{{rr}}{}^{{(0)}}\right)+\frac{1}{2} W^{{(0)}} \left(\partial _{\phi \phi } T_{{rr}}{}^{{(0)}}\right)-\\
& 2 T_{{r\phi }}{}^{{(2)}}-2 T_{{u\phi }}{}^{{(2)}}+\frac{1}{6} \left(T_{\theta \phi }{}^{{(3)}} (-\cot (\theta ))-\partial _{\theta } T_{\theta \phi }{}^{{(3)}}-\csc ^2(\theta ) \left(\partial _{\phi } T_{\phi \phi }{}^{{(3)}}\right)\right),
\end{aligned}
\end{equation}

\begin{equation}\label{Truu}
\begin{aligned}
\partial_{u}T_{ru}{}^{(0)}=\;& 
\frac{1}{8} \left(6 U^{{(0)}} \csc ^2(\theta ) \left(\partial _{\phi \phi } T_{{r\theta }}{}^{{(1)}}\right)-2 U^{{(0)}} T_{{r\theta }}{}^{{(1)}} (9 \cos (2 \theta )-4) \csc ^2(\theta )+\right .\\
& 10 U^{{(0)}} \cot (\theta ) \left(\partial _{\theta } T_{{r\theta }}{}^{{(1)}}\right)+10 U^{{(0)}} \left(\partial _{\theta \theta } T_{{r\theta }}{}^{{(1)}}\right)+16 W^{{(0)}} \cot (\theta ) \left(\partial _{\phi } T_{{r\theta }}{}^{{(1)}}\right)+\\
& 4 W^{{(0)}} \partial _{\theta \phi } T_{{r\theta }}{}^{{(1)}}+4 U^{{(0)}} \csc ^2(\theta ) \partial _{\theta \phi } T_{{r\phi }}{}^{{(1)}}-20 U^{{(0)}} \cot (\theta ) \csc ^2(\theta ) \partial _{\phi } T_{{r\phi }}{}^{{(1)}}+\\
& 10 W^{{(0)}} \csc ^2(\theta ) \left(\partial _{\phi \phi } T_{{r\phi }}{}^{{(1)}}\right)-6 W^{{(0)}} \cot (\theta ) \left(\partial _{\theta } T_{{r\phi }}{}^{{(1)}}\right)+6 W^{{(0)}} \left(\partial _{\theta \theta } T_{{r\phi }}{}^{{(1)}}\right)+\\
& 36 W^{{(0)}} T_{{r\phi }}{}^{{(1)}}-48 T_{{ur}}{}^{{(0)}} \Phi ^{{(1)}}-U^{{(0)}} \cot (\theta ) \left(\partial _{\theta \theta } T_{{rr}}{}^{{(0)}}\right)-\\
& U^{{(0)}} \csc ^2(\theta ) \left(\partial _{\theta \phi \phi } T_{{rr}}{}^{{(0)}}\right)+2 U^{{(0)}} \cot (\theta ) \csc ^2(\theta ) \left(\partial _{\phi \phi } T_{{rr}}{}^{{(0)}}\right)+\\
& U^{{(0)}} \left(2 \cot ^2(\theta )-\csc ^2(\theta )+8\right) \left(\partial _{\theta } T_{{rr}}{}^{{(0)}}\right)-U^{{(0)}} \left(\partial _{\theta \theta \theta } T_{{rr}}{}^{{(0)}}\right)-\\
& W^{{(0)}} \cot (\theta ) \left(\partial _{\theta \phi } T_{{rr}}{}^{{(0)}}\right)-W^{{(0)}} \csc ^2(\theta ) \left(\partial _{\phi \phi \phi } T_{{rr}}{}^{{(0)}}\right)-W^{{(0)}} \left(\partial _{\theta \theta \phi } T_{{rr}}{}^{{(0)}}\right)+\\
& 6 W^{{(0)}} \left(\partial _{\phi } T_{{rr}}{}^{{(0)}}\right)-8 T_{{ur}}{}^{{(0)}} U^{{(0)}} \cot (\theta )-8 T_{{ur}}{}^{{(0)}} \left(\partial _{\theta } U^{{(0)}}\right)-8 T_{{ur}}{}^{{(0)}} \left(\partial _{\phi } W^{{(0)}}\right)-\\
&\left . 24 T_{{ur}}{}^{{(1)}}-24 T_{{uu}}{}^{{(1)}}-4 T_{{u\theta }}{}^{{(2)}} \cot (\theta )-4 \left(\partial _{\theta } T_{{u\theta }}{}^{{(2)}}\right)-4 \csc ^2(\theta ) \left(\partial _{\phi } T_{{u\phi }}{}^{{(2)}}\right)\right)+\\
& \frac{3}{32 \pi } \left(U^{{(0)}} \left(-3 \cot (\theta ) \partial _{\theta \theta } \gamma ^{{(2)}}-\csc ^2(\theta ) \partial _{\theta \phi \phi } \gamma ^{{(2)}}-4 \cot (\theta ) \csc ^2(\theta ) \partial _{\phi \phi } \gamma ^{{(2)}}-\right .\right .\\
& 12 \gamma ^{{(2)}} \cot (\theta )-4 \left(\partial _{\theta } \gamma ^{{(2)}}\right)+3 \csc ^2(\theta ) \left(\partial _{\theta } \gamma ^{{(2)}}\right)-\partial _{\theta \theta \theta } \gamma ^{{(2)}}-\csc (\theta ) \left(\partial _{\theta \theta \phi } \delta ^{{(2)}}\right)+\\
& \left .3 \cot (\theta ) \csc (\theta ) \partial _{\theta \phi } \delta ^{{(2)}}+(5 \cos (2 \theta )-1) \csc ^3(\theta ) \partial _{\phi } \delta ^{{(2)}}-\csc ^3(\theta ) \partial _{\phi \phi \phi } \delta ^{{(2)}}\right)+\\
& W^{{(0)}} \left(-3 \cot (\theta ) \left(\partial _{\theta \phi } \gamma ^{{(2)}}\right)-(5 \cos (2 \theta )-1) \csc ^2(\theta ) \left(\partial _{\phi } \gamma ^{{(2)}}\right)+\right .\\
& \csc ^2(\theta ) \left(\partial _{\phi \phi \phi } \gamma ^{{(2)}}\right)+\partial _{\theta \theta \phi } \gamma ^{{(2)}}-3 \cos (\theta ) \left(\partial _{\theta \theta } \delta ^{{(2)}}\right)-\sin (\theta ) \left(\partial _{\theta \theta \theta } \delta ^{{(2)}}\right)-\\
& \csc (\theta ) \left(\partial _{\theta \phi \phi } \delta ^{{(2)}}\right)-4 \cot (\theta ) \csc (\theta ) \left(\partial _{\phi \phi } \delta ^{{(2)}}\right)-12 \delta ^{{(2)}} \cos (\theta )+\csc (\theta ) \left(\partial _{\theta } \delta ^{{(2)}}\right)+\\
&\left .\left . 2 \cos (2 \theta ) \csc (\theta ) \left(\partial _{\theta } \delta ^{{(2)}}\right)\right)\right).
\end{aligned}
\end{equation}

Initial-value constraints:

\begin{equation}\label{Tthetatheta2}
\begin{aligned}
T_{\theta\theta}{}^{(2)}=\;&
\frac{1}{4} \left(\csc ^2(\theta ) \left(\partial _{\phi \phi } T_{{rr}}{}^{{(0)}}\right)+\cot (\theta ) \left(\partial _{\theta } T_{{rr}}{}^{{(0)}}\right)-\partial _{\theta \theta } T_{{rr}}{}^{{(0)}}+8 T_{{rr}}{}^{{(0)}}+8 T_{{ur}}{}^{{(0)}}-\right .\\
&\left . 2 T_{{r\theta }}{}^{{(1)}} \cot (\theta )+10 \partial _{\theta } T_{{r\theta }}{}^{{(1)}}-2 \csc ^2(\theta ) \partial _{\phi } T_{{r\phi }}{}^{{(1)}}\right)+\frac{3}{16 \pi } \left(-\csc ^2(\theta ) \partial _{\phi \phi } \gamma ^{{(2)}}+\right .\\
& \left .\gamma ^{{(2)}} (3 \cos (2 \theta )+1) \csc ^2(\theta )-\cot (\theta ) \left(\partial _{\theta } \gamma ^{{(2)}}\right)-\partial _{\theta \theta } \gamma ^{{(2)}}+4 \cot (\theta ) \csc (\theta ) \left(\partial _{\phi } \delta ^{{(2)}}\right)\right),
\end{aligned}
\end{equation}

\begin{equation}\label{Tphiphi2}
\begin{aligned}
T_{\phi\phi}{}^{(2)}=\;&
\frac{1}{8} \left(2 \sin ^2(\theta ) \left(\partial _{\theta \theta } T_{{rr}}{}^{{(0)}}\right)-\sin (2 \theta ) \left(\partial _{\theta } T_{{rr}}{}^{{(0)}}\right)+16 \sin ^2(\theta ) \left(T_{{rr}}{}^{{(0)}}+T_{{ur}}{}^{{(0)}}\right)-\right .\\
&\left . 2 \left(\partial _{\phi \phi } T_{{rr}}{}^{{(0)}}\right)+10 T_{{r\theta }}{}^{{(1)}} \sin (2 \theta )-4 \sin ^2(\theta ) \left(\partial _{\theta } T_{{r\theta }}{}^{{(1)}}\right)+20 \left(\partial _{\phi } T_{{r\phi }}{}^{{(1)}}\right)\right)-\\
& \frac{3}{32 \pi } \left(-2 \sin ^2(\theta ) \left(\partial _{\theta \theta } \gamma ^{{(2)}}\right)+\gamma ^{{(2)}} (6 \cos (2 \theta )+2)-\sin (2 \theta ) \left(\partial _{\theta } \gamma ^{{(2)}}\right)-2 \left(\partial _{\phi \phi } \gamma ^{{(2)}}\right)+\right .\\
&\left .8 \cos (\theta ) \left(\partial _{\phi } \delta ^{{(2)}}\right)\right),
\end{aligned}
\end{equation}

\begin{equation}\label{Tthetaphi2}
\begin{aligned}
T_{\theta\phi}{}^{(2)}=\;&
\frac{1}{2} \left(\cot (\theta ) \left(\partial _{\phi } T_{{rr}}{}^{{(0)}}\right)-\partial _{\theta \phi } T_{{rr}}{}^{{(0)}}+3 \left(\partial _{\phi } T_{{r\theta }}{}^{{(1)}}\right)-6 T_{{r\phi }}{}^{{(1)}} \cot (\theta )+3 \left(\partial _{\theta } T_{{r\phi }}{}^{{(1)}}\right)\right)+\\
& \frac{3}{16 \pi } \left(-4 \cot (\theta ) \left(\partial _{\phi } \gamma ^{{(2)}}\right)-\sin (\theta ) \left(\partial _{\theta \theta } \delta ^{{(2)}}\right)-\csc (\theta ) \left(\partial _{\phi \phi } \delta ^{{(2)}}\right)+\right .\\
& \left .\delta ^{{(2)}} (3 \cos (2 \theta )+1) \csc (\theta )-\cos (\theta ) \left(\partial _{\theta } \delta ^{{(2)}}\right)\right),
\end{aligned}
\end{equation}

\begin{equation}\label{Tutheta1}
\begin{aligned}
T_{u\theta}{}^{(1)}=\;&
\frac{1}{24} \left(\cot (\theta ) \left(\partial _{\theta \theta } T_{{rr}}{}^{{(0)}}\right)+\csc ^2(\theta ) \left(\partial _{\theta \phi \phi } T_{{rr}}{}^{{(0)}}\right)-2 \cot (\theta ) \csc ^2(\theta ) \left(\partial _{\phi \phi } T_{{rr}}{}^{{(0)}}\right)+\right .\\
& (3 \cos (2 \theta )-4) \csc ^2(\theta ) \left(\partial _{\theta } T_{{rr}}{}^{{(0)}}\right)+\partial _{\theta \theta \theta } T_{{rr}}{}^{{(0)}}-8 \left(\partial _{\theta } T_{{ur}}{}^{{(0)}}\right)-\\
& 6 \csc ^2(\theta ) \left(\partial _{\phi \phi } T_{{r\theta }}{}^{{(1)}}\right)+2 T_{{r\theta }}{}^{{(1)}} (9 \cos (2 \theta )-4) \csc ^2(\theta )-10 \cot (\theta ) \left(\partial _{\theta } T_{{r\theta }}{}^{{(1)}}\right)-\\
& \left .10 \left(\partial _{\theta \theta } T_{{r\theta }}{}^{{(1)}}\right)-4 \csc ^2(\theta ) \left(\partial _{\theta \phi } T_{{r\phi }}{}^{{(1)}}\right)+20 \cot (\theta ) \csc ^2(\theta ) \left(\partial _{\phi } T_{{r\phi }}{}^{{(1)}}\right)\right)+\\
& \frac{1}{32 \pi }\left(3 \cot (\theta ) \left(\partial _{\theta \theta } \gamma ^{{(2)}}\right)+\csc ^2(\theta ) \left(\partial _{\theta \phi \phi } \gamma ^{{(2)}}\right)+4 \cot (\theta ) \csc ^2(\theta ) \left(\partial _{\phi \phi } \gamma ^{{(2)}}\right)+\right .\\
& 12 \gamma ^{{(2)}} \cot (\theta )+\left(4-3 \csc ^2(\theta )\right) \left(\partial _{\theta } \gamma ^{{(2)}}\right)+\partial _{\theta \theta \theta } \gamma ^{{(2)}}+\csc (\theta ) \left(\partial _{\theta \theta \phi } \delta ^{{(2)}}\right)-\\
&\left . 3 \cot (\theta ) \csc (\theta ) \left(\partial _{\theta \phi } \delta ^{{(2)}}\right)+(1-5 \cos (2 \theta )) \csc ^3(\theta ) \left(\partial _{\phi } \delta ^{{(2)}}\right)+\csc ^3(\theta ) \left(\partial _{\phi \phi \phi } \delta ^{{(2)}}\right)\right),
\end{aligned}
\end{equation}

\begin{equation}\label{Tuphi1}
\begin{aligned}
T_{u\phi}{}^{(1)}=\;&
\frac{1}{24} \left(\cot (\theta ) \left(\partial _{\theta \phi } T_{{rr}}{}^{{(0)}}\right)+\csc ^2(\theta ) \left(\partial _{\phi \phi \phi } T_{{rr}}{}^{{(0)}}\right)+\partial _{\theta \theta \phi } T_{{rr}}{}^{{(0)}}-6 \left(\partial _{\phi } T_{{rr}}{}^{{(0)}}\right)-\right .\\
& 8 \left(\partial _{\phi } T_{{ur}}{}^{{(0)}}\right)-16 \cot (\theta ) \left(\partial _{\phi } T_{{r\theta }}{}^{{(1)}}\right)-4 \left(\partial _{\theta \phi } T_{{r\theta }}{}^{{(1)}}\right)-10 \csc ^2(\theta ) \left(\partial _{\phi \phi } T_{{r\phi }}{}^{{(1)}}\right)+\\
& \left .6 \cot (\theta ) \left(\partial _{\theta } T_{{r\phi }}{}^{{(1)}}\right)-6 \left(\partial _{\theta \theta } T_{{r\phi }}{}^{{(1)}}\right)-36 T_{{r\phi }}{}^{{(1)}}\right)+\\
& \frac{1}{32 \pi }\left(3 \cot (\theta ) \left(\partial _{\theta \phi } \gamma ^{{(2)}}\right)+(5 \cos (2 \theta )-1) \csc ^2(\theta ) \left(\partial _{\phi } \gamma ^{{(2)}}\right)-\csc ^2(\theta ) \left(\partial _{\phi \phi \phi } \gamma ^{{(2)}}\right)-\right .\\
& \partial _{\theta \theta \phi } \gamma ^{{(2)}}+3 \cos (\theta ) \left(\partial _{\theta \theta } \delta ^{{(2)}}\right)+\sin (\theta ) \left(\partial _{\theta \theta \theta } \delta ^{{(2)}}\right)+\csc (\theta ) \left(\partial _{\theta \phi \phi } \delta ^{{(2)}}\right)+\\
& \left .4 \cot (\theta ) \csc (\theta ) \left(\partial _{\phi \phi } \delta ^{{(2)}}\right)+12 \delta ^{{(2)}} \cos (\theta )+(-2 \cos (2 \theta )-1) \csc (\theta ) \left(\partial _{\theta } \delta ^{{(2)}}\right)\right),
\end{aligned}
\end{equation}

\begin{equation}\label{Tuu0}
\begin{aligned}
T_{uu}{}^{(0)}=\;&
\frac{1}{48} \left(\left(\cot ^2(\theta )+8\right) \left(\partial _{\theta \theta } T_{{rr}}{}^{{(0)}}\right)-2 \cot (\theta ) \left(\partial _{\theta \theta \theta } T_{{rr}}{}^{{(0)}}\right)-2 \csc ^2(\theta ) \left(\partial _{\theta \theta \phi \phi } T_{{rr}}{}^{{(0)}}\right)+\right .\\
& 2 \cot (\theta ) \csc ^2(\theta ) \left(\partial _{\theta \phi \phi } T_{{rr}}{}^{{(0)}}\right)-4 \cos (2 \theta ) \csc ^4(\theta ) \left(\partial _{\phi \phi } T_{{rr}}{}^{{(0)}}\right)-\\
& \csc ^4(\theta ) \left(\partial _{\phi \phi \phi \phi } T_{{rr}}{}^{{(0)}}\right)-(3 \cos (2 \theta )-2) \cot (\theta ) \csc ^2(\theta ) \left(\partial _{\theta } T_{{rr}}{}^{{(0)}}\right)-\partial _{\theta \theta \theta \theta } T_{{rr}}{}^{{(0)}}+\\
& 8 \csc ^2(\theta ) \left(\partial _{\phi \phi } T_{{ur}}{}^{{(0)}}\right)+8 \cot (\theta ) \left(\partial _{\theta } T_{{ur}}{}^{{(0)}}\right)+8 \left(\partial _{\theta \theta } T_{{ur}}{}^{{(0)}}\right)-48 T_{{ur}}{}^{{(0)}}+\\
& 20 \cot (\theta ) \left(\partial _{\theta \theta } T_{{r\theta }}{}^{{(1)}}\right)+10 \csc ^2(\theta ) \left(\partial _{\theta \phi \phi } T_{{r\theta }}{}^{{(1)}}\right)+10 \cot (\theta ) \csc ^2(\theta ) \left(\partial _{\phi \phi } T_{{r\theta }}{}^{{(1)}}\right)-\\
& 2 T_{{r\theta }}{}^{{(1)}} (9 \cos (2 \theta )-14) \cot (\theta ) \csc ^2(\theta )-(13 \cos (2 \theta )-3) \csc ^2(\theta ) \left(\partial _{\theta } T_{{r\theta }}{}^{{(1)}}\right)+\\
& 10 \left(\partial _{\theta \theta \theta } T_{{r\theta }}{}^{{(1)}}\right)+10 \csc ^2(\theta ) \left(\partial _{\theta \theta \phi } T_{{r\phi }}{}^{{(1)}}\right)-30 \cot (\theta ) \csc ^2(\theta ) \left(\partial _{\theta \phi } T_{{r\phi }}{}^{{(1)}}\right)-\\
&\left . 8 (\cos (2 \theta )-6) \csc ^4(\theta ) \left(\partial _{\phi } T_{{r\phi }}{}^{{(1)}}\right)+10 \csc ^4(\theta ) \left(\partial _{\phi \phi \phi } T_{{r\phi }}{}^{{(1)}}\right)\right)+\\
& \frac{1}{128 \pi }\left((\cos (2 \theta )+5) \csc ^2(\theta ) \left(\partial _{\theta \theta } \gamma ^{{(2)}}\right)-8 \cot (\theta ) \left(\partial _{\theta \theta \theta } \gamma ^{{(2)}}\right)-\right .\\
& 12 \cot (\theta ) \csc ^2(\theta ) \left(\partial _{\theta \phi \phi } \gamma ^{{(2)}}\right)+2 (7-3 \cos (2 \theta )) \csc ^4(\theta ) \left(\partial _{\phi \phi } \gamma ^{{(2)}}\right)+\\
& 2 \csc ^4(\theta ) \left(\partial _{\phi \phi \phi \phi } \gamma ^{{(2)}}\right)+2 (8 \cos (2 \theta )-11) \cot (\theta ) \csc ^2(\theta ) \left(\partial _{\theta } \gamma ^{{(2)}}\right)-2 \left(\partial _{\theta \theta \theta \theta } \gamma ^{{(2)}}\right)+\\
& 24 \gamma ^{{(2)}}-4 \csc (\theta ) \left(\partial _{\theta \theta \theta \phi } \delta ^{{(2)}}\right)+2 (7 \cos (2 \theta )-3) \csc ^3(\theta ) \left(\partial _{\theta \phi } \delta ^{{(2)}}\right)-\\
& 4 \csc ^3(\theta ) \left(\partial _{\theta \phi \phi \phi } \delta ^{{(2)}}\right)+4 (3 \cos (2 \theta )-7) \cot (\theta ) \csc ^3(\theta ) \left(\partial _{\phi } \delta ^{{(2)}}\right)-\\
&\left . 4 \cot (\theta ) \csc ^3(\theta ) \left(\partial _{\phi \phi \phi } \delta ^{{(2)}}\right)\right).
\end{aligned}
\end{equation}

\section{Matter evolution equations}\label{intermediate}

From the definitions of the covariant derivative and of the Christoffel symbol and using the formula for its contracted form, we have, for a general symmetric tensor $H^{ab}$, that
\begin{equation*}
\begin{split}
\nabla_a H^a{}_b & = \partial_a H^a{}_b + \Gamma^a{}_{ac} H^c{}_b - \Gamma^c{}_{ab} H^a{}_c \\
& = \partial_a H^a{}_b + (-g)^{-1/2} \left(\partial_c (-g)^{1/2} \right) H^c{}_b - \frac{1}{2} g^{cd}\left( \partial_a g_{bd} + \partial_b g_{ad} - \partial_d g_{ab} \right) H^a{}_c \\
& = (-g)^{-1/2} \partial_a \left( (-g)^{1/2}  H^a{}_b \right) - \frac{1}{2} \left( \partial_b g_{ad} \right) H^{ad} \\
& = (-g)^{-1/2}\partial_a \left( (-g)^{1/2} H^a{}_b \right) - \frac{1}{2} \left( \partial_b g_{ad} \right) g^{ac}g^{de}H_{ce} \\ 
& = (-g)^{-1/2} \partial_a \left( (-g)^{1/2}  H^a{}_b \right) - \frac{1}{2} \left( \partial_b \left( g_{ad} g^{ac}g^{de} \right) -  g_{ad} g^{de}  \partial_b g^{ac}    -  g_{ad} g^{ac}  \partial_b g^{de} \right) H_{ce} \\
& = (-g)^{-1/2} \partial_a \left( (-g)^{1/2}  H^a{}_b \right) - \frac{1}{2} \left( \partial_b  g^{ce}  - \partial_b \left( g^{ac} g_{a}{}^e \right)   -   \partial_b \left( g^{de} g_{d}{}^{c} \right) \right) H_{ce} \\
& = (-g)^{-1/2} \partial_a \left( (-g)^{1/2}  H^a{}_b \right) + \frac{1}{2} \left( \partial_b  g^{cd} \right) H_{cd}.
\end{split}
\end{equation*}

\section{Kinematics of light cones}\label{kinematics}

\begin{equation}\label{l}
l^\mu = \left( e^{-2 \beta },\frac{1}{2} e^{2 (\Phi - \beta) },e^{-2 \beta } U,e^{-2 \beta } W\right),
\end{equation}

\[ q^{\mu}{}_{\nu}= \left(
\begin{array}{cccc}
0 & 0 & 0 & 0 \\
0 & 0 & 0 & 0 \\
-U & 0 & 1 & 0 \\
-W & 0 & 0 & 1 \\
\end{array}
\right), \]

\begin{equation}\label{shear2}
\sigma^2\coloneqq\sigma^{a b} \sigma_{a b}=2\left[\cosh ^2(2 \delta)\left(\partial_r \gamma\right)^2+\left(\partial_r \delta\right)^2\right],
\end{equation}

\[ \Omega^H{}_{\theta} =\partial _{\theta } \beta -\frac{1}{2} e^{-2 \beta } r^2 \left(e^{2 \gamma } \cosh (2 \delta ) \left(\partial _r U\right)+\sinh (2 \delta ) \sin (\theta ) \left(\partial _r W\right)\right), \]

\[ \Omega^H{}_{\phi} =  \partial _{\phi } \beta -\frac{1}{2}  e^{-2 \beta } r^2\sin (\theta ) \left( \sinh (2 \delta ) \left(\partial _r U\right)+e^{-2 \gamma }\cosh (2 \delta ) \sin (\theta ) \left(\partial _r W\right)\right), \]

\begin{equation}\label{Omegatheta}
 \Omega^H{}_{\theta}{}^{(2)} = \pi  \partial _{\theta } T_{{rr}}{}^{{(0)}}+ 2 \pi  T_{{r\theta }}{}^{{(1)}} -\frac{1}{4} \left(2 \gamma ^{{(2)}} \cot (\theta )+\partial _{\theta } \gamma ^{{(2)}}+\csc (\theta ) \partial _{\phi } \delta ^{{(2)}}\right),
\end{equation}

\begin{equation}\label{Omegaphi}
\Omega^H{}_{\phi}{}^{(2)} = \pi  \partial _{\phi } T_{{rr}}{}^{{(0)}}+2 \pi  T_{{r\phi }}{}^{{(1)}}+\frac{1}{4} \left(\partial _{\phi } \gamma ^{{(2)}}-2 \delta ^{{(2)}} \cos (\theta )-\sin (\theta ) \partial _{\theta } \delta ^{{(2)}}\right),
\end{equation}

\begin{equation}\label{x}
\xi = \frac{2}{r} \left( e^{2( \Phi - \beta) }-1\right)+2 e^{-2 \beta } \left(\partial _{\theta } U+U \cot (\theta )+\partial _{\phi } W\right),
\end{equation}

\begin{equation}\label{xi1}
\begin{aligned}
\xi^{(1)} =\;& \frac{1}{3} \left(16 \pi  T_{{ur}}{}^{{(0)}}-\csc ^2(\theta ) \left(\partial _{\phi \phi } \gamma ^{{(2)}}\right)+3 \cot (\theta ) \left(\partial _{\theta } \gamma ^{{(2)}}\right)+\partial _{\theta \theta } \gamma ^{{(2)}}-2 \gamma ^{{(2)}}+\right .\\
& 2 \csc (\theta ) \left(\partial _{\theta \phi } \delta ^{{(2)}}\right)+2 \cot (\theta ) \csc (\theta ) \left(\partial _{\phi } \delta ^{{(2)}}\right)-2 {\Omega^H{} _\theta }^{{(2)}} \cot (\theta )-2 \left(\partial _{\theta } {\Omega^H{}_\theta }^{{(2)}}\right)-\\
&\left . 2 \csc ^2(\theta ) \left(\partial _{\phi } {\Omega^H{} _\phi }^{{(2)}}\right)-2 \Lambda \right).
\end{aligned}
\end{equation}

From definition \eqref{kappa}, one finds that
\begin{equation}\label{kappadelrbeta}
\kappa=\Gamma^r{}_{rr} = g^{ru}\partial_r g_{ru} = 2 \partial_r \beta.
\end{equation}
Also, we have that \cite[p. 49]{Wald1984}
\[ \nabla_a \partial_{(r)}{}^a = \frac{1}{\sqrt{\abs{g}}} \partial_\mu \left(  \sqrt{\abs{g}} \partial_{(r)}{}^\mu \right) = 2\partial_r \beta + \frac{2}{r},\]
where \eqref{detg} was used. Given that \cite[eq. 5.69]{Gourgoulhon2006}
\[ \nabla_a \partial_{(r)}{}^a = \kappa + \Theta, \]
it follows that
\begin{equation}\label{Theta}
\Theta =  \frac{2}{r}.
\end{equation}

Now, consider the relation between the affine parameter $ \lambda' $ and the areal radius $ r $. Since they parametrize the same generators of each light cone, their corresponding null vector fields, $ k^a $ and $ \partial_{(r)}{}^a $, are collinear. Thus, $ k^a = \alpha \partial_{(r)}{}^a $ and the geodesic equation for $ k^a $ implies that \cite[eq. 2.25]{Gourgoulhon2006}
\[ \nabla_r \ln(\alpha) = -\kappa = -2\partial_r \beta, \]
\[\qor* \alpha = \zeta e^{-2\beta},  \]
for some function $ \zeta $ that does not depend on $ r $. By noting that, along $ r=0 $,
\[ 1= k^a v_a = \alpha \partial_{(r)}{}^a  \partial_{(u)}{}_a = \alpha  e^{2\beta}, \]
one finds that $ \zeta =1 $ and
\[ \alpha = e^{-2\beta}. \]
Then,
\[ k^\mu= \partial_{(\lambda')}{}^\mu= \pdv{x^\mu}{\lambda'} = \alpha \partial_{(r)}{}^\mu = \alpha \pdv{x^\mu}{r} = \alpha \pdv{x^\mu}{y^{\mu'}}\pdv{y^{\mu'}}{r} = \alpha \pdv{x^\mu}{\lambda'}\pdv{\lambda'}{r}  \]
implies that \cite{Maedler2016}
\begin{equation}\label{delrlambda}
\partial_r \lambda' = \alpha^{-1} = e^{2\beta} \qand \partial_{\lambda'}r = e^{-2\beta} . 
\end{equation}

Another important relation is given by the product
\begin{equation}\label{Ray}
\kappa \Theta = \frac{4\partial_r \beta}{r}=  \sigma^2+8 \pi  {T}_{{rr}},
\end{equation}
where the second equality follows from equations \eqref{beta} and \eqref{shear2}. In fact, the above equation is the Raychaudhuri equation \cite[eq. 6.9]{Gourgoulhon2006}, which takes a simpler form in Bondi-Sachs coordinates, since the expansion scalar becomes a trivial function of the areal radius and
\[ \partial_r \Theta =  -\frac{\Theta^2}{2}. \]
Given the central conditions \eqref{ccg}, we have that
\begin{equation}\label{kT0}
\left(\kappa \Theta \right)^{(0)} = 8 \pi  {T}_{{rr}}{}^{(0)}.
\end{equation}

One can invert equations \eqref{xi1} and \eqref{kT0} to get an expression for $ T_{{ur}}{}^{{(0)}} $ and $ {T}_{{rr}}{}^{(0)} $, respectively. Similarly, equations
\eqref{Omegatheta} and \eqref{Omegaphi} provide expressions for $ T_{{r\theta }}{}^{{(1)}} $ and $ T_{{r\phi}}{}^{{(1)}} $, from which we can calculate $\partial_\theta T_{{r\theta }}{}^{{(1)}} $ and $\partial_\phi T_{{r\phi}}{}^{{(1)}} $. Using these results, it is possible to get equation \eqref{rhokine}.

\section{Spherically symmetric spacetimes}\label{SSS}

Isotropic observations from a center of symmetry are not consistent with the deformations induced by shear, as represented in Figure \ref{shear}. As a result, spherically symmetric spacetimes require a vanishing shear and, by equation \eqref{shear2}, it implies that $ \partial_r \gamma=\partial_r \delta=0 $ \cite{Ellis1985}. Together with the central conditions \eqref{ccg}, this enforces that $ \gamma= \delta=0 $. Then, in analogy to the definition of a static spacetime \cite[p. 119]{Wald1984}, we shall say that the metric \eqref{g} is \textit{nonrotating} if $ \partial_{(u)}{}^a \partial_{(A)}{}_a = 0 $. Combined with the condition on $ \gamma $ and $ \delta $, this leads to $ U=W=0 $, and it follows from the central conditions that $ a^{\bar{i}}=\Omega^{\bar{i}}=0 $.

In addition, there can be no net momentum in angular directions, meaning that $ T_{rA}=g_{ru}g_{AA}T^{uA}=0 $. Shear in the matter content would also imply preferred angular directions and, consequently, we must have $ T_{uA}=g_{uu}g_{AA}T^{uA}+g_{ur}g_{AA}T^{rA}=0 $ and $ T_{\theta\phi}=g_{\theta\theta}g_{\phi\phi}T^{\theta\phi}=0 $. Given that the norm of $ \partial_{(\phi)}{}^a $ depends on $ \theta $, $ T_{\phi\phi} $ may depend on $ \theta $, but not on $ \phi $. The remaining components will encode information about energy density, radial velocity, and pressures, and they must be independent of $ \theta $ and $ \phi $:
\[  T_{uu}=T_{uu}(u,r), \quad T_{ur}=T_{ur}(u,r), \quad T_{rr}=T_{rr}(u,r), \qand  T_{\theta\theta}=T_{\theta\theta}(u,r).\]

By plugging these initial data and $ \Lambda $ into the hypersurface equations \eqref{beta}, \eqref{UW1}, \eqref{UW2}, and \eqref{Phi}, we recover $ U=W=0 $ and find that
\begin{equation}\label{betasphe}
\beta(u,r)= \int_{0}^{r} 2\pi r' T_{rr}(u,r') \dd{r'}  
\end{equation}
and
\begin{equation}\label{phisphe}
\partial_r e^{2\Phi} + \frac{1-4\pi r^2 T_{rr}}{r}e^{2\Phi} = \frac{1-r^2 \Lambda}{r}e^{2\beta}+ 8\pi r T_{ur}.
\end{equation}
Multiplying the last equation by the integrating factor \eqref{FPhi} and integrating yields
\[ e^{2\Phi}= F_\Phi^{-1}\int_{0}^{r} F_\Phi \left(\frac{1-{r'}^2 \Lambda}{r'}e^{2\beta}+ 8\pi r' T_{ur}\right) \dd{r'}.\]
From \eqref{betasphe}, one finds that
\[ F_\Phi(u,r)= \frac{r}{r^*}e^{2\beta(u,r^*)}e^{-2\beta(u,r)} \]
and
\begin{equation}\label{e2P}
e^{2\Phi} = e^{2\beta}\left(  1-\frac{\Lambda}{3}r^2+\frac{1}{r}   \int_{0}^{r} 8\pi {r'}^2 T_{ur} e^{-2\beta} \dd{r'} \right).
\end{equation}

Then, the metric \eqref{g} becomes
\begin{equation}\label{gsphe}
\dd{s}^2=  -e^{2(\Phi+\beta)} \dd{u}^2 +2 e^{2\beta}\dd{u}\dd{r}+r^2\dd{\theta}^2+ r^2\sin[2](\theta)\dd{\phi}^2. 
\end{equation}
By computing the Jacobian using equations \eqref{ytoz} and \eqref{delrlambda}, one finds that the metric written in $ y^{\mu'} $ coordinates is
\begin{multline}
\dd{s}^2=  \left[-e^{2(\Phi+\beta)} + 2 e^{2\beta}\partial_{u'}r(u',\lambda')\right] \dd{u'}^2 +2\dd{u'}\dd{\lambda'}+\\
r^2(u',\lambda')\dd{\theta'}^2+r^2(u',\lambda')\sin[2](\theta')\dd{\phi'}^
2,
\end{multline}
with determinant $ g'=-r^4(u',\lambda')\sin[2](\theta') $. The inverse metric is given by
\begin{equation}\label{invgprime}
\begin{matrix}
g^{\mu'\nu'} = \mqty(0 & 1 & 0 & 0 \\ 1 & e^{2(\Phi+\beta)} - 2 e^{2\beta}\partial_{u'}r(u',\lambda') & 0 & 0 \\ 0 & 0 & r^{-2} & 0 \\ 0 & 0 & 0 & r^{-2} \csc[2](\theta') ) .  
\end{matrix}
\end{equation}

It is convenient to define, from \eqref{l}, a new transversal null vector field that is normalized with respect to $ \partial_{(\lambda')}{}^a $: 
\begin{equation}\label{lprime}
{l'}^a \coloneqq e^{2\beta}l^a, \qq{with}  {l'}^{\mu} = \left( 1,\frac{e^{2 \Phi }}{2} ,0,0\right).
\end{equation}
Since $ \partial_{u'}\lambda'=0 $ implies that $ \partial_{u}\lambda' = -\partial_{r}\lambda'\partial_{u'}r= -e^{2\beta}\partial_{u'}r$, one can compute the components of the above vector field in the $ y^{\mu'} $ system and find that
\begin{equation}\label{lprimeprime}
{l'}^{\mu'} = \left( 1,\frac{g^{\lambda'\lambda'}}{2} ,0,0\right).
\end{equation}
Equations \eqref{delrlambda} and \eqref{ytoz} imply that $ \partial_{(\lambda')}{}^a = e^{-2\beta}\partial_{(r)}{}^a $, so the definition \eqref{ldef} leads to $ l'_a\partial_{(\lambda')}{}^a = 1 $.
By calculating Christoffel symbols from the metric \eqref{gsphe}, one finds that $ {l'}^a $ is the generator of non-affinely parametrized geodesics, since $ {l'}^\mu\nabla_\mu {l'}^\nu=\Gamma^{u}{}_{uu} {l'}^\nu$. What is interesting in defining this new vector field is that the Lie derivative $ \mathcal{L}_{l'}u=1 $, meaning that the Lie dragging produced by $ {l'}^a $ is compatible with the one given by $ \partial_{(u)}{}^a $. In other words, the parametrization of the geodesics generated by $ {l'}^a $ is compatible with the observer's perception of time evolution, mapping a past light cone into another successive one. Therefore, commutation with $ \partial_{(r)}{}^a $ is restricted only by the radial condition
\begin{equation}\label{commucond}
\partial_r e^{2 \Phi }=0 ,
\end{equation}
since
\[ \left[l',\partial_{(r)}\right]^{\mu} = \left( 0,-\frac{\partial_r e^{2 \Phi }}{2},0,0 \right) = \left( 0, -e^{2\Phi}\partial_r \Phi,0,0 \right). \]

Lastly, we note that the non-trivial initial-value constraints of Appendix \ref{firstterms} that result from the restrictions on $ T_{\mu\nu},\ \gamma, $ and $ \delta $ are
\begin{equation}\label{ivc}
T_{\theta\theta}{}^{(2)}=2\left(T_{rr}{}^{(0)}+T_{ur}{}^{(0)}\right), \quad T_{\phi\phi}{}^{(2)}=2\sin[2](\theta)\left(T_{rr}{}^{(0)}+T_{ur}{}^{(0)}\right) \qand T_{uu}{}^{(0)}=-T_{ur}{}^{(0)}.
\end{equation}

\subsection{Exterior solution: Schwarzschild-de Sitter spacetime}

In case the matter content is restricted to a maximum radius $ R $, it is useful to define (this is the past light cone, quasi-local version of the Bondi mass in equation (5.22) of Reference \citeonline{Siebel2002}) 
\begin{equation}\label{M}
M(u,R)\coloneqq -\int_{0}^{R} 4\pi r^2 T_{ur}(u,r)e^{-2\beta(u,r)} \dd{r}
\end{equation}
and
\begin{equation}\label{betaR}
\beta_R(u)\coloneqq\beta(u,R)= \int_{0}^{R} 2\pi r T_{rr}(u,r) \dd{r}  .
\end{equation}
Then, one has that, for the region $ r\geq R $,
\begin{equation}\label{e2phi}
e^{2\Phi} = e^{2\beta_R}\left(  1-\frac{\Lambda}{3}r^2-\frac{2M}{r}   \right)
\end{equation}
and
\[ \Phi= \beta_R + \frac{1}{2}\ln\left(  1-\frac{\Lambda}{3}r^2-\frac{2M}{r}  \right). \]

Expanding the last logarithm around $ 1 $, one recovers the Newtonian gravitational potential,
\[ \Phi\approx-\frac{1}{r}\left(M +\frac{4\pi r^3}{3}\frac{\Lambda}{8\pi }\right)+ \text{constant}, \]
with an extra mass term due to the cosmological constant. Moreover, the dominant energy condition \cite{Poisson2004} implies that $ -T^a{}_b \partial_{(u)}{}^b $ is a future-directed, timelike or null, vector field (as long as the coordinate system remains well defined and $ \partial_{(u)}{}^a $ timelike). For $ \partial_{(r)}{}^a $ is past-directed and null, it follows that $ -T_{ab} \partial_{(u)}{}^a \partial_{(r)}{}^b = -T_{ur}\geq 0 $, and, from definition \eqref{M}, that
\begin{equation}\label{Mpositive}
M\geq0,
\end{equation}
with equality holding only for $ T_{ab}=0 $ or $ T^a{}_b \partial_{(u)}{}^b \propto \partial_{(r)}{}^a$. We also note that the geodesic equation with null initial velocities and $ \beta_R $ and $ M $ constants leads to
\begin{equation}\label{force}
\dv[2]{r}{u} =-\frac{\partial_{r}e^{4\Phi}}{4}=  -e^{4\Phi}\partial_{r}\Phi.
\end{equation}
In a weak-field approximation with $ \Phi \ll 1 $, it follows that
\[\dv[2]{r}{u}  \approx-\partial_{r}\Phi. \]
Therefore, $ M $ may be seen as a generalization of the Newtonian active gravitational mass due to the matter content and $ \Phi $ as a generalization of the Newtonian gravitational potential.

The case where $ \beta_R $ and $ M $ are independent of $ u $ and $ \Lambda $ is positive corresponds to the Schwarzschild-de Sitter spacetime.
Our interest in this solution resides in the existence of a radius where the commutation condition \eqref{commucond} is satisfied. From equation \eqref{e2phi}, one finds that this radius is
\begin{equation}\label{r1}
r_1= \left(\frac{3M}{\Lambda}\right)^{\frac{1}{3}}.
\end{equation}

\subsection{Interior solution: dusty spacetime}

Now, consider a pressureless perfect fluid with proper energy density $ \mu(u,r) $, also known as dust. Spherical symmetry requires that its four-velocity, $ v^a $, be orthogonal to $ \partial_{(\theta)}{}^a $ and $ \partial_{(\phi)}{}^a $. Then, $ T_{\theta\theta}=T_{\phi\phi}=0 $ and
\[ T_{ab}= \mu(u,r)v_av_b.\]
For $ v^av_a=-1 $, contraction of the last equality using the inverse metric \eqref{g} leads to
\begin{equation}\label{Tur}
T_{ur} = -\frac{e^{2\beta}}{2}\mu  -\frac{e^{2\Phi}}{2}T_{rr}.
\end{equation}

Following Reference \citeonline{Ellis1985}, we write the four-velocity of the fluid in terms of the observed redshift $ z $:
\[ v^{u'}=1+z(u',\lambda'). \]
The coordinate transformation \eqref{ytoz} leads then to
\begin{equation}\label{vu}
v^u=1+z(u,r)>0,
\end{equation}
where the inequality is required for a future-directed, timelike flow of dust. The normalization of this vector field implies that
\begin{equation}\label{vr}
 v^r=\frac{e^{2\Phi}(1+z)}{2}-\frac{e^{-2\beta}}{2(1+z)},
\end{equation}
and we also have that
\[v_u=-\frac{1+e^{2(\Phi+\beta)}(1+z)^2}{2(1+z)} \qand v_r= e^{2\beta}(1+z).\]
Therefore,
\begin{equation}\label{Trr}
T_{rr}=\mu e^{4\beta}(1+z)^2
\end{equation}
and
\begin{equation}\label{Tuu}
T_{uu}= \frac{\mu}{4(1+z)^2} \left[1+e^{2(\Phi+\beta)}(1+z)^2\right]^2 = \frac{e^{-4\beta}}{\mu(1+z)^2}T_{ur}{}^2.
\end{equation}

Now, we note that, given the way we are modeling matter, $ T_{\mu\nu} $ is not prescribed from the beginning, and we must alter the hierarchical scheme of integration. Equation \eqref{betasphe} becomes
\[ \partial_r \beta= 2\pi r \mu (1+z)^2 e^{4\beta}, \]
and can be solved by defining $ x\coloneqq e^{4\beta} $. This yields a Bernoulli differential equation,
\[ \partial_r x= 8\pi r \mu (1+z)^2 x^2, \]
which is reduced to a linear differential equation by $ y\coloneqq x^{-1} $. Straightforward integration and substitutions result in
\begin{equation}\label{betasol}
\beta=-\frac{1}{4} \ln\left( 1- \int_{0}^{r} 8\pi r' \mu (1+z)^2 \dd{r'}   \right).
\end{equation}
By inserting this into equation \eqref{Trr}, one finds that
\begin{equation}\label{Trrsol}
T_{rr} = \frac{\mu(1+z)^2}{1- \int_{0}^{r} 8\pi r' \mu (1+z)^2 \dd{r'}}. 
\end{equation}

Then, multiplying equation \eqref{phisphe} by $ r $ and using equation \eqref{Tur} leads to
\begin{equation}\label{delrre2Phi}
\partial_r \left(r e^{2\Phi}\right)= e^{2\beta}\left[1-(4\pi \mu + \Lambda)r^2\right],
\end{equation}
which, together with equation \eqref{betasol}, results in
\begin{equation}\label{phisol}
e^{2\Phi} = \frac{1}{r} \int_{0}^{r}\frac{1-(4\pi \mu + \Lambda){r'}^2}{\sqrt{1- \int_{0}^{r'} 8\pi r'' \mu (1+z)^2 \dd{r''}}} \dd{r'}.
\end{equation}
Finally, one can compute $ T_{ur} $ from equations \eqref{Tur}, \eqref{betasol}, \eqref{phisol} and \eqref{Trrsol}:
\begin{multline}\label{Tursol}
T_{ur}= -\frac{\mu}{2\sqrt{1- \int_{0}^{r} 8\pi r' \mu (1+z)^2 \dd{r'}}} -\\
\frac{\mu(1+z)^2}{2r\left[1- \int_{0}^{r} 8\pi r' \mu (1+z)^2 \dd{r'}\right]} \int_{0}^{r}\frac{1-(4\pi \mu + \Lambda){r'}^2}{\sqrt{1- \int_{0}^{r'} 8\pi r'' \mu (1+z)^2 \dd{r''}}} \dd{r'}.
\end{multline}

The time evolution of this spacetime is given by equations \eqref{deluTrr}, \eqref{Phiur} and \eqref{deluTur}. After inserting the components of the energy-momentum tensor as given above, expressing $ \partial_r \beta $ and $ \partial_r \Phi $ using equations \eqref{betasphe} and \eqref{phisphe} and using equations \eqref{Tur} and \eqref{Tuu}, one gets
\begin{equation}\label{deluTrrsphe}
\partial_u T_{rr} = \frac{e^{2\beta}}{2r^2}\partial_r(r^2\mu) -\frac{e^{2\Phi}}{2}\partial_r T_{rr} - \frac{e^{2\beta}}{r}T_{rr}\left[1-(4\pi \mu+\Lambda)r^2\right].
\end{equation}
Then, one can use this result and equations \eqref{Trr} and \eqref{vr} in \eqref{Phiur} to find that
\begin{equation}\label{delue2Phi}
\partial_u e^{2\Phi} = \frac{1}{r}\int_{0}^{r} \frac{4\pi}{1+z} \left[\frac{{r'}^2\mu\partial_{r'} z}{(1+z)^2}+e^{2\beta}v^r\partial_{r'}({r'}^2\mu)\right]+ \left[1-(4\pi\mu+\Lambda){r'}^2\right]\partial_u e^{2\beta} \dd{r'},
\end{equation}
where $ \partial_u \beta $ can be computed taking the $ u $-derivative of equation \eqref{betasphe} and using \eqref{deluTrrsphe}.

The last equation is then used to calculate $ \partial_u T_{ur} $, which in turn can be used to find $ \partial_u \mu $ from equation \eqref{Tur}:  
\begin{equation}\label{delumu}
\partial_u \mu=-\frac{1}{1+z}\left[r^2v^r\partial_r\left(\frac{\mu}{r^2}\right)+\frac{\mu e^{-2\beta}}{(1+z)^2}\partial_r z\right] .
\end{equation}
Finally, $ \partial_u z $ can be calculated using equation \eqref{Trr}.

The initial-value constraints can be evaluated by expanding $ \mu $ and $ z $ as in equation \eqref{f}. Using the central conditions \eqref{ccg} and equations \eqref{Trr}, \eqref{Tuu}, and \eqref{Tur}, all three constraints \eqref{ivc} become equivalent to
\begin{equation}\label{ivcTrrTur}
T_{rr}{}^{(0)}=-T_{ur}{}^{(0)},
\end{equation}
which is satisfied if
\begin{equation}\label{ivcmuz}
\mu^{(0)}=0 \qor z^{(0)}=0.
\end{equation}
In case $ z^{(0)}=0 $, then $ v^{u(0)}=1 $ and $ v^{r(0)}=0 $. In other words, $ v^a $ corresponds to the observer's four-velocity along their worldline, justifying the use of the same notation for both.

\section{Data from supernova simulations}\label{super}

\begin{table}[h]
	\centering
	\caption{From Figure 7.3 of Reference \citeonline{Siebel2002}, we first read the snapshot times measured in Bondi time, $ u_B $, which corresponds to the proper time of inertial observers at future null infinity. At each snapshot, we identified the minimum of the radial velocity profile, collecting the value of the velocity, $ v^r(u_B,r) $, and the corresponding radius, $ r $. With $ u_B $ and $ r $, we can use Figure 7.2 to identify which mass shells are immediately above and below the point of minimum velocity and estimate the enclosed Bondi mass, $ M(u_B,r) $. The values of $ v^r $ and $ M $ are given in units in which $ c=M_\odot=1 $}
	\begin{tabular}{|c|c|c|c|}
		\hline
		\textbf{$ u_B (\operatorname{ms}) $} & \textbf{$ r (\operatorname{km}) $} & \textbf{$ -v^r/c $} & \textbf{$ M/M_\odot $} \\ 
		\hline
			30 & 210 & 0.035 & 0.5 \\
			\hline
			31 & 200 & 0.038 & 0.5 \\
			\hline
			32 & 190 & 0.040 & 0.5 \\
			\hline
			33 & 180 & 0.043 & 0.5 \\
			\hline
			34 & 170 & 0.047 & 0.5 \\
			\hline
			35 & 160 & 0.051 & 0.5 \\
			\hline
			36 & 140 & 0.056 & 0.5 \\
			\hline
			37 & 120 & 0.063 & 0.5 \\
			\hline
			38 &  92 & 0.072 & 0.4 \\
			\hline
			39 &  67 & 0.088 & 0.4 \\
			\hline
			40 &  20 & 0.140 & 0.4 \\
			\hline
			41 &  45 & 0.147 & 0.5 \\
			\hline
			42 &  78 & 0.118 & 0.6 \\
			\hline
			43 & 105 & 0.105 & 0.7 \\
			\hline
			44 & 134 & 0.098 & 0.8 \\
			\hline
			45 & 155 & 0.092 & 0.8 \\
			\hline
	\end{tabular}
		\label{table}
\end{table}
\begin{table}[h]
	\centering
	\caption{We read the infall velocities at bounce for 11 values of baryon mass coordinate for the ``model SH'' shown in Figure 6 of Reference \citeonline{Sumiyoshi2005}. The first one represents the boundary of the inner core, and the rest are convenient values of mass coordinate inside the infall region. Then, one can read the radius corresponding to the mass coordinate at bounce from Figure 1 of Reference \citeonline{Sumiyoshi2005}. The values of $ v^r $ and $ M $ are given in units in which $ c=M_\odot=1 $}
\begin{tabular}{|c|c|c|}
	\hline
	$ M/M_\odot $ & $ -v^r/c $ & $ r (\operatorname{km}) $ \\
	\hline
		0.61 & 0.228 & 15 \\
		\hline
		0.70 & 0.177 & 19 \\
		\hline
		0.80 & 0.135 & 31 \\
		\hline
		0.90 & 0.098 & 44 \\
		\hline
		1.00 & 0.077 & 60 \\
		\hline
		1.10 & 0.057 & 78 \\
		\hline
		1.20 & 0.042 & 130 \\
		\hline
		1.30 & 0.027 & 700 \\
		\hline
		1.40 & 0.012 & 1460 \\
		\hline
		1.50 & 0.002 & 2600 \\
		\hline
		1.60 & 0.001 & 3900 \\
	\hline
\end{tabular}
\label{table2}
\end{table}

\clearpage
\section{Landau-Lifshitz pseudotensor}\label{LLP}

We define the Landau-Lifshitz pseudotensor similarly to \cite[p. 466]{Misner1973}, but including the cosmological constant term in it, as a contribution of the gravitational field:

\begin{equation*}
\begin{aligned}
t^{\mu\nu}\coloneqq & -\frac{\Lambda}{8\pi}g^{\mu\nu} - \frac{1}{16\pi g}\left[  \partial_{\sigma}\mathfrak{g}^{\mu\nu} \partial_{\rho}\mathfrak{g}^{\sigma\rho}  - \partial_{\sigma}\mathfrak{g}^{\mu\sigma} \partial_{\rho}\mathfrak{g}^{\nu\rho} + \frac{1}{2}g^{\mu\nu}g_{\sigma\rho}\partial_{\omega}\mathfrak{g}^{\sigma\tau} \partial_{\tau}\mathfrak{g}^{\omega\rho} - \right.\\
&\left( g^{\mu\sigma}g_{\rho\tau}\partial_{\omega}\mathfrak{g}^{\nu\tau} \partial_{\sigma}\mathfrak{g}^{\rho\omega} +  g^{\nu\sigma}g_{\rho\tau}\partial_{\omega}\mathfrak{g}^{\mu\tau} \partial_{\sigma}\mathfrak{g}^{\rho\omega} \right) +  g_{\sigma\rho}g^{\tau\omega}\partial_{\tau}\mathfrak{g}^{\mu\sigma} \partial_{\omega}\mathfrak{g}^{\nu\rho} + \\
&\left.\frac{1}{8}  \left(2g^{\mu\sigma}g^{\nu\rho}-g^{\mu\nu}g^{\sigma\rho}\right)\left(2g_{\tau\omega}g_{\zeta\chi}-g_{\omega\zeta}g_{\tau\chi}\right)\partial_{\sigma}\mathfrak{g}^{\tau\chi} \partial_{\rho}\mathfrak{g}^{\omega\zeta}     \right], 
\end{aligned}
\end{equation*}
where $\mathfrak{g}^{\mu\nu}\coloneqq (-g)^{1/2}g^{\mu\nu} $, with $ g_{\mu\nu}  $, $ g^{\mu\nu}  $ and $ g $ given by equations \eqref{g} and \eqref{detg}.

It is defined in such a way that the effective total energy-momentum (including matter and gravitational degrees of freedom),
\[ T_{\text{eff}}^{\mu\nu}\coloneqq-g(T^{\mu\nu}+t^{\mu\nu}), \]
is divergenceless:
\[ \partial_{\nu}T_{\text{eff}}^{\mu\nu}=0. \]
Therefore, one can obtain conserved charges by considering the integral
\[ \int_{0}^{2\pi}\int_{0}^{\pi}\int_{0}^{R}\int_{u_1}^{u_2} \partial_{\nu}T_{\text{eff}}^{\mu\nu} \dd{u}\dd{r}\dd{\theta}\dd{\phi}=0,\]
which implies that
\begin{multline*}
\int_{0}^{2\pi}\int_{0}^{\pi}\int_{0}^{R} \eval{T_{\text{eff}}^{\mu u}}_{u_1}^{u_2} \dd{r}\dd{\theta}\dd{\phi}   +   \int_{0}^{2\pi}\int_{0}^{\pi}\int_{u_1}^{u_2} \eval{T_{\text{eff}}^{\mu r}}_{0}^{R} \dd{u}\dd{\theta}\dd{\phi}   + \\ \int_{0}^{2\pi}\int_{0}^{R}\int_{u_1}^{u_2} \eval{T_{\text{eff}}^{\mu \theta}}_{0}^{\pi} \dd{u}\dd{r}\dd{\phi}   +   \int_{0}^{\pi}\int_{0}^{R}\int_{u_1}^{u_2} \eval{T_{\text{eff}}^{\mu \phi}}_{0}^{2\pi} \dd{u}\dd{r}\dd{\theta}=0.
\end{multline*}
The last integral vanishes, since $ T_{\text{eff}}^{\mu \phi}(u,r,\theta,0) =T_{\text{eff}}^{\mu \phi}(u,r, \theta,2\pi) $. Also, equation \eqref{deltheta02pi} implies that $ T_{\text{eff}}^{\mu \theta}(u,r,0/\pi,\phi) =-T_{\text{eff}}^{\mu \theta}(u,r,0/\pi,\phi\pm\pi) $, so that its integral over $ \phi $ yields 0. Moreover, equations \eqref{dbarbondi} and the tensor transformation below it imply that $ T^{\mu\nu} $ diverges at most as $ \mathcal{O}(1/r^2) $ as $ r\to0 $. Explicit calculations of $ t^{\mu r} $ using the central conditions \eqref{ccg} show that these components diverge at most as $ \mathcal{O}(1/r^3) $. Given that $ g = \mathcal{O}(r^4) $, it follows that $ T_{\text{eff}}^{\mu r}=0 $ at $ r=0 $. As a result, we have that
\begin{multline*}
\int_{0}^{2\pi}\int_{0}^{\pi}\int_{0}^{R} T_{\text{eff}}^{\mu u}(u_2,r,\theta,\phi) \dd{r}\dd{\theta}\dd{\phi} -\\ \int_{0}^{2\pi}\int_{0}^{\pi}\int_{0}^{R} T_{\text{eff}}^{\mu u}(u_1,r,\theta,\phi) \dd{r}\dd{\theta}\dd{\phi}  =   -\int_{0}^{2\pi}\int_{0}^{\pi}\int_{u_1}^{u_2} T_{\text{eff}}^{\mu r}(u,R,\theta,\phi) \dd{u}\dd{\theta}\dd{\phi}
\end{multline*}
and the total four-momentum of matter plus gravitational field,
\[ P^\mu(u,R)\coloneqq\int_{0}^{2\pi}\int_{0}^{\pi}\int_{0}^{R} T_{\text{eff}}^{\mu u}(u,r,\theta,\phi) \dd{r}\dd{\theta}\dd{\phi}, \]
will be conserved between $ u_1 $ and $ u_2 $ in case
\[ \int_{0}^{2\pi}\int_{0}^{\pi}\int_{u_1}^{u_2} T_{\text{eff}}^{\mu r}(u,R,\theta,\phi) \dd{u}\dd{\theta}\dd{\phi}=0. \]

Most of the components of the Landau-Lifshitz pseudotensor are quite complicated when calculated using the metric written in Bondi-Sachs coordinates $ z^\mu $ for general spacetimes. However, its energy component assumes an interesting simple form. Using equations \eqref{kappadelrbeta}, \eqref{Theta}, \eqref{shear2} and \eqref{Ray} and raising the indices of $ T_{rr} $ with the metric \eqref{g}, one finds that
\[t^{uu}=e^{-4\beta}\left(-\frac{3}{4\pi r^2}-\frac{\kappa\Theta}{8\pi}+\frac{\sigma^2}{8\pi}\right) = e^{-4\beta}\left(-\frac{3}{4\pi r^2}-T_{rr}\right)= -\frac{3e^{-4\beta}}{4\pi r^2}-T^{uu}.\]
Therefore,
\[T_{\text{eff}}^{uu}= -\frac{3}{4\pi}r^2\sin[2](\theta) \]
and
\begin{equation}\label{Pu}
P^u=- \frac{3}{4\pi}\int_{0}^{2\pi}\int_{0}^{\pi}\int_{0}^{R}r^2\sin[2](\theta) \dd{r}\dd{\theta}\dd{\phi}=-\frac{\pi}{4}R^3.
\end{equation}

\end{appendices}

\phantomsection
\addcontentsline{toc}{section}{\refname} 
\bibliography{refs}

\end{document}